\newcommand*{\ATLASLATEXPATH}{latex/}
\documentclass[cernpreprint,texlive=2016,UKenglish]{\ATLASLATEXPATH atlasdoc}
\usepackage[full, backend=bibtex, maxbibnames=5]{\ATLASLATEXPATH atlaspackage}
\usepackage[articletitle]{\ATLASLATEXPATH atlasbiblatex}
\usepackage[unit=false, journal=false]{\ATLASLATEXPATH atlasphysics}
\usepackage{multirow}
\graphicspath{{logos/}{images/}}

\usepackage{Pflow-defs}

\AtlasTitle{Jet reconstruction and performance using particle flow with the ATLAS Detector}

\author{The ATLAS Collaboration}

\date{\today}

\AtlasRefCode{PERF-2015-09}

\PreprintIdNumber{CERN-EP-2017-024}

\arXivId{1703.10485}

\AtlasJournalRef{Eur.\ Phys.\ J.\ C \textbf{77} (2017) 466}
\AtlasDOI{10.1140/epjc/s10052-017-5031-2}

\AtlasAbstract{%
This paper describes the implementation and performance of a particle flow algorithm applied to \SI{20.2}{\per\fb} of ATLAS data from 8 TeV proton--proton collisions in Run 1 of the LHC.
The algorithm removes calorimeter energy deposits due to charged hadrons from consideration during jet reconstruction,
instead using measurements of their momenta from the inner tracker.
This improves the accuracy of the charged-hadron measurement,
while retaining the calorimeter measurements of neutral-particle energies.
The paper places emphasis on how this is achieved,
while minimising double-counting of charged-hadron signals between the inner tracker and calorimeter.
The performance of particle flow jets,
formed from the ensemble of signals from the calorimeter and the inner tracker,
is compared to that of jets reconstructed from calorimeter energy deposits alone,
demonstrating improvements in resolution and pile-up stability.
}

\begin{document}

\maketitle

\tableofcontents

\clearpage
%-------------------------------------------------------------------------------
% Introduction
%-------------------------------------------------------------------------------
% !TeX root = Pflow.tex
%-------------------------------------------------------------------------------
\section{Introduction}
\label{sect:intro}
%-------------------------------------------------------------------------------

Jets are a key element in many analyses of the data collected by the experiments 
at the Large Hadron Collider (LHC)~\cite{LHC}.
The jet calibration procedure should correctly determine the jet energy scale and additionally the best possible energy and angular resolution should be achieved.
Good jet reconstruction and calibration facilitates the identification of known resonances that decay to hadronic jets,
as well as the search for new particles.
A complication, at the high luminosities encountered by the ATLAS detector~\cite{PERF-2007-01},
is that multiple interactions can contribute to the detector signals associated with a single bunch-crossing (pile-up).
These interactions, which are mostly soft, have to be separated from the hard interaction that is of interest.

Pile-up contributes to the detector signals from the collision environment,
and is especially important for higher-intensity operations of the LHC.
One contribution arises from particle emissions produced by the additional proton--proton (\pp) collisions
occurring in the same bunch crossing as the hard-scatter interaction (in-time pile-up).
Further pile-up influences on the signal are from signal remnants in the ATLAS calorimeters from the energy deposits in other bunch crossings (out-of-time pile-up).

In Run~1 of the LHC, the ATLAS experiment used either solely the calorimeter
or solely the tracker to reconstruct hadronic jets and soft particle activity.
The vast majority of analyses utilised jets that were built from topological clusters of calorimeter cells (\topoclusters)~\cite{topoclustering}.
These jets were then calibrated to the particle level using a jet energy scale (JES) correction factor~\cite{ATLAS-CONF-2015-002, ATLAS-CONF-2015-017, ATLAS-CONF-2015-037,ATLAS-CONF-2015-057}.
For the final Run~1 jet calibration,
this correction factor also took into account the tracks associated with the jet,
as this was found to greatly improve the jet resolution~\cite{ATLAS-CONF-2015-002}.
\enquote{Particle flow} introduces an alternative approach,
in which measurements from both the tracker and the calorimeter are combined to form the signals,
which ideally represent individual particles.
The energy deposited in the calorimeter by all the charged particles is removed.
Jet reconstruction is then performed on an ensemble of \enquote{particle flow objects}
consisting of the remaining calorimeter energy and tracks which are matched to the hard interaction.

The chief advantages of integrating tracking and calorimetric information into one hadronic reconstruction step are as follows:
\begin{itemize}
\item The design of the ATLAS detector~\cite{ATLAS-TDR} specifies a calorimeter energy resolution for single charged pions in the centre of the detector of
  \begin{equation}
    \frac{\sigma(E)}{E}=\frac{\SI{50}{\%}}{\sqrt{E}}\oplus\SI{3.4}{\%}\oplus\frac{\SI{1}{\%}}{E}\,,
  \end{equation}
while the design inverse transverse momentum resolution for the tracker is
  \begin{equation}
    \sigma\left(\frac{1}{\pt}\right)\cdot\pt=\SI{0.036}{\%} \cdot \pt\oplus\SI{1.3}{\%}\,,
  \end{equation}
where energies and transverse momenta are measured in GeV.
  Thus for low-energy charged particles, the momentum resolution of the tracker is significantly better than the energy resolution of the calorimeter. Furthermore, the acceptance of the detector is extended to softer particles, as tracks are reconstructed for charged particles with a minimum transverse momentum $\pT>\SI{400}{\MeV}$, whose energy deposits often do not pass the noise thresholds required to seed topo-clusters~\cite{ATL-PHYS-PUB-2014-002}.
\item The angular resolution of a single charged particle, reconstructed using the tracker is much better than that of the calorimeter.
\item Low-\pT charged particles originating within a hadronic jet are swept out of the jet cone by the magnetic field by the time they reach the calorimeter.
  By using the track's azimuthal coordinate\footnote{%
ATLAS uses a right-handed coordinate system with its origin at the nominal interaction point (IP) in the centre of the detector and the $z$-axis along the beam direction. The $x$-axis points from the IP to the centre of the LHC ring, and the $y$-axis points upward. Cylindrical coordinates $(r,\phi)$ are used in the transverse plane, $\phi$ being the azimuthal angle around the $z$-axis. The pseudorapidity is defined in terms of the polar angle $\theta$ as $\eta=-\ln\tan(\theta/2)$. Angular distance is measured in units of $\Delta R=\sqrt{(\Delta\phi)^2+(\Delta\eta)^2}$.} at the perigee, these particles are clustered into the jet.
\item When a track is reconstructed, one can ascertain whether it is associated with a vertex, and if so the vertex from which it originates.
  Therefore, in the presence of multiple in-time pile-up interactions, 
  the effect of additional particles on the hard-scatter interaction signal can be mitigated by rejecting signals
  originating from pile-up vertices.\footnote{%
    The standard ATLAS reconstruction defines the hard-scatter primary vertex to be the primary vertex
    with the largest $\sum \pT^2$ of the associated tracks.
    All other primary vertices are considered to be contributed by pile-up.}
\end{itemize}
The capabilities of the tracker in reconstructing charged particles are complemented by the calorimeter's ability to reconstruct both the charged and neutral particles.
At high energies, the calorimeter's energy resolution is superior to the tracker's momentum resolution.
Thus a combination of the two subsystems is preferred for optimal event reconstruction.
Outside the geometrical acceptance of the tracker, only the calorimeter information is available.
Hence, in the forward region the \topoclusters alone are used as inputs to the particle flow jet reconstruction.

However, particle flow introduces a complication.
For any particle whose track measurement ought to be used,
it is necessary to correctly identify its signal in the calorimeter,
to avoid double-counting its energy in the reconstruction.
In the particle flow algorithm described herein,
a Boolean decision is made as to whether to use the tracker or calorimeter measurement.
If a particle's track measurement is to be used,
the corresponding energy must be subtracted from the calorimeter measurement.
The ability to accurately subtract all of a single particle's energy,
without removing any energy deposited by any other particle,
forms the key performance criterion upon which the algorithm is optimised.

Particle flow algorithms were pioneered in the ALEPH experiment at LEP~\cite{Buskulic:272484}.
They have also been used in the H1~\cite{PeezH1PFlow}, ZEUS~\cite{ZeusPFlowRefA,ZeusPFlowRefB} and  DELPHI~\cite{Abreu:1995uz} experiments.
Subsequently, they were used for the reconstruction of hadronic $\tau$-lepton decays in the 
CDF~\cite{Abulencia:2007iy}, D0~\cite{Abazov2009292} and ATLAS~\cite{PERF-2014-06} experiments.
In the CMS experiment at the LHC, large gains in the performance of the reconstruction of hadronic jets and $\tau$ leptons have been seen from the use of particle flow algorithms~\cite{CMS-PAS-PFT-09-001, CMS-PAS-PFT-10-001, CMS-JME-10-011}.
Particle flow is a key ingredient in the design of detectors for the planned International Linear Collider~\cite{Thomson:2009rp}
and the proposed calorimeters are being optimised for its use~\cite{CALICE-testbeam}.
While the ATLAS calorimeter already measures jet energies precisely \cite{ATLAS-CONF-2015-037}, %under LHC conditions,
it is desirable to explore the extent to which particle flow is able to further improve the ATLAS hadronic-jet reconstruction, in particular in the presence of pile-up interactions.

This paper is organised as follows.
A description of the detector is given in \Sect{\ref{sec:det}}, the Monte Carlo (MC) simulated event samples and the dataset used are described in \Sects{\ref{sec:MC}}{\ref{sec:dataset}}, 
while \Sect{\ref{sec:topocluster}} outlines the relevant properties of \topoclusters.
The particle flow algorithm is described in \Sect{\ref{sec:alg}}.
Section~\ref{sec:calHits} details the algorithm's performance in energy subtraction at the level of individual particles in a variety of cases, starting from a single pion through to dijet events.
The building and calibration of reconstructed jets is covered in \Sect{\ref{sec:jet:cal}}.
The improvement in jet energy and angular resolution is shown in \Sect{\ref{sec:jet:res}} and the sensitivity to pile-up is detailed in \Sect{\ref{sec:jet:PU}}.
A comparison between data and MC simulation is shown in \Sect{\ref{sec:DataMC}} before the conclusions are presented in \Sect{\ref{sec:conc}}.

%-------------------------------------------------------------------------------
% The ATLAS detector
%-------------------------------------------------------------------------------
% !TeX root = Pflow.tex
%-------------------------------------------------------------------------------
\section{ATLAS detector}
\label{sec:det}
%-------------------------------------------------------------------------------

The ATLAS experiment features a multi-purpose detector
designed to precisely measure jets, leptons and photons produced in the \pp collisions at the LHC. 
From the inside out, the detector consists of a tracking system called the inner detector (ID), surrounded by electromagnetic (EM) sampling calorimeters.
These are in turn surrounded by hadronic sampling calorimeters and an air-core toroid muon spectrometer (MS).
A detailed description of the ATLAS detector can be found in \Ref{\cite{PERF-2007-01}}.

The high-granularity silicon pixel detector covers the vertex region and typically provides three measurements per track.
%the first hit being normally in the innermost layer.
It is followed by the silicon microstrip tracker which usually provides eight hits, corresponding to four two-dimensional measurement points, per track.
These silicon detectors are complemented by the transition radiation tracker,
which enables radially extended track reconstruction up to $|\eta| = 2.0$. 
The ID is immersed in a \SI{2}{\tesla} axial magnetic field and can reconstruct tracks within the pseudorapidity range $|\eta|<2.5$.
For tracks with transverse momentum $\pT<\SI{100}{\GeV}$, the fractional inverse momentum resolution $\sigma({1/\pT}) \cdot \pT$
measured using 2012 data, ranges from approximately \SI{2}{\%} to \SI{12}{\%} depending on pseudorapidity and \pT~\cite{ATLAS-CONF-2014-047}.

\begin{figure}[htbp]
  \centering
  \includegraphics[width=12cm]{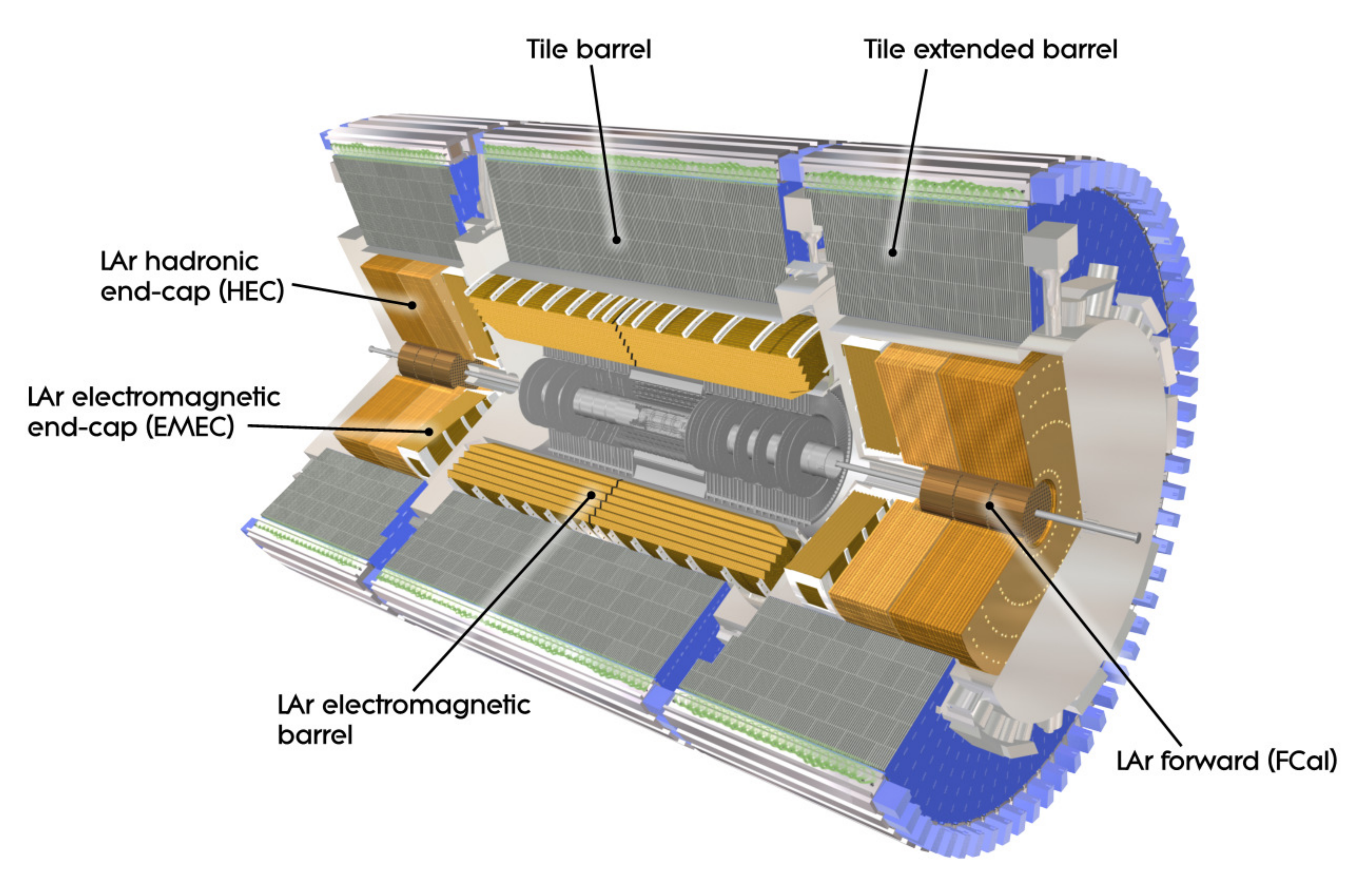}
  \caption{Cut-away view of the ATLAS calorimeter system.}
  \label{fig:atlas:cutout}
\end{figure}

The calorimeters provide hermetic azimuthal coverage in the range $|\eta|<4.9$.
The detailed structure of the calorimeters within the tracker acceptance strongly influences the development of the shower subtraction algorithm described in this paper.
In the central barrel region of the detector, a high-granularity liquid-argon (LAr) electromagnetic calorimeter with lead absorbers is surrounded by a hadronic sampling calorimeter (Tile) with steel absorbers and active scintillator tiles.
The same LAr technology is used in the calorimeter endcaps, with fine granularity and lead absorbers for the EM endcap (EMEC),
while the hadronic endcap (HEC) utilises copper absorbers with reduced granularity.
The solid angle coverage is completed with forward copper/LAr and tungsten/LAr calorimeter modules (FCal)
optimised for electromagnetic and hadronic measurements respectively.
Figure~\ref{fig:atlas:cutout} shows the physical location of the different calorimeters.
To achieve a high spatial resolution, the calorimeter cells are arranged in a projective geometry with fine segmentation in $\phi$ and $\eta$.
Additionally, each of the calorimeters is longitudinally segmented into multiple layers, capturing the shower development in depth.
In the region $|\eta| < 1.8$, a presampler detector is used to correct for the energy lost by electrons and photons upstream of the calorimeter.
The presampler consists of an active LAr layer of thickness \SI{1.1}{\cm} (\SI{0.5}{\cm}) in the barrel (endcap) region.
The granularity of all the calorimeter layers within the tracker acceptance is given in \Tab{\ref{table:atlas:granularity}}.

The EM calorimeter is over 22 radiation lengths in depth, ensuring that there is little leakage of EM showers into the hadronic calorimeter.
The total depth of the complete calorimeter is over 9 interaction lengths in the barrel and over 10 interaction lengths in the endcap,
such that good containment of hadronic showers is obtained.
Signals in the MS are used to correct the jet energy if the hadronic shower is not completely contained.
In both the EM and Tile calorimeters, most of the absorber material is in the second layer.
In the hadronic endcap, the material is more evenly spread between the layers.

\begin{table}[htbp]
  \centering
  \begin{tabular}{l|l r|l r}
  \toprule
  \multicolumn{5}{c}{\textbf{EM LAr calorimeter}} \\ \midrule
   & \multicolumn{2}{c|}{\textbf{Barrel}} & \multicolumn{2}{c}{\textbf{Endcap}} \\
   \midrule
  Presampler &  $0.025\times \pi/32$ & $|\eta|<1.52$ & $0.025\times \pi/32$ & $1.5<|\eta|<1.8$ \\
  {\small\texttt{PreSamplerB/E}} & & & & \\
  \midrule
1st layer &  $0.025/8\times \pi/32$ & $|\eta|<1.4$ & $0.050\times \pi/32$ & $1.375<|\eta|<1.425$ \\
{\small\texttt{EMB1/EME1}} & $0.025\times \pi/128$ & $1.4<|\eta|<1.475$ & $0.025\times \pi/32$ & $1.425<|\eta|<1.5$ \\
 & & & $0.025/8\times \pi/32$ & $1.5<|\eta|<1.8$ \\
 & & & $0.025/6\times \pi/32$ & $1.8<|\eta|<2.0$ \\
 & & & $0.025/4\times \pi/32$ & $2.0<|\eta|<2.4$ \\
 & & & $0.025\times \pi/32$ & $2.4<|\eta|<2.5$ \\
 & & & $0.1\times \pi/32$ & $2.5<|\eta|<3.2$ \\ \midrule
2nd layer &  $0.025\times \pi/128$ & $|\eta|<1.4$ & $0.050\times \pi/128$ & $1.375<|\eta|<1.425$ \\
{\small\texttt{EMB2/EME2}} & $0.075\times \pi/128$ & $1.4<|\eta|<1.475$ & $0.025\times \pi/128$ & $1.425<|\eta|<2.5$ \\
 & & & $0.1\times \pi/32$ & $2.5<|\eta|<3.2$ \\ \midrule
3rd layer &  $0.050\times \pi/128$ & $|\eta|<1.35$ & $0.050\times \pi/128$ & $1.5<|\eta|<2.5$ \\
{\small\texttt{EMB3/EME3}} & & & & \\
\bottomrule
\multicolumn{5}{c}{\textbf{Tile calorimeter}} \\ \midrule
 & \multicolumn{2}{c|}{\textbf{Barrel}} & \multicolumn{2}{c}{\textbf{Extended barrel}} \\ \midrule
1st layer &  $0.1\times \pi/32$ & $|\eta|<1.0$ & $0.1\times \pi/32$ & $0.8<|\eta|<1.7$ \\
{\small\texttt{TileBar0/TileExt0}} & & & & \\
2nd layer &  $0.1\times \pi/32$ & $|\eta|<1.0$ & $0.1\times \pi/32$ & $0.8<|\eta|<1.7$ \\
{\small\texttt{TileBar1/TileExt1}} & & & & \\
3rd layer &  $0.2\times \pi/32$ & $|\eta|<1.0$ & $0.2\times \pi/32$ & $0.8<|\eta|<1.7$ \\
{\small\texttt{TileBar2/TileExt2}} & & & & \\
\bottomrule
\multicolumn{5}{c}{\textbf{Hadronic LAr calorimeter}} \\
\midrule
 & \multicolumn{2}{c|}{\mbox{}} & \multicolumn{2}{c}{\textbf{Endcap}} \\
\midrule
1st layer &   &  & $0.1\times \pi/32$ & $1.5<|\eta|<2.5$ \\ 
{\small\texttt{HEC0}} &   &  & $0.2\times \pi/16$ & $2.5<|\eta|<3.2$ \\ \midrule
2nd layer &   &  & $0.1\times \pi/32$ & $1.5<|\eta|<2.5$ \\ 
{\small\texttt{HEC1}} &   &  & $0.2\times \pi/16$ & $2.5<|\eta|<3.2$ \\ \midrule
3rd layer &   &  & $0.1\times \pi/32$ & $1.5<|\eta|<2.5$ \\ 
{\small\texttt{HEC2}} &   &  & $0.2\times \pi/16$ & $2.5<|\eta|<3.2$ \\ \midrule
4th layer &   &  & $0.1\times \pi/32$ & $1.5<|\eta|<2.5$ \\ 
{\small\texttt{HEC3}} &   &  & $0.2\times \pi/16$ & $2.5<|\eta|<3.2$ \\
\bottomrule
\end{tabular}
\caption{The granularity in $\Delta\eta\times\Delta\phi$ of all the different ATLAS calorimeter layers relevant to the tracking coverage of the inner detector.
}
\label{table:atlas:granularity}
\end{table}

The muon spectrometer surrounds the calorimeters and is based on
three large air-core toroid superconducting magnets with eight coils each.
The field integral of the toroids ranges from \num{2.0} to \SI{6.0}{\tesla\metre} across most of the detector.
It includes a system of precision tracking chambers and fast detectors for triggering.

%-------------------------------------------------------------------------------
% Monte Carlo samples
%-------------------------------------------------------------------------------
% !TeX root = Pflow.tex
%-------------------------------------------------------------------------------
\section{Simulated event samples}
\label{sec:MC}
%-------------------------------------------------------------------------------

A variety of MC samples are used in the optimisation and performance evaluation of the particle flow algorithm.
The simplest samples consist of a single charged pion generated with a uniform spectrum in the logarithm of the generated pion energy and in the generated $\eta$.
Dijet samples generated with \PYTHIAV{8} (v8.160)~\cite{pythia8,pythiapartonshower},
with parameter values set to the ATLAS AU2 tune~\cite{ATL-PHYS-PUB-2012-003} and the CT10 parton distribution functions (PDF) set~\cite{Lai:2010vv},
form the main samples used to derive the jet energy scale and determine the jet energy resolution in simulation.
The dijet samples are generated with a series of jet \pT thresholds applied to the leading jet,
reconstructed from all stable final-state particles excluding muons and neutrinos,
using the \akt algorithm~\cite{Cacciari:2008gp} with radius parameter 0.6 using \textsc{FastJet}~(v3.0.3)~\cite{Fastjet,Cacciari200657}.

For comparison with collision data, $Z\rightarrow\mu\mu$ events are generated with \POWHEGBOX (r1556)~\cite{Nason:2004rx} using the CT10 PDF and are showered with \PYTHIAV{8}, with the ATLAS AU2 tune.
Additionally, top quark pair production is simulated with \MCatNLO~(v4.03)~\cite{FRI-0201,FRI-0301} using the CT10 PDF set,
interfaced with \HERWIG~(v6.520)~\cite{COR-0001} for parton showering, and the underlying event is modelled by \JIMMY~(v4.31)~\cite{Butterworth:1996zw}.
The top quark samples are normalised using the cross-section calculated at next-to-next-to-leading order (NNLO) in QCD including resummation of next-to-next-to-leading logarithmic soft gluon terms with top++2.0 \cite{Aliev:2010zk,Beneke:2011mq,Czakon:2013goa,Czakon:2012pz,Czakon:2012zr,Baernreuther:2012ws,Cacciari:2011hy,Czakon:2011xx}, assuming a top quark mass of \SI{172.5}{\GeV}.
Single-top-quark production processes contributing to the distributions shown are also simulated, but their contributions are negligible.

\subsection{Detector simulation and pile-up modelling}

All samples are simulated using \GEANTFour~\cite{Geant4} within the ATLAS simulation framework~\cite{SOFT-2010-01}
and are reconstructed using the noise threshold criteria used in 2012 data-taking~\cite{topoclustering}.
Single-pion samples are simulated without pile-up,
while dijet samples are simulated under three conditions: with no pile-up; with pile-up conditions similar to those in the 2012 data;
and with a mean number of interactions per bunch crossing $\langle \mu \rangle = 40$, 
where $\mu$ follows a Poisson distribution.
In 2012, the mean value of $\mu$ was \num{20.7} and the actual number of interactions per bunch crossing ranged from around 10 to 35 depending on the luminosity.
The bunch spacing was \SI{50}{\ns}.
When compared to data, the MC samples are reweighted to have the same distribution of $\mu$ as present in the data.
In all the samples simulated including pile-up,
effects from both the same bunch crossing and previous/subsequent crossings are simulated by overlaying additional generated minimum-bias events on the hard-scatter event prior to reconstruction.
The minimum-bias samples are generated using \PYTHIAV{8} with the ATLAS AM2 tune~\cite{ATL-PHYS-PUB-2011-009}
and the MSTW2009 PDF set~\cite{PDF-MRST}, and are simulated using the same software as the hard-scatter event.

\subsection{Truth calorimeter energy and tracking information}
\label{sec:MC_calhits}

For some samples the full \GEANTFour hit information~\cite{Geant4} is retained for each calorimeter cell such that the true amount of hadronic and electromagnetic energy deposited by each generated particle is known. 
Only the measurable hadronic and electromagnetic energy deposits are counted, while the energy lost due to nuclear capture and particles escaping from the detector is not included. For a given charged pion the sum of these hits in a given cluster $i$ originating from this particle is denoted by $E^\text{clus $i$}_\text{true, $\pi$}$.

Reconstructed \topocluster energy is assigned to a given truth particle according to the proportion of \GEANTFour hits supplied to that \topocluster by that particle.
Using the \GEANTFour hit information in the inner detector
a track is matched to a generated particle based on the fraction of hits on the track which originate
from that particle~\cite{STDM-2015-02}.

%-------------------------------------------------------------------------------
% Data sample
%-------------------------------------------------------------------------------
% !TeX root = Pflow.tex
%-------------------------------------------------------------------------------
\section{Data sample}
\label{sec:dataset}
%-------------------------------------------------------------------------------

Data acquired during the period from March to December 2012 with the LHC operating at a $pp$ centre-of-mass energy of \SI{8}{\TeV}
are used to evaluate the level of agreement between data and Monte Carlo simulation of different outputs of the algorithm.
Two samples with a looser preselection of events are reconstructed using the particle flow algorithm.
A tighter selection is then used to evaluate its performance.

First, a $Z\rightarrow\mu\mu$ enhanced sample is extracted from the 2012 dataset by selecting events containing two reconstructed muons~\cite{PERF-2014-05},
each with $\pT > \SI{25}{\GeV}$ and $|\eta|<2.4$, where the invariant mass of the dimuon pair is greater than \SI{55}{\GeV},
and the \pT of the dimuon pair is greater than \SI{30}{\GeV}.

Similarly, a sample enhanced in $\ttbar\rightarrow\bbbar\qqbar\mu\nu$ events is obtained from events with an isolated muon and at least one hadronic jet which is required to be identified as a jet
containing $b$-hadrons ($b$-jet).
Events are selected that pass single-muon triggers and include one reconstructed muon satisfying $\pT > \SI{25}{\GeV}$, $|\eta|<2.4$, for which the sum of additional track momenta in a cone of size $\dR = 0.2$ around the muon track is less than \SI{1.8}{\GeV}.
Additionally, a reconstructed calorimeter jet is required to be present with $\pT > \SI{30}{\GeV}$, $|\eta|<2.5$, and pass the \SI{70}{\percent} working point of the MV1 $b$-tagging algorithm~\cite{ATLAS-CONF-2011-102}.

For both datasets, all ATLAS subdetectors are required to be operational with good data quality.
Each dataset corresponds to an integrated luminosity of \SI{20.2}{\per\fb}.
To remove events suffering from significant electronic noise issues, cosmic rays or beam background,
the analysis excludes events that contain calorimeter jets with $\pT > \SI{20}{\GeV}$
which fail to satisfy the \enquote{looser} ATLAS jet quality criteria~\cite{DAPR-2012-01,ATLAS-CONF-2012-020}.

%-------------------------------------------------------------------------------
% Topological clusters
%-------------------------------------------------------------------------------
% !TeX root = Pflow.tex
%-------------------------------------------------------------------------------
\section{Topological clusters}
\label{sec:topocluster}
%-------------------------------------------------------------------------------

The lateral and longitudinal segmentation of the calorimeters permits three-dimensional reconstruction of particle showers,
implemented in the topological clustering algorithm~\cite{topoclustering}.
\Topoclusters of calorimeter cells are seeded by cells whose absolute energy measurements $|E|$ exceed the expected noise
by four times its standard deviation.
The expected noise includes both electronic noise and the average contribution from pile-up, which depends on the run conditions.
The \topoclusters are then expanded both laterally and longitudinally in two steps,
first by iteratively adding all adjacent cells with absolute energies two standard deviations above noise,
and finally adding all cells neighbouring the previous set.
A splitting step follows, separating at most two local energy maxima into separate \topoclusters.
Together with the ID tracks, these \topoclusters form the basic inputs to the particle flow algorithm.

The topological clustering algorithm employed in ATLAS is not designed to separate energy deposits from different particles,
but rather to separate continuous energy showers of different nature,
i.e.\ electromagnetic and hadronic, and also to suppress noise.
The cluster-seeding threshold in the topo-clustering algorithm results
in a large fraction of low-energy particles being unable to seed their own clusters.
For example, in the central barrel $\sim$\SI{25}{\%} of \SI{1}{\GeV} charged pions do not seed their own cluster~\cite{ATL-PHYS-PUB-2014-002}.

While the granularity, noise thresholds and employed technologies vary across the different ATLAS calorimeters,
they are initially calibrated to the electromagnetic scale (EM scale) to give the same response for electromagnetic showers from electrons or photons.
Hadronic interactions produce responses that are lower than the EM scale, by amounts depending on where the showers develop.
To account for this, the mean ratio of the energy deposited by a particle
to the momentum of the particle is determined based on the position of the particle's shower in the detector,
as described in \Sect{\ref{sec:eflowRec:EoverP}}.

A local cluster (LC) weighting scheme is used to calibrate
hadronic clusters to the correct scale~\cite{topoclustering}.
Further development is needed to combine this with particle flow;
therefore, in this work the \topoclusters used in the particle flow algorithm are calibrated at the EM scale.

%-------------------------------------------------------------------------------
% Particle flow algorithm
%-------------------------------------------------------------------------------
% !TeX root = Pflow.tex
%-------------------------------------------------------------------------------
\section{Particle flow algorithm}
\label{sec:alg}
%-------------------------------------------------------------------------------

A cell-based energy subtraction algorithm is employed to remove overlaps between the momentum and energy measurements made in the inner detector and calorimeters, respectively.
Tracking and calorimetric information is combined for the reconstruction of hadronic jets and
soft activity (additional hadronic recoil below the threshold used in jet reconstruction) in the event.
The reconstruction of the soft activity is important for the calculation of the missing transverse momentum in the event~\cite{PERF-2014-04},
whose magnitude is denoted by \MET.

\begin{figure}[htbp]
  \centering
  \includegraphics[width=\textwidth]{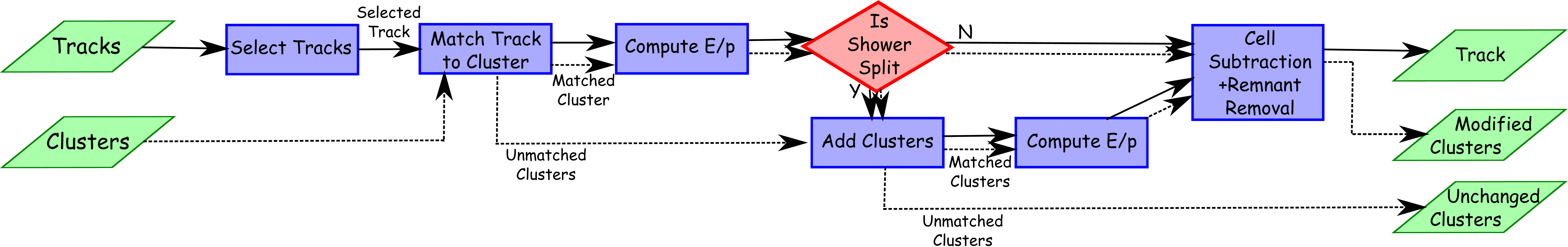}
  \caption{A flow chart of how the particle flow algorithm proceeds,
    starting with track selection and continuing until the energy associated with the selected tracks has been removed from the calorimeter.
    At the end, charged particles, \topoclusters which have not been modified by the algorithm, and remnants of \topoclusters which have had part of their energy removed remain.}
  \label{fig:eflowRec:flowChart}
\end{figure}

The particle flow algorithm provides a list of tracks and a list of \topoclusters containing both the unmodified \topoclusters and a set of new \topoclusters resulting from the energy subtraction procedure.
This algorithm is sketched in \Fig{\ref{fig:eflowRec:flowChart}}.
First, well-measured tracks are selected following the criteria discussed in \Sect{\ref{sec:trksel}}.
The algorithm then attempts to match each track to a single \topocluster in the calorimeter (\Sect{\ref{sec:trkclus}}).
The expected energy in the calorimeter, deposited by the particle that also created the track, is computed based on the \topocluster position and the track momentum (\Sect{\ref{sec:eflowRec:EoverP}}).
It is relatively common for a single particle to deposit energy in multiple \topoclusters.
For each track/\topocluster system, the algorithm evaluates the probability that the particle energy was deposited in more than one \topocluster. 
On this basis it decides if it is necessary
to add more \topoclusters to the track/\topocluster system to recover the full shower energy (\Sect{\ref{sec:eflowRec_ssr}}).
The expected energy deposited in the calorimeter by the particle that produced the track  is subtracted cell by cell
from the set of matched \topoclusters (\Sect{\ref{sec:eflowRec_cellsub}}).
Finally, if the remaining energy in the system is consistent with the expected shower fluctuations of a single particle's signal, the \topocluster remnants are removed (\Sect{\ref{sec:eflowRec_remnant}}).

\begin{figure}[htbp]
\centering
\begin{tabular}{|@{}c@{}| @{}c@{}| @{}c@{}| @{}c@{}| @{}c@{}|}
\hline
 & Track/\topocluster & Split shower & Cell subtraction & Remnant removal \\
 & matching & recovery & & \\ \hline
$\begin{array}{l}\text{1 particle,}\\ \text{1 \topocluster}\end{array}$ & \,\includegraphics[trim=0cm 1.5cm 0cm -0.25cm,clip,width=2.75cm]{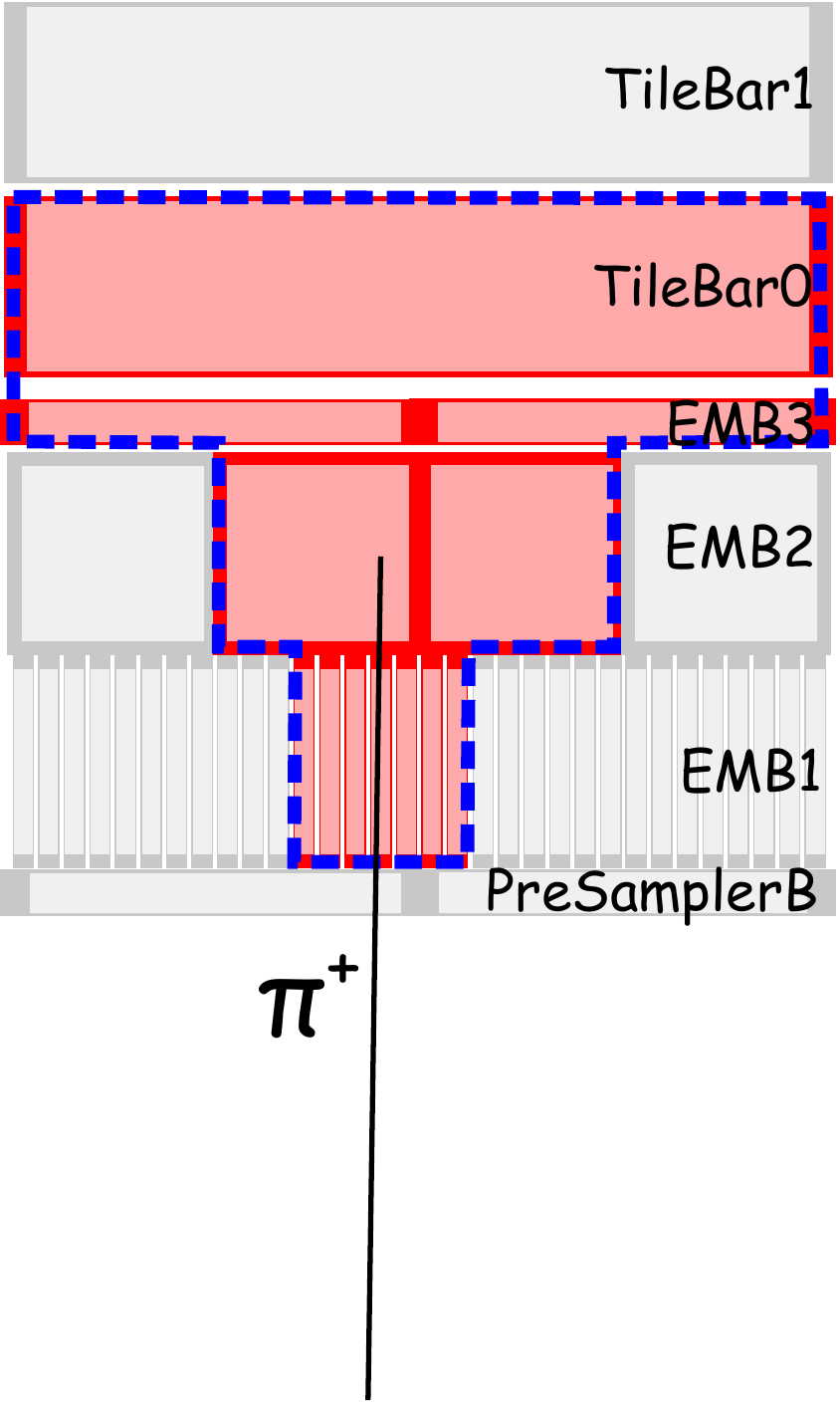}\, & \,\includegraphics[trim=0cm 1.5cm 0cm -0.25cm,clip,width=2.75cm]{flowchart_calo-a}\, & \,\includegraphics[trim=0cm 1.5cm 0cm -0.25cm,clip,width=2.75cm]{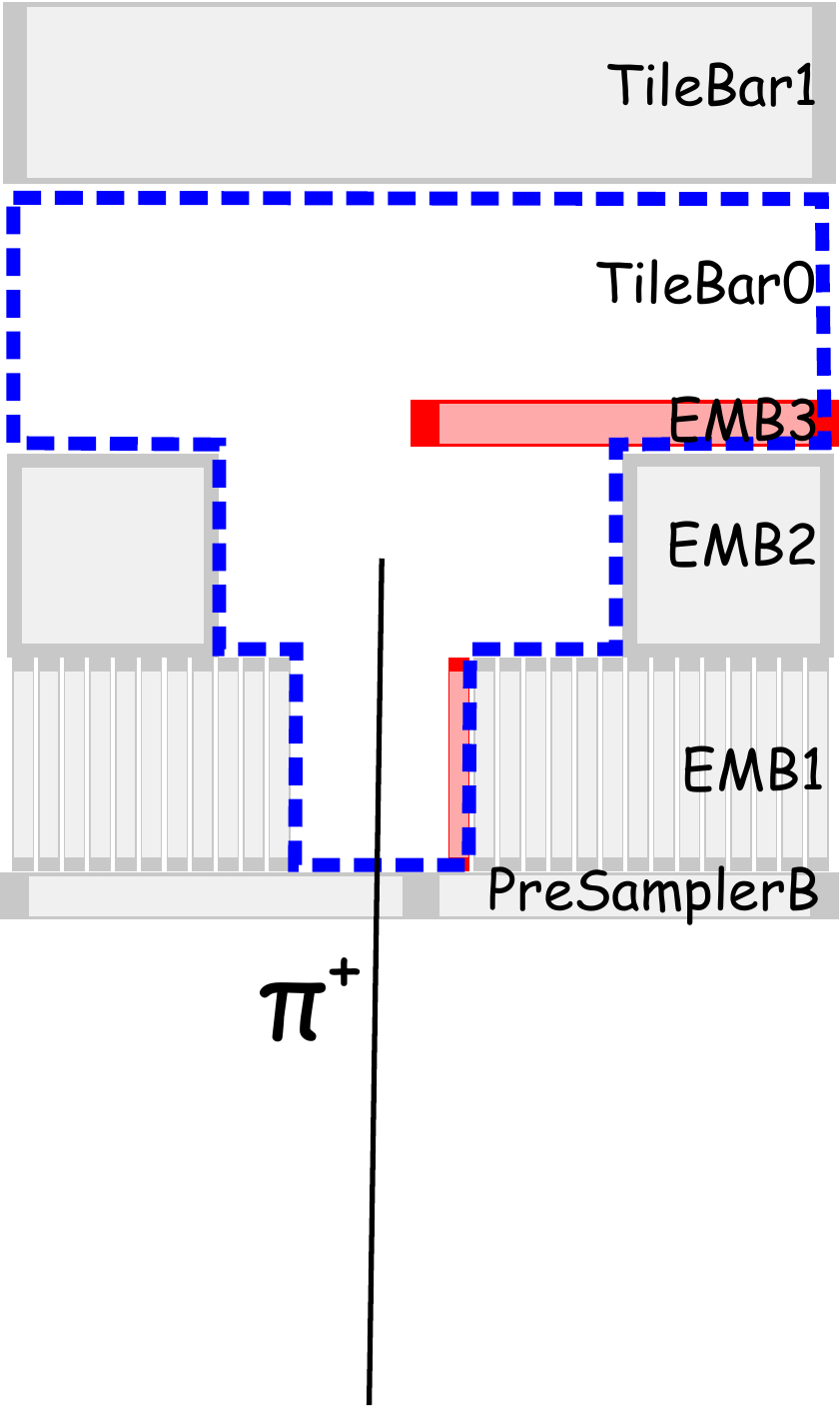}\, & \,\includegraphics[trim=0cm 1.5cm 0cm -0.25cm,clip,width=2.75cm]{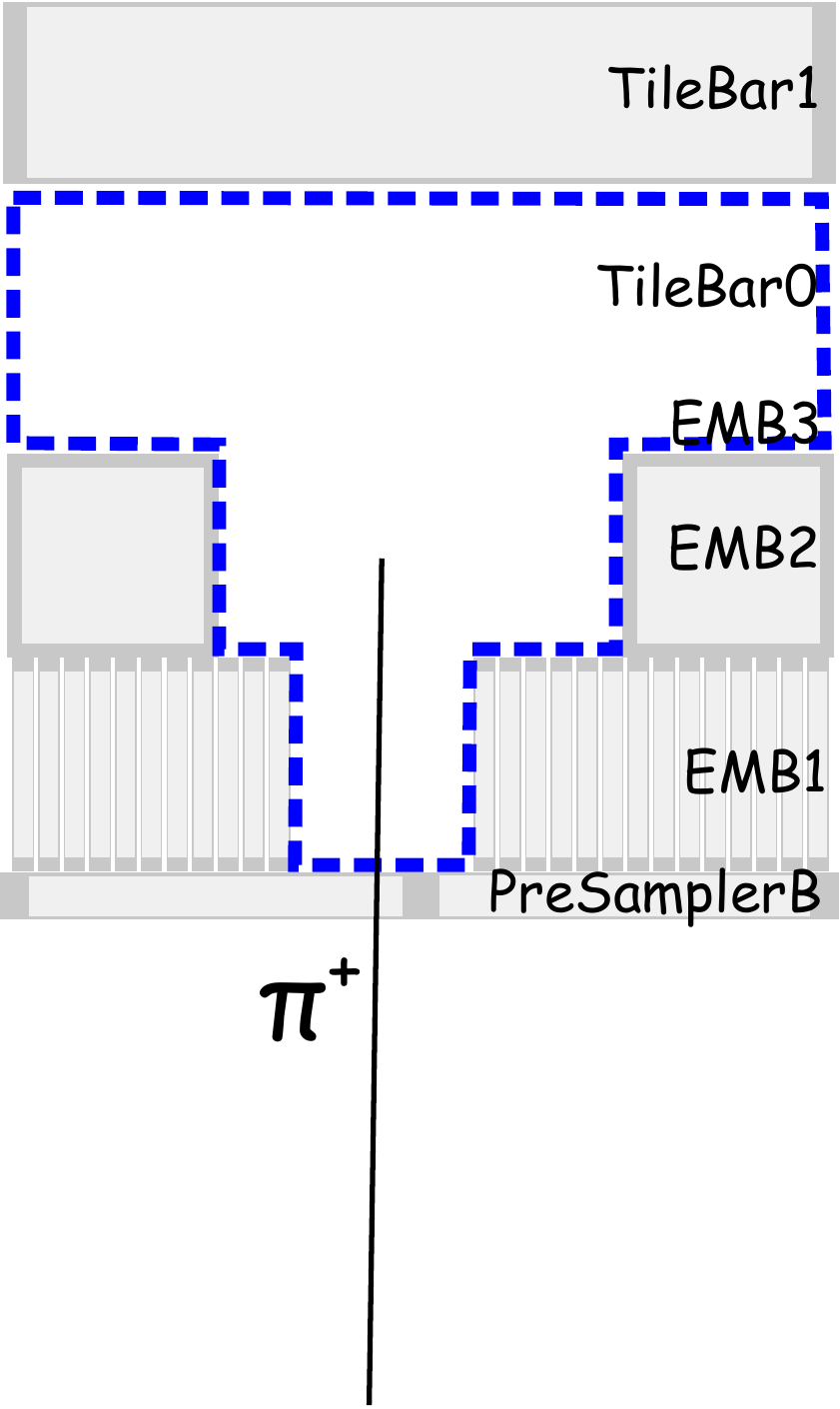}\,\\ \hline
$\begin{array}{l}\text{1 particle,}\\ \text{2 \topoclusters}\end{array}$ & \includegraphics[trim=0cm 1.5cm 0cm -0.25cm,clip,width=2.75cm]{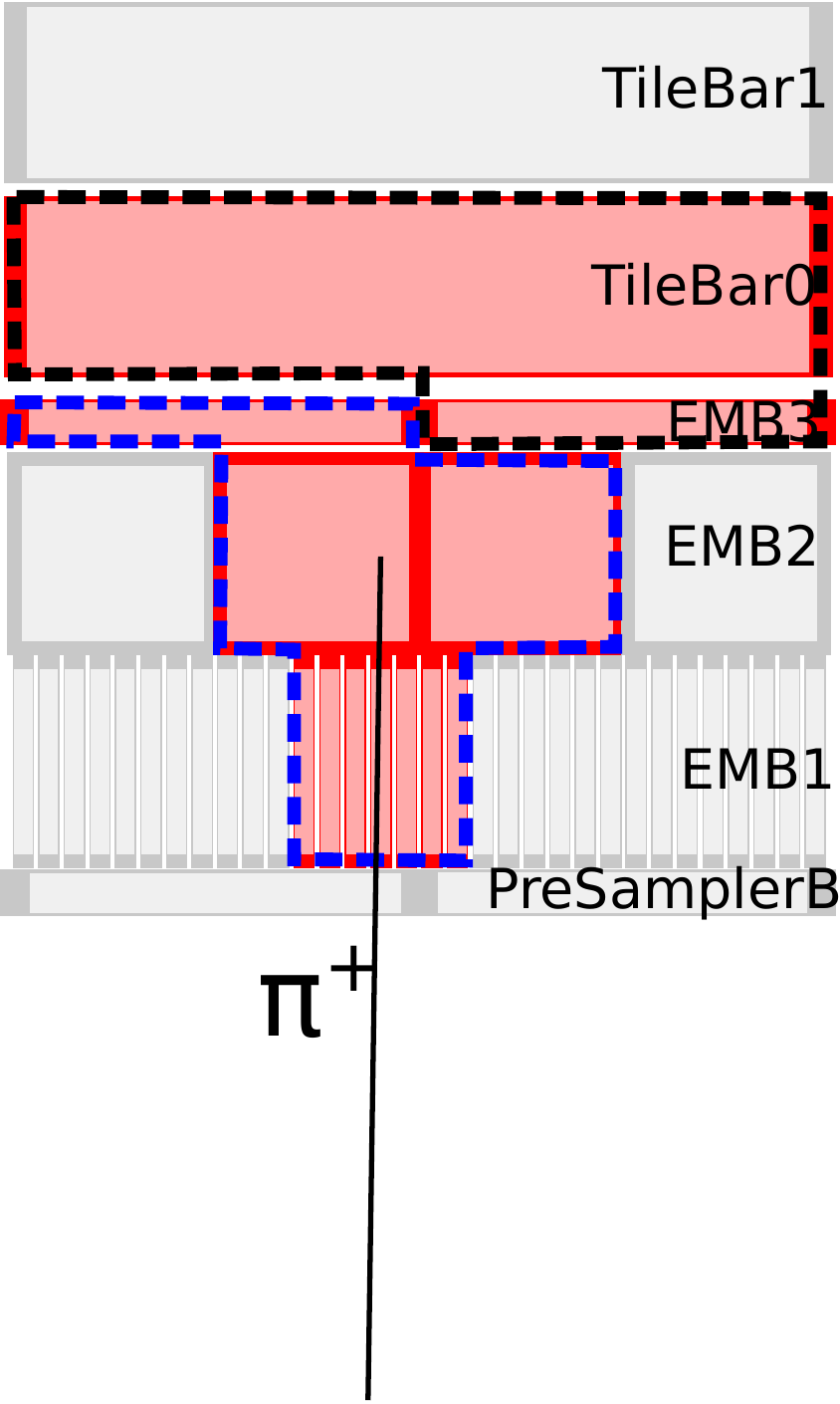} & \includegraphics[trim=0cm 1.5cm 0cm -0.25cm,clip,width=2.75cm]{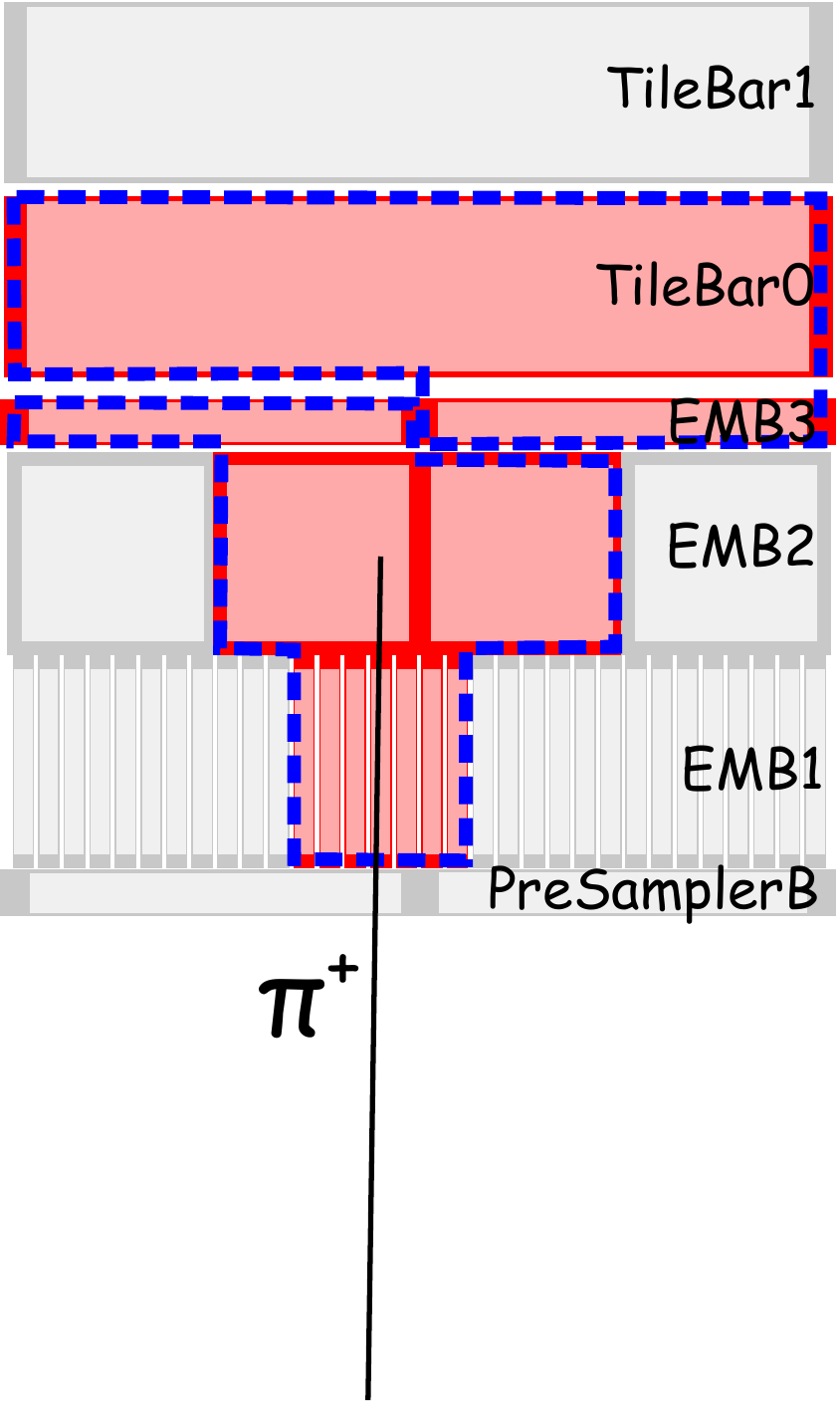} & \includegraphics[trim=0cm 1.5cm 0cm -0.25cm,clip,width=2.75cm]{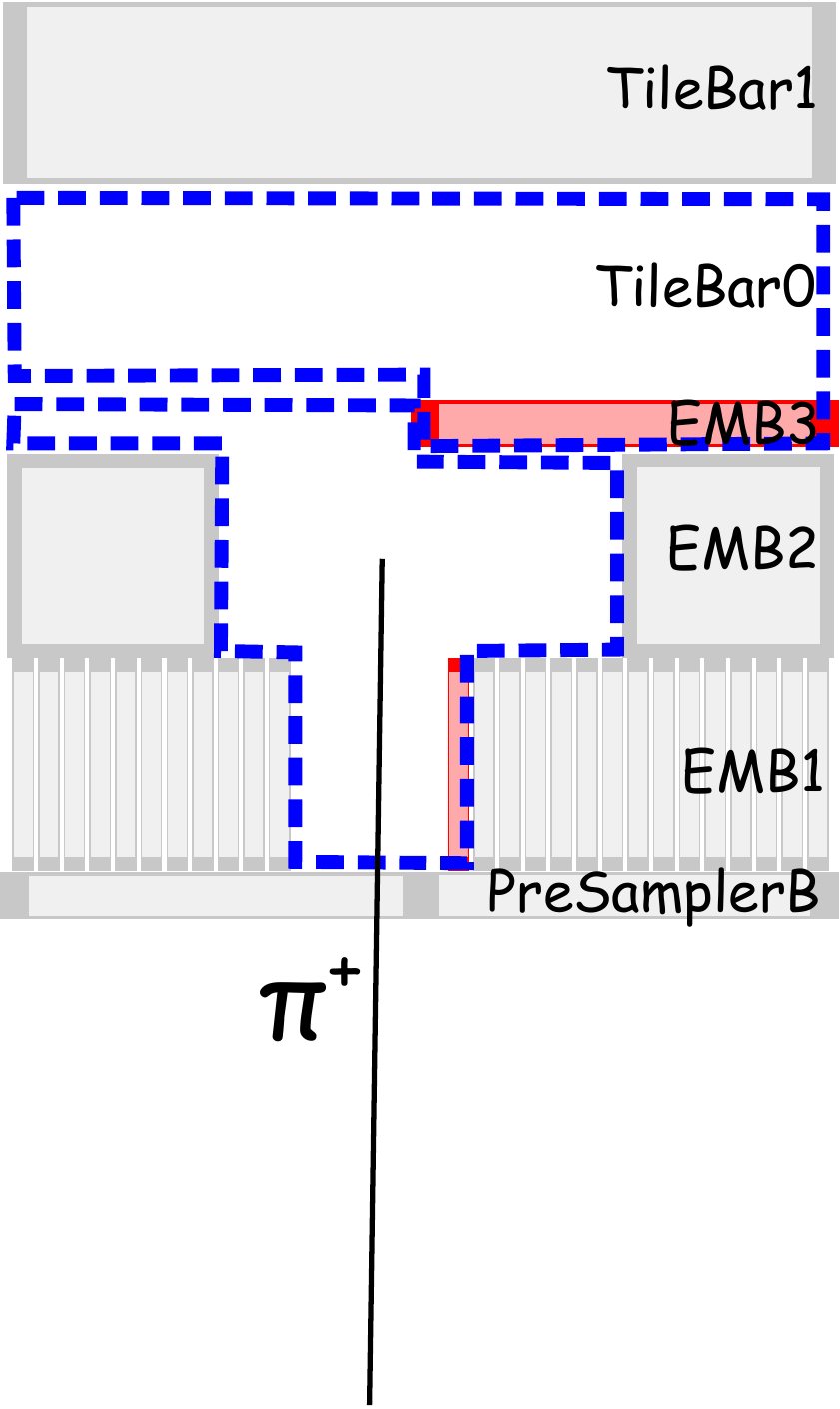} & \includegraphics[trim=0cm 1.5cm 0cm -0.25cm,clip,width=2.75cm]{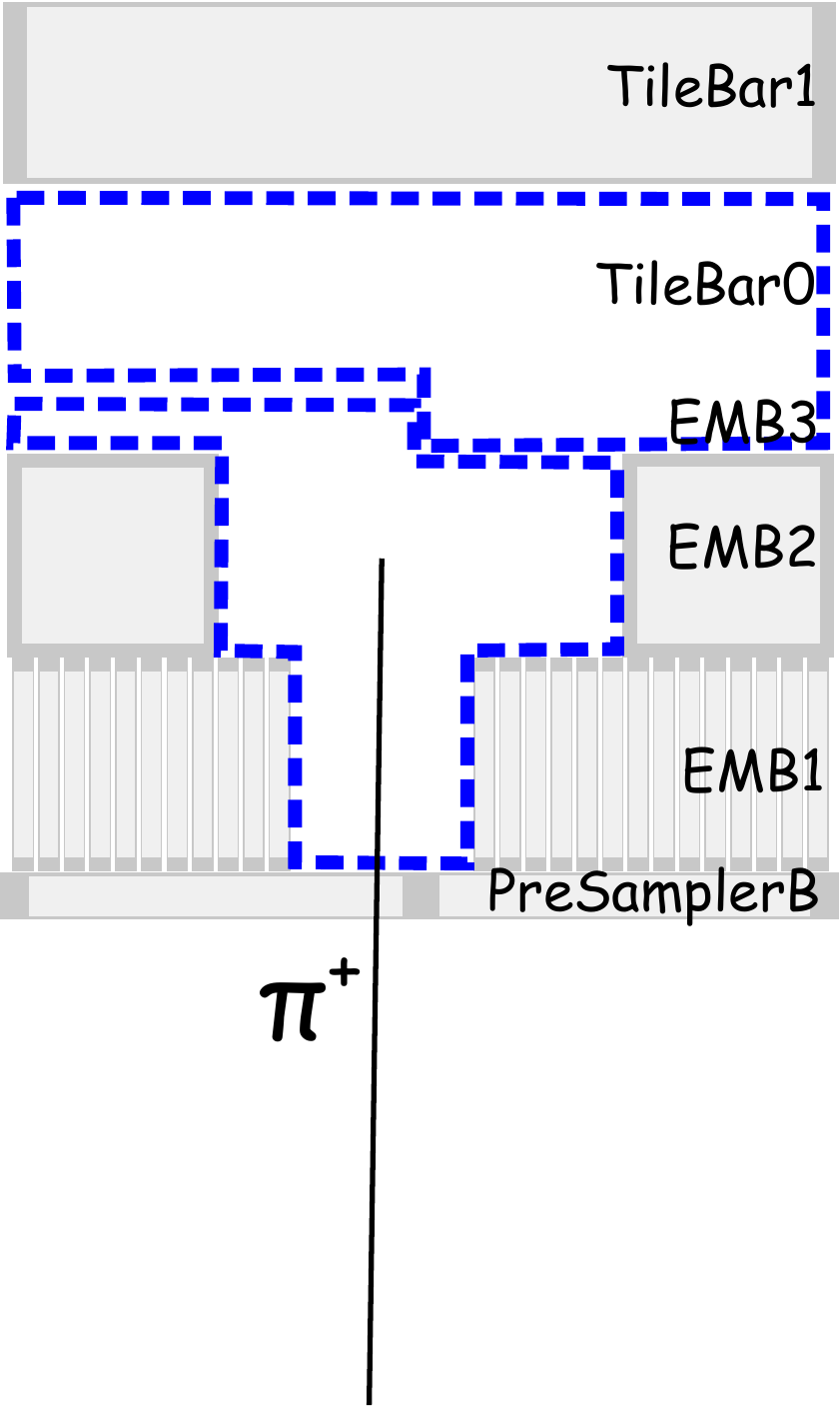}\\ \hline
$\begin{array}{l}\text{2 particles,}\\ \text{2 \topoclusters}\end{array}$ & \includegraphics[trim=0cm 1.5cm 0cm -0.25cm,clip,width=2.75cm]{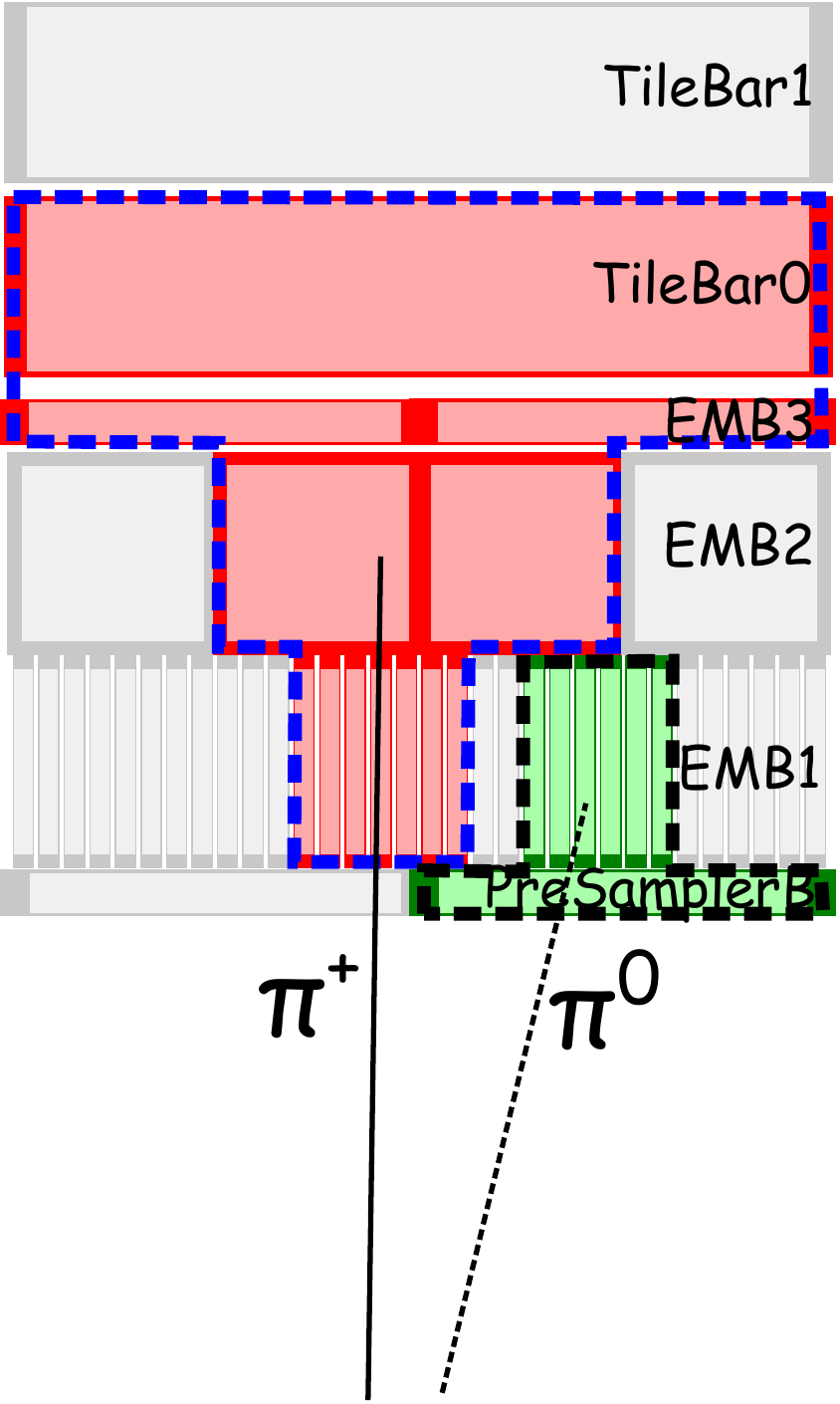} & \includegraphics[trim=0cm 1.5cm 0cm -0.25cm,clip,width=2.75cm]{flowchart_calo-i} & \includegraphics[trim=0cm 1.5cm 0cm -0.25cm,clip,width=2.75cm]{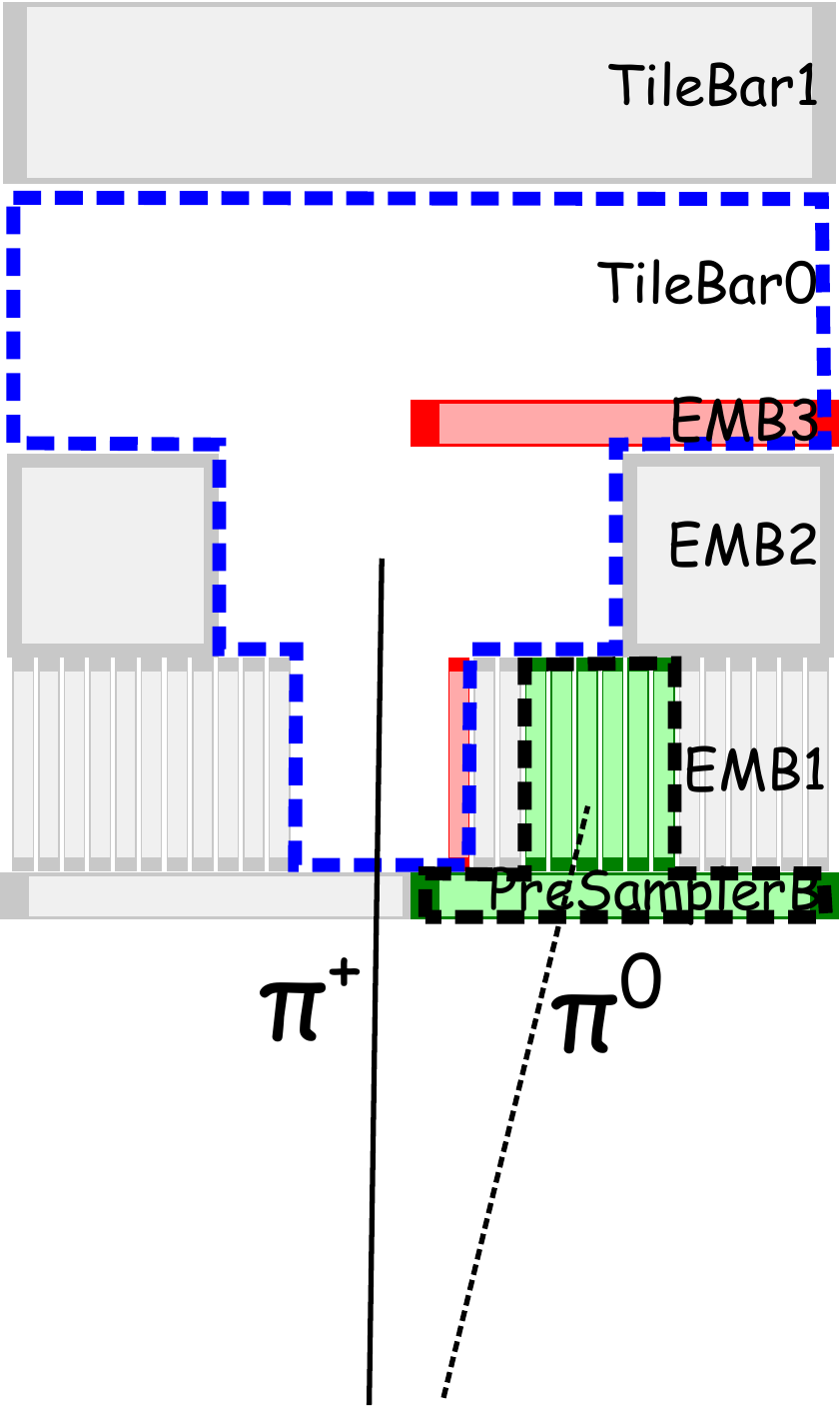} & \includegraphics[trim=0cm 1.5cm 0cm -0.25cm,clip,width=2.75cm]{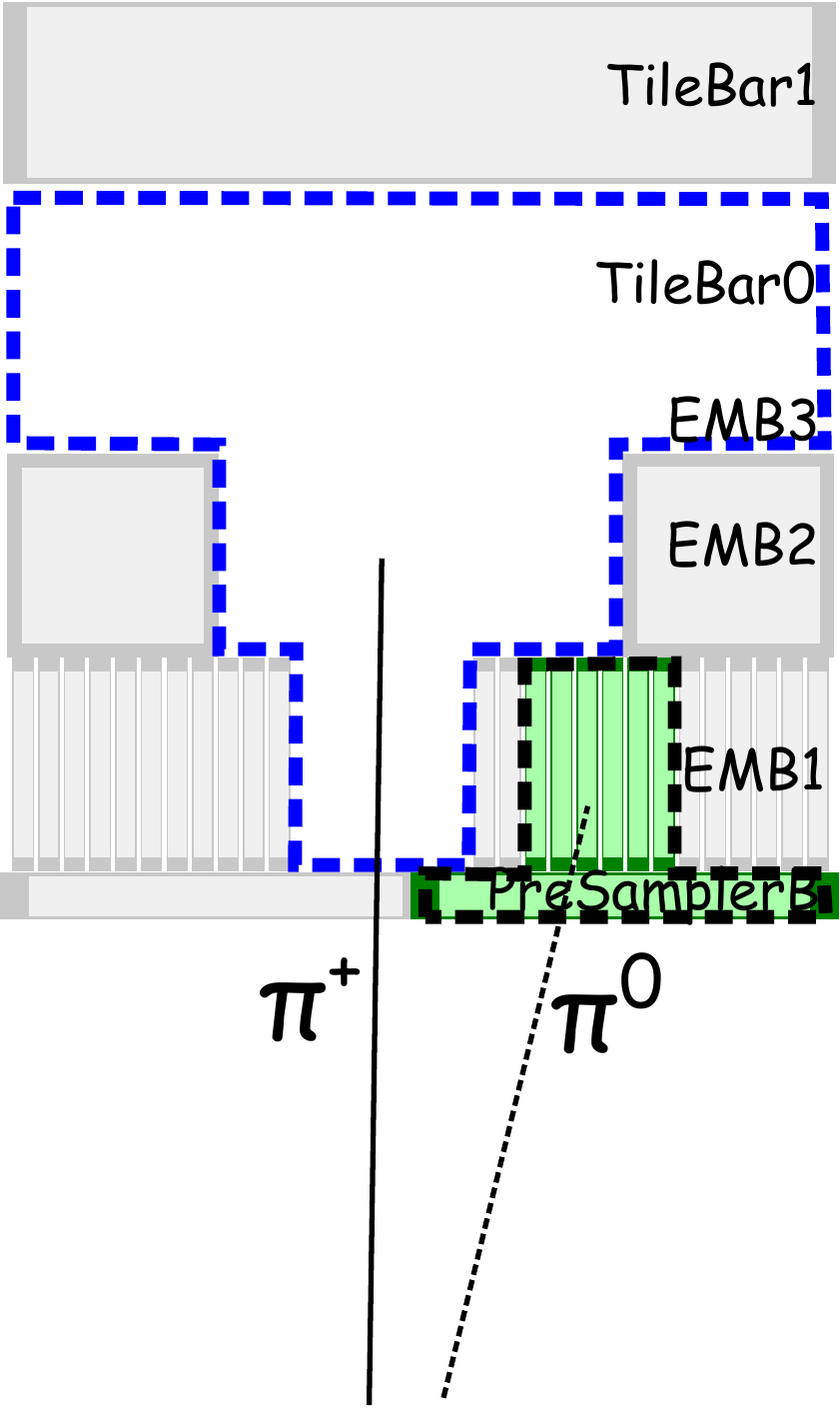}\\ \hline
$\begin{array}{l}\text{2 particles,}\\ \text{1 \topocluster}\end{array}$ & \includegraphics[trim=0cm 1.5cm 0cm -0.25cm,clip,width=2.75cm]{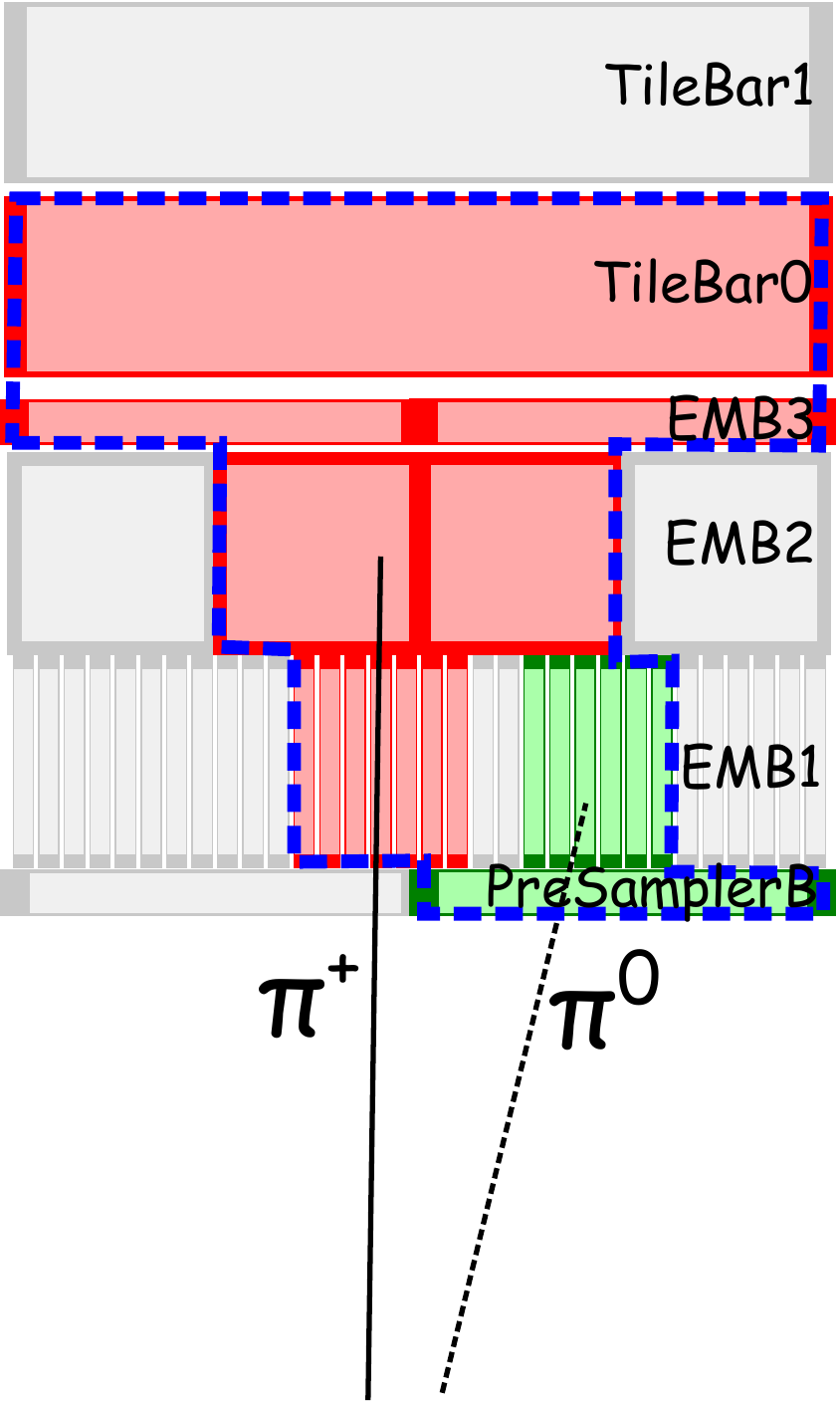} & \includegraphics[trim=0cm 1.5cm 0cm -0.25cm,clip,width=2.75cm]{flowchart_calo-m} & \includegraphics[trim=0cm 1.5cm 0cm -0.25cm,clip,width=2.75cm]{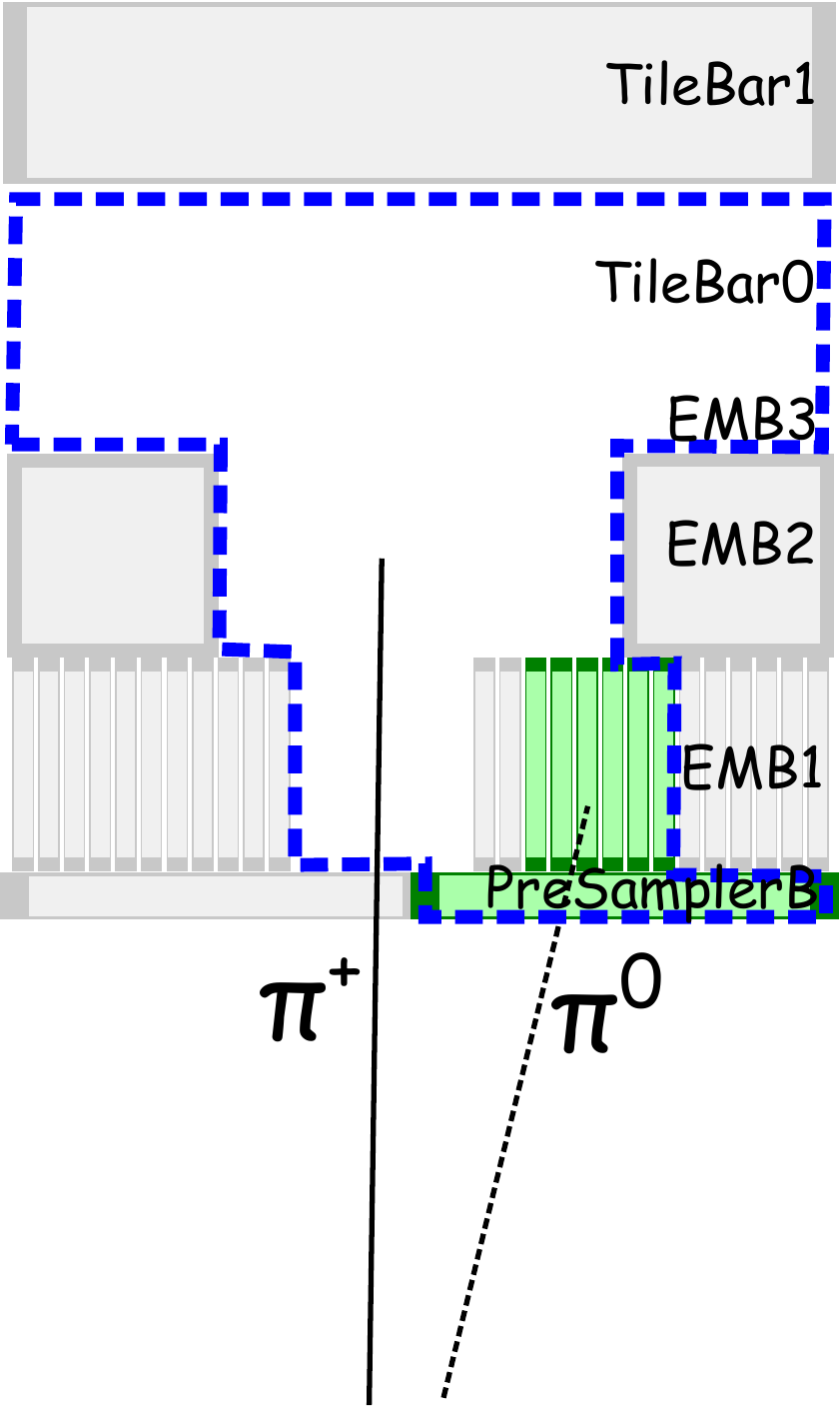} & \includegraphics[trim=0cm 1.5cm 0cm -0.25cm,clip,width=2.75cm]{flowchart_calo-o}\\ \hline
\end{tabular}
\caption{Idealised examples of how the algorithm is designed to deal with several different cases.
  The red cells are those which have energy from the $\pi^+$,
  the green cells energy from the photons from the $\pi^0$ decay,
  the dotted lines represent the original \topocluster boundaries with those outlined in blue having been matched by the algorithm to the $\pi^+$, while those in black are yet to be selected.
  The different layers in the electromagnetic calorimeter (Presampler, EMB1, EMB2, EMB3) are indicated.
  In this sketch only the first two layers of the Tile calorimeter are shown (TileBar0 and TileBar1).
}
\label{fig:eflowRec:cartoon}
\end{figure}

This procedure is applied to tracks sorted in descending \pT-order, firstly to the cases where only a single \topocluster is matched to the track, and then to the other selected tracks.
This methodology is illustrated in \Fig{\ref{fig:eflowRec:cartoon}}.

Details about each step of the procedure are given in the rest of this section.
After some general discussion of the properties of \topoclusters in the calorimeter,
the energy subtraction procedure for each track is described.
The procedure is  accompanied by illustrations of performance metrics used to validate the configuration of the algorithm. 
The samples used for the validation are single-pion and dijet MC samples without pile-up, as described in the previous section.
Charged pions dominate the charged component of the jet,
which on average makes up two-thirds of the visible jet energy~\cite{Knowles:1997dk,Green:357258}.
Another quarter of the jet energy is contributed by photons from neutral hadron decays, and the remainder is carried by neutral hadrons that reach the calorimeter.
Because the majority of tracks are generated by charged pions~\cite{ATLAS-CONF-2011-016}, particularly at low \pT,
the pion mass hypothesis is assumed for all tracks used by the particle flow algorithm to reconstruct jets.
Likewise the energy subtraction is based on the calorimeter's response to charged pions.

In the following sections, the values for the parameter set and the performance obtained for the 2012 dataset are discussed.
These parameter values are not necessarily the product of a full optimisation,
but it has been checked that the performance is not easily improved by variations of these choices.
Details of the optimisation are beyond the scope of the paper.

%-------------------------------------------------------------------------------
\subsection{Containment of showers within a single \topocluster}

The performance of the particle flow  algorithm, especially the shower subtraction procedure,
strongly relies on the topological clustering algorithm.
Hence, it is important to quantify the extent to which the clustering algorithm distinguishes individual particles' showers
and how often it splits a single particle's shower into more than one \topocluster.
The different configurations of \topoclusters containing energy from a given single pion are classified using two variables.

For a given \topocluster $i$, the fraction of the particle's true energy contained in the \topocluster (see \Sect{\ref{sec:MC_calhits}}),
with respect to the total true energy deposited by the particle in all clustered cells, is defined as
\begin{equation}
  \effi = \frac
    {E^\text{clus $i$}_\text{true, $\pi$}}
    {E^\text{all \topoclusters}_\text{true, $\pi$}}\,,
\end{equation}
where ${E^\text{clus $i$}_\text{true, $\pi$}}$ is the true energy deposited in topo-cluster $i$ by the generated particle under consideration and 
${E^\text{all \topoclusters}_\text{true, $\pi$}}$ is the true energy deposited in all topo-clusters by that truth particle.
For each particle, the \topocluster with the highest value of \effi is designated the leading \topocluster, for which $\efflead = \effi$.
The minimum number of \topoclusters needed to capture at least \SI{90}{\%} of the particle's true energy, i.e.\ 
such that $\sum_{i=0}^{n}\effi > \SI{90}{\%}$, is denoted by $n_\text{clus}^{90}$.

\Topoclusters can contain contributions from multiple particles, affecting the ability of the subtraction algorithm to separate the energy deposits of different particles.
The purity \puri for a \topocluster $i$ is defined as the fraction of true energy within the \topocluster which originates from the particle of interest:
\begin{equation}
  \puri = \frac
    {E^\text{clus $i$}_\text{true, $\pi$} }
    {E^\text{clus $i$}_\text{true, all particles}}\,.
\end{equation}
For the leading \topocluster, defined by having the highest \effi, the purity value is denoted by \purlead. 

Only charged particles depositing significant energy (at least \SI{20}{\%} of their true energy) in clustered cells are considered in the following plots,
as in these cases there is significant energy in the calorimeter to remove.
%to avoid double counting of signals between tracks and \topoclusters.
This also avoids the case where insufficient energy is present in any cell to form a cluster,
which happens frequently for very low-energy particles \cite{topoclustering}.

Figure \ref{fig:eflowRec:cartoon} illustrates how the subtraction procedure is designed to deal with cases of different complexity. Four different scenarios are shown covering cases where the charged pion deposits its energy in one cluster, in two clusters, and where there is a nearby neutral pion which either deposits its energy in a separate cluster or the same cluster as the charged pion.

\begin{figure}[htbp]
\centering
\subfloat[$|\etatrue|<1.0$]{\includegraphics[width=0.31\textwidth]{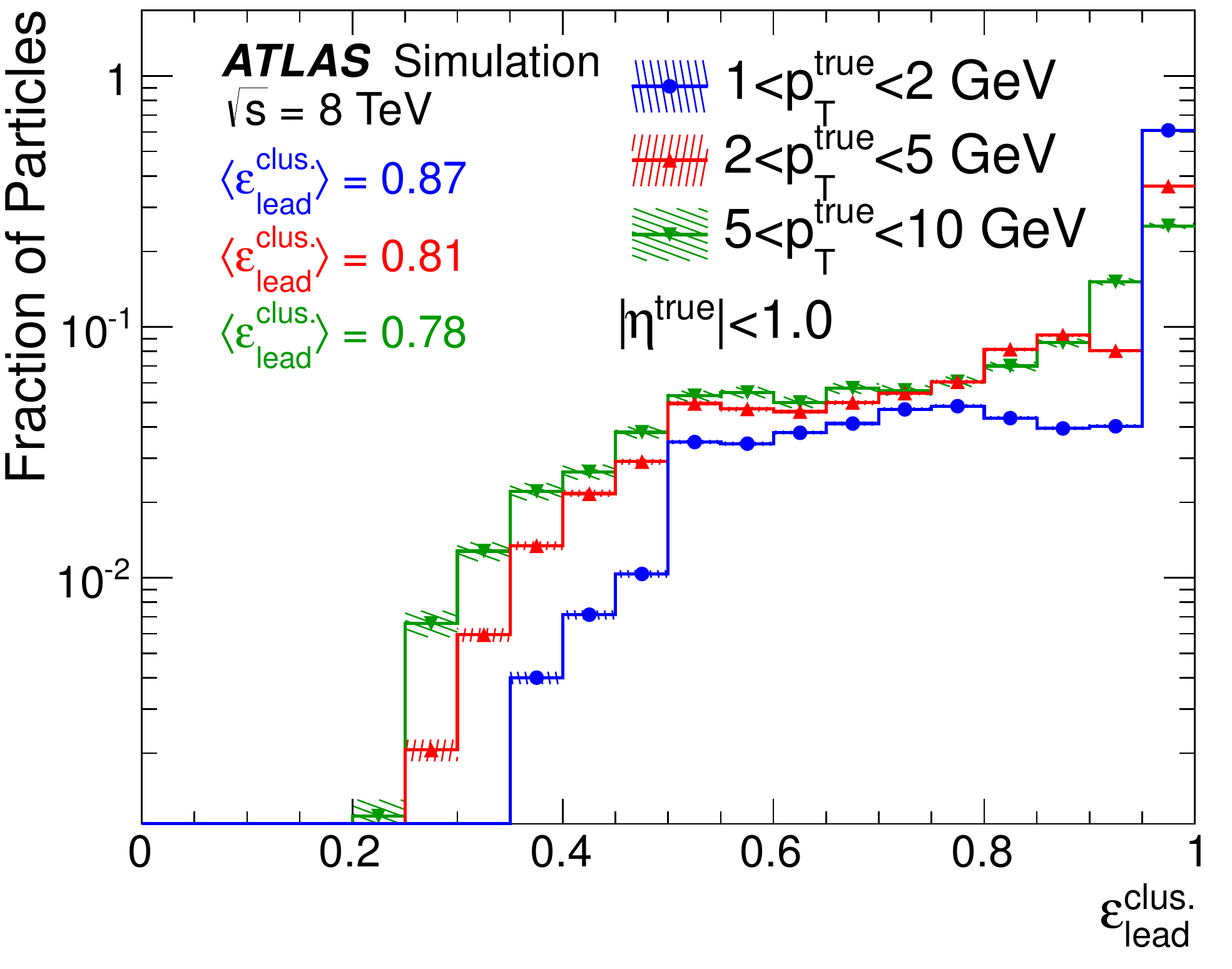}}\quad
\subfloat[$1.0<|\etatrue|<2.0$]{\includegraphics[width=0.31\textwidth]{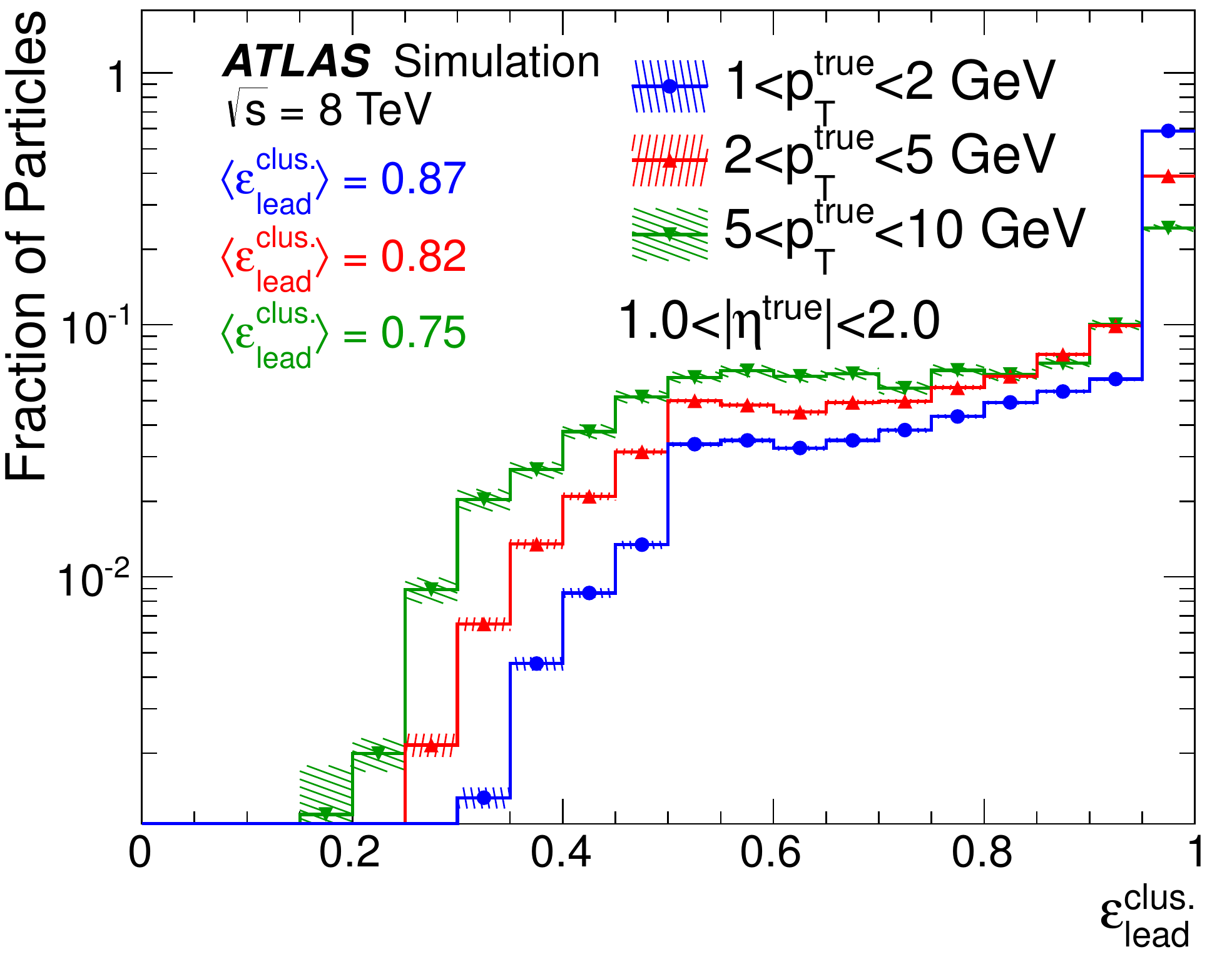}}\quad
\subfloat[$2.0<|\etatrue|<2.5$]{\includegraphics[width=0.31\textwidth]{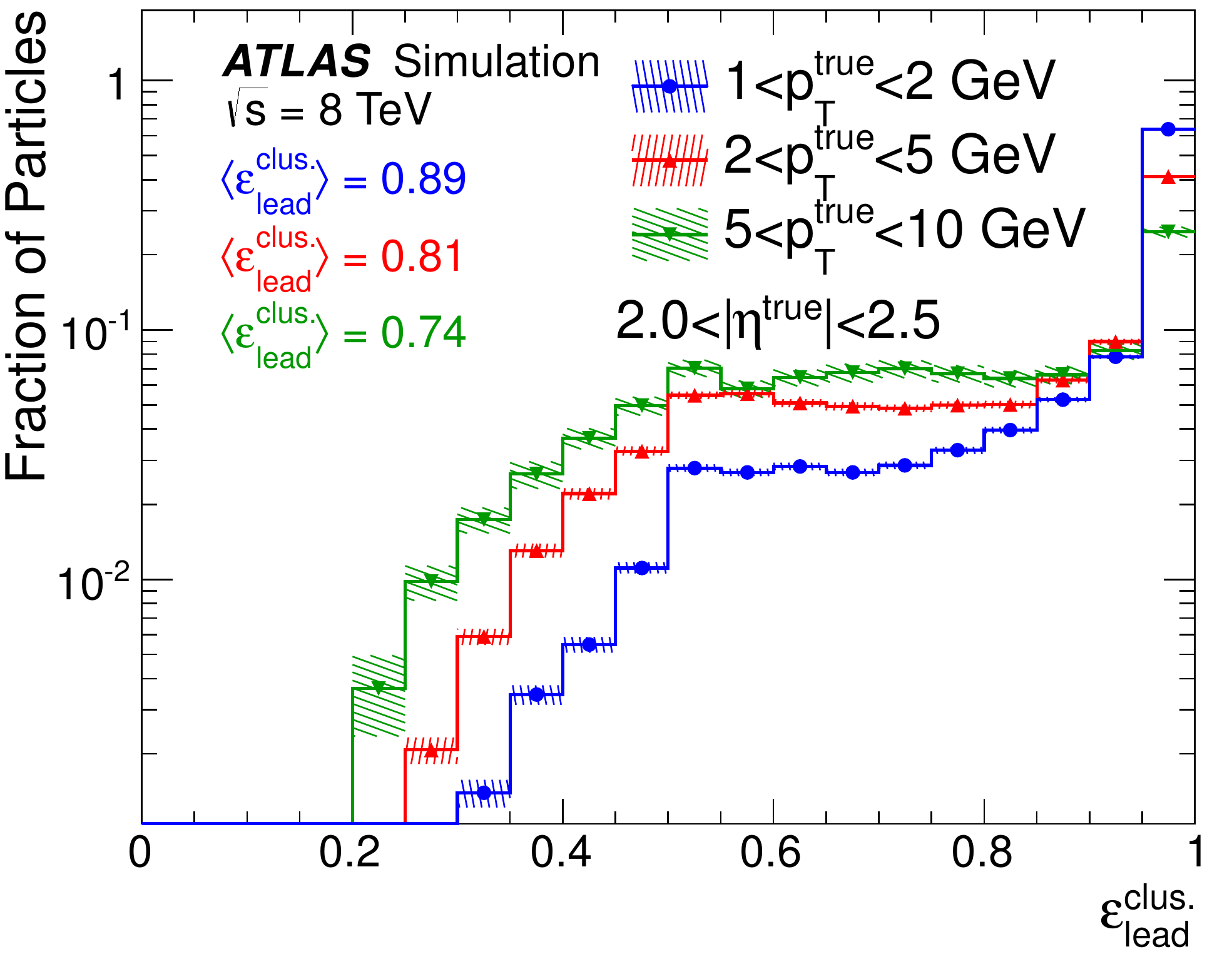}}
\caption{Distribution of the fraction of the total true energy in the leading \topocluster, \efflead, for charged pions
  which deposit significant energy (\SI{20}{\%} of the particle's energy) in the clustered cells for three different $\pTtrue$ bins in three $|\etatrue|$ regions.
  \DijetSample
}
\label{fig:eflowRec:eff}
\end{figure}

\begin{figure}[htbp]
\centering
\subfloat[$|\etatrue|<1.0$]{\includegraphics[width=0.31\textwidth]{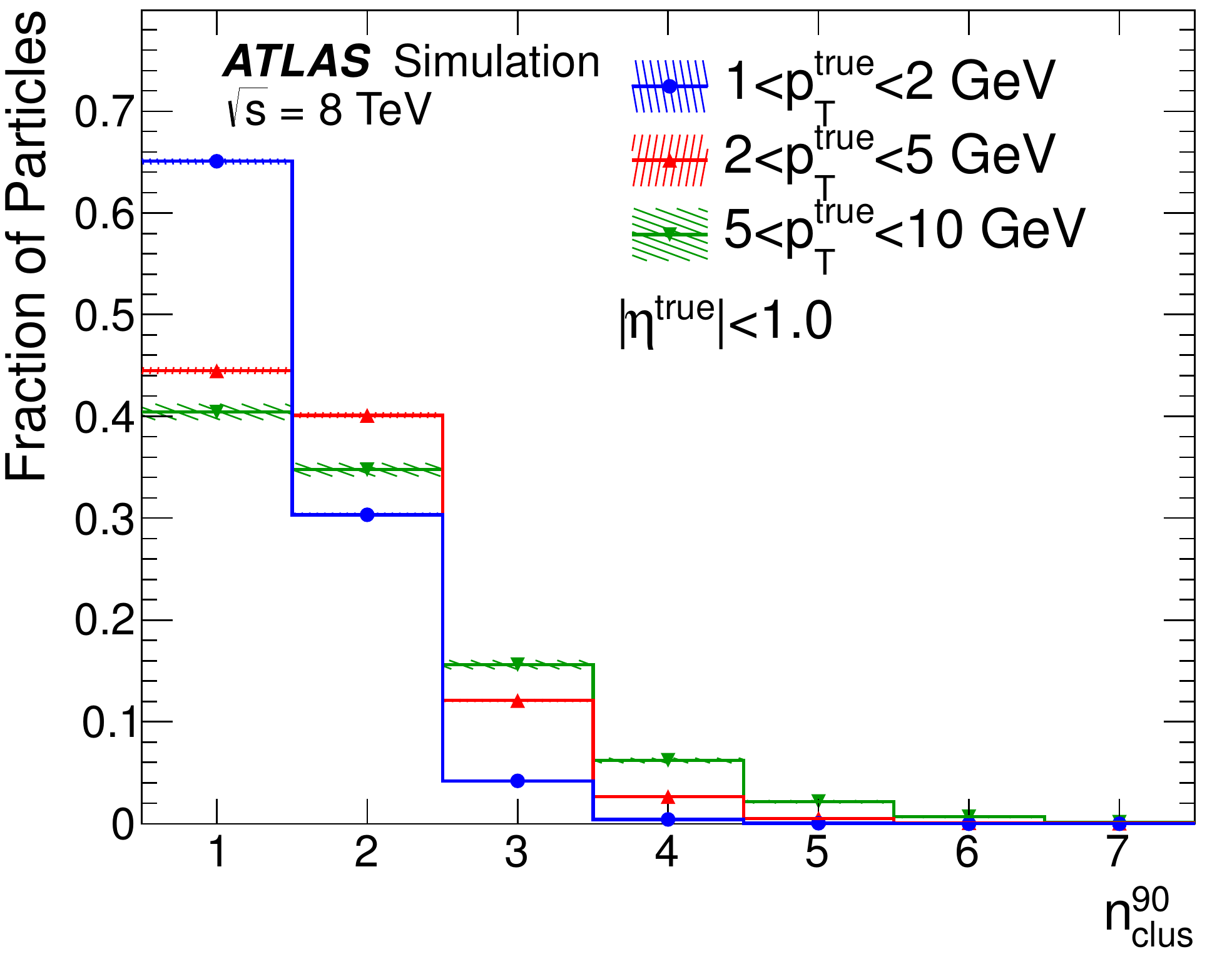}}\quad
\subfloat[$1.0<|\etatrue|<2.0$]{\includegraphics[width=0.31\textwidth]{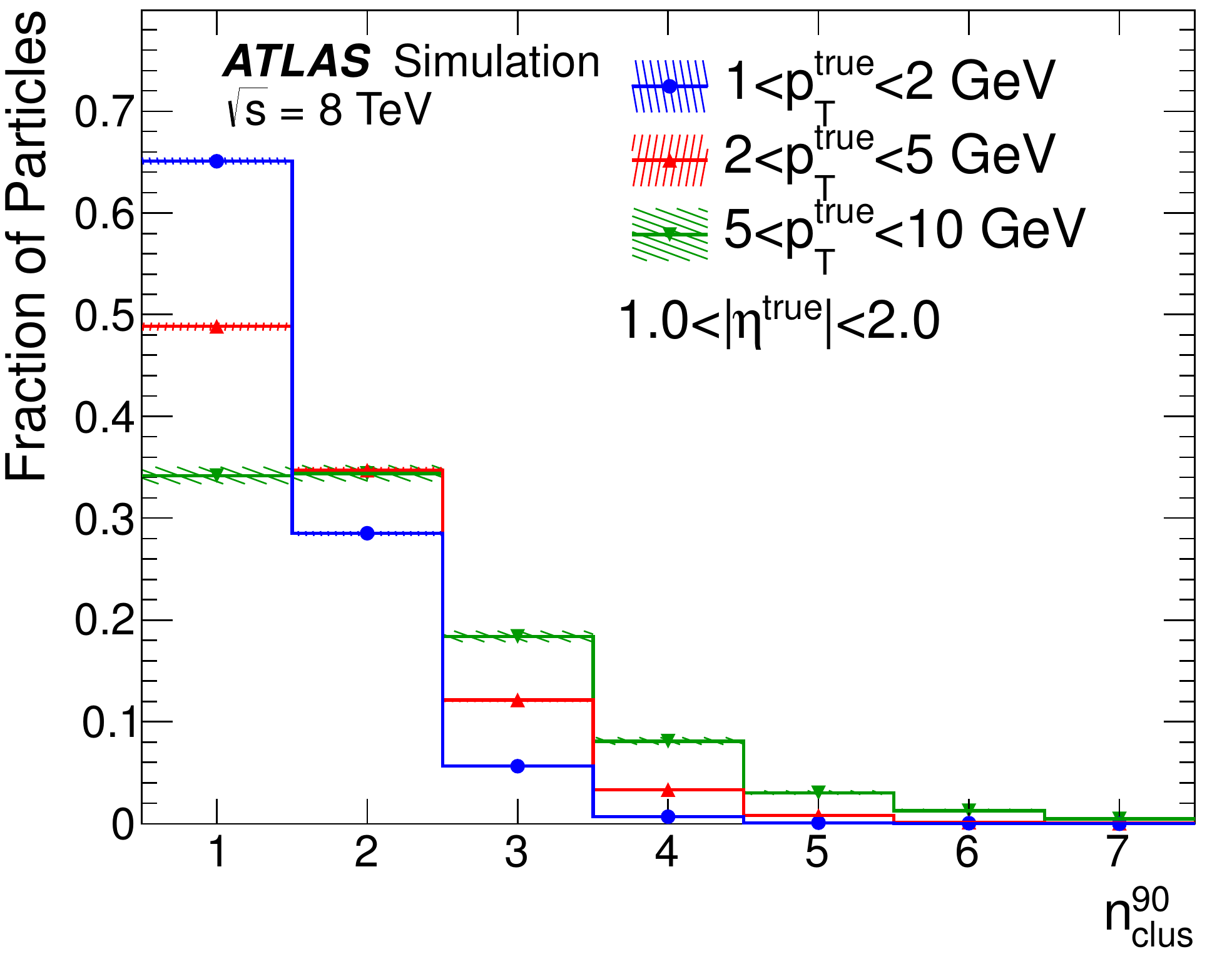}}\quad
\subfloat[$2.0<|\etatrue|<2.5$]{\includegraphics[width=0.31\textwidth]{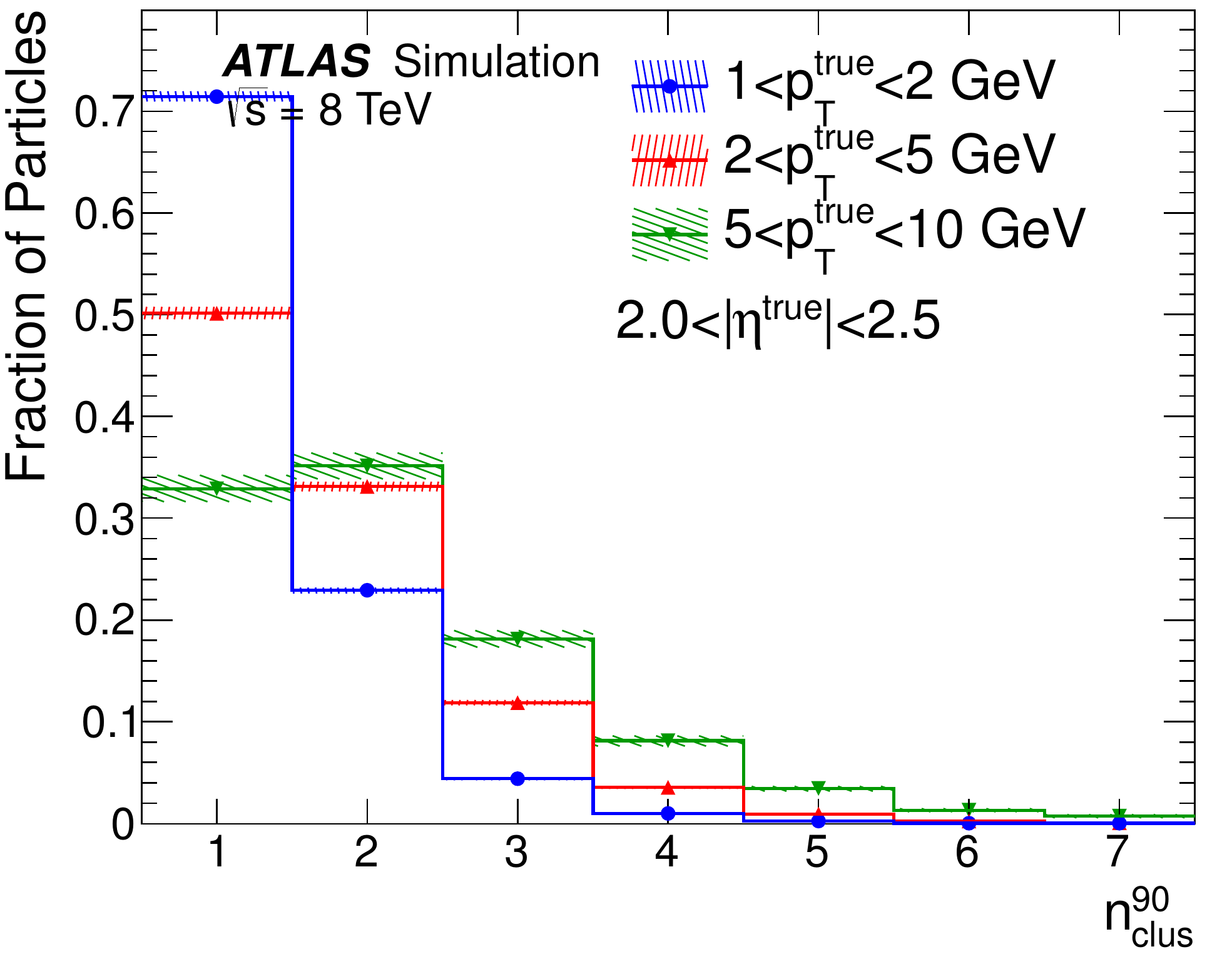}}
\caption{Distributions of the number of \topoclusters required to contain $> \SI{90}{\%}$ of the true deposited energy of a single charged pion
  which deposits significant energy (\SI{20}{\%} of the particle's energy) in the clustered cells.
  The distributions are shown for three $\pTtrue$ bins in three $|\etatrue|$ regions.
  \DijetSample
}
\label{fig:eflowRec:nClus}
\end{figure}

\begin{figure}[htbp]
\centering
\includegraphics[width=0.48\textwidth]{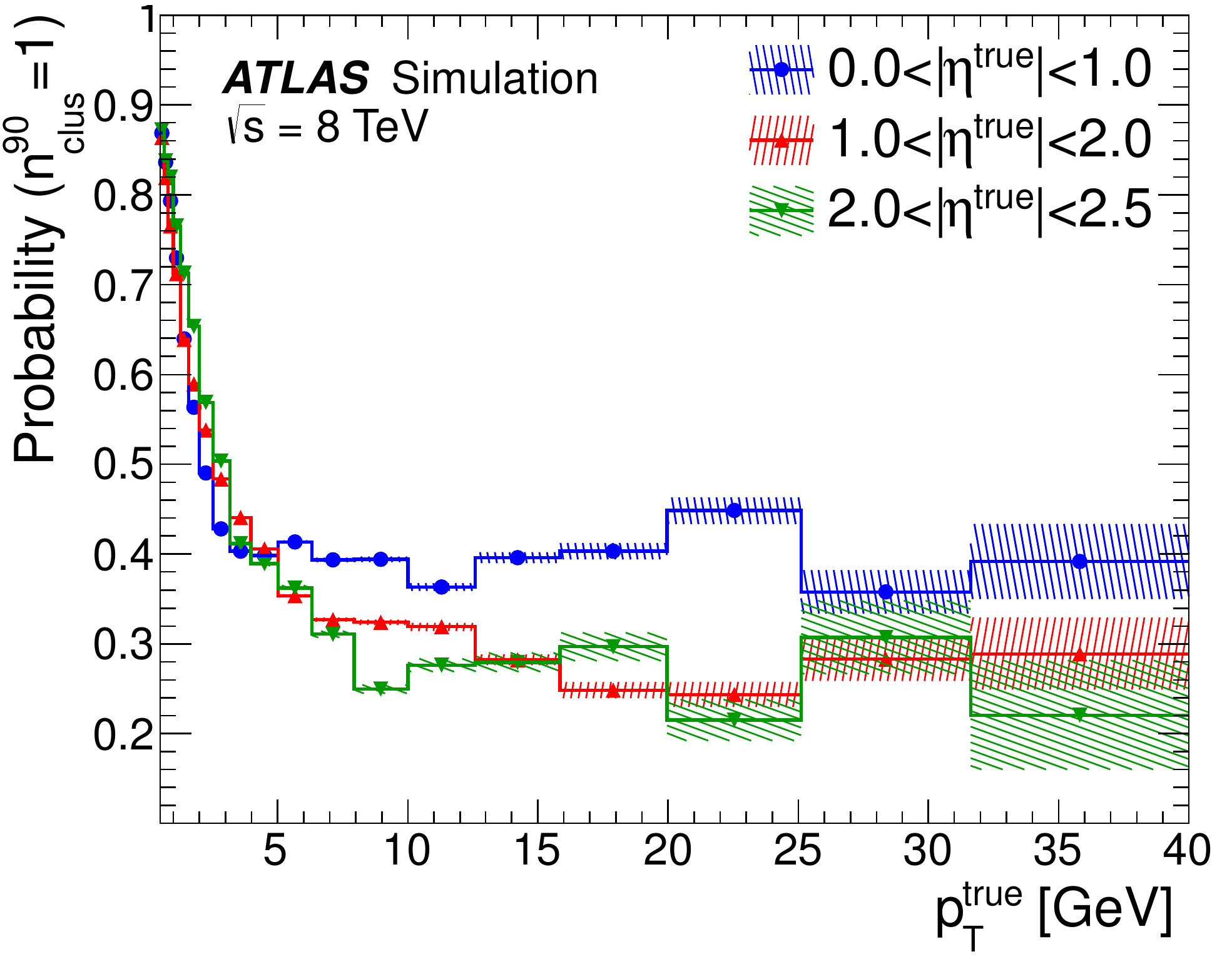}
\caption{The probability that a single \topocluster contains $> \SI{90}{\%}$ of the true deposited energy of a single charged pion,
  which deposits significant energy (\SI{20}{\%} of the particle's energy) in the clustered cells.
  The distributions are shown as a function of $\pTtrue$ in three $|\etatrue|$ regions.
  \DijetSample
}
\label{fig:eflowRec:nClusTProfile}
\end{figure}

\begin{figure}[htbp]
\centering
\subfloat[$|\etatrue|<1.0$]{\includegraphics[width=0.31\textwidth]{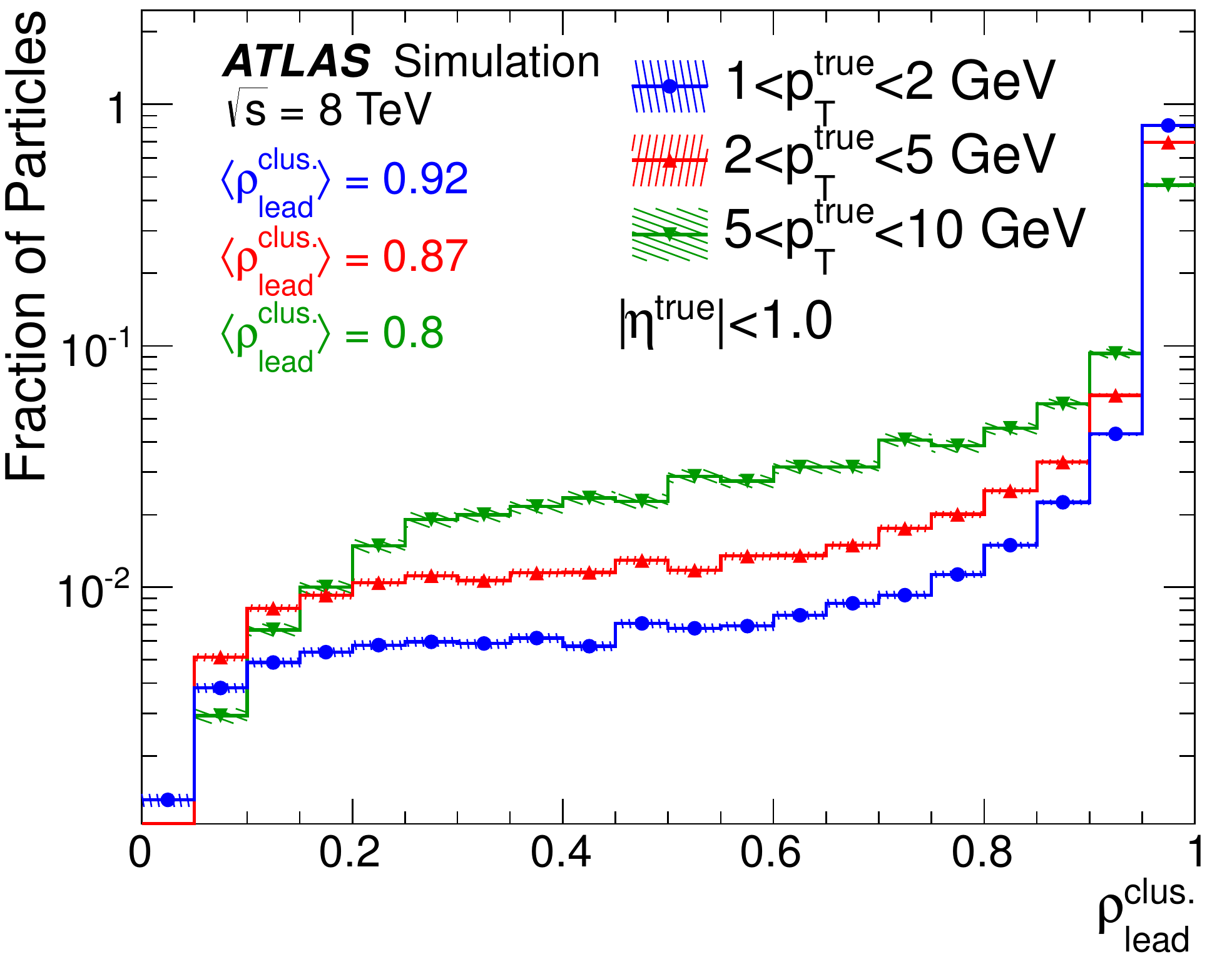}}\quad
\subfloat[$1.0<|\etatrue|<2.0$]{\includegraphics[width=0.31\textwidth]{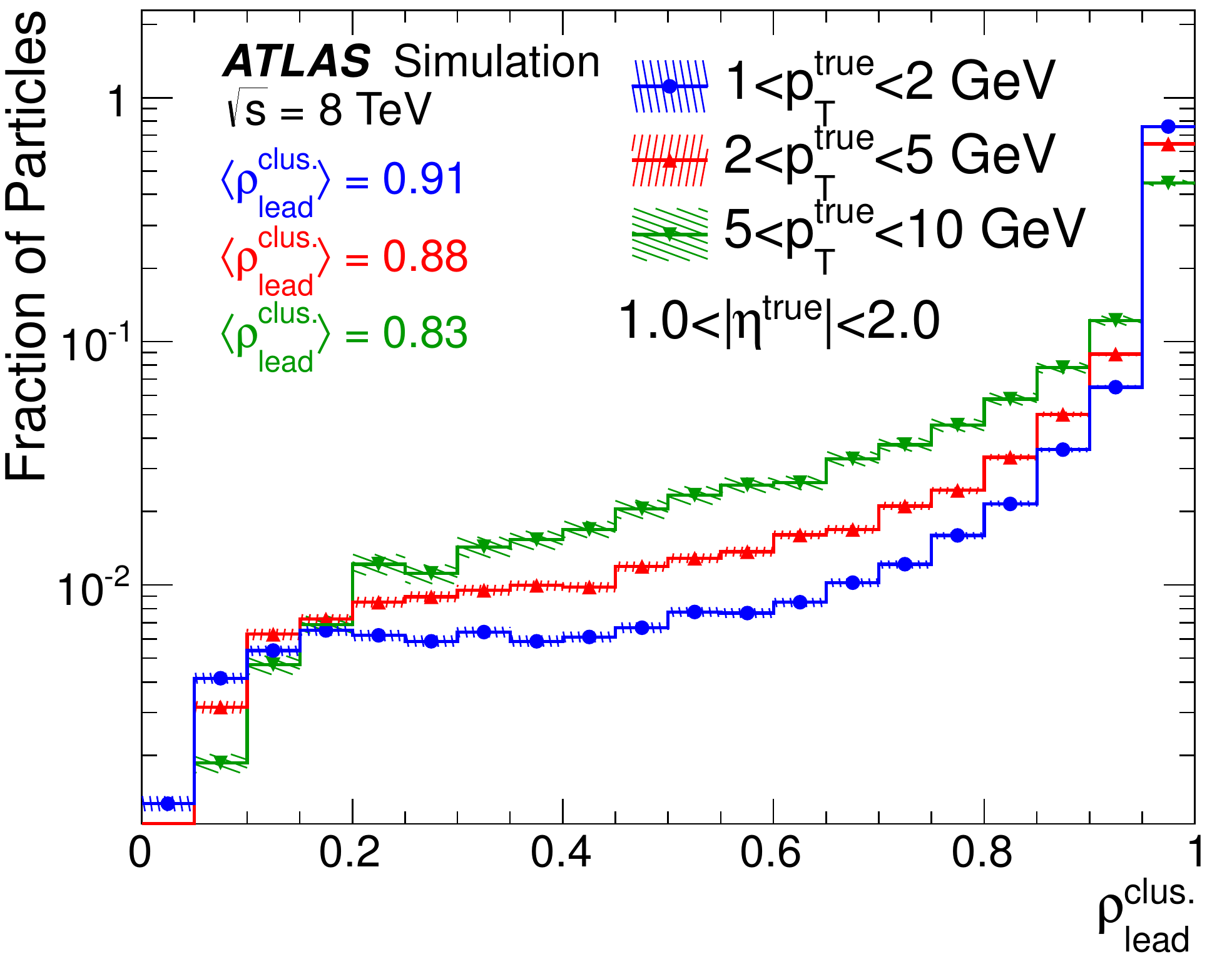}}\quad
\subfloat[$2.0<|\etatrue|<2.5$]{\includegraphics[width=0.31\textwidth]{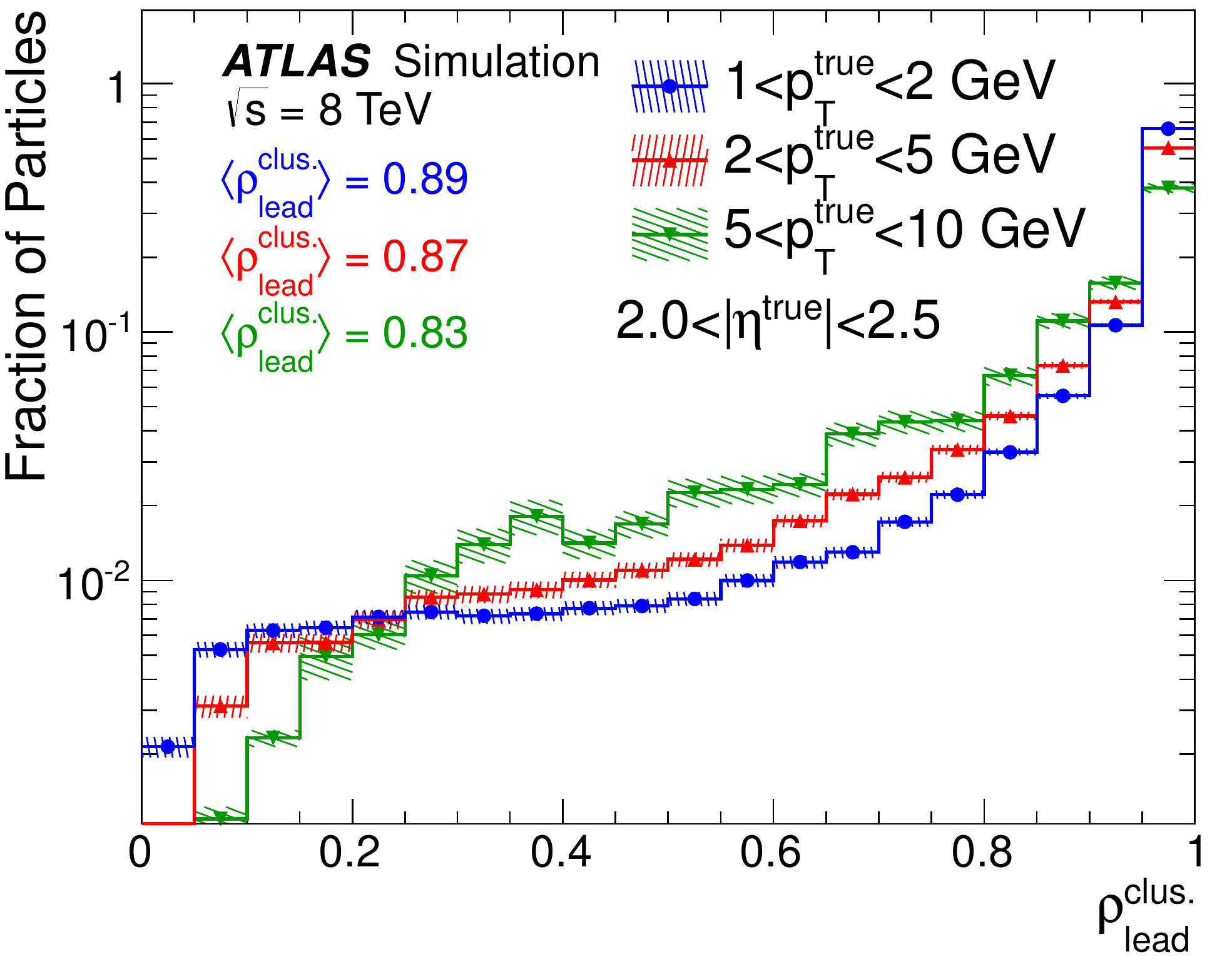}}
\caption{The purity \purlead, defined for a selected charged pion as the fractional contribution of the chosen particle to the total true energy in the leading \topocluster, shown for pions with \efflead>\SI{50}{\%}.
  Distributions are shown for several $\pTtrue$ bins and in three $|\etatrue|$ regions.
  \DijetSample
}
\label{fig:eflowRec:purity}
\end{figure}

Several distributions are plotted for the dijet sample in which the energy of the leading jet, measured at truth level,
is in the range $20 < \pTlead < \SI{500}{\GeV}$.
The distribution of \efflead is shown in \Fig{\ref{fig:eflowRec:eff}} for different $\pTtrue$ and $\etatrue$ bins.
It can be seen that \efflead decreases as the \pT of the particle increases and
very little dependence on $\eta$ is observed.
Figure ~\ref{fig:eflowRec:nClus} shows the distribution of $n_\text{clus}^{90}$.
As expected, $n_\text{clus}^{90}$ increases with particle \pT.
It is particularly interesting to know the fraction of particles for which at least \SI{90}{\%} of the true energy is contained in a single \topocluster ($n_\text{clus}^{90}=1$)
and this is shown in \Fig{\ref{fig:eflowRec:nClusTProfile}}.
Lastly, \Fig{\ref{fig:eflowRec:purity}} shows the distribution of \purlead.
This decreases as \pTtrue increases and has little dependence on $|\eta^\mathrm{true}|$.

For more than $\SI{60}{\%}$ of particles with $1 < \pTtrue < \SI{2}\GeV$, the shower is entirely contained within a single \topocluster ($\efflead\sim 1$).
This fraction falls rapidly with particle \pT, reaching $\sim\SI{25}{\%}$ for particles in the range $5<\pTtrue<\SI{10}{\GeV}$.
For particles with $\pTtrue<\SI{2}\GeV$, $\SI{90}{\%}$ of the particle energy can be captured within two \topoclusters in $\sim\SI{95}{\%}$ of cases.
The \topocluster purity also falls as the pion \pT increases,
with the target particle only contributing between \SI{38}{\%} and \SI{45}{\%} of the \topocluster energy when $5<\pTtrue<\SI{10}{\GeV}$.
This is in part due to the tendency for high-\pT particles to be produced in dense jets,
while softer particles from the underlying event tend to be isolated from nearby activity.

In general, the subtraction of the hadronic shower is easier for cases with \topoclusters with high \puri, and high \effi,
since in this configuration the \topocluster{}ing algorithm has separated out the contributions from different particles.

%-------------------------------------------------------------------------------
\subsection{Track selection}
\label{sec:trksel}

Tracks are selected which pass stringent quality criteria: at least nine hits in the silicon detectors are required, and tracks must have no missing Pixel hits when such hits would be expected~\cite{ATLAS-CONF-2012-042}.
This selection is designed such that the number of badly measured tracks is minimised and is referred to as \enquote{tight selection}.
No selection cuts are made on the association to the hard scatter vertex at this stage
Additionally, tracks are required to be within $|\eta|<2.5$ and have $\pT > \SI{0.5}{\GeV}$.
These criteria remain efficient for tracks from particles which are expected to deposit energy below the threshold needed to seed a \topocluster or particles that do not reach the calorimeter.
Including additional tracks by reducing the \pT requirement to $\SI{0.4}{\GeV}$ leads to a substantial increase in computing time
without any corresponding improvement in jet resolution. This is due to their small contribution to the total jet \pT.

Tracks with $\pT$ $>$ $\SI{40}{\GeV}$ are excluded from the algorithm,
as such energetic particles are often poorly isolated from nearby activity,
compromising the accurate removal of the calorimeter energy associated with the track.
In such cases, with the current subtraction scheme, there is no advantage in using the tracker measurement.
This requirement was tuned both by monitoring the effectiveness of the energy subtraction using the true energy deposited in dijet MC events, and by measuring the jet resolution in MC simulation.
The majority of tracks in jets with \pT between 40 and 60 GeV have \pT below 40 GeV, as shown later in \Sect{\ref{sec:DataMC}}.

In addition, any tracks matched to candidate electrons~\cite{PERF-2016-01} or muons~\cite{PERF-2014-05},
without any isolation requirements, identified with medium quality criteria, are not selected and therefore are not considered for subtraction, 
as the algorithm is optimised for the subtraction of hadronic showers.
The energy deposited in the calorimeter by electrons and muons is hence taken into account in the 
particle flow algorithm and any resulting \topoclusters are generally left unsubtracted.

\begin{figure}[htbp]
  \centering
  \subfloat[]{
    \includegraphics[width=0.48\textwidth]{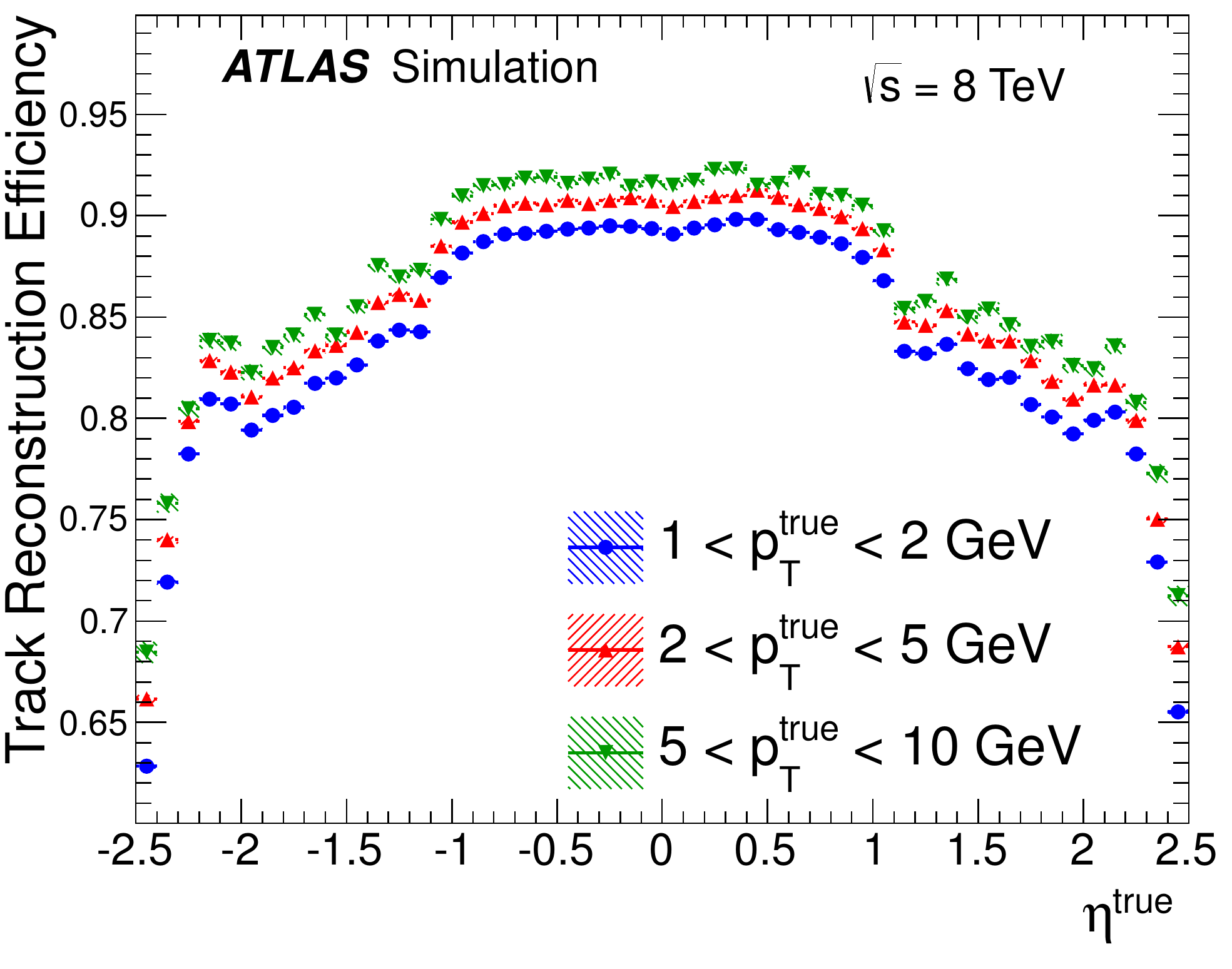}}\quad
  \subfloat[]{
    \includegraphics[width=0.48\textwidth]{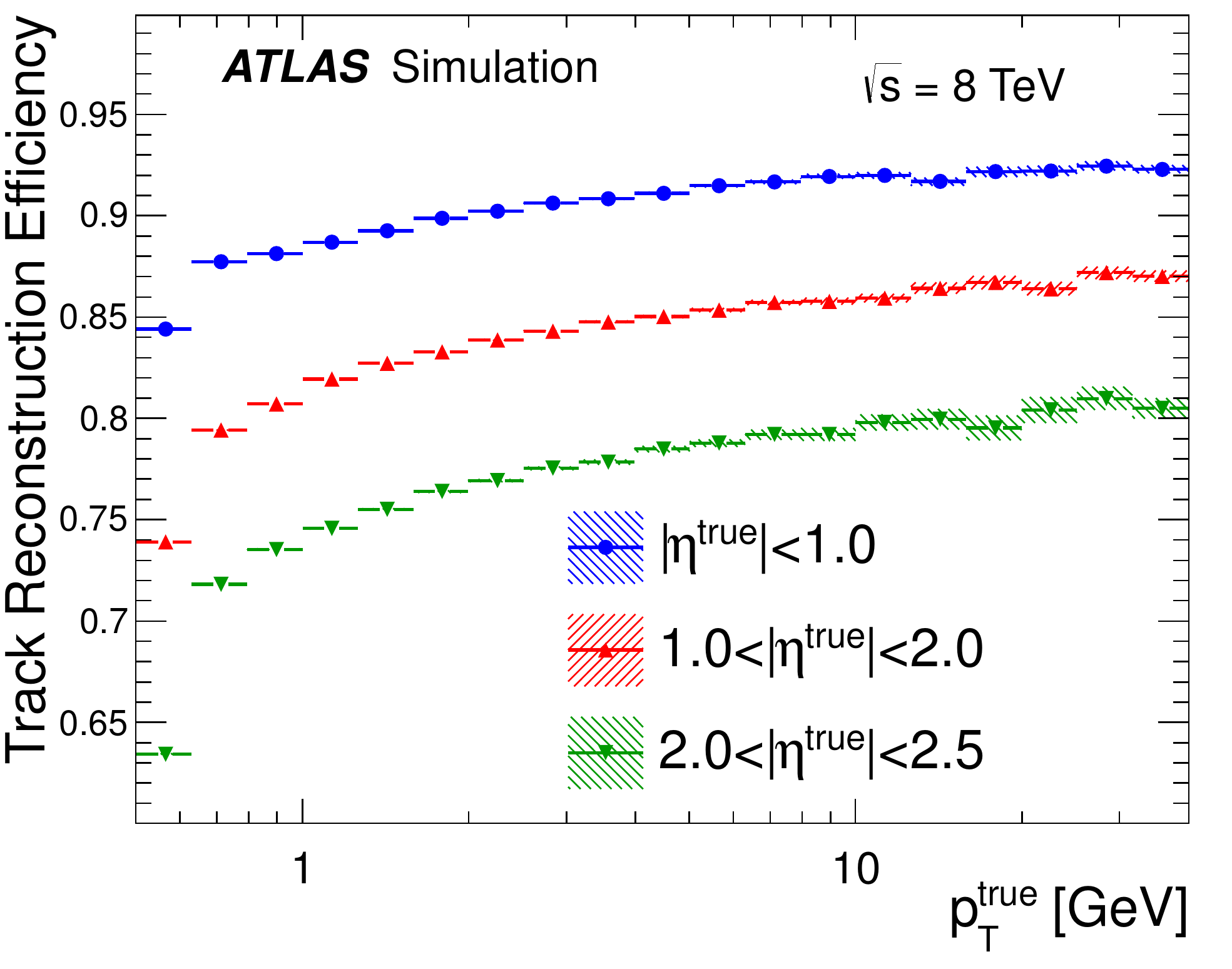}}
  \caption{The track reconstruction efficiency for charged pions after applying the tight quality selection criteria to the tracks.
   Subfigure (a) shows the efficiency for 1--2\,\si{\GeV}, 2--5\,\si{\GeV} and 5--10\,\si{\GeV} particles as a function of $\eta$, while
   (b) shows the track reconstruction efficiency as a function of \pT in three $|\eta|$ bins.
   A simulated dijet sample is used, with similar pile-up to that in the 2012 data, and for which $20 < \pTlead < \SI{1000}{\GeV}$.
   The statistical uncertainties in the number of MC simulated events are shown in a darker shading.
  }
\label{fig:eflowRec:trkEff}
\end{figure}

\begin{figure}[htbp]
  \centering
  \captionsetup[subfigure]{justification=centering}
  \subfloat[$1<\pTtrue< \SI{2}{\GeV}$,\protect\\ $|\etatrue|<1.0$.]{%
    \includegraphics[width=0.48\textwidth]{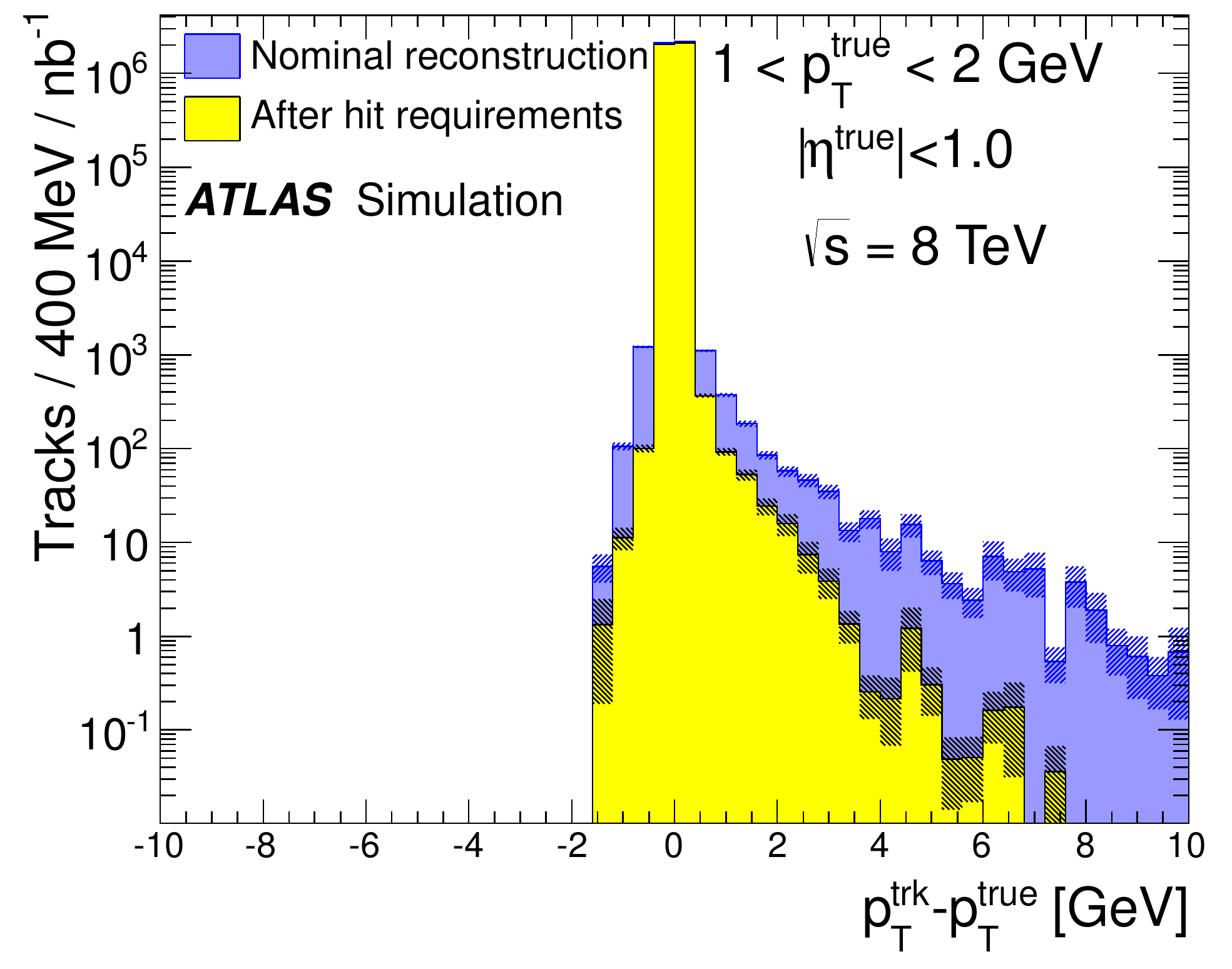}
  }\quad
  \subfloat[$5<\pTtrue< \SI{10}{\GeV}$,\protect\\ $2.0<|\etatrue|<2.5$.]{%
    \includegraphics[width=0.48\textwidth]{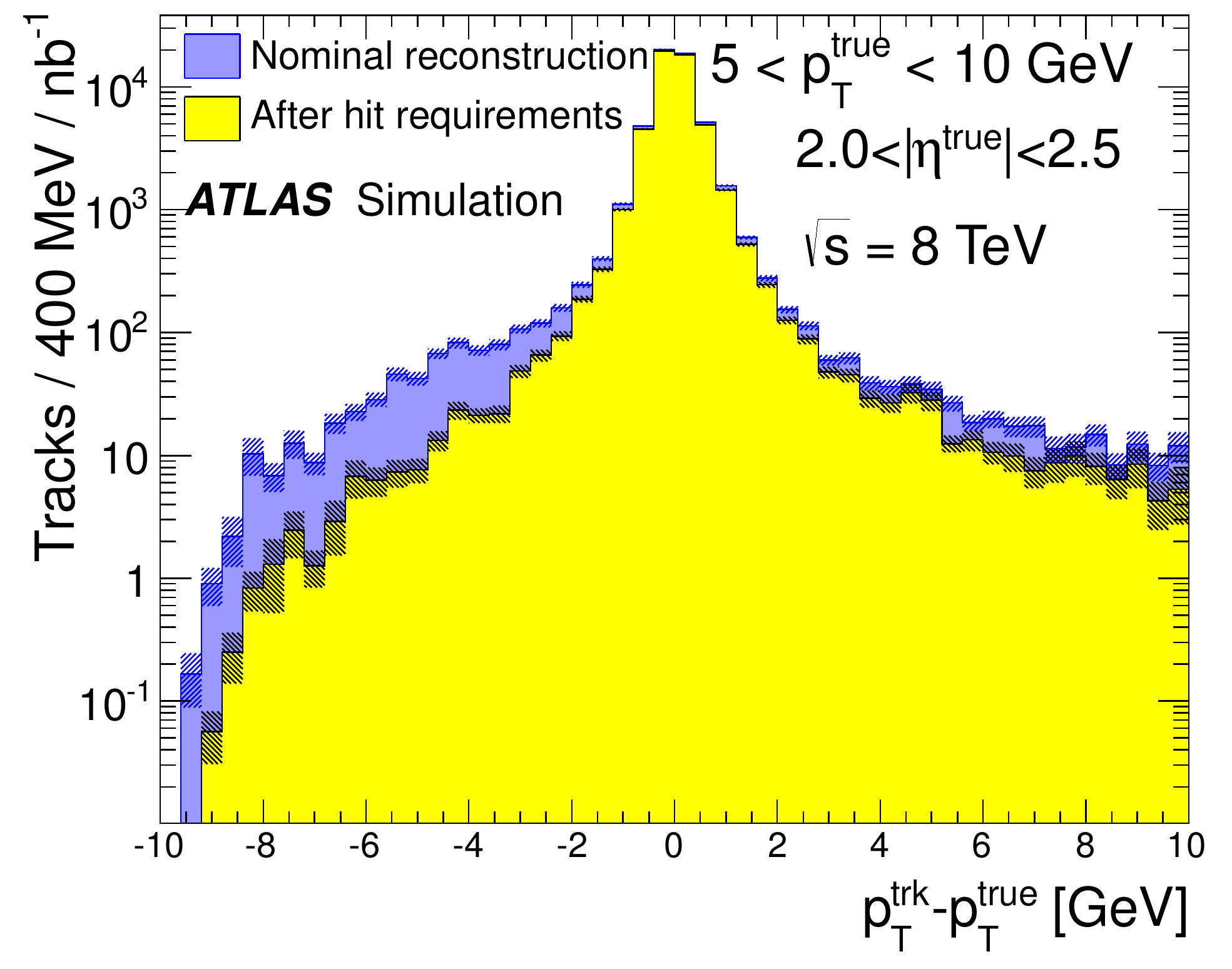}
  }
  \caption{The difference between the reconstructed \pT of the track from a charged pion and the particle's true \pT
    for two bins in truth particle \pT and $|\eta|$, 
    determined in dijet MC simulation with similar pile-up to that in the 2012 data.
    The shaded bands represent the statistical uncertainty.
    The tails in the residuals are substantially diminished upon the application of the more stringent 
    silicon detector hit requirements.
    A simulated dijet sample with $20 < \pTlead < \SI{1000}{\GeV}$ is used,
    and the statistical uncertainties in the number of MC simulated events are shown
    as a hatched band.}
\label{fig:eflowRec:trkQual}
\end{figure}

Figure~\ref{fig:eflowRec:trkEff} shows the charged-pion track reconstruction efficiency,
for the tracks selected with the criteria described above, as a function of $\etatrue$ and $\pTtrue$ in the dijet MC sample,
with leading jets in the range $20<\pTlead <\SI{1000}{\GeV}$ and with similar pile-up to that in the 2012 data.
The Monte Carlo generator information is used to match the reconstructed tracks to the
generated particles~\cite{STDM-2015-02}.
The application of the tight quality criteria substantially reduces the rate of poorly measured tracks, as shown in \Fig{\ref{fig:eflowRec:trkQual}}.
Additionally, using the above selection, the fraction of combinatorial fake tracks arising from combining ID hits from different particles is negligible~\cite{STDM-2015-02}.

%-------------------------------------------------------------------------------
\subsection{Matching tracks to \topoclusters}
\label{sec:trkclus}

To remove the calorimeter energy where a particle has formed a single \topocluster, the algorithm first attempts to match each selected track to one \topocluster.
The distances $\Delta\phi$ and $\Delta\eta$ 
between the barycentre of the \topocluster and the track, extrapolated to the second layer of the EM calorimeter,
are computed for each \topocluster.
The \topoclusters are ranked based on the distance metric 
\begin{equation}
\label{eq:dRprime}
  \dRprime = \sqrt{ \left( \frac{\Delta\phi}{\sigma_\phi} \right)^ 2+ \left( \frac{\Delta\eta}{\sigma_\eta} \right)^2 },
\end{equation}
where $\sigma_\eta$ and $\sigma_\phi$ represent the angular \topocluster widths, computed as the standard deviation of the displacements of the \topocluster's constituent cells in $\eta$ and $\phi$ with respect to the \topocluster barycentre.
This accounts for the spatial extent of the \topoclusters, which may contain energy deposits from multiple particles.

\begin{figure}[htbp]
\centering
\subfloat[$\sigma_\eta, |\etatrue|<1.0$]{\includegraphics[width=0.31\textwidth]{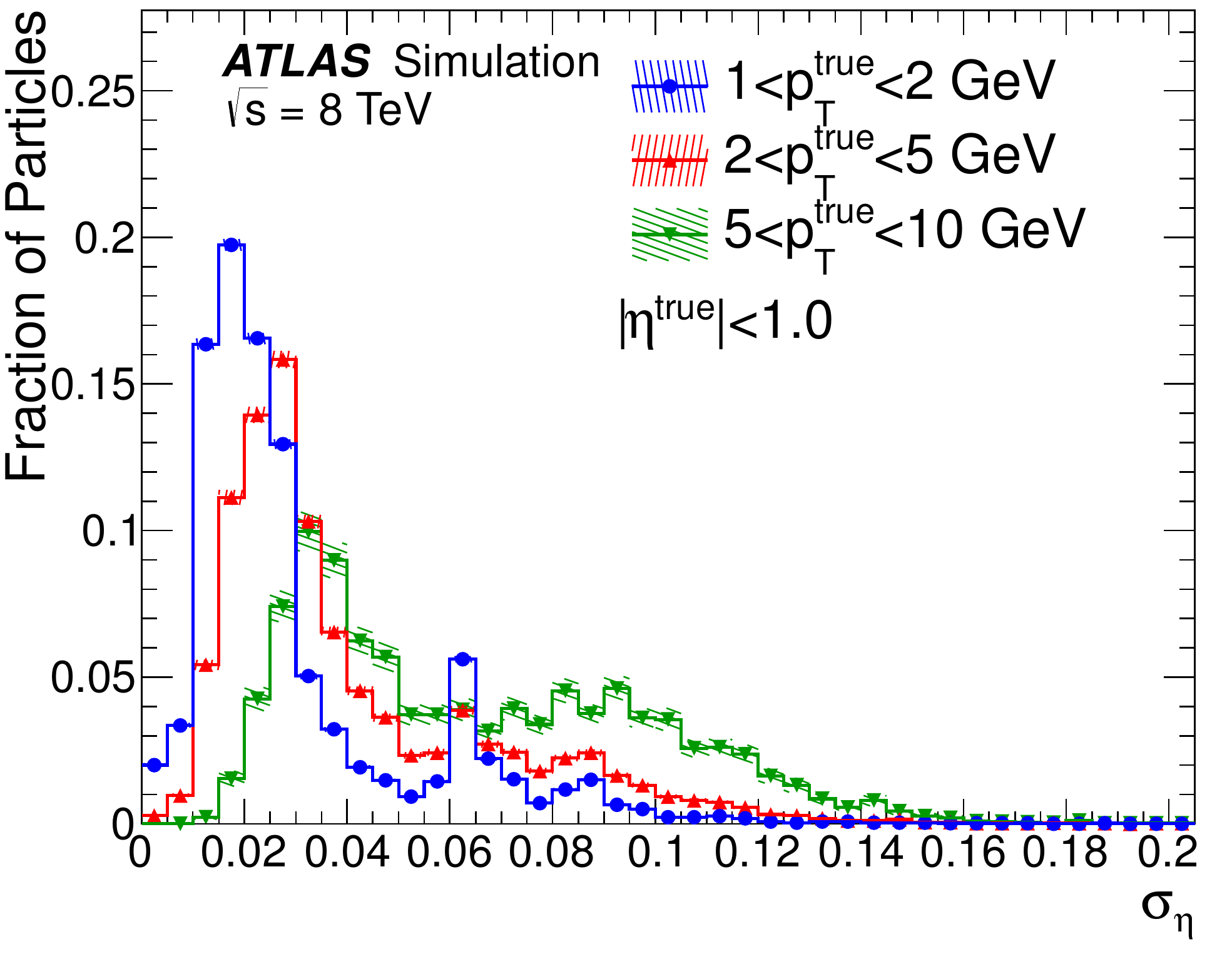}}\quad
\subfloat[$\sigma_\eta, 1.0<|\etatrue|<2.0$]{\includegraphics[width=0.31\textwidth]{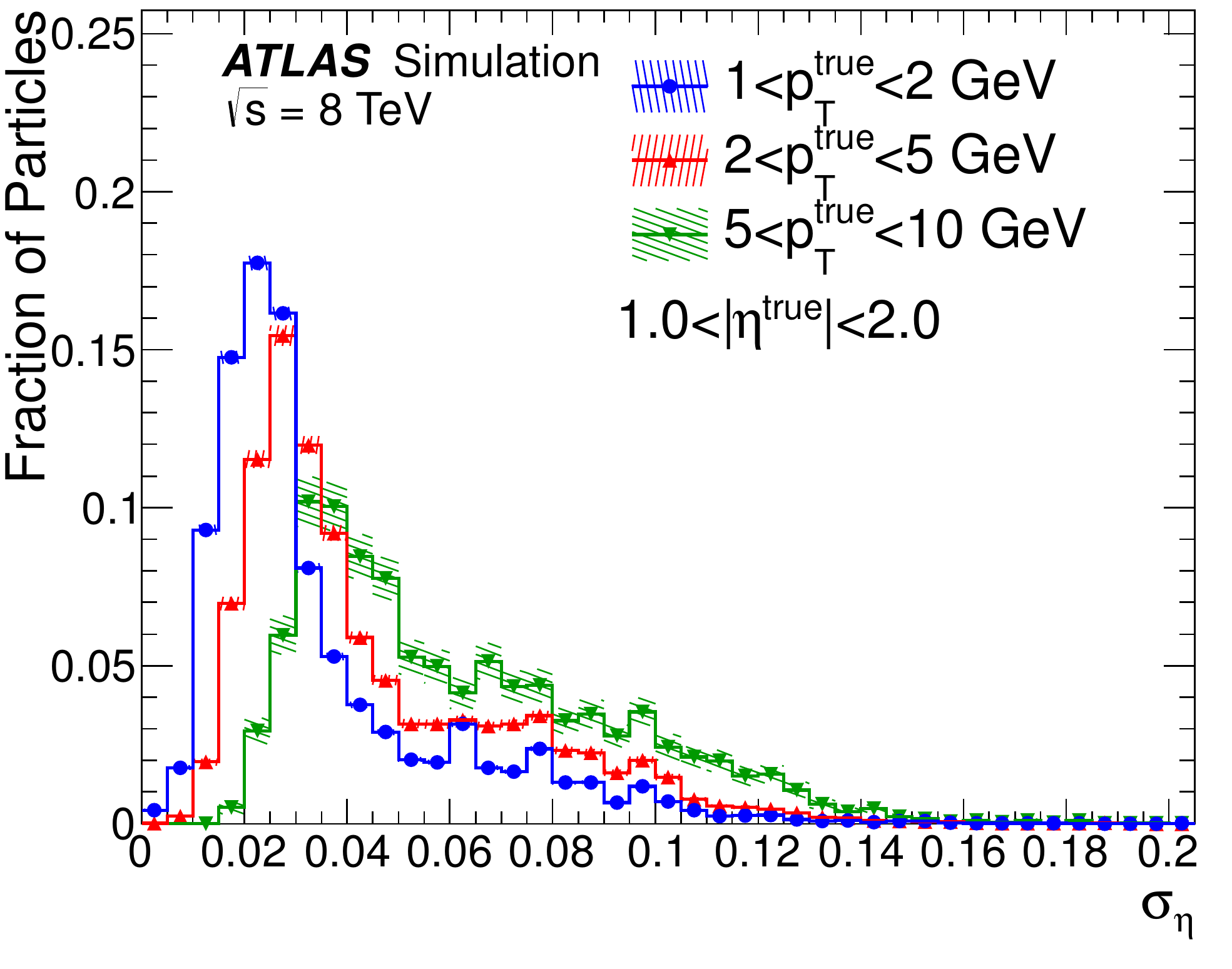}}\quad
\subfloat[$\sigma_\eta, 2.0<|\etatrue|<2.5$]{\includegraphics[width=0.31\textwidth]{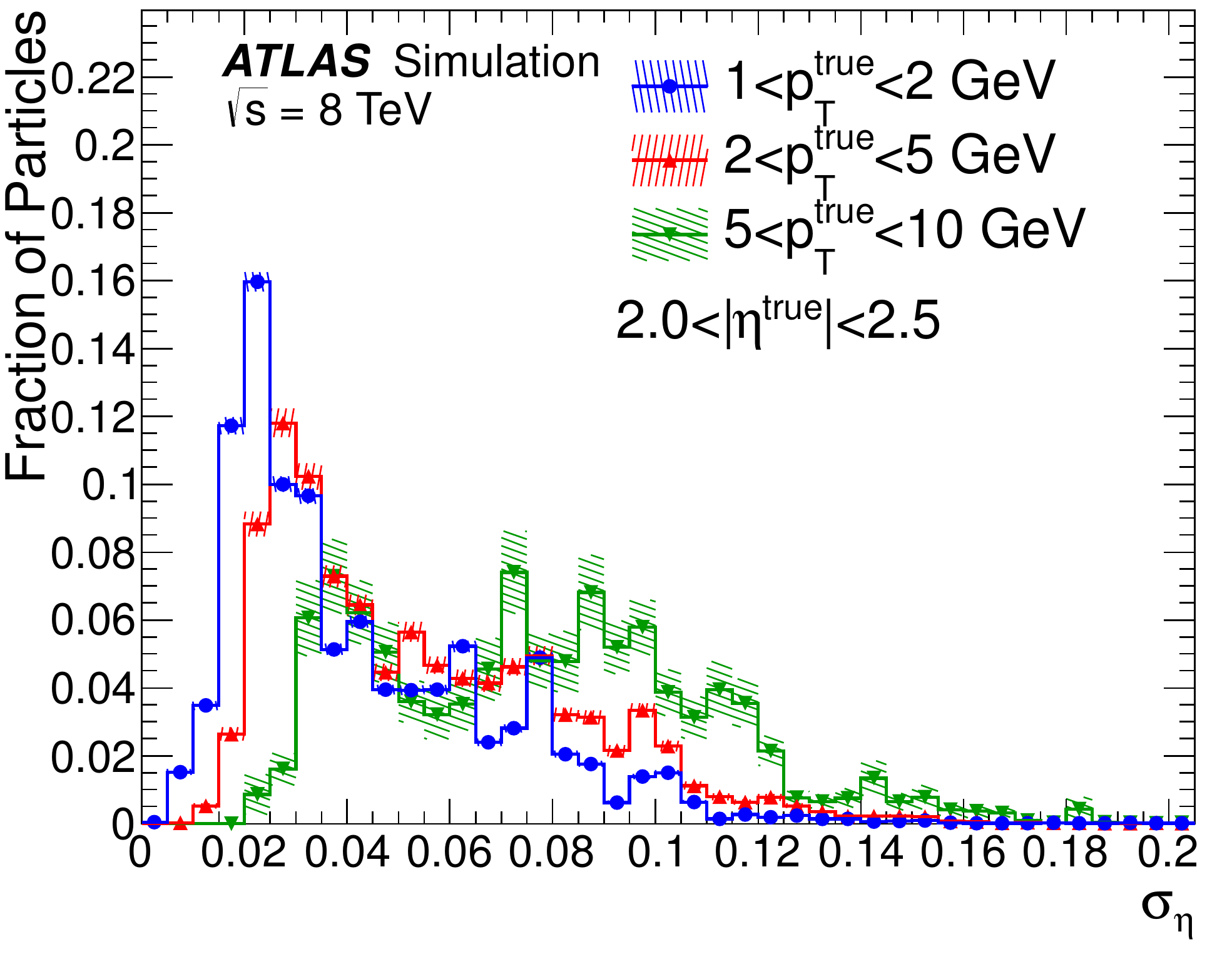}}\\
\subfloat[$\sigma_\phi, |\etatrue|<1.0$]{\includegraphics[width=0.31\textwidth]{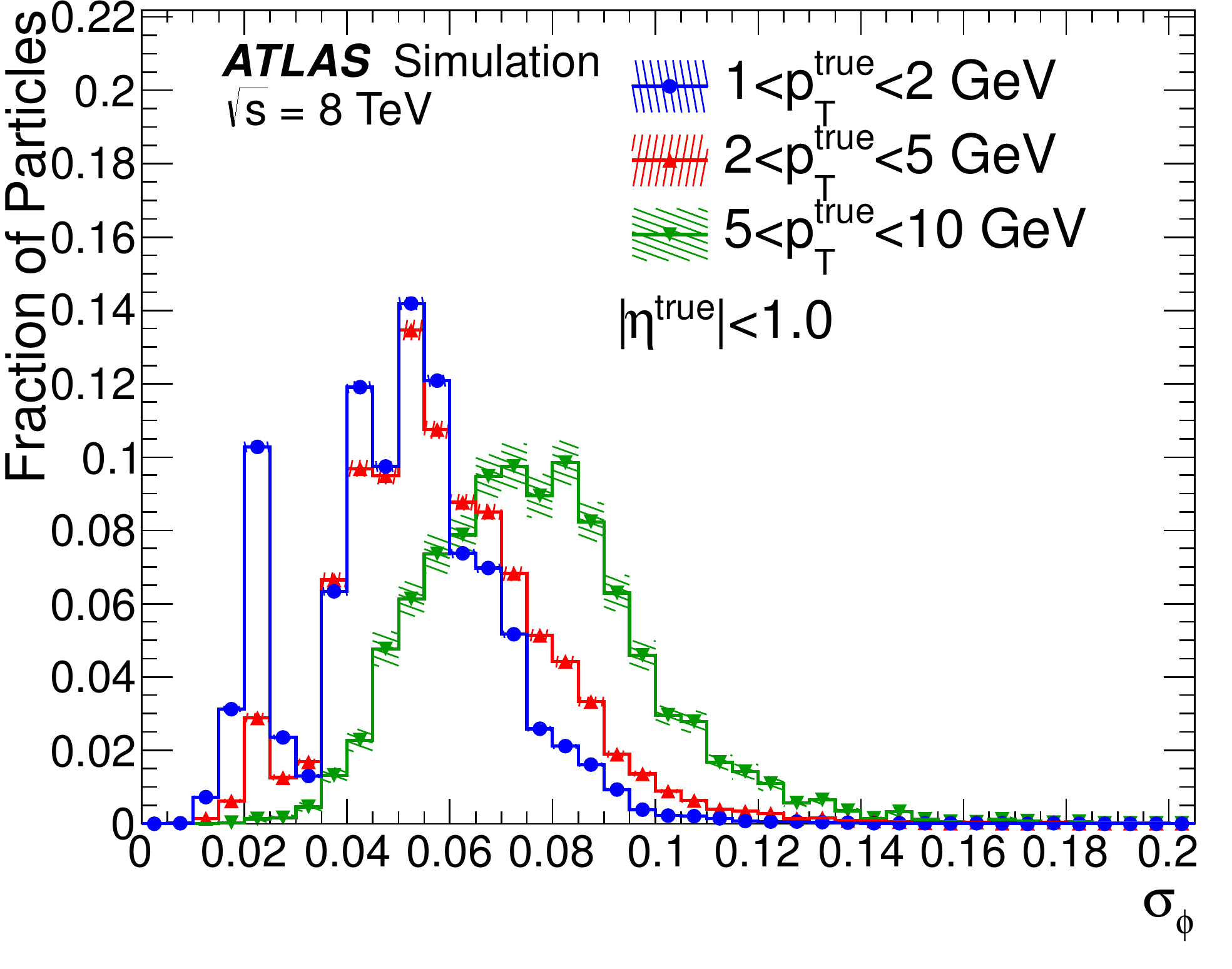}}\quad
\subfloat[$\sigma_\phi, 1.0<|\etatrue|<2.0$]{\includegraphics[width=0.31\textwidth]{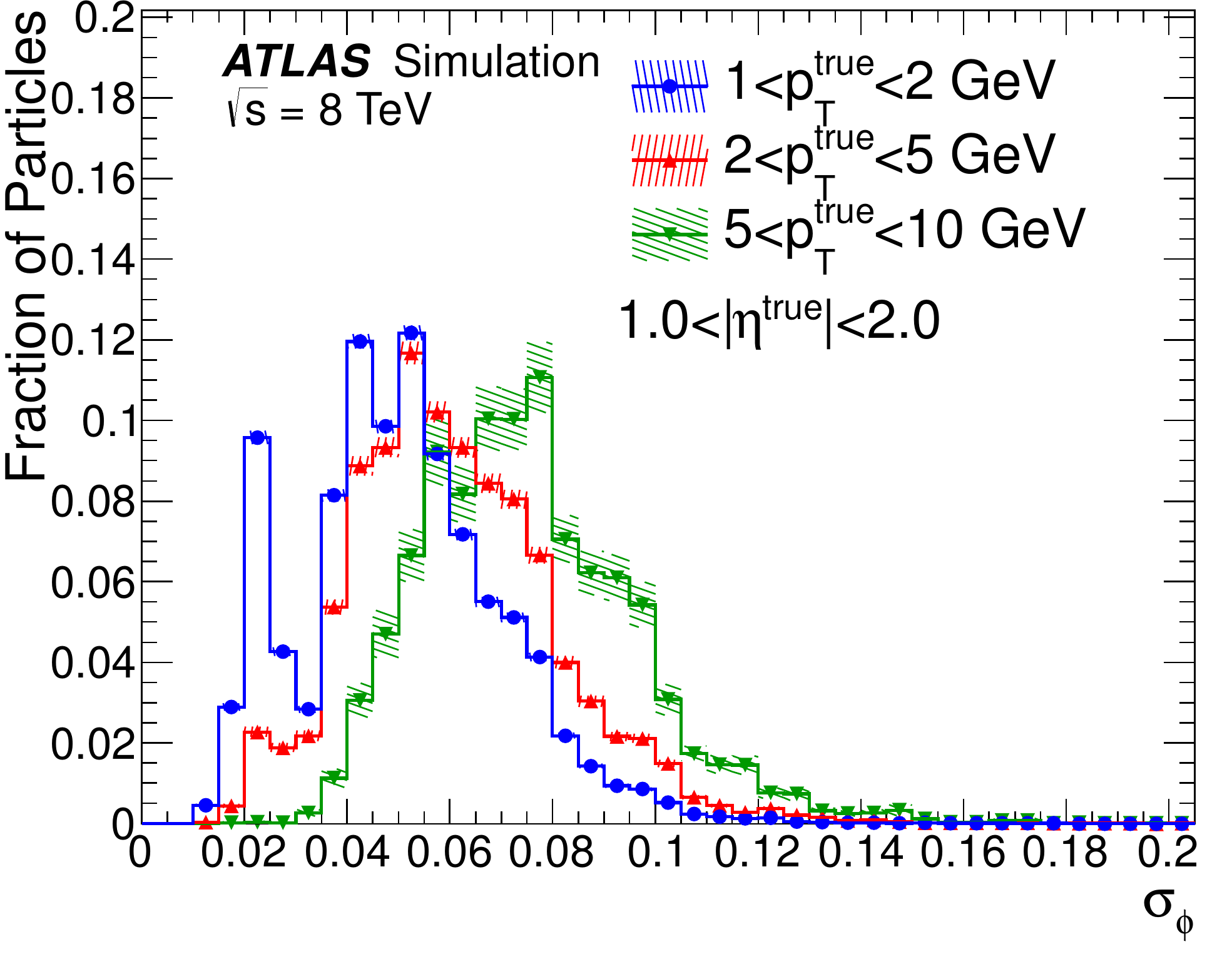}}\quad
\subfloat[$\sigma_\phi, 2.0<|\etatrue|<2.5$]{\includegraphics[width=0.31\textwidth]{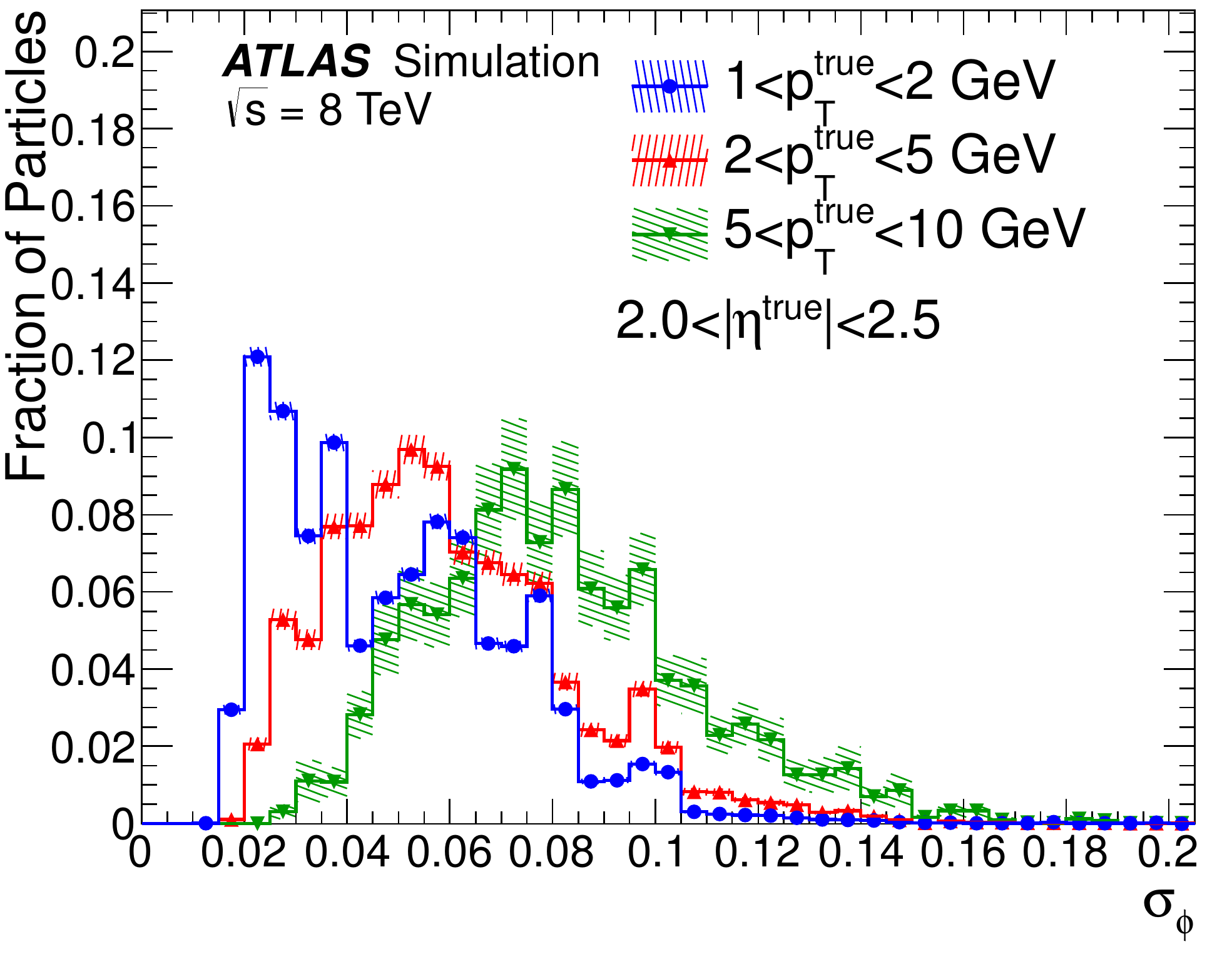}}
\caption{The distribution of $\sigma_\eta$ and $\sigma_\phi$, for charged pions, in three different regions of the detector for three particle \pT ranges.
  \DijetSample
}
\label{fig:eflowRec:sigmaEtaPhi}
\end{figure}

The distributions of $\sigma_\eta$ and $\sigma_\phi$ for single-particle samples are shown in \Fig{\ref{fig:eflowRec:sigmaEtaPhi}}.
The structure seen in these distributions is related to the calorimeter geometry.
Each calorimeter layer has a different cell granularity in both dimensions, and this sets the minimum \topocluster size.
In particular, the granularity is significantly finer in the electromagnetic calorimeter, thus particles that primarily deposit their energy in either the electromagnetic and hadronic calorimeters form distinct populations.
High-energy showers typically spread over more cells, broadening the corresponding \topoclusters.
If the computed value of $\sigma_{\eta}$ or $\sigma_{\phi}$ is smaller than 0.05,
it is set to 0.05.

\begin{figure}[htbp]
\centering
\captionsetup[subfigure]{justification=centering}
\subfloat[$1<\pTtrue< \SI{2}{\GeV}$,\protect\\ $|\eta|<1.0$]{\includegraphics[width=0.31\textwidth]{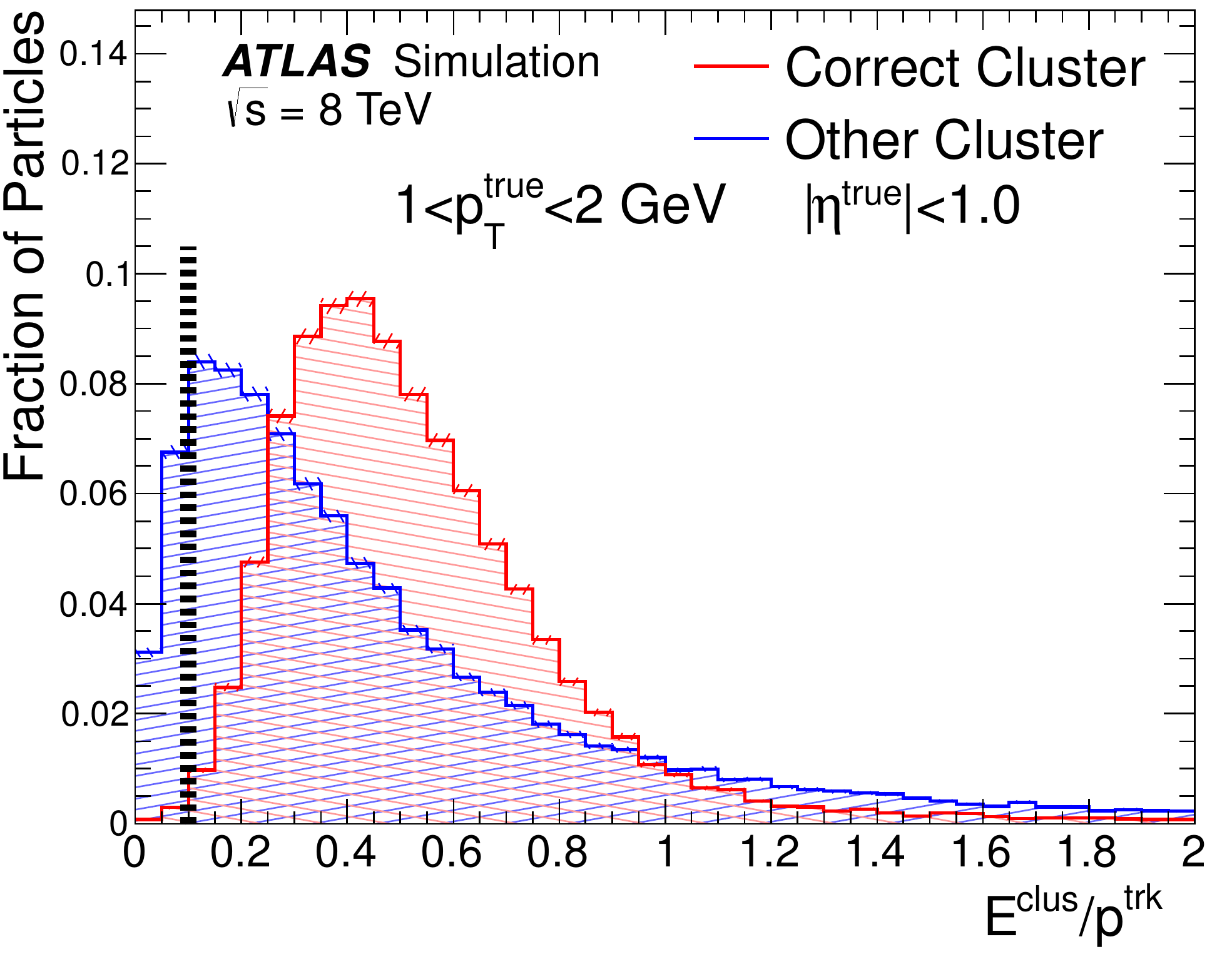}}\quad
\subfloat[$1<\pTtrue< \SI{2}{\GeV}$,\protect\\ $1.0<|\eta|< 2.0$]{\includegraphics[width=0.31\textwidth]{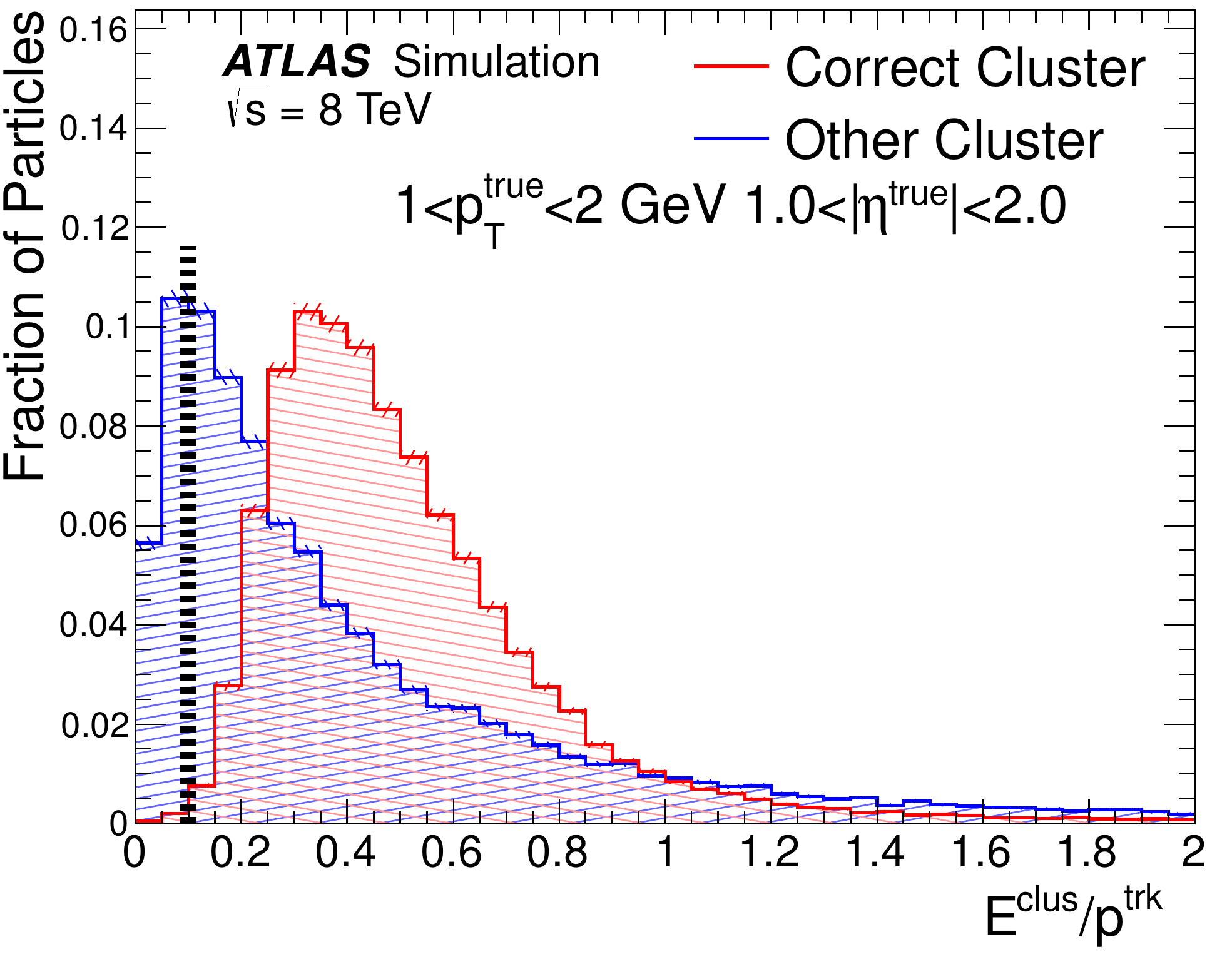}}\quad
\subfloat[$1<\pTtrue< \SI{2}{\GeV}$,\protect\\ $2.0<|\eta|<2.5$]{\includegraphics[width=0.31\textwidth]{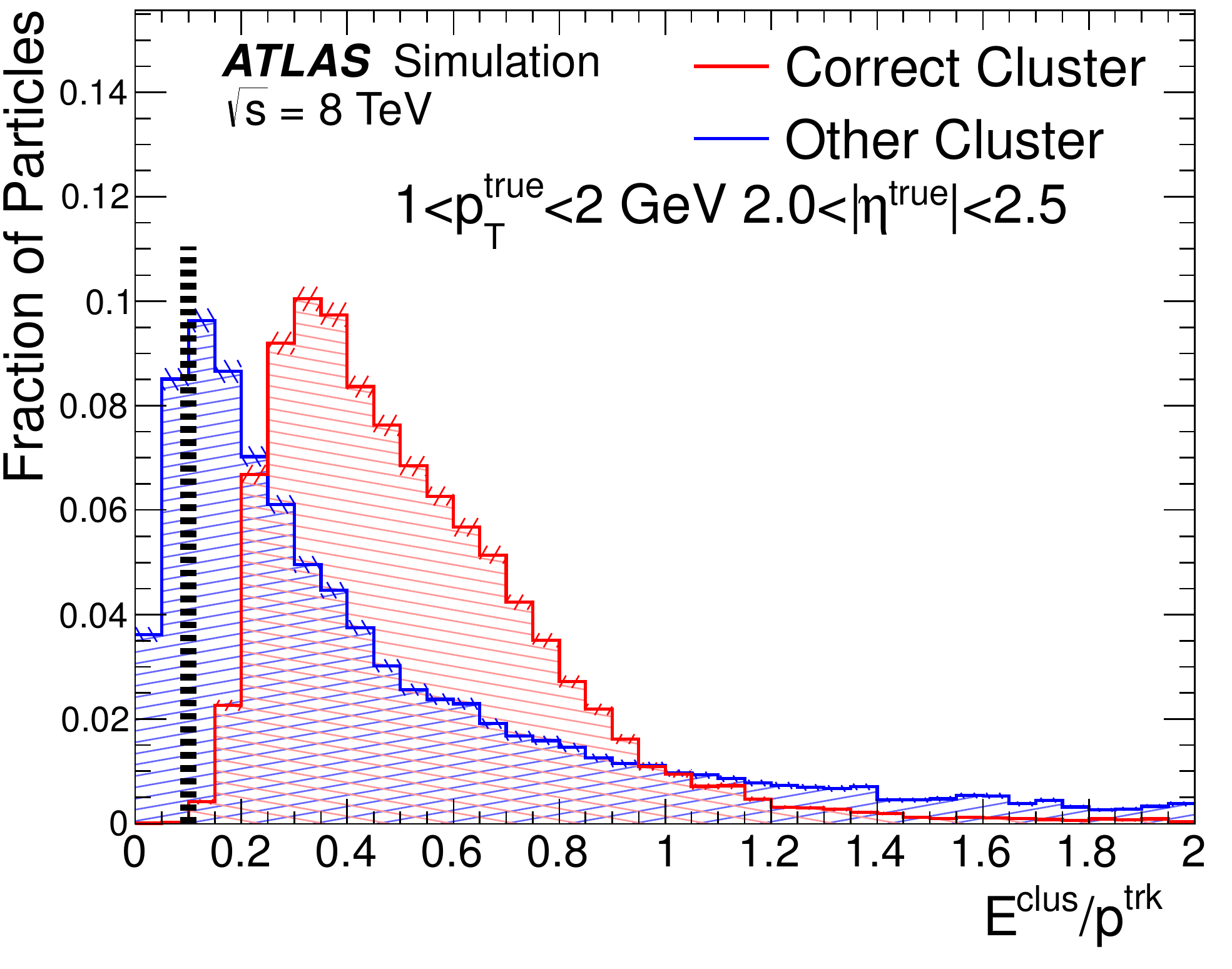}}\\
\subfloat[$5<\pTtrue< \SI{10}{\GeV}$,\protect\\ $|\eta|<1.0$]{\includegraphics[width=0.31\textwidth]{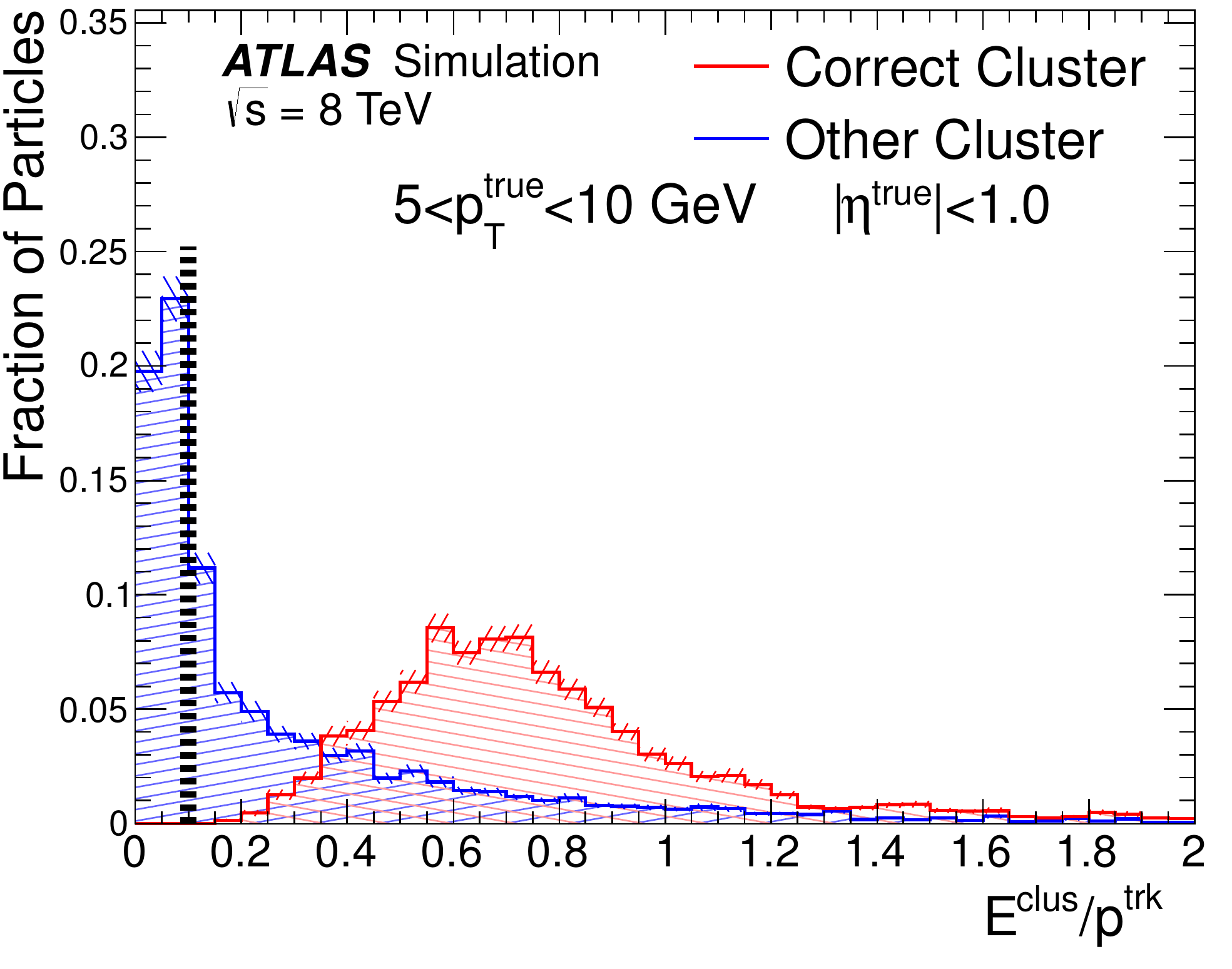}}\quad
\subfloat[$5<\pTtrue< \SI{10}{\GeV}$,\protect\\ $1.0<|\eta|<2.0$]{\includegraphics[width=0.31\textwidth]{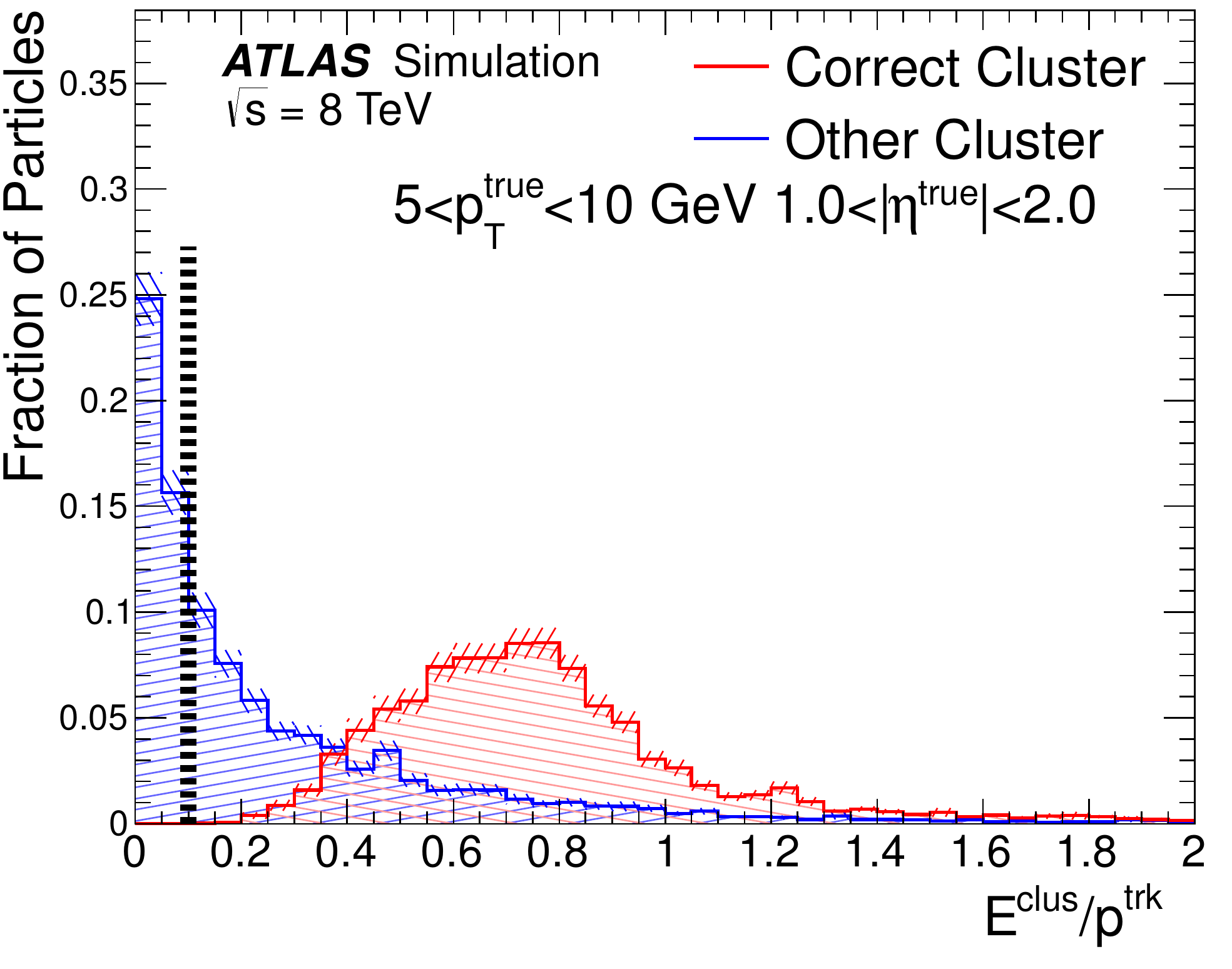}}\quad
\subfloat[$5<\pTtrue< \SI{10}{\GeV}$,\protect\\ $2.0<|\eta|<2.5$]{\includegraphics[width=0.31\textwidth]{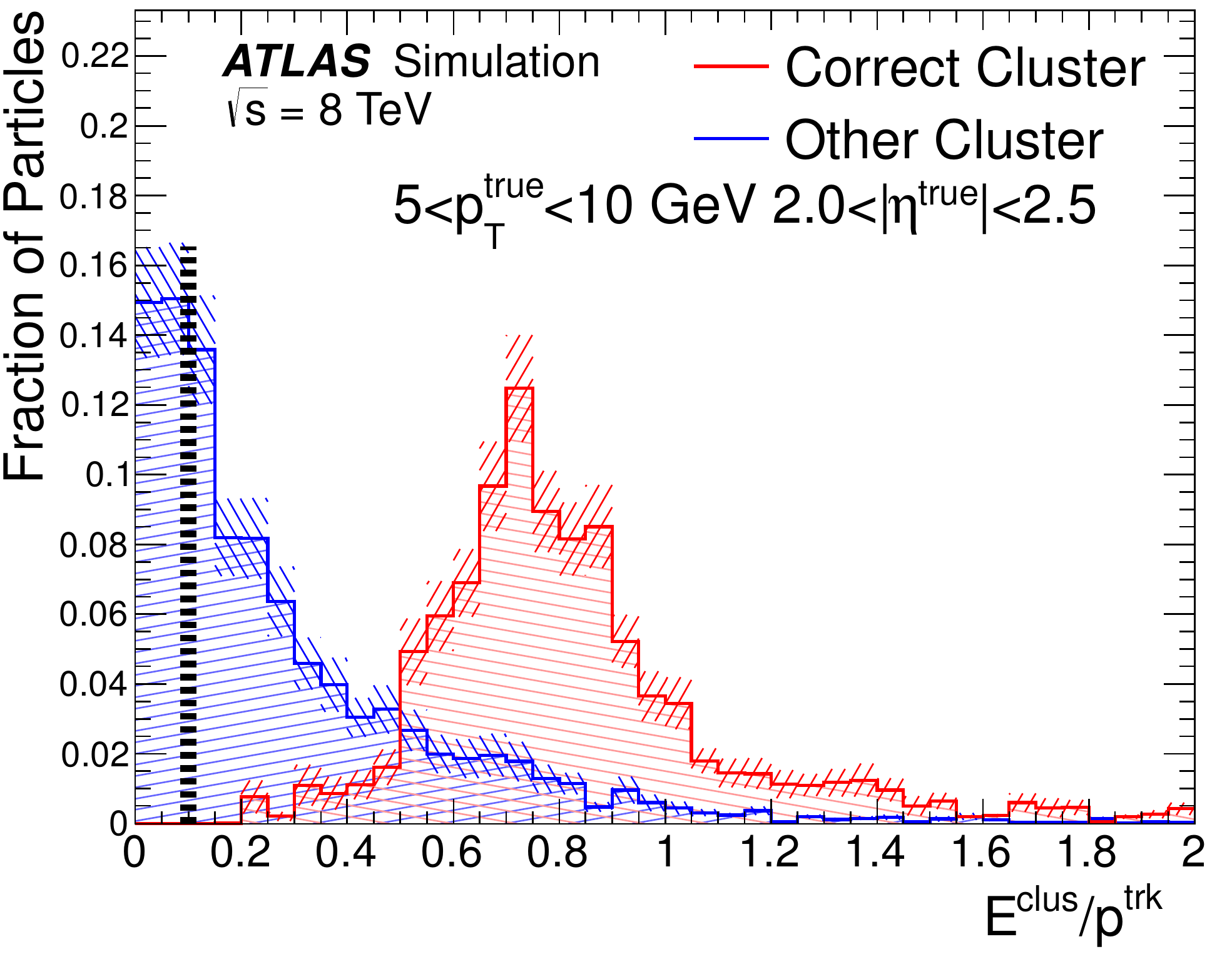}}
\caption{The distributions of $\EoPtrk$ for the \topocluster with $> \SI{90}{\%}$ of the true energy of the particle and the closest other \topocluster in \dRprime.
  \DijetSample
    A track is only used for energy subtraction if a \topocluster is found inside a cone of $\dRprime = 1.64$ for which $\EoPtrk > 0.1$, as indicated by the vertical dashed line.
}
\label{fig:eflowRec:matchEoP}
\end{figure}

A preliminary selection of \topoclusters to be matched to the tracks is performed by requiring that $\EoPtrk>0.1$,
where $\Eclus$ is the energy of the \topocluster and \ptrk is the track momentum.
The distribution of $\EoPtrk$ for the \topocluster with at least \SI{90}{\%} of the true energy from the particle matched to the track -- the ``correct" one to match to -- and for the closest other \topocluster in \dRprime is shown in \Fig{\ref{fig:eflowRec:matchEoP}}.
For very soft particles, it is common that the closest other \topocluster carries \EoPtrk comparable to (although smaller than) the correct \topocluster.
About \SI{10}{\%} of incorrect \topoclusters are rejected by the $\EoPtrk$ cut for particles with $1<\pT<\SI{2}{\GeV}$.
The difference in $\EoPtrk$ becomes much more pronounced for particles with $\pT>\SI{5}{\GeV}$,
for which there is a very clear separation between the correct and incorrect \topocluster matches, resulting in a 30--\SI{40}{\%} rejection rate for the incorrect \topoclusters. 
This is because at lower $\pT$ clusters
come from both signal and electronic or pile-up noise. Furthermore, the particle $\pT$ spectrum is peaked towards lower values, and thus higher-$\pT$ \topoclusters are rarer.
The $\EoPtrk>0.1$ requirement rejects the correct cluster for far less than \SI{1}{\%} of particles.

\begin{figure}[htbp]
\centering
\captionsetup[subfigure]{justification=centering}
\subfloat[$1<\pTtrue< \SI{2}{\GeV}$,\protect\\ $|\eta|<1.0$]{\includegraphics[width=0.31\textwidth]{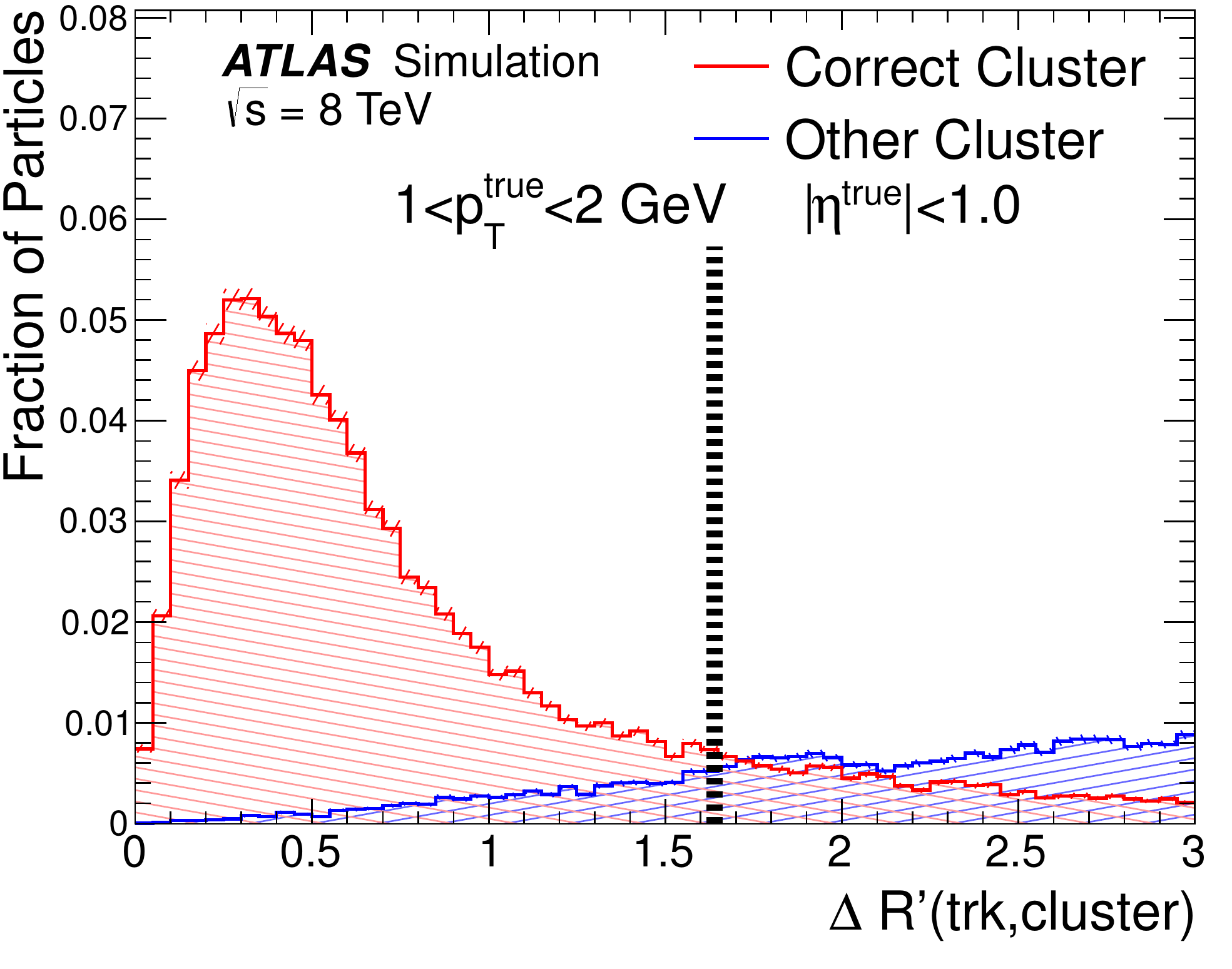}}\quad
\subfloat[$1<\pTtrue< \SI{2}{\GeV}$,\protect\\ $1.0<|\eta|<2.0$]{\includegraphics[width=0.31\textwidth]{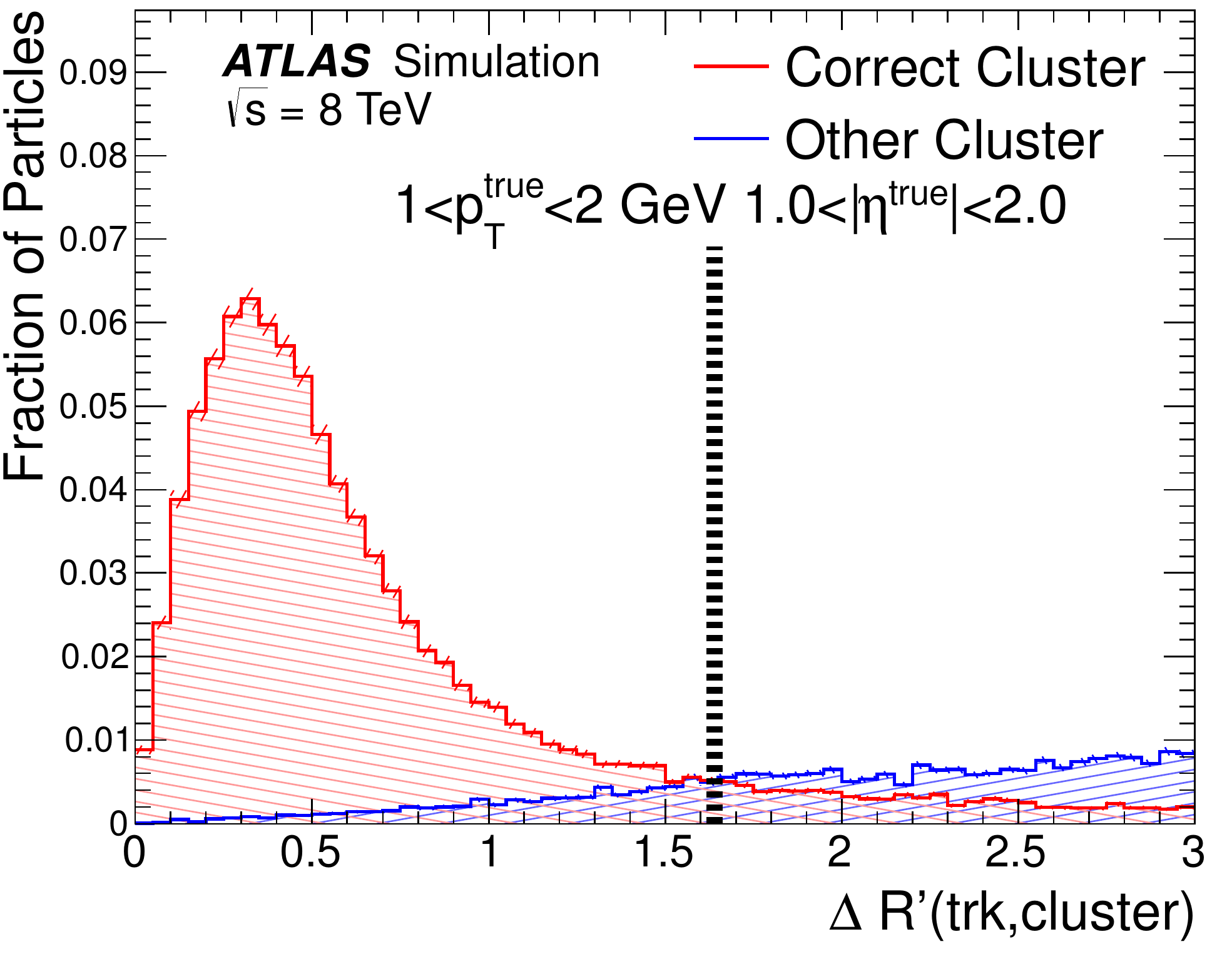}}\quad
\subfloat[$1<\pTtrue< \SI{2}{\GeV}$,\protect\\ $2.0<|\eta|<2.5$]{\includegraphics[width=0.31\textwidth]{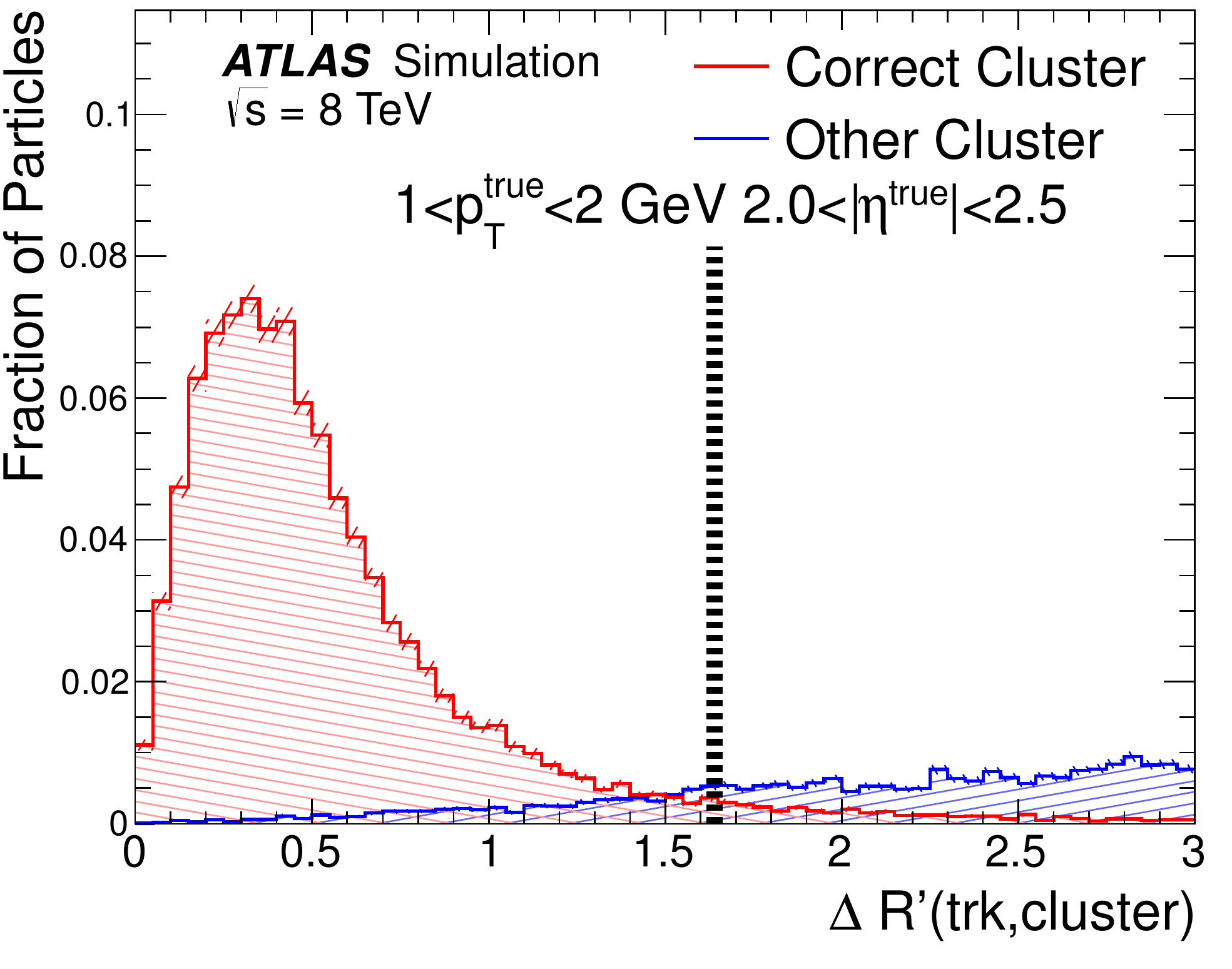}}\\
\subfloat[$5<\pTtrue< \SI{10}{\GeV}$,\protect\\ $|\eta|<1.0$]{\includegraphics[width=0.31\textwidth]{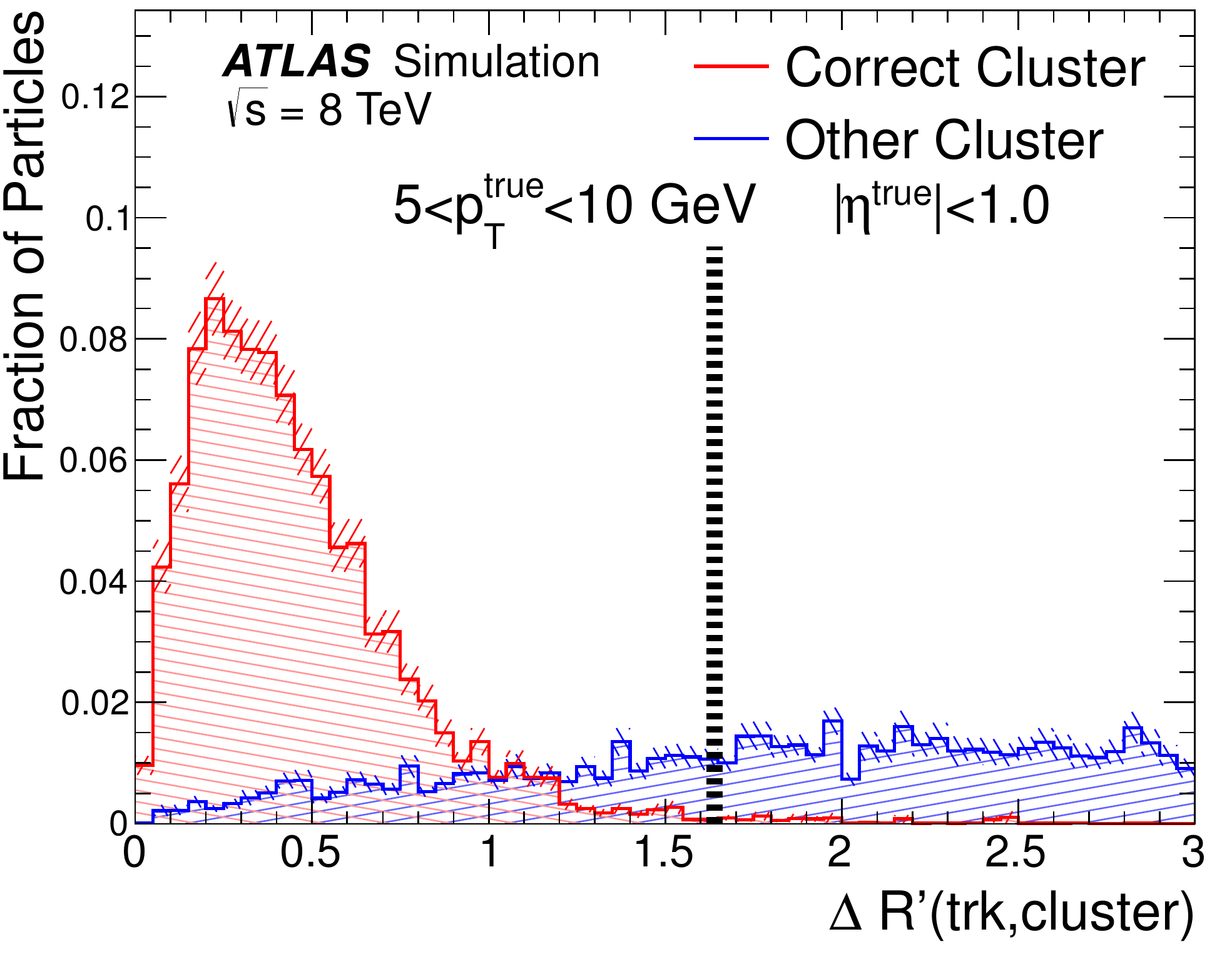}}\quad
\subfloat[$5<\pTtrue< \SI{10}{\GeV}$,\protect\\ $1.0<|\eta|<2.0$]{\includegraphics[width=0.31\textwidth]{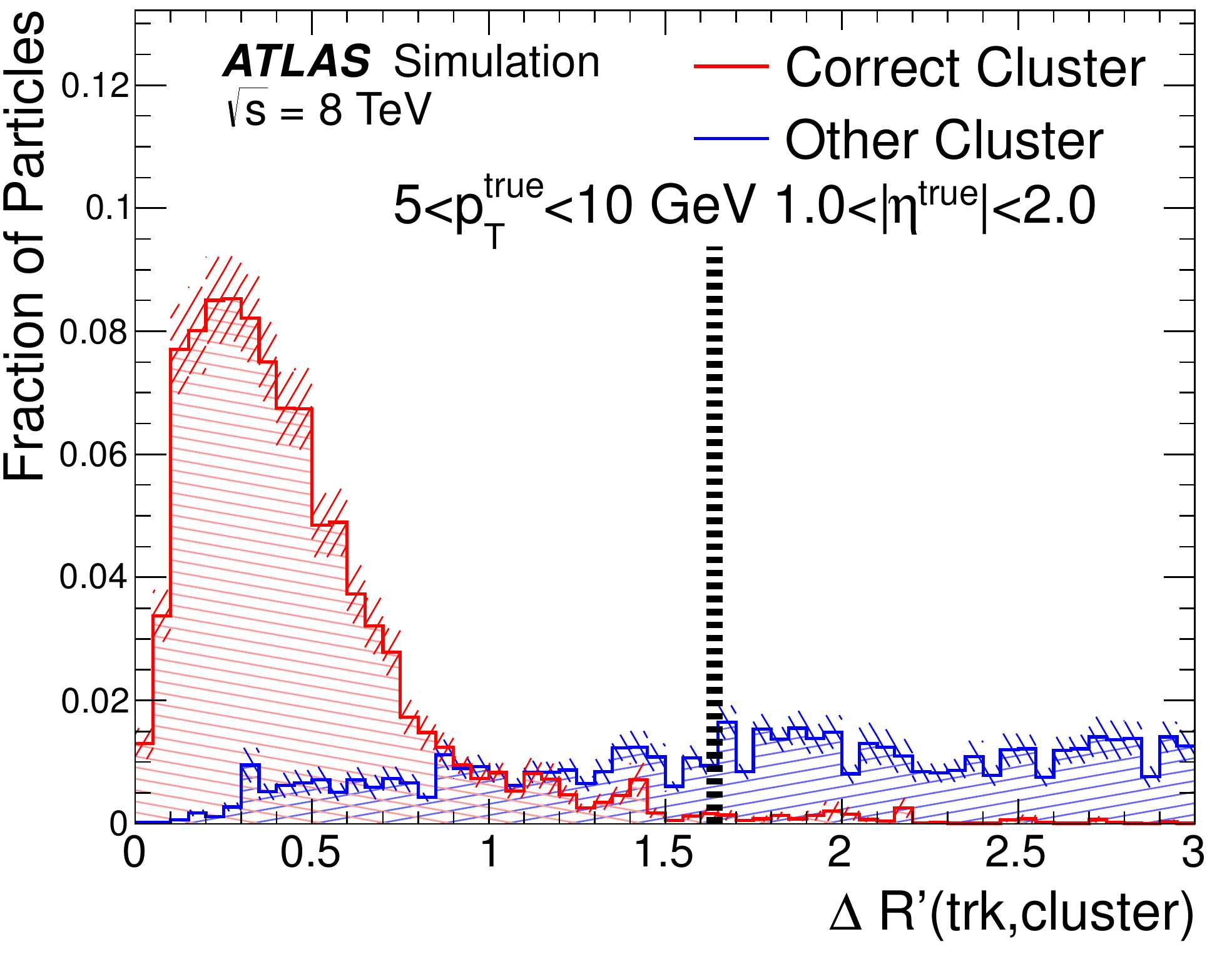}}\quad
\subfloat[$5<\pTtrue< \SI{10}{\GeV}$,\protect\\ $2.0<|\eta|<2.5$]{\includegraphics[width=0.31\textwidth]{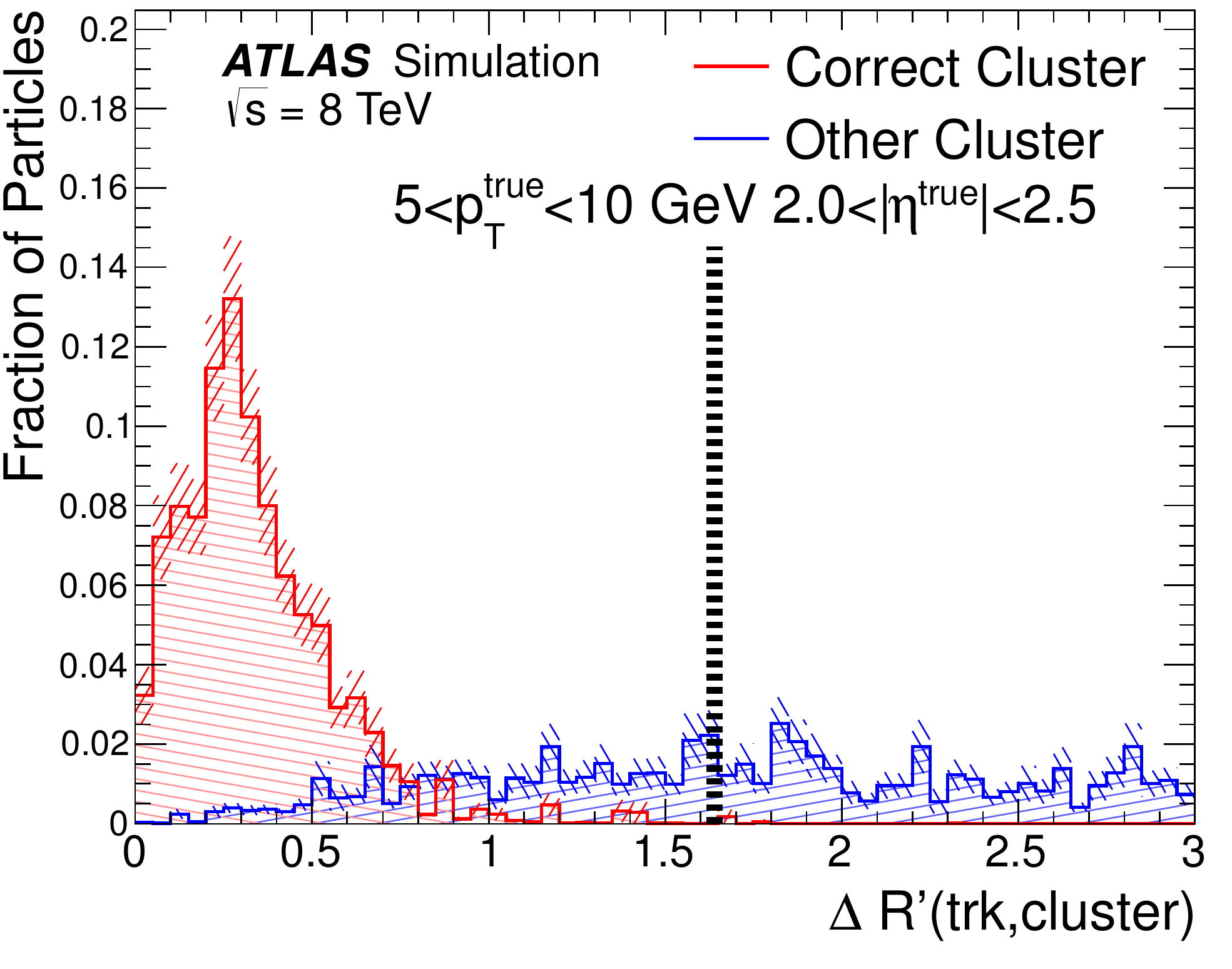}}
\caption{The distributions of $\dRprime$ for the \topocluster with $>\SI{90}{\%}$ of the true energy of the particle and the closest other \topocluster, both satisfying $\EoPtrk > 0.1$.
    \DijetSample
    A track is only used for energy subtraction if a \topocluster is found with $\EoPtrk > 0.1$ inside a cone of $\dRprime < 1.64$, as indicated by the vertical dashed line.
}
\label{fig:eflowRec:matchDR}
\end{figure}

Next, an attempt is made to match the track to one of the preselected \topoclusters using the distance metric \dRprime defined in \Eqn{\ref{eq:dRprime}}.
The distribution of \dRprime between the track and the \topocluster with $> \SI{90}{\%}$ of the truth particle energy and to the closest other preselected \topocluster
is shown in \Fig{\ref{fig:eflowRec:matchDR}} for the dijet MC sample.
From this figure, it is seen that the correct \topocluster almost always lies at a small \dRprime relative to other clusters.
Hence, the closest preselected \topocluster in $\dRprime$ is taken to be the matched \topocluster.
This criterion selects the correct \topocluster with a high probability, succeeding for virtually all particles with $\pT>\SI{5}{\GeV}$.
If no preselected \topocluster is found in a cone of size $\dRprime=1.64$, it is assumed that this particle did not form a \topocluster in the calorimeter.
In such cases the track is retained in the list of tracks and no subtraction is performed.
The numerical value corresponds to a one-sided Gaussian confidence interval of \SI{95}{\%}, and has not been optimised.
However, as seen in \Fig{\ref{fig:eflowRec:matchDR}}, this cone size almost always includes the correct \topocluster, while rejecting the bulk of incorrect clusters.

%-------------------------------------------------------------------------------
\subsection{Evaluation of the expected deposited particle energy through $\MeanEoP$ determination}
\label{sec:eflowRec:EoverP}

\begin{figure}[htbp]
\centering
\subfloat[$|\eta|<1.0$]{\includegraphics[width=0.31\textwidth]{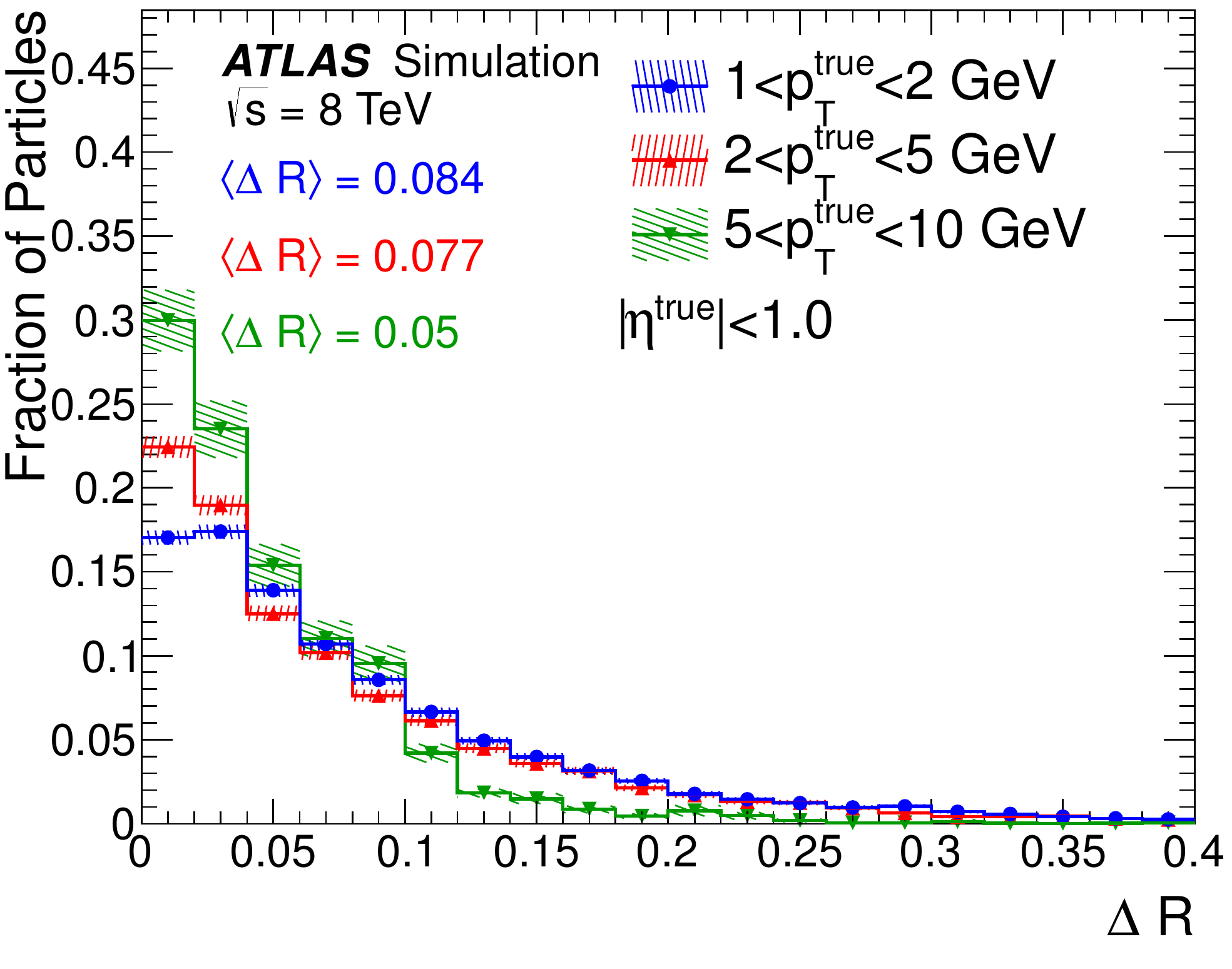}}\quad
\subfloat[$1.0<|\eta|<2.0$]{\includegraphics[width=0.31\textwidth]{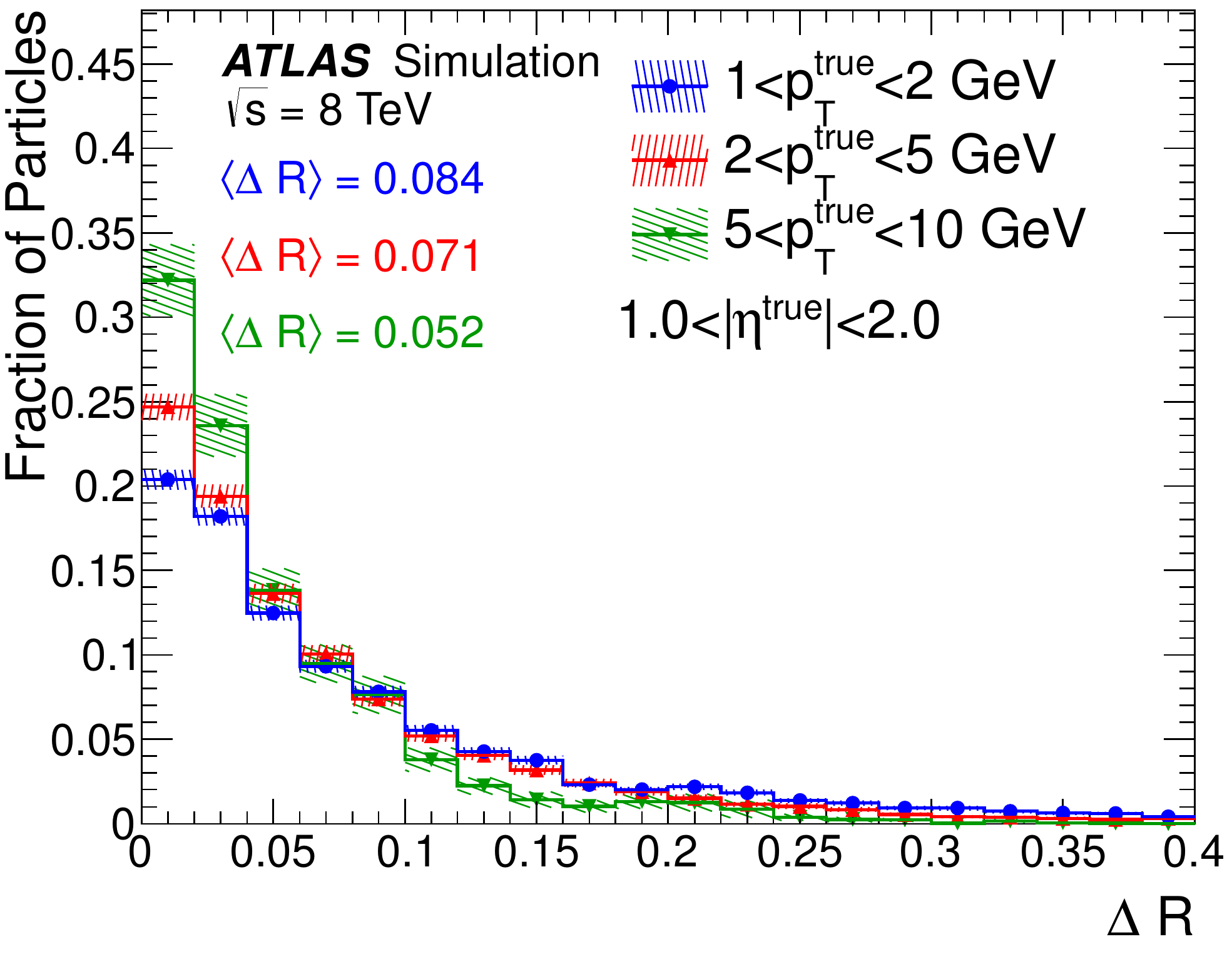}}\quad
\subfloat[$2.0<|\eta|<2.5$]{\includegraphics[width=0.31\textwidth]{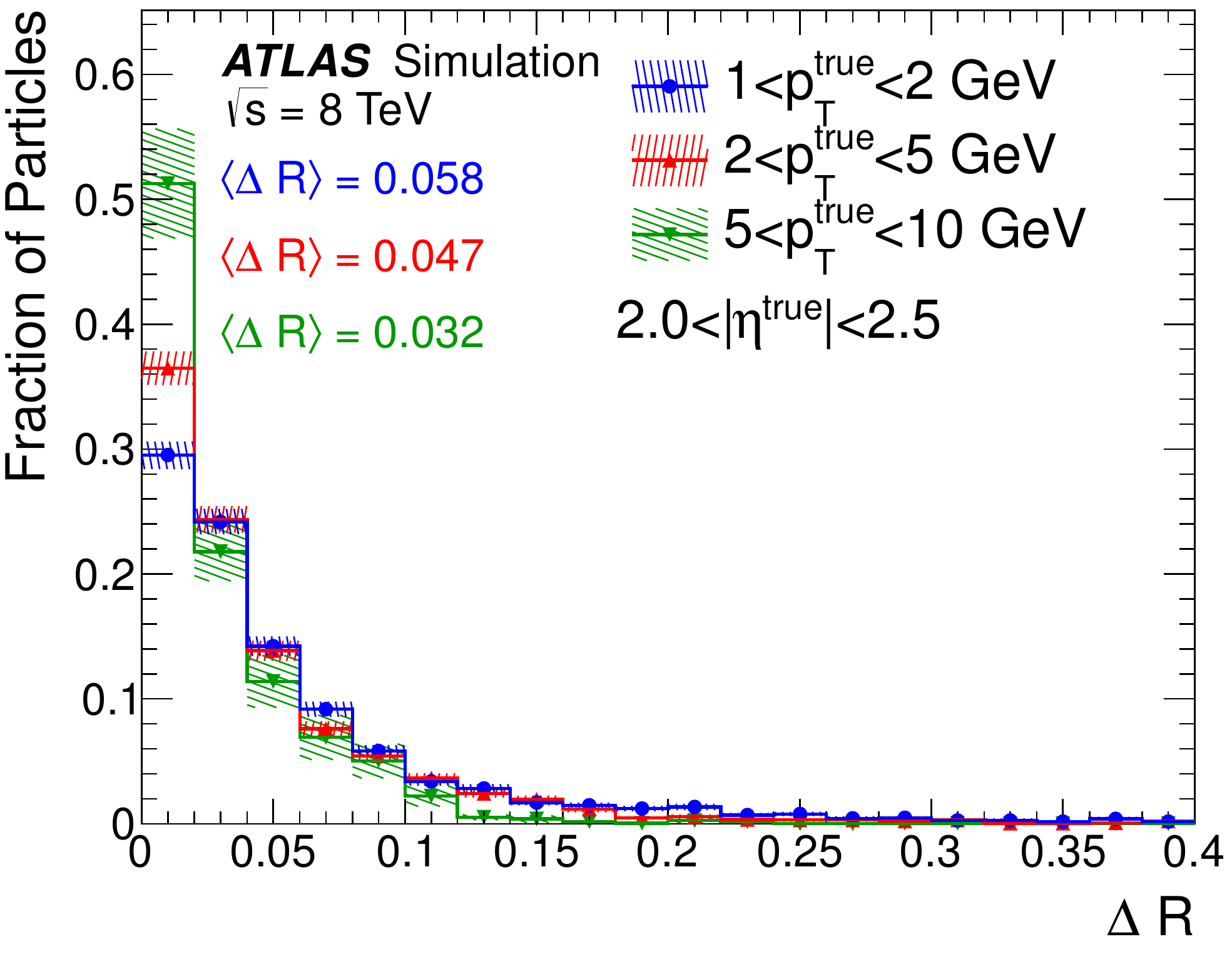}}\qquad
\caption{The cone size $\Delta R$ around the extrapolated track required to encompass both the leading and sub-leading \topoclusters,
  for $\pi^\pm$ when $< \SI{70}{\%}$ of their true deposited energy in \topoclusters is contained in the leading \topocluster,
  but $> \SI{90}{\%}$ of the energy is contained in the two leading \topoclusters.
  \DijetSample
}
\label{fig:eflowRec:RSSdR}
\end{figure}

It is necessary to know how much energy a particle with measured momentum \ptrk deposits on average, given by
$\MeanEdep = \ptrk \,\MeanEoP$,
in order to correctly subtract the energy from the calorimeter for a particle whose track has been reconstructed.
The expectation value $\MeanEoP$ (which is also a measure of the mean response) is determined using single-particle samples without pile-up
by summing the energies of \topoclusters in a \dR cone of size 0.4 around the track position,
extrapolated to the second layer of the EM calorimeter.
This cone size is large enough to entirely capture the energy of the majority of particle showers.
This is also sufficient in dijet events, as demonstrated in \Fig{\ref{fig:eflowRec:RSSdR}},
where one might expect the clusters to be broader due to the presence of other particles.
The subscript \enquote{ref} is used here and in the following to indicate \EoPtrk values determined from single-pion samples.

Variations in $\MeanEoP$ due to detector geometry and shower development are captured by binning the measurement in
the $\pt$ and $\eta$ of the track as well as the layer of highest energy density (LHED), defined in the next section.
The LHED is also used to determine the order in which cells are subtracted in subsequent stages of the algorithm.

The spread of the expected energy deposition, denoted by $\sigmaEdep$, is determined from the standard deviation
of the \EoPtrkref distribution in single-pion samples.
It is used in order to quantify the consistency of the measured $\EoPtrk$ with the expectation from $\MeanEoP$ in both
the split-shower recovery (\Sect{\ref{sec:eflowRec_ssr}}) and remnant removal (\Sect{\ref{sec:eflowRec_remnant}}).

%-------------------------------------------------------------------------------
\subsubsection{Layer of highest energy density}
\label{sec:eflowRec:LHED}

The dense electromagnetic shower core has a well-defined ellipsoidal shape in $\eta$--$\phi$.
It is therefore desirable to locate this core,
such that the energy subtraction may be performed first in this region before progressing to the less regular shower periphery.
The LHED is taken to be the layer which shows the largest rate of increase in energy density, as a function of the number of interaction lengths from the front face of the calorimeter.
This is determined as follows:
\begin{itemize}
\item The energy density is calculated for the $j$th cell in the $i$th layer of the calorimeter as
\begin{equation}
  \rho_{ij} = \frac{E_{ij}}{V_{ij}} \left(\si{\GeV}/X_{0}^{3}\right)\,,
\end{equation}
with $E_{ij}$ being the energy in and $V_{ij}$ the volume of the cell expressed in radiation lengths.
The energy measured in the Presampler is added to that of the first layer in the EM calorimeter.
In addition, the Tile and HEC calorimeters are treated as single layers.
Thus, the procedure takes into account four layers -- three in the EM calorimeter and one in the hadronic calorimeter.
Only cells in the \topoclusters matched to the track under consideration are used.

\item Cells are then weighted based on their proximity to the extrapolated track position in the layer, favouring cells that are closer to the track and hence more likely to contain energy from the selected particle.
The weight for each cell, $w_{ij}$, is computed from the integral over the cell area in $\eta$--$\phi$ of a Gaussian distribution centred on the extrapolated track position with a width in $\Delta R$ of 0.035, similar to the Molière radius of the LAr calorimeter.

\item A weighted average energy density for each layer is calculated as
\begin{equation}
  \langle \rho' \rangle_{i} = \sum_{j} w_{ij} \rho_{ij}\,.
\end{equation}

\item Finally, the rate of increase in $\langle\rho'\rangle_i$ in each layer is determined.
Taking $d_i$ to be the depth of layer $i$ in interaction lengths, the rate of increase is defined as
\begin{equation}
  \Delta\rho'_{i} = \frac{ \langle\rho'\rangle_i - \langle\rho'\rangle_{i-1} } {d_i - d_{i-1}}\,,
\end{equation}
where the values $\langle\rho'\rangle_0=0$ and $d_0=0$ are assigned, and the first calorimeter layer has the index $i=1$.
\end{itemize}
The layer for which $\Delta\rho'$ is maximal is identified as the LHED.

%-------------------------------------------------------------------------------
\subsection{Recovering split showers}
\label{sec:eflowRec_ssr}

Particles do not always deposit all their energy in a single \topocluster, as seen in \Fig{\ref{fig:eflowRec:nClus}}.
Clearly, handling the multiple \topocluster case is crucial,
particularly the two \topocluster case, which is very common.
The next stages of the algorithm are therefore firstly to determine if the shower is split across several clusters,
and then to add further clusters for consideration when this is the case.

\begin{figure}[htbp]
\centering
\captionsetup[subfigure]{justification=centering}
\subfloat[$2<\pTtrue<\SI{5}{\GeV}$,\protect\\ $|\etatrue|<1.0$]{\includegraphics[width=0.31\textwidth]{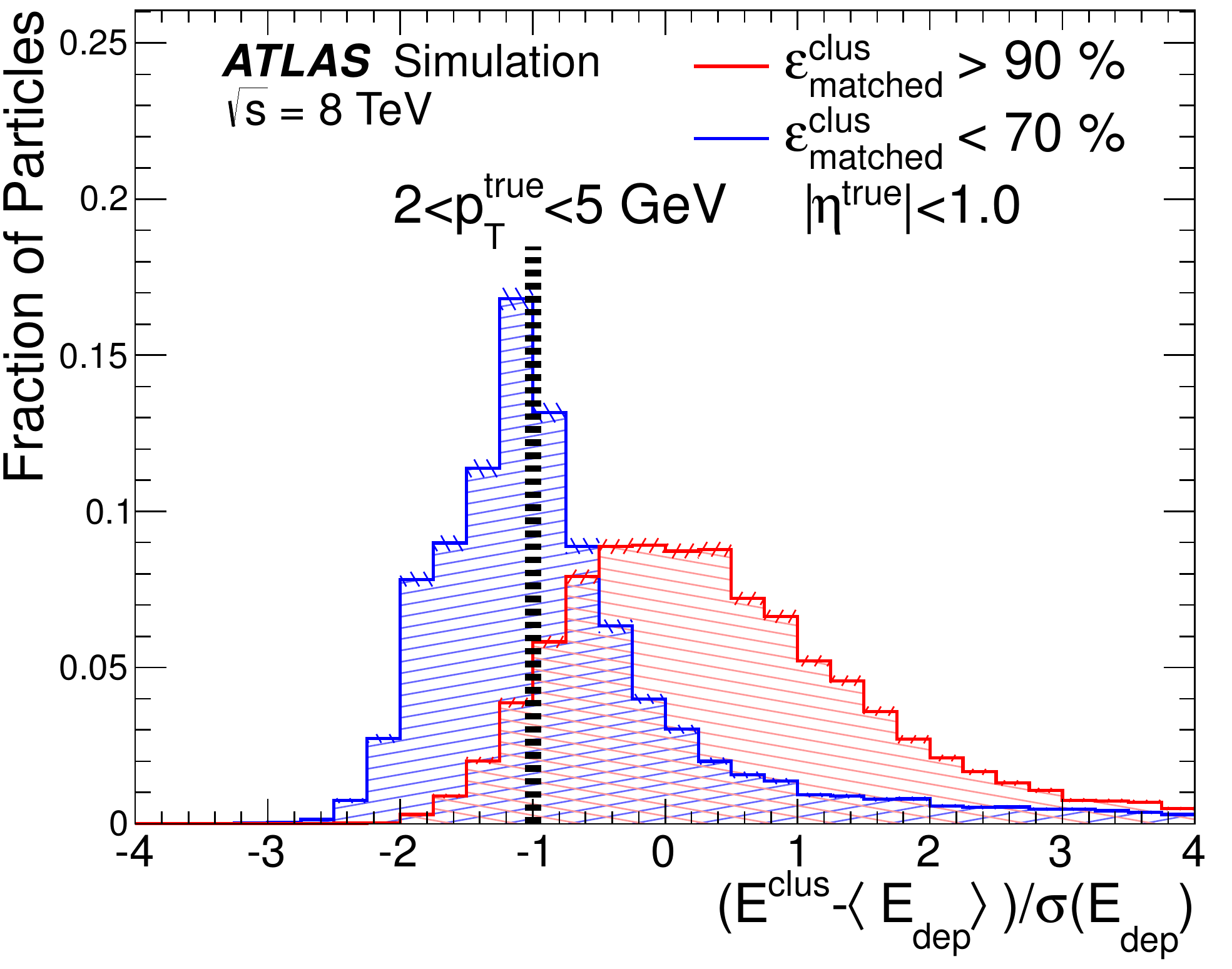}}\quad
\subfloat[$2<\pTtrue<\SI{5}{\GeV}$,\protect\\ $1.0<|\etatrue|<2.0$]{\includegraphics[width=0.31\textwidth]{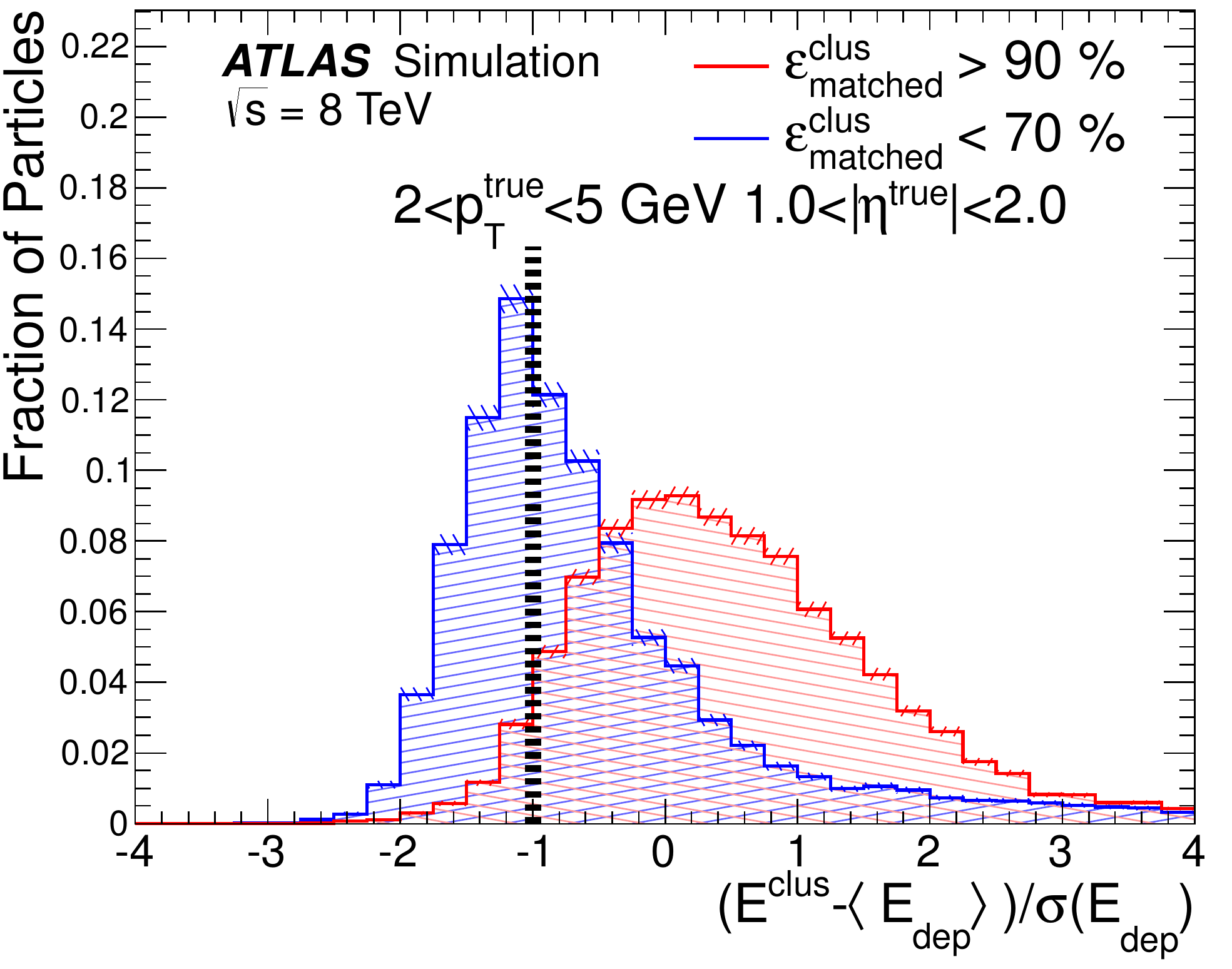}}\quad
\subfloat[$2<\pTtrue<\SI{5}{\GeV}$,\protect\\ $2.0<|\etatrue|<2.5$]{\includegraphics[width=0.31\textwidth]{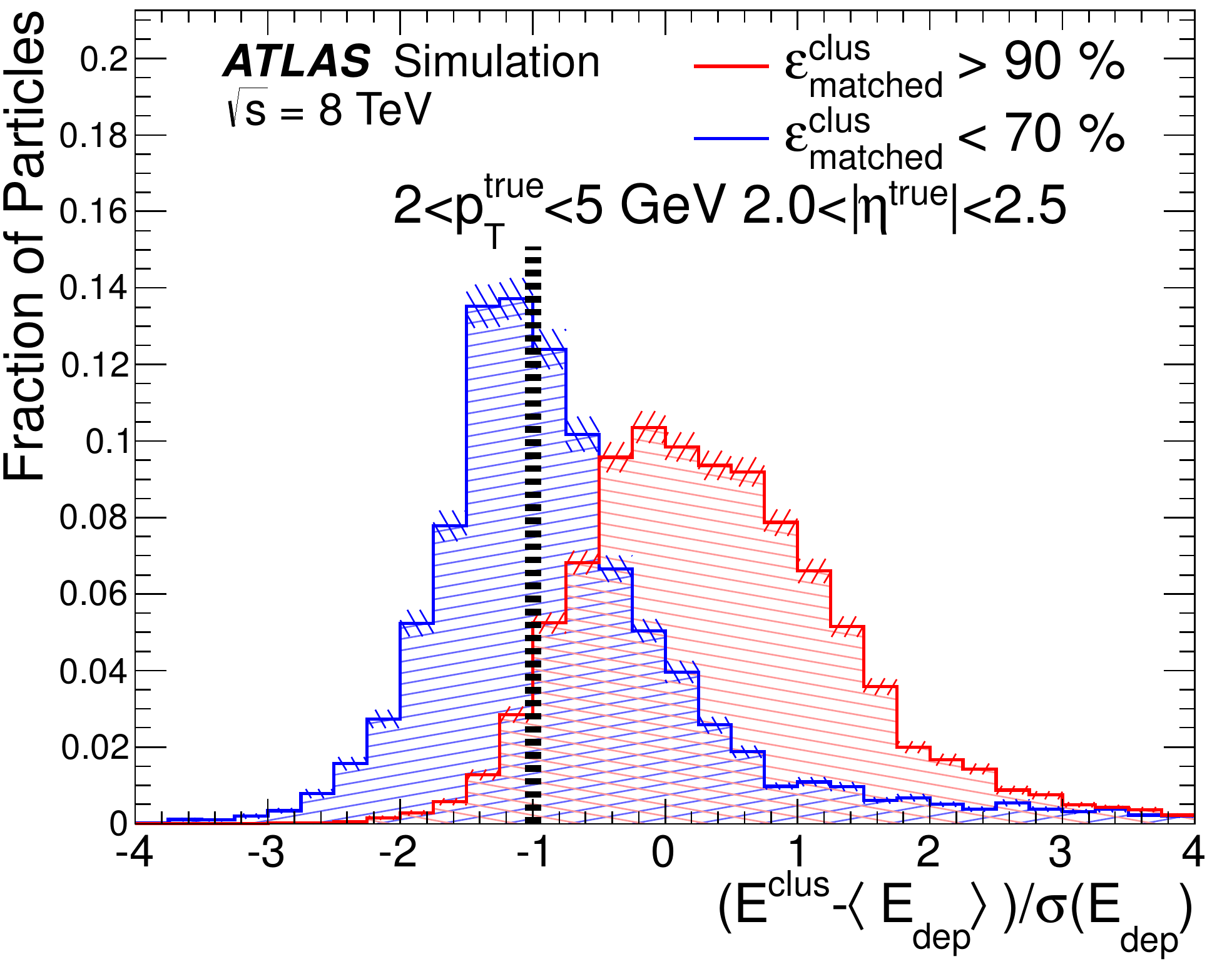}}\qquad
\subfloat[$5<\pTtrue<\SI{10}{\GeV}$,\protect\\ $|\etatrue|<1.0$]{\includegraphics[width=0.31\textwidth]{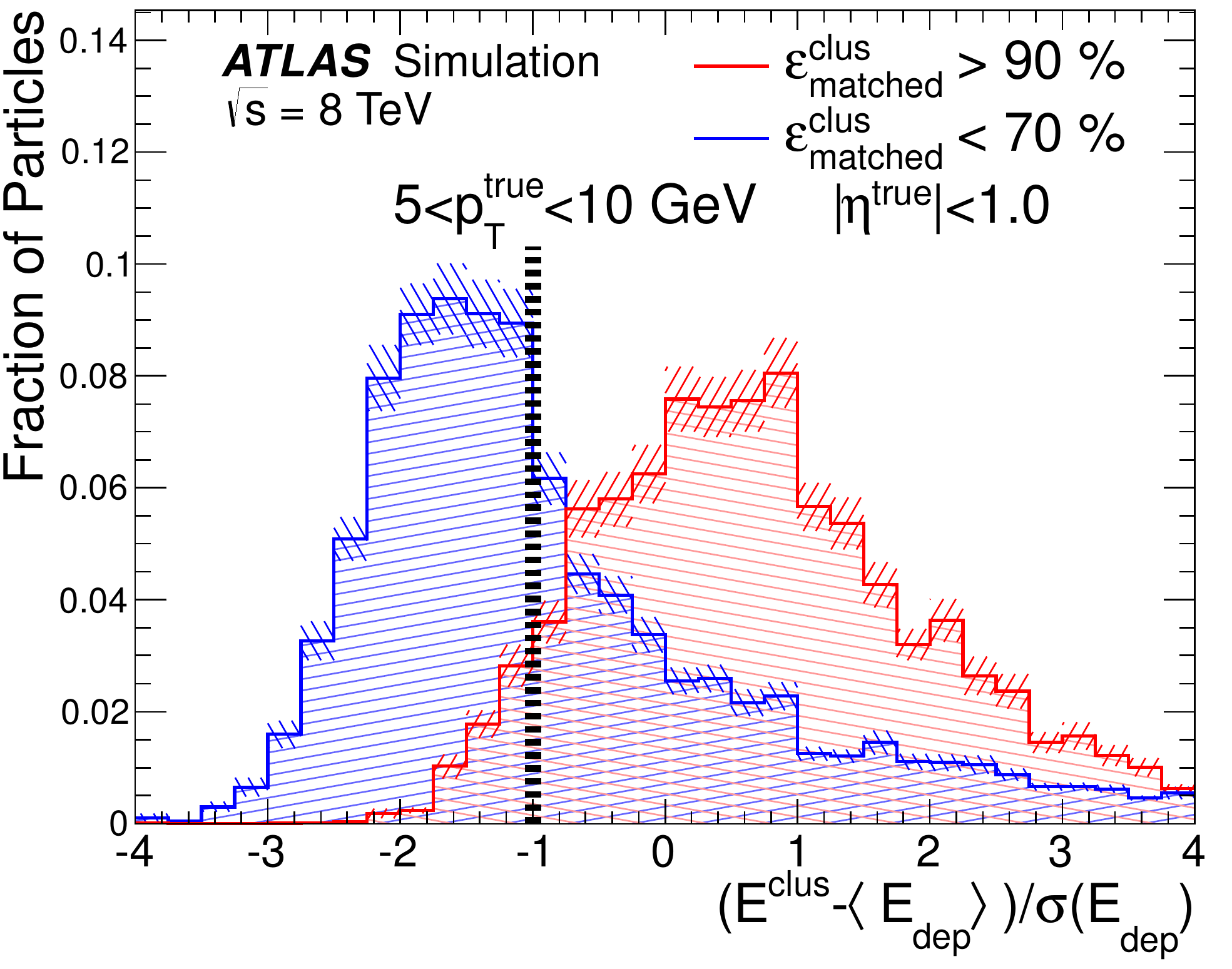}}\quad
\subfloat[$5<\pTtrue<\SI{10}{\GeV}$,\protect\\ $1.0<|\etatrue|<2.0$]{\includegraphics[width=0.31\textwidth]{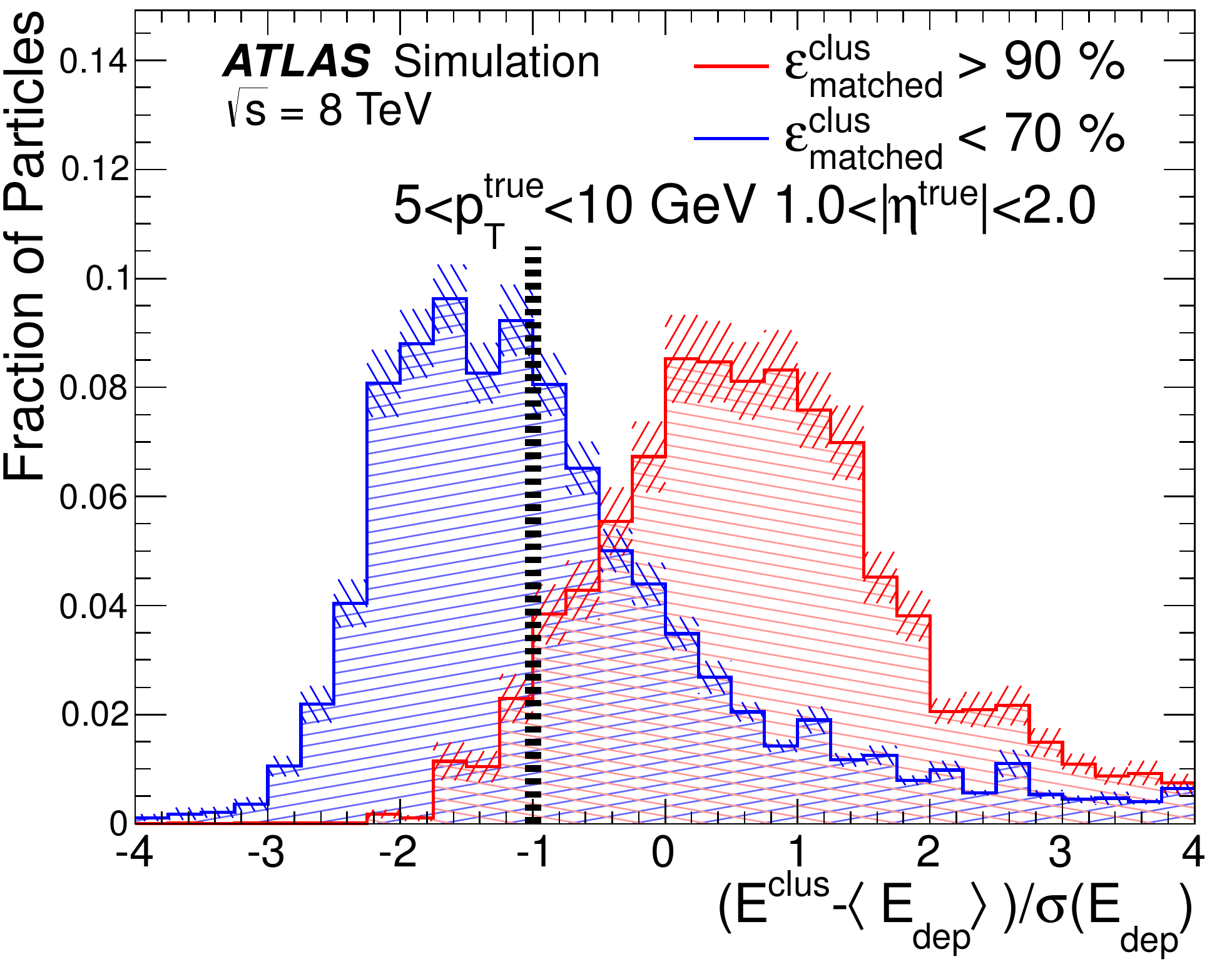}}\quad
\subfloat[$5<\pTtrue<\SI{10}{\GeV}$,\protect\\ $2.0<|\etatrue|<2.5$]{\includegraphics[width=0.31\textwidth]{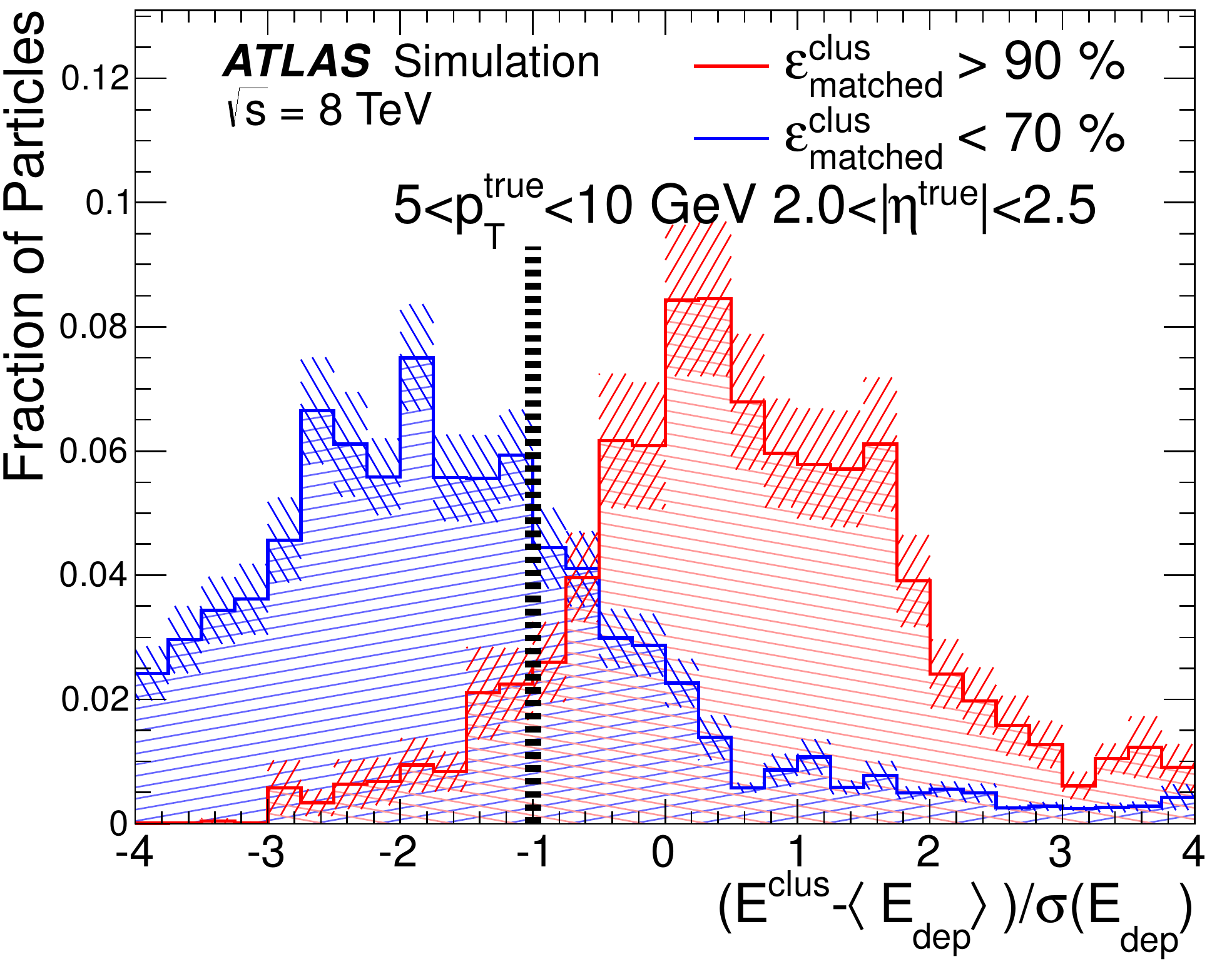}}\qquad
\caption{The significance of the difference between the energy of the matched \topocluster and the expected deposited energy $\MeanEdep$ and that of the matched \topocluster,
  for $\pi^\pm$ when $< \SI{70}{\%}$ and $> \SI{90}{\%}$ of the true deposited energy in \topoclusters is contained in the matched \topocluster
  for different \pTtrue and $|\etatrue|$ ranges.
  The vertical line indicates the value below which additional \topoclusters are matched to the track for cell subtraction.
  Subfigures (a)--(f) indicate that a single cluster is considered $(93,95,95,94,95,91)\,\%$ of the time when $\effmatch > \SI{90}{\%}$;
  while additional \topoclusters are considered $(49,39,46,56,52,60)\,\%$ of the time when $\effmatch < \SI{70}{\%}$.
  \DijetSample
}
\label{fig:eflowRec:RSSpull}
\end{figure}

The discriminant used to distinguish the single and multiple \topocluster cases is the significance of the difference between the expected energy
and that of the matched \topocluster (defined using the algorithm in \Sect{\ref{sec:trkclus}}),
\begin{equation}
  S(\Eclus) = \frac{\Eclus - \MeanEdep}{\sigmaEdep}\,.
\end{equation}
The distribution of $S(\Eclus)$ is shown in \Fig{\ref{fig:eflowRec:RSSpull}} 
for two categories of matched \topoclusters:
those with $\effi > \SI{90}{\%}$ and those with $\effi < \SI{70}{\%}$.
A clear difference is observed between the $S(\Eclus)$ distributions for the two categories,
demonstrating the separation between showers that are and are not contained in a single cluster.
More than \SI{90}{\%} of clusters with $\effi > \SI{90}{\%}$ have $S(\Eclus)>-1$.
Based on this observation a \SSR procedure is run if $S(\Eclus)<-1$:
\topoclusters within a cone of $\dR=0.2$ around the track position
extrapolated to the second EM calorimeter layer are considered to be matched to the track.
As can be seen in the figure, the \SSR procedure is typically run \SI{50}{\%}
of the time when $\effmatch < \SI{70}{\%}$.
The full set of matched clusters is then considered when the energy is subtracted from the calorimeter.

%-------------------------------------------------------------------------------
\subsection{Cell-by-cell subtraction}
\label{sec:eflowRec_cellsub}

Once a set of \topoclusters corresponding to the track has been selected,
the subtraction step is executed.
If $\MeanEdep$ exceeds the total energy of the set of matched \topoclusters, then the \topoclusters are simply removed.
Otherwise, subtraction is performed cell by cell.

\begin{figure}[htbp]
\centering
\subfloat[]{\includegraphics[width=0.23\textwidth]{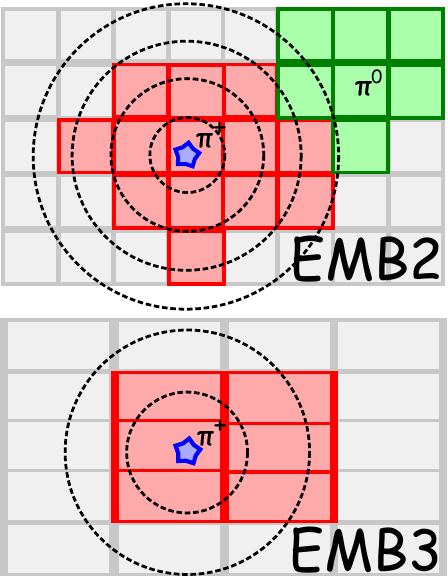}}\quad
\subfloat[]{\includegraphics[width=0.23\textwidth]{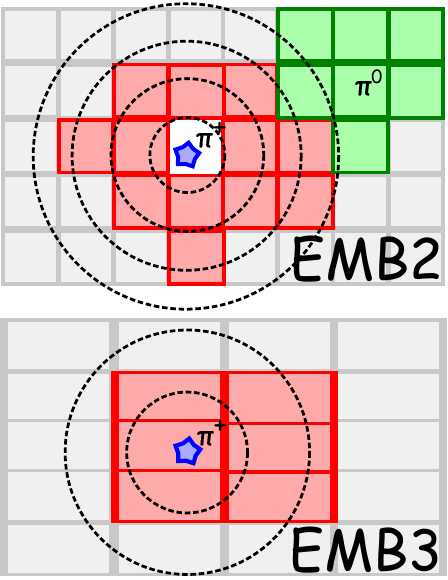}}\quad
\subfloat[]{\includegraphics[width=0.23\textwidth]{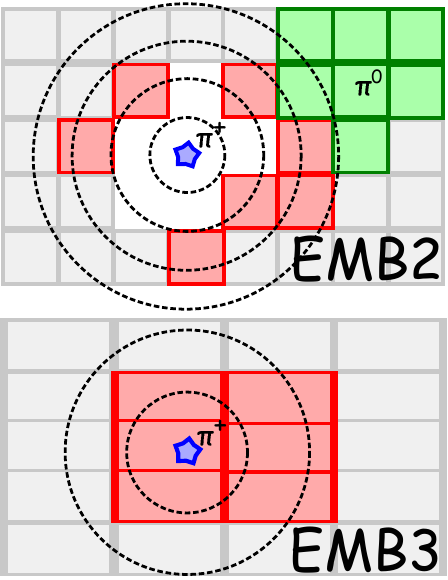}}\quad
\subfloat[]{\includegraphics[width=0.23\textwidth]{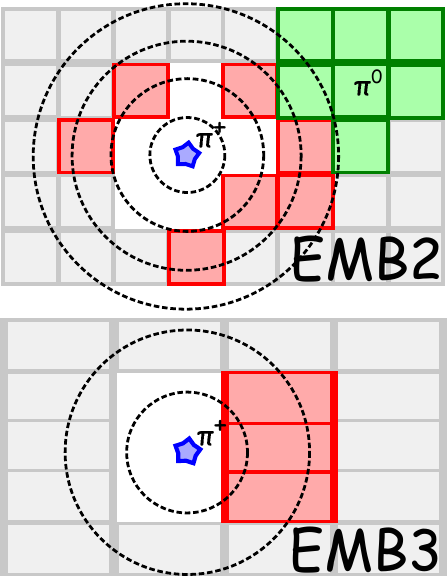}}\\
\subfloat[]{\includegraphics[width=0.23\textwidth]{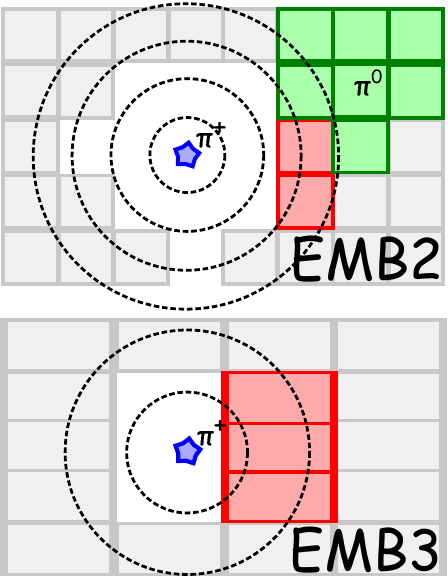}}\quad
\subfloat[]{\includegraphics[width=0.23\textwidth]{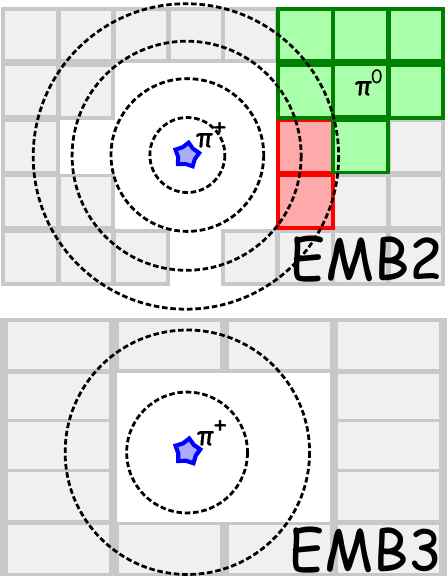}}\quad
\subfloat[]{\includegraphics[width=0.23\textwidth]{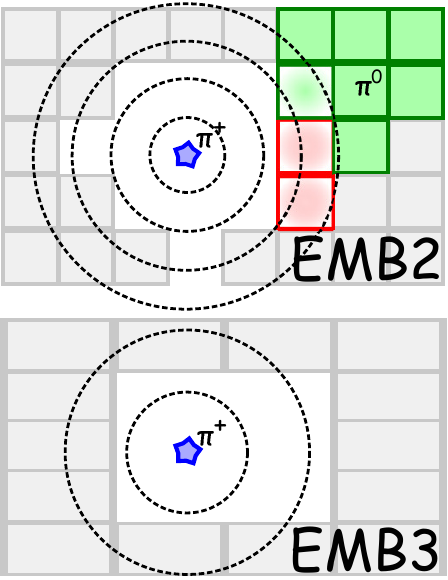}}
\caption{An idealised example of how the cell-by-cell subtraction works.
  Cells in two adjacent calorimeter layers (EMB2 and EMB3) are shown in grey if they are not in clusters, red if they belong to a $\pi^+$ cluster and in green if contributed by a $\pi^0$ meson.
  Rings are placed around the extrapolated track (represented by a star) and then the cells in these are removed ring by ring starting with the centre of the shower, (a),  where the expected energy density is highest and moving outwards, and between layers. This sequence of ring subtraction is shown in subfigures (a) through (g).
  The final ring contains more energy than the expected energy, hence this is only partially subtracted (g), indicated by a lighter shading.}
\label{fig:eflowRec:cellSub}
\end{figure}

Starting from the extrapolated track position in the LHED, a parameterised shower shape is used to map out the most likely energy density profile in each layer.
This profile is determined from a single $\pi^\pm$ MC sample and is dependent on the track momentum and pseudorapidity, as well as on the LHED for the set of considered \topoclusters.
Rings are formed in $\eta$,$\phi$ space around the extrapolated track.
The rings are just wide enough to always contain at least one calorimeter cell, independently of the extrapolated position, and are confined to a single calorimeter layer.
Rings within a single layer are equally spaced in radius.
The average energy density in each ring is then computed, and the rings are ranked in descending order of energy density, irrespective of which layer each ring is in.
Subtraction starts from the ring with the highest energy density (the innermost ring of the LHED)
and proceeds successively to the lower-density rings.
If the energy in the cells in the current ring is less than the remaining energy required to reach \MeanEdep,
these cells are simply removed and the energy still to be subtracted is reduced by the total energy of the ring.
If instead the ring has more energy than is still to be removed,
each cell in the ring is scaled down in energy by the fraction needed to reach the expected energy from the particle, 
then the process halts.
Figure~\ref{fig:eflowRec:cellSub} shows a cartoon of how this subtraction works, removing cells in different rings from different layers until the expected energy deposit is reached.

\subsection{Remnant removal}
\label{sec:eflowRec_remnant}

\begin{figure}[htbp]
\centering
\captionsetup[subfigure]{justification=centering}
\subfloat[$2<\pTtrue<\SI{5}{\GeV}$,\protect\\ $|\etatrue|<1.0$]{\includegraphics[width=0.31\textwidth]{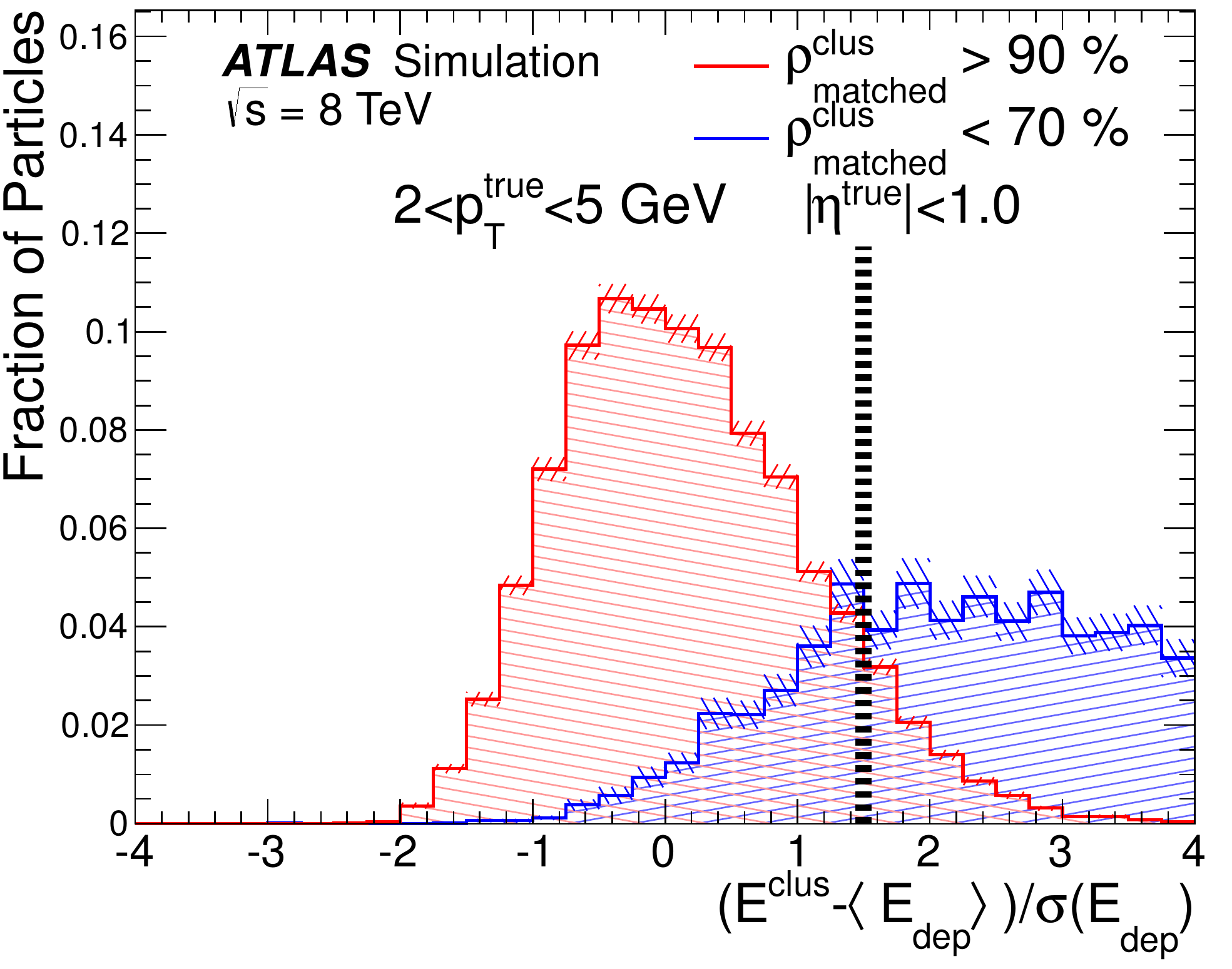}}\quad
\subfloat[$2<\pTtrue<\SI{5}{\GeV}$,\protect\\ $1.0<|\etatrue|<2.0$]{\includegraphics[width=0.31\textwidth]{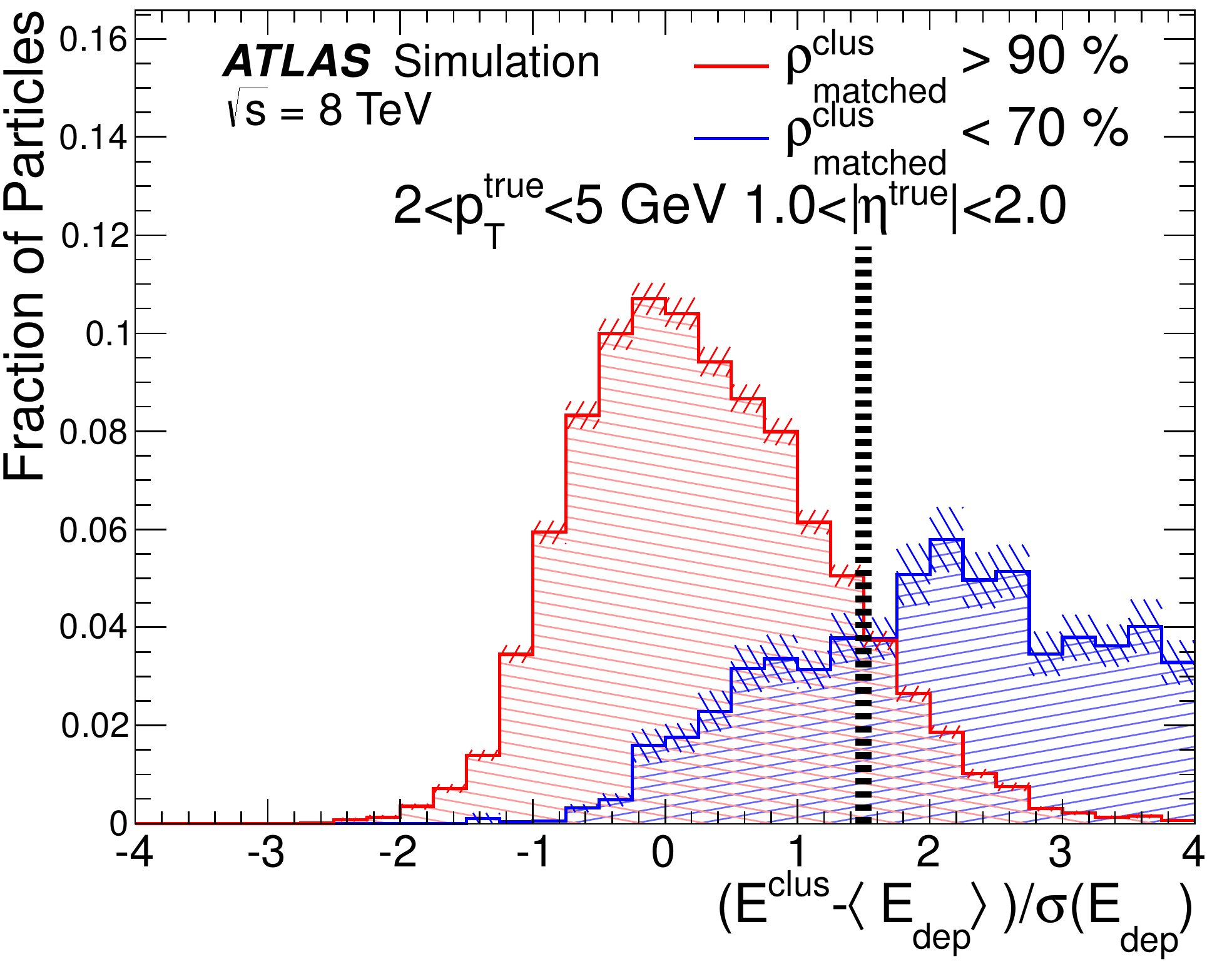}}\quad
\subfloat[$2<\pTtrue<\SI{5}{\GeV}$,\protect\\ $2.0<|\etatrue|<2.5$]{\includegraphics[width=0.31\textwidth]{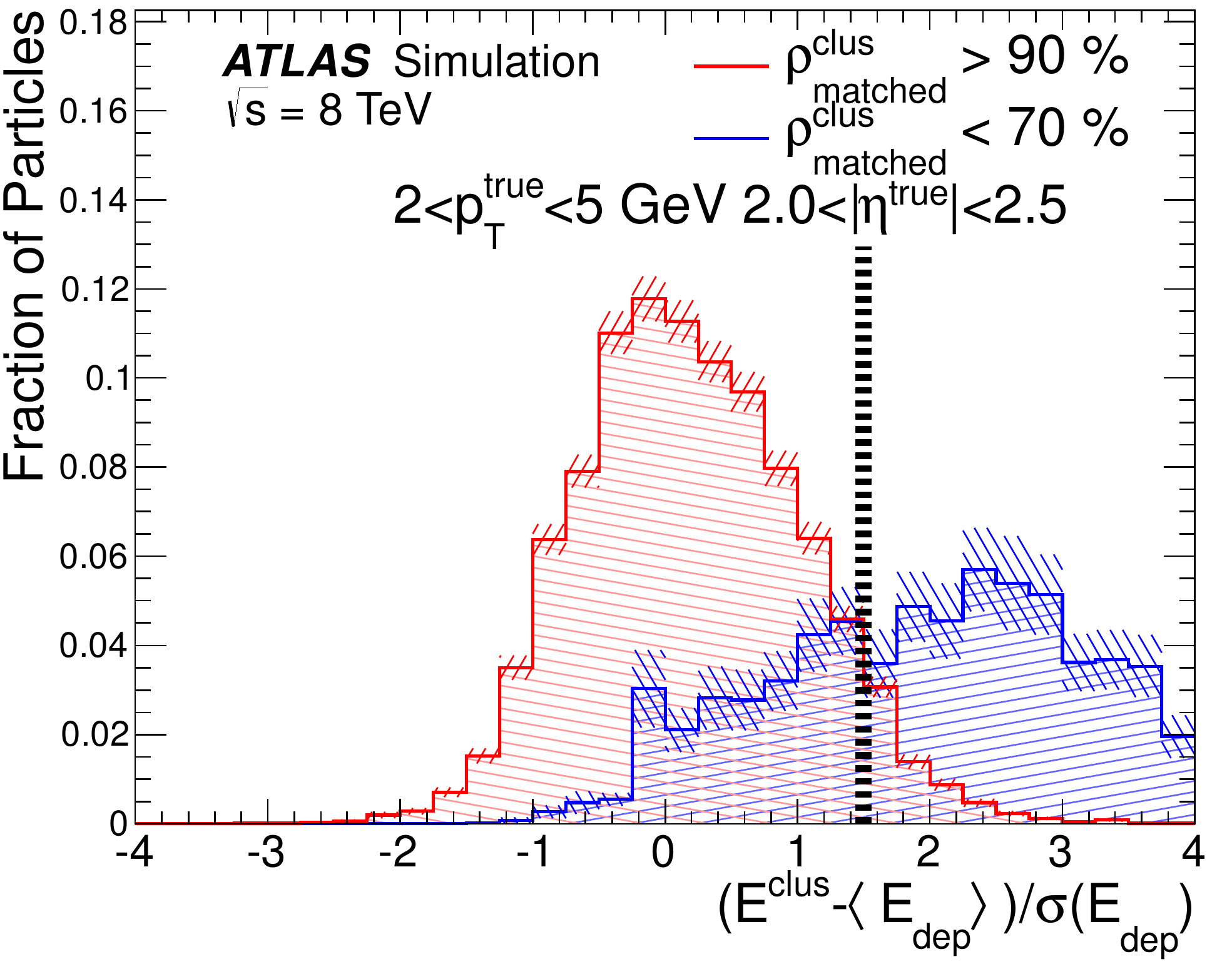}}\\
\subfloat[$5<\pTtrue<\SI{10}{\GeV}$,\protect\\ $|\etatrue|<1.0$]{\includegraphics[width=0.31\textwidth]{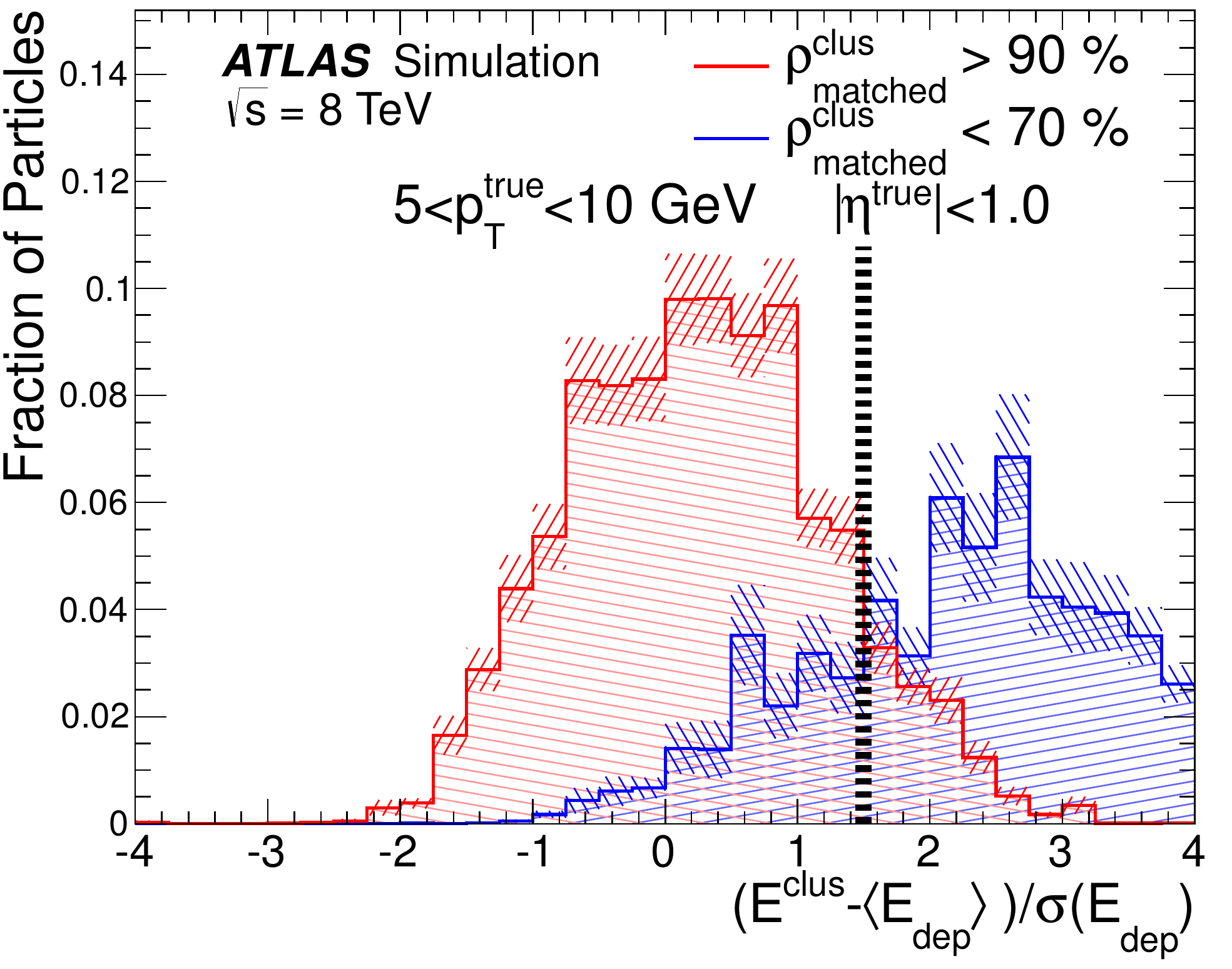}}\quad
\subfloat[$5<\pTtrue<\SI{10}{\GeV}$,\protect\\ $1.0<|\etatrue|<2.0$]{\includegraphics[width=0.31\textwidth]{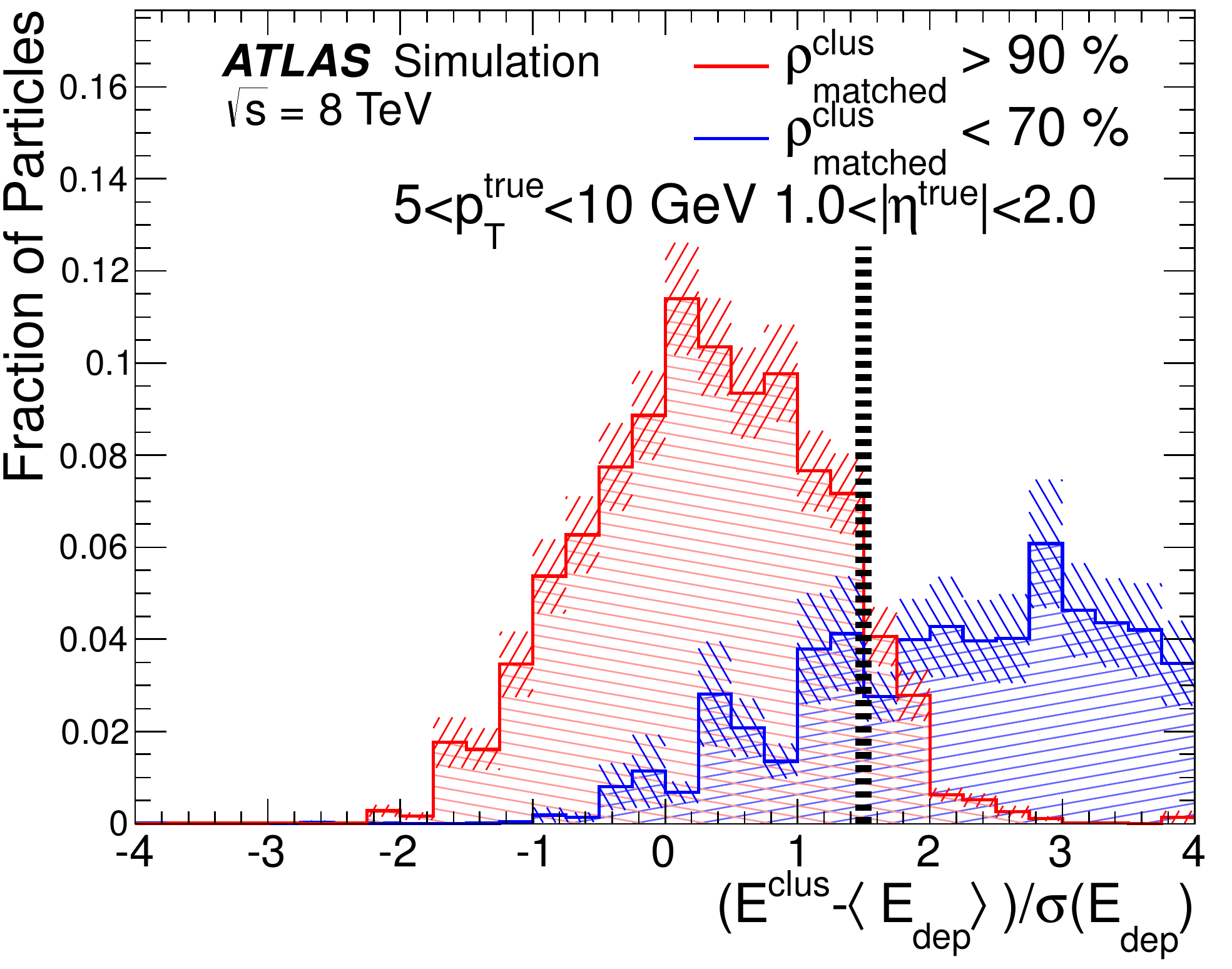}}\quad
\subfloat[$5<\pTtrue<\SI{10}{\GeV}$,\protect\\ $2.0<|\etatrue|<2.5$]{\includegraphics[width=0.31\textwidth]{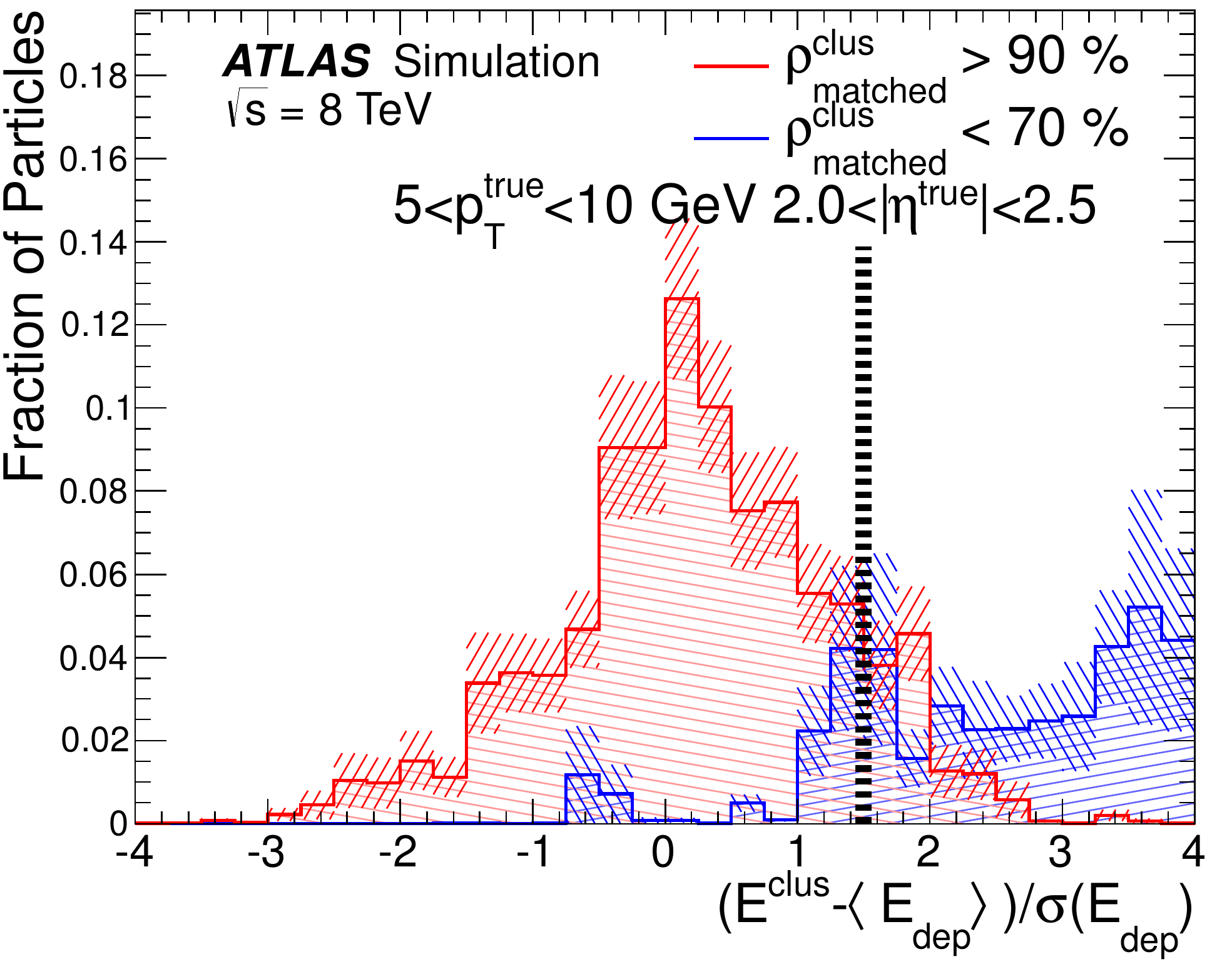}}
\caption{The significance of the difference between the energy of the matched \topocluster and the expected deposited energy $\MeanEdep$
  for $\pi^\pm$ with either $< \SI{70}{\%}$ or $> \SI{90}{\%}$ of the total true energy in the matched \topocluster originating from the $\pi^{\pm}$
  for different \pTtrue and $|\etatrue|$ ranges.
  The vertical line indicates the value below which the remnant \topocluster is removed,
  as it is assumed that in this case no other particles contribute to the \topocluster.
  Subfigures (a)--(f) indicate that when $\purmatch > \SI{90}{\%}$ the remnant is successfully removed $(91,89,94,89,91,88)\,\%$ of the time;
  while when $\purmatch < \SI{70}{\%}$ the remnant is retained $(81,80,76,84,83,91)\,\%$ of the time.
  \DijetSample
}
\label{fig:eflowRec:Remnantpull}
\end{figure}

If the energy remaining in the set of cells and/or \topoclusters that survive the energy subtraction is consistent with the width of the $\EoPtrkref$ distribution,
specifically if this energy is less than $1.5 \sigmaEdep$,
it is assumed that the \topocluster system was produced by a single particle.
The remnant energy therefore originates purely from shower fluctuations and so the energy in the remaining cells is removed.
Conversely, if the remaining energy is above this threshold,
the remnant \topocluster{}(s) are retained -- it being likely that multiple particles deposited energy in the vicinity.
Figure~\ref{fig:eflowRec:Remnantpull} shows how this criterion is able to separate cases
where the matched \topocluster has true deposited energy only from a single particle
from those where there are multiple contributing particles.

After this final step,
the set of selected tracks and the remaining \topoclusters in the calorimeter together should ideally represent the reconstructed event
with no double counting of energy between the subdetectors.

%-------------------------------------------------------------------------------
% Performance of the subtraction algorithm at truth level
%-------------------------------------------------------------------------------
% !TeX root = Pflow.tex
%-------------------------------------------------------------------------------
\section{Performance of the subtraction algorithm at truth level}
\label{sec:calHits}
%-------------------------------------------------------------------------------

The performance of each step of the particle flow algorithm is evaluated exploiting the detailed energy information
at truth level available in Monte Carlo generated events. 
For these studies a dijet sample with leading truth jet \pT between 20 and \SI{500}{\GeV} without pile-up is used.

%-------------------------------------------------------------------------------
\subsection{Track--cluster matching performance}

\begin{figure}[htbp]
\centering
\subfloat[$\efflead > \SI{90}{\%}$]{\includegraphics[width=6cm]{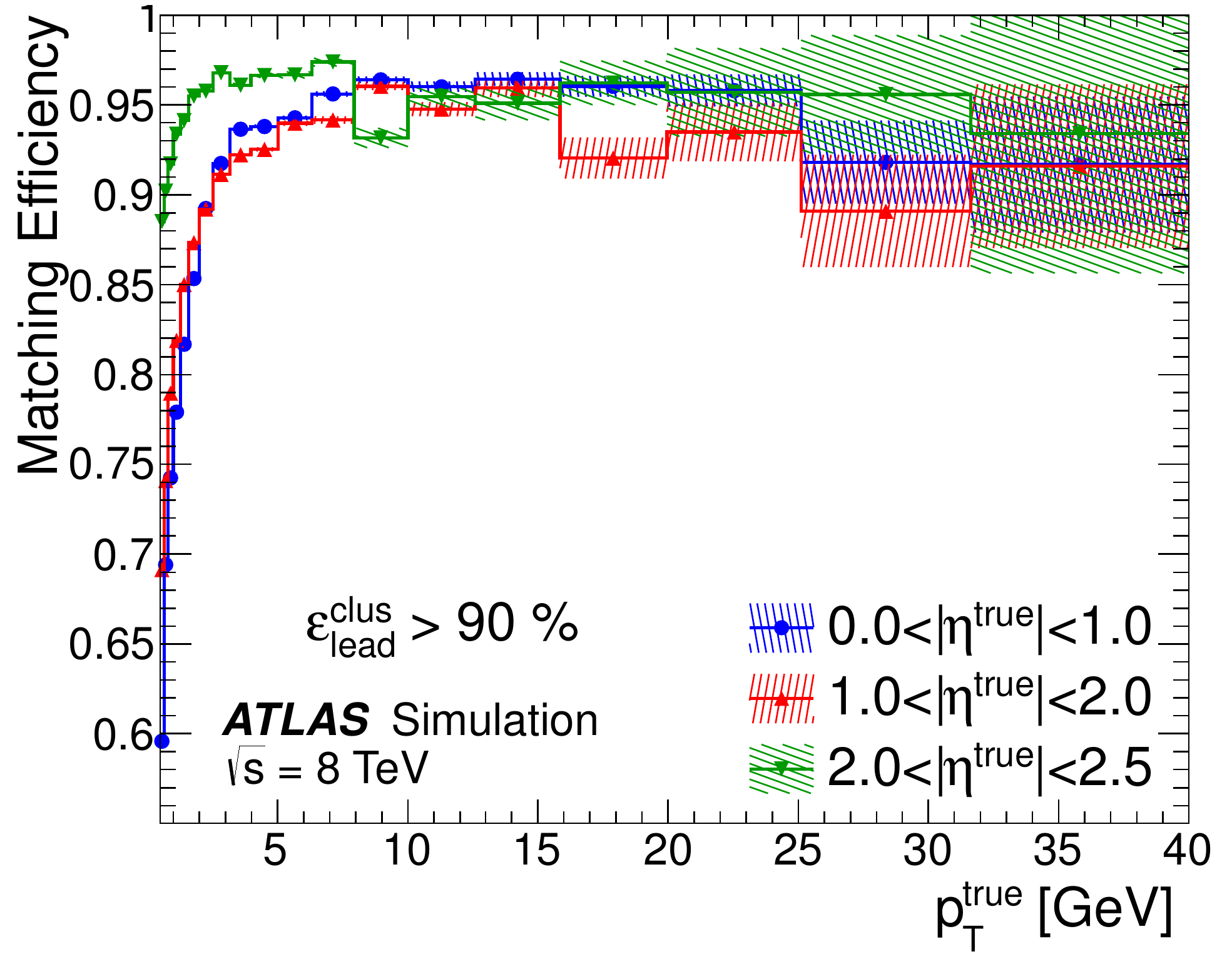}}\quad
\subfloat[$\efflead > \SI{50}{\%}$]{\includegraphics[width=6cm]{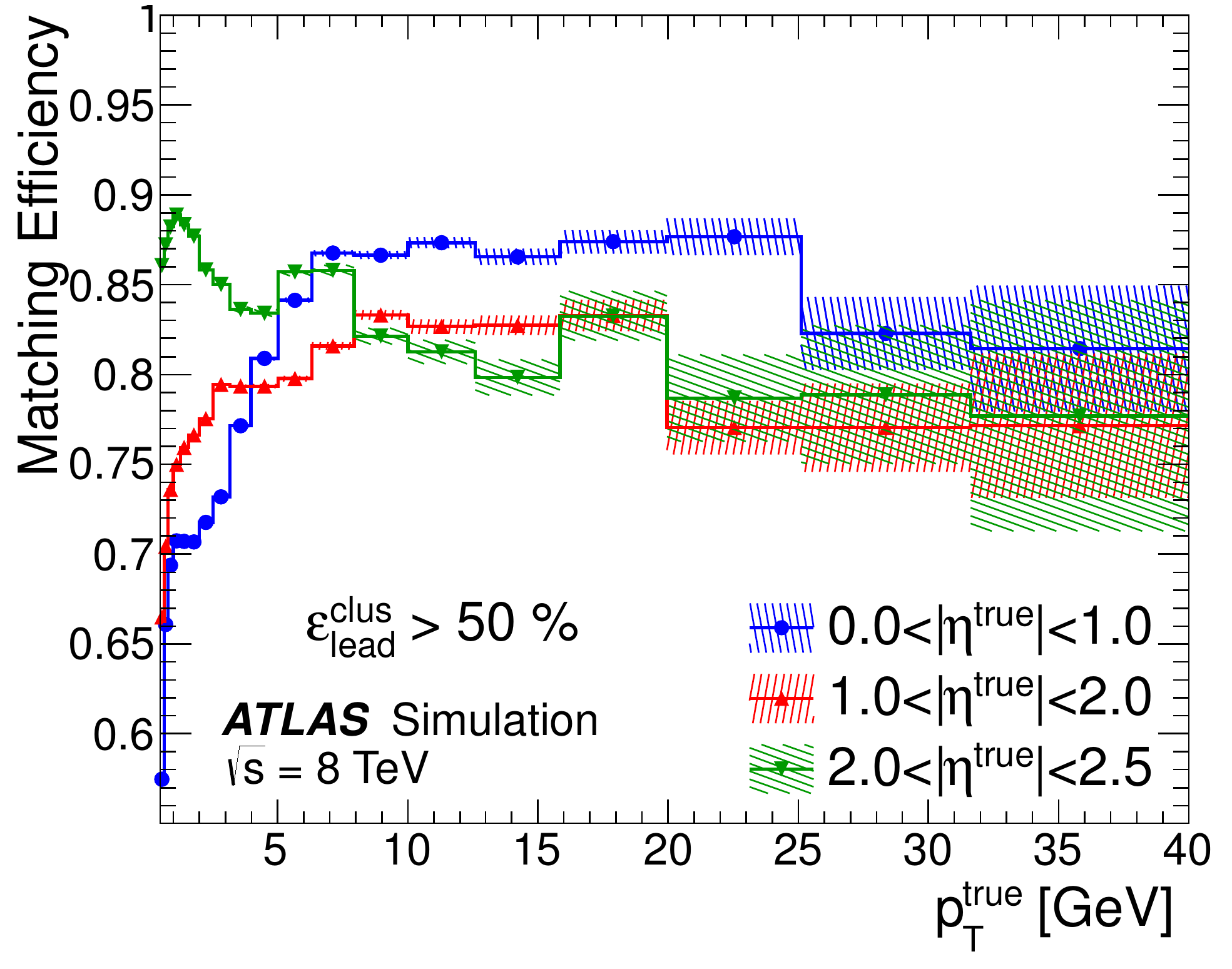}}
\caption{The probability to match the track to the leading \topocluster
  (a) when $\efflead > \SI{90}{\%}$ and 
  (b) when $\efflead > \SI{50}{\%}$.
  \DijetSample
}
\label{fig:eflowRec:matchEff}
\end{figure}

Initially, the algorithm attempts to match the track to a single \topocluster containing the full particle energy.
Figure~\ref{fig:eflowRec:matchEff} shows the fraction of tracks whose matched cluster has $\efflead>\SI{90}{\%}$ or $\efflead>\SI{50}{\%}$.
When almost all of the deposited energy is contained within a single \topocluster,
the probability to match a track to this \topocluster (matching probability)
is above \SI{90}{\%} in all $\eta$ regions, for particles with $\pT > \SI{2}{\GeV}$.
The matching probability falls to between $\SI{70}{\%}$ and $\SI{90}{\%}$ when up to half the particle's energy is permitted to fall in other \topoclusters.
Due to changes in the calorimeter geometry, the splitting rate and hence the matching probability vary significantly for particles in different pseudorapidity regions.
In particular, the larger cell size at higher $|\eta|$ enhances the likelihood of capturing soft particle showers in a single \topocluster, as seen in \Figs{\ref{fig:eflowRec:eff}}{\ref{fig:eflowRec:nClus}}, which results in the matching efficiency increasing at low \pT\ for $|\eta|>2$.

%-------------------------------------------------------------------------------
\subsection{Split-shower recovery performance}

Frequently, a particle's energy is not completely contained within the single best-match \topocluster,
in which case the \SSR procedure is applied.
The effectiveness of the recovery can be judged based on whether the procedure is correctly triggered,
and on the extent to which the energy subtraction is improved by its execution.

\begin{figure}[htbp]
  \centering
  \captionsetup[subfigure]{justification=centering}
  \subfloat[$5<\pTtrue<\SI{10}{\GeV}$,\protect\\ $|\etatrue|<1.0$]{\includegraphics[width=0.32\textwidth]{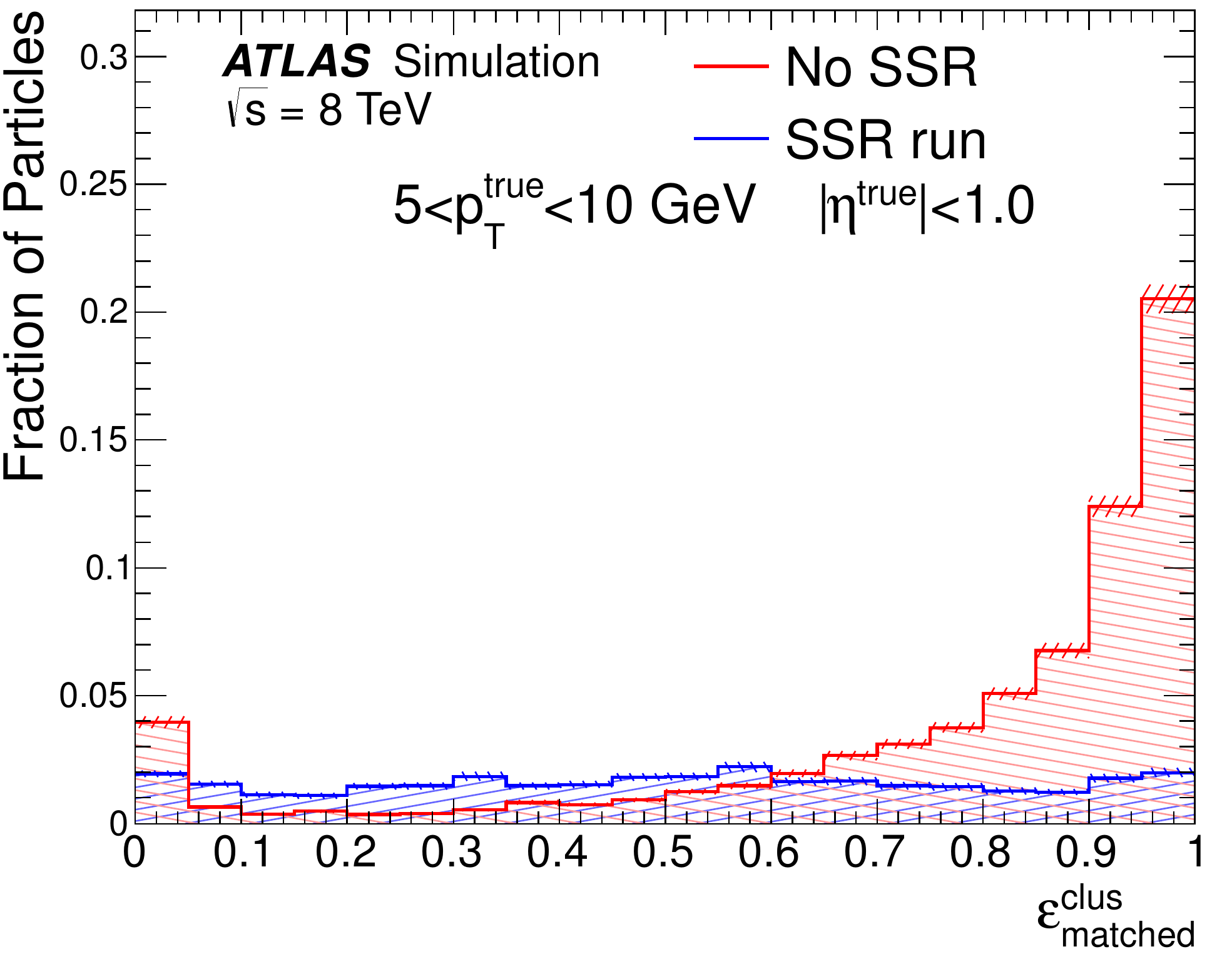}}
  \subfloat[$10<\pTtrue<\SI{20}{\GeV}$,\protect\\ $|\etatrue|<1.0$]{\includegraphics[width=0.32\textwidth]{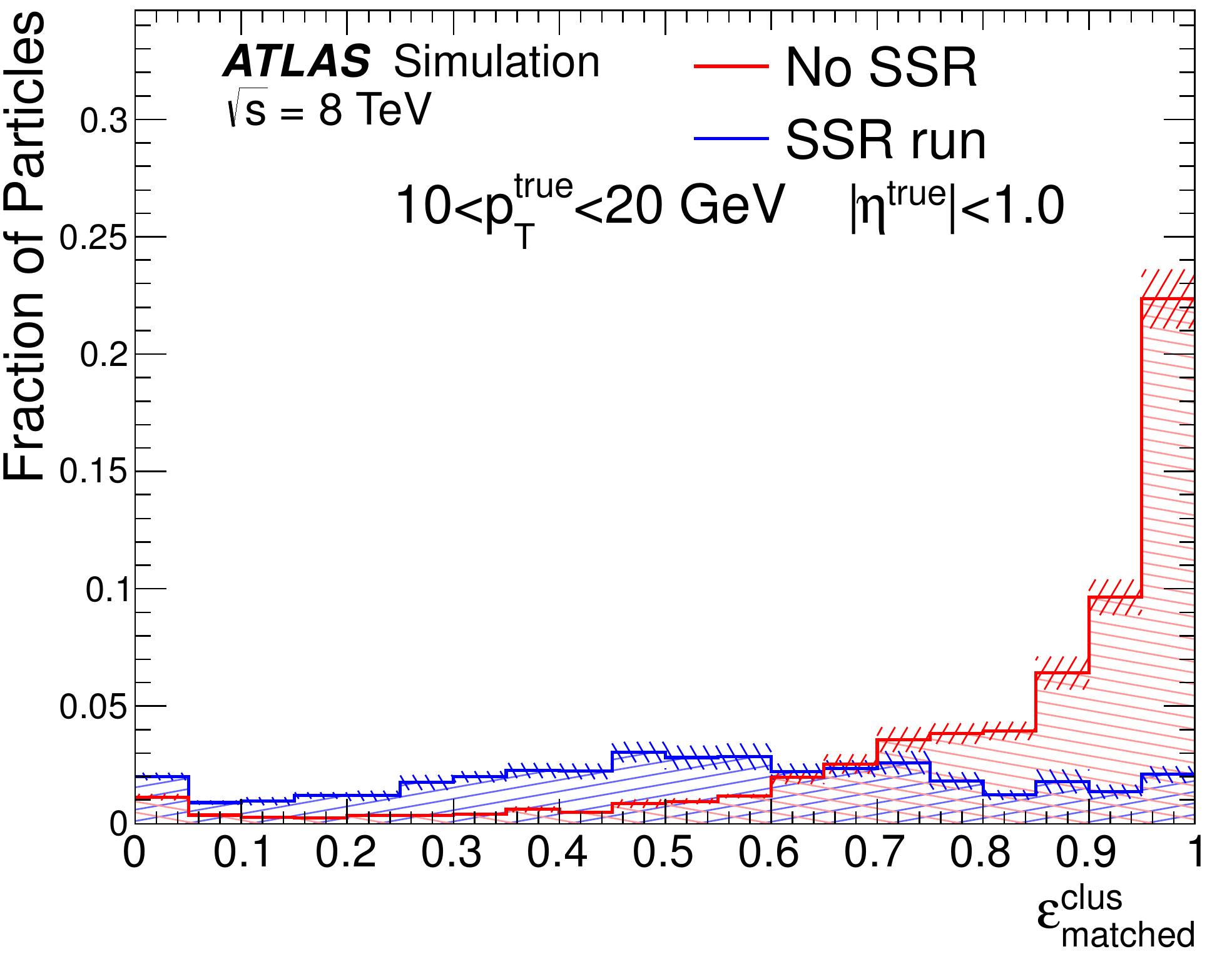}}
  \subfloat[$10<\pTtrue<\SI{20}{\GeV}$,\protect\\ $1.0<|\etatrue|<2.0$]{\includegraphics[width=0.32\textwidth]{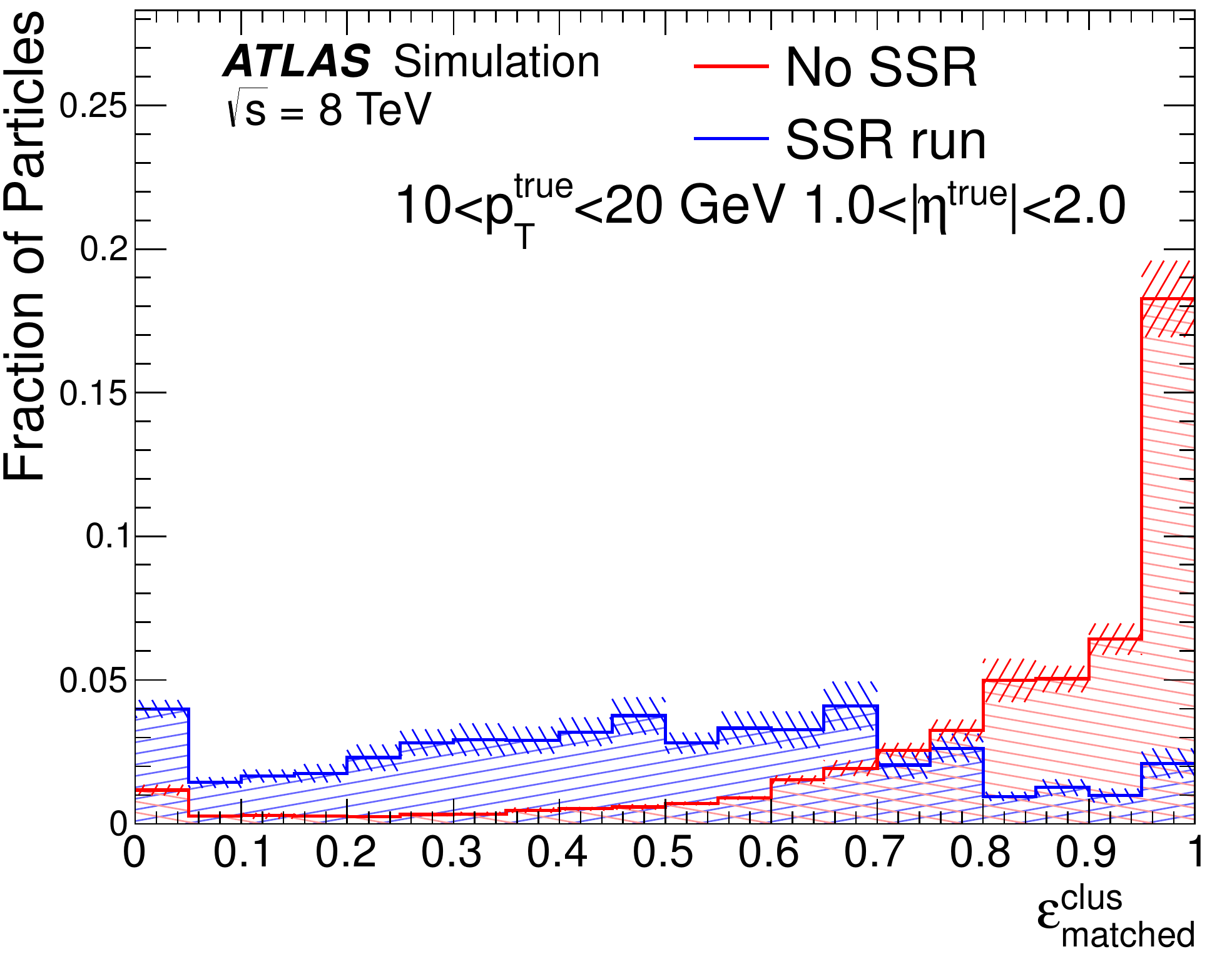}}
  \caption{The fraction of the true energy of a given particle contained within the initially matched \topocluster for particles where the \SSR procedure is run (SSR run) and where it is not (No SSR).
  For cases where most of the energy is contained in the initially matched \topocluster the procedure is less likely to be run.
  \DijetSample
}
  \label{fig:perf:RSSexample}
\end{figure}

Figure~\ref{fig:perf:RSSexample} shows the fraction $\varepsilon^\text{clus}_\text{matched}$ of the true deposited energy contained within the matched \topocluster,
separately for cases where the \SSR procedure is and is not triggered,
as determined by the criteria described in \Sect{\ref{sec:eflowRec_ssr}}.
In the cases where the \SSR procedure is not run, $\varepsilon^\text{clus}_\text{matched}$ is found to be high, 
confirming that the comparison of \topocluster energy and $\MeanEoP$ is successfully identifying good \topocluster matches.
Conversely, the \SSR procedure is activated when $\varepsilon^\text{clus}_\text{matched}$ is low,
particularly for higher-\pT particles, which are expected to split their energy between multiple \topoclusters more often.
Furthermore, as the particle \pT rises, the width of the calorimeter response distribution decreases,
making it easier to distinguish the different cases.

\begin{figure}[htbp]
  \centering
  \includegraphics[width=0.45\textwidth]{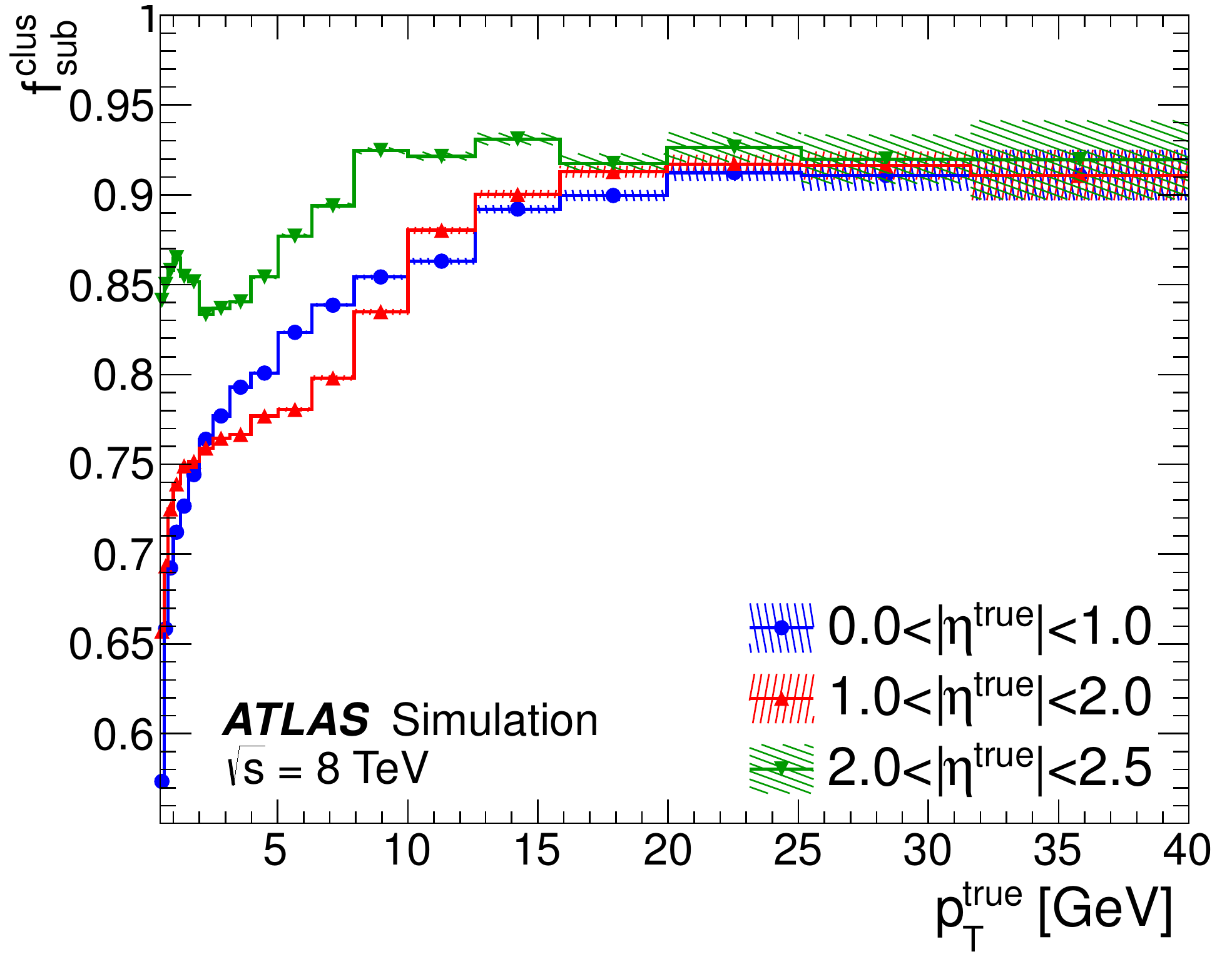}
  \caption{The fraction of the true energy of a given particle considered in the subtraction procedure $f^\text{clus}_\text{sub}$ after the inclusion of the \SSR algorithm.
  \DijetSample}
  \label{fig:perf:RSS}
\end{figure}

Figure~\ref{fig:perf:RSS} shows the fraction $f^\text{clus}_\text{sub}$ of the true deposited energy of the pions considered for subtraction,
in the set of clusters matched to the track, as a function of true \pT.
For particles with $\pT>\SI{20}{\GeV}$, with \SSR active, $f^\text{clus}_\text{sub}$ is greater than $\SI{90}{\%}$ on average.
The subtraction algorithm misses more energy for softer showers, which are harder to capture completely.
While $f^\text{clus}_\text{sub}$ could be increased by simply attempting recovery more frequently,
expanding the \topocluster matching procedure in this fashion increases the risk of incorrectly subtracting neutral energy; hence the \SSR procedure cannot be applied indiscriminately.
The settings used in the studies presented in this paper are a reasonable compromise between these two cases.

%-------------------------------------------------------------------------------
\subsection{Accuracy of cell subtraction}

The cell subtraction procedure removes the expected calorimeter energy contribution based on the track properties.
It is instructive to identify the energy that is incorrectly subtracted from the calorimeter,
to properly understand and optimise the performance of the algorithm.

Truth particles are assigned reconstructed energy in \topoclusters as described in \Sect{\ref{sec:MC_calhits}},
and then classified depending on whether or not a track was reconstructed for the particle.
The reconstructed energy assigned to each particle is computed both before subtraction and after the subtraction has been performed, using the remaining cells.
In the ideal case, the subtraction should remove all the energy in the calorimeter assigned to stable truth particles which have reconstructed tracks, and should not remove any energy assigned to other particles.
The total transverse momentum of clusters associated with particles in a truth jet where a track was reconstructed before (after) subtraction is defined as $p^{\pm}_{\text{T,pre-sub}} (p^{\pm}_{\text{T,post-sub}})$.
Similarly, the transverse momentum of clusters associated with the other particles in a truth jet, neutral particles and those that did not create selected, reconstructed tracks, before (after) subtraction as $p^{0}_{\text{T,pre-sub}} (p^{0}_{\text{T,post-sub}})$.
The corresponding transverse momentum fractions are defined as 
$f^{\pm} = p^{\pm}_{\text{T,pre-sub}}/\pT^{\text{jet, true}}$ ($f^0 = p^{0}_{\text{T,pre-sub}}/\pT^{\text{jet, true}}$).

Three measures are established, to quantify the degree to which the energy is incorrectly subtracted.
The incorrectly subtracted fractions for the two classes of particles are:
\begin{equation}
  R^\pm = \frac{p^{\pm}_{\text{T,post-sub}}}{\pT^\text{jet, true}}
\end{equation}
and
\begin{equation}
  R^0 = \frac{p^0_\text{T,pre-sub} - p^0_\text{T,post-sub}}{\pT^\text{jet, true}}\,,
\end{equation}
such that $R^\pm$ corresponds to the fraction of surviving momentum associated with particles where the track measurement is used,
which should have been removed,
while $R^0$ gives the fraction of momentum removed that should have been retained as it is associated with particles where the calorimeter measurement is being used.
These two variables are combined into the confusion term
\begin{equation}
  C = R^\pm-R^0\,,
\end{equation}
which is equivalent to the net effect of both mistakes on the final jet transverse momentum, as there is a potential cancellation between the two effects. An ideal subtraction algorithm would give zero for all three quantities.

\begin{figure}[htbp]
  \centering
  \subfloat[]{\includegraphics[width=0.32\textwidth]{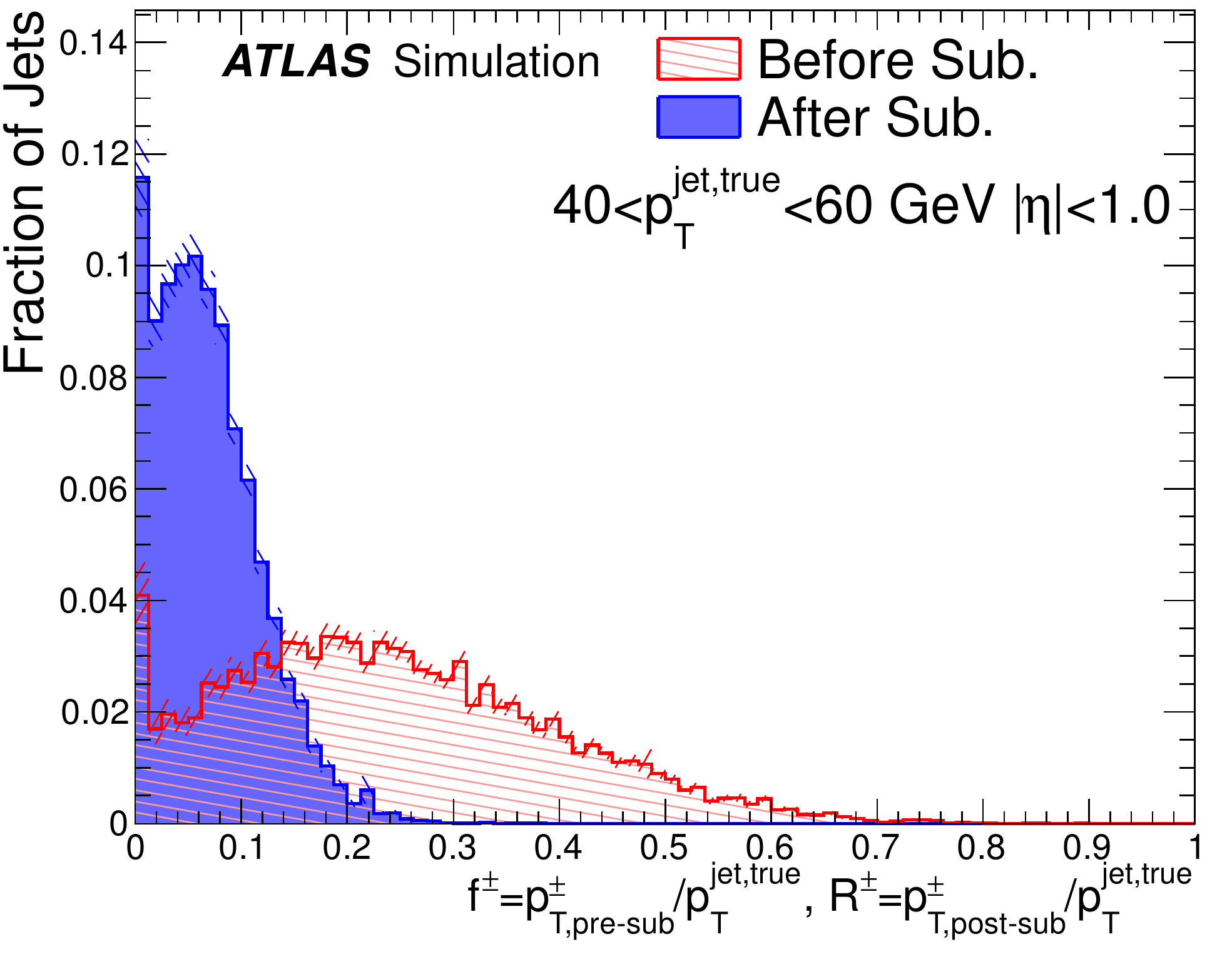}\label{fig:perf:sub_a}}
  \subfloat[]{\includegraphics[width=0.32\textwidth]{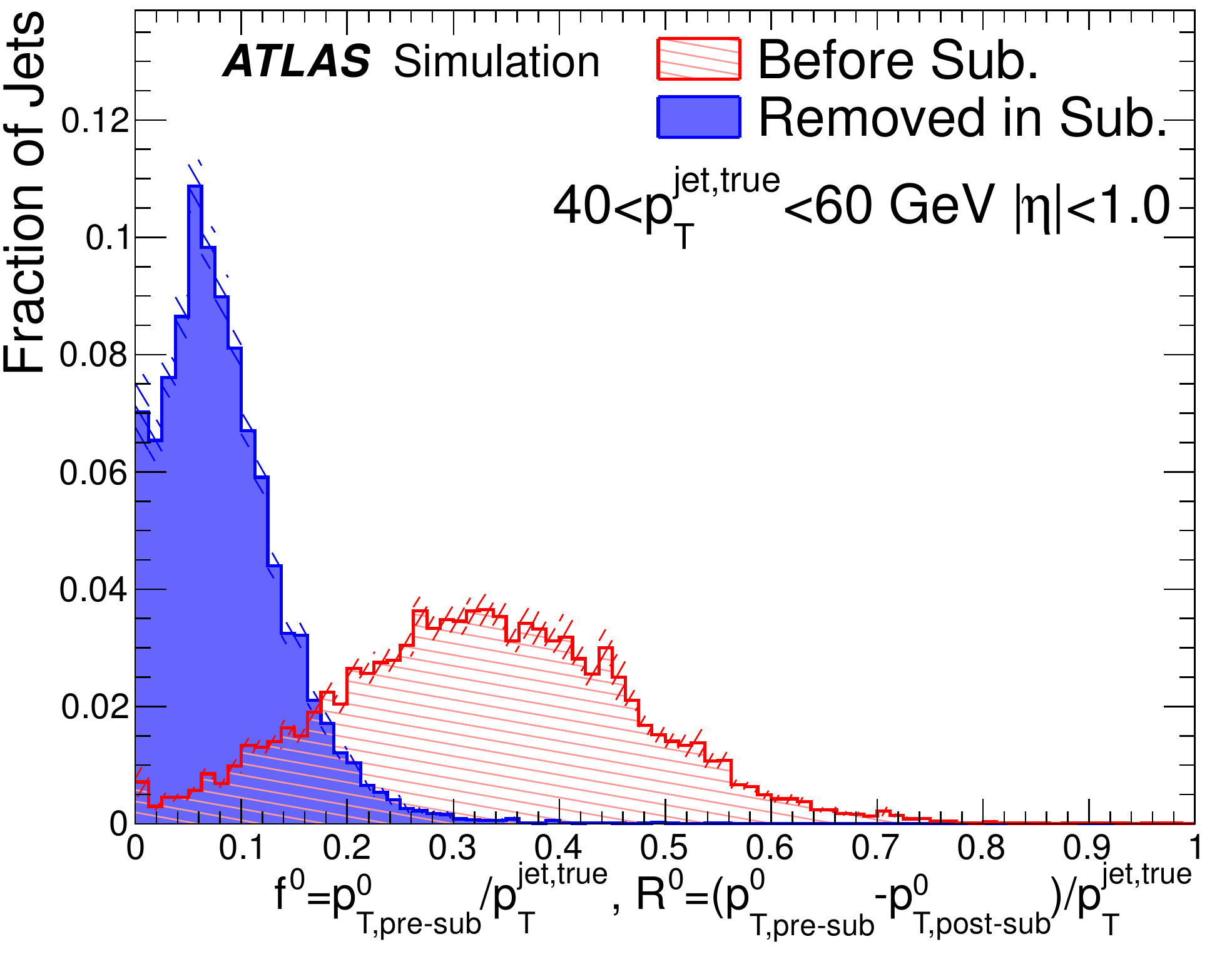}\label{fig:perf:sub_b}}
  \subfloat[]{\includegraphics[width=0.32\textwidth]{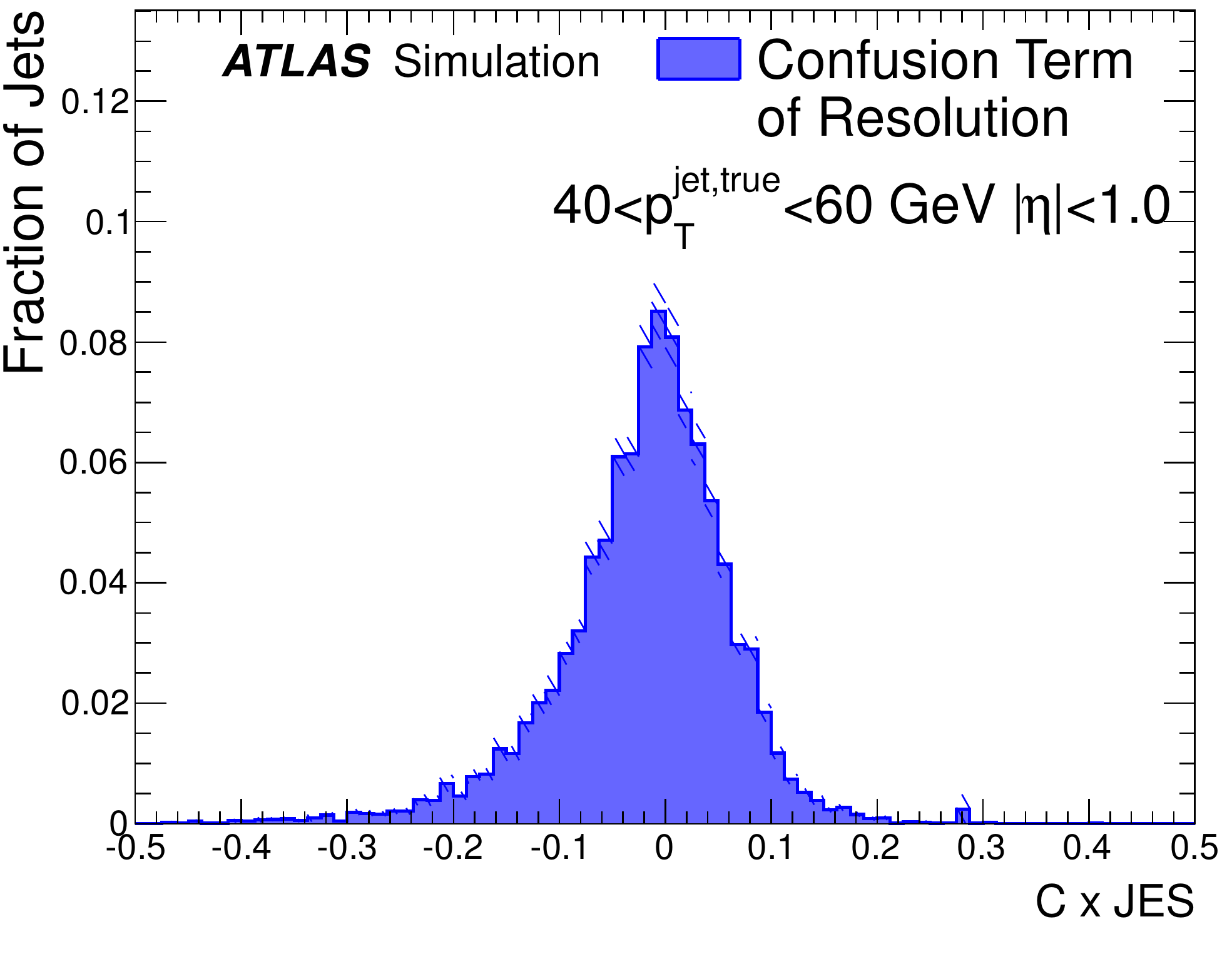}\label{fig:perf:sub_c}}
  \caption{The fractions of the jet calorimeter energy that have been incorrectly subtracted by the cell subtraction algorithm, 
    for jets with $40 < \pTtrue < \SI{60}{\GeV}$ and $|\eta|<1.0$ in dijet MC simulation without pile-up.
    The statistical uncertainty is indicated by the hatched bands.
    Subfigure (a) shows the fraction of jet transverse momentum carried by reconstructed tracks before subtraction $f^\pm$ (hashed) and the corresponding fraction after subtraction $R^\pm$ (solid);
    (b) shows the fraction of jet transverse momentum carried by particles without reconstructed tracks before subtraction $f^0$ (hashed) and the corresponding fraction after subtraction $R^0$ (solid); and
    (c) shows the confusion $C = R^\pm-R^0$, scaled up by the jet energy scale, derived as discussed in \Sect{\ref{sec:jet:cal}}.
  }
  \label{fig:perf:sub}
\end{figure}

Figure~\ref{fig:perf:sub} shows the fractions associated with the different classes of particle,
before and after the subtraction algorithm has been executed for jets with a true energy in the range \SIrange{40}{60}{\GeV}.
The confusion term is also shown,
multiplied by the jet energy scale factor that would be applied to these reconstructed jets,
such that its magnitude ($C \times \text{JES}$) is directly comparable to the reconstructed jet resolution.

Clearly, the subtraction does not perform perfectly, but most of the correct energy is removed -- the mean value of the confusion is \SI{-1}{\%}, with an RMS of \SI{7.6}{\%}.
The slight bias towards negative values suggests that the subtraction algorithm is more likely to remove additional neutral energy rather than to miss charged energy and the RMS gives an indication of the contribution from this confusion to the overall jet resolution.

\begin{figure}[htbp]
  \centering
  \subfloat[]{\includegraphics[width=0.45\textwidth]{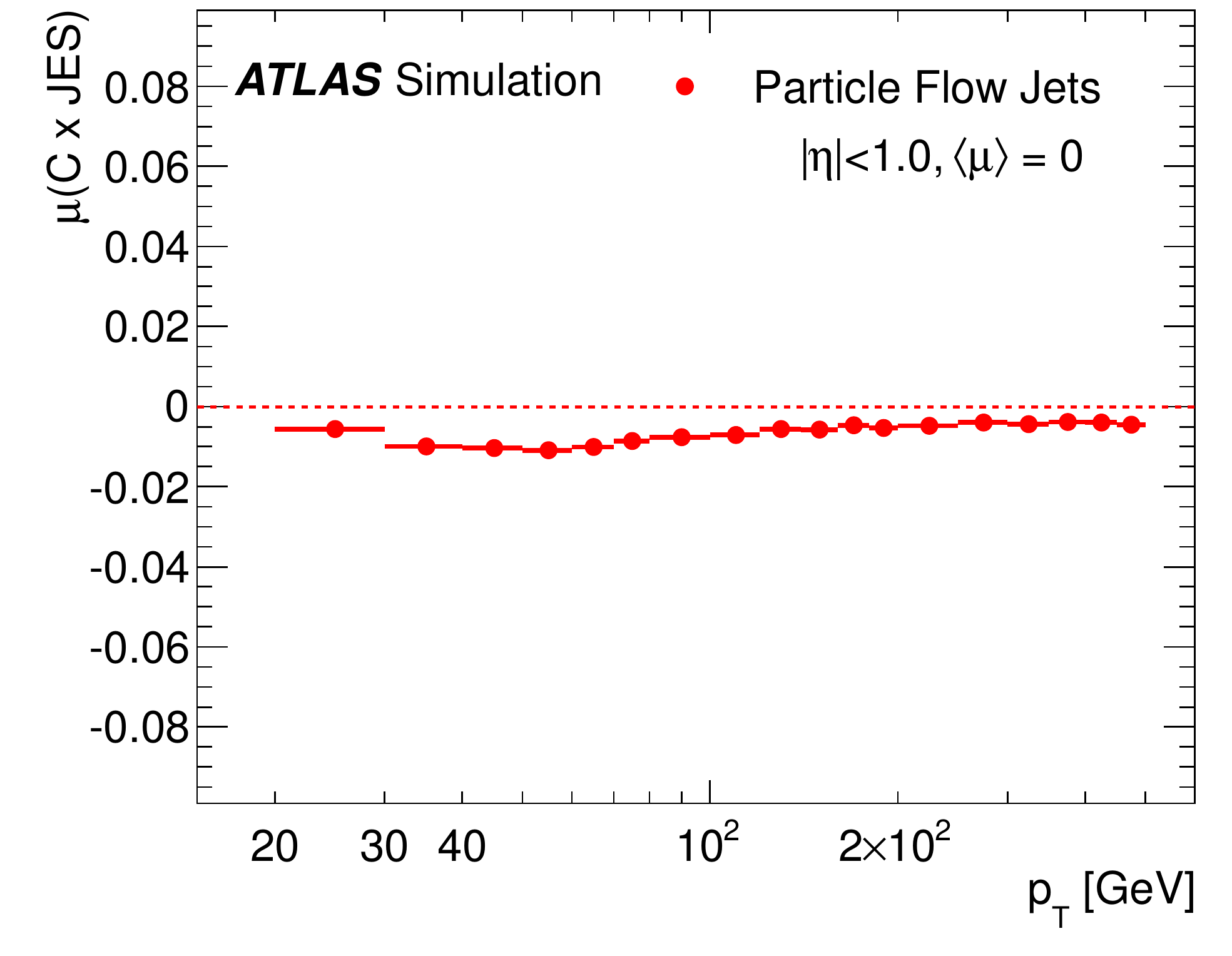}\label{fig:perf:pT:a}}
  \subfloat[]{\includegraphics[width=0.45\textwidth]{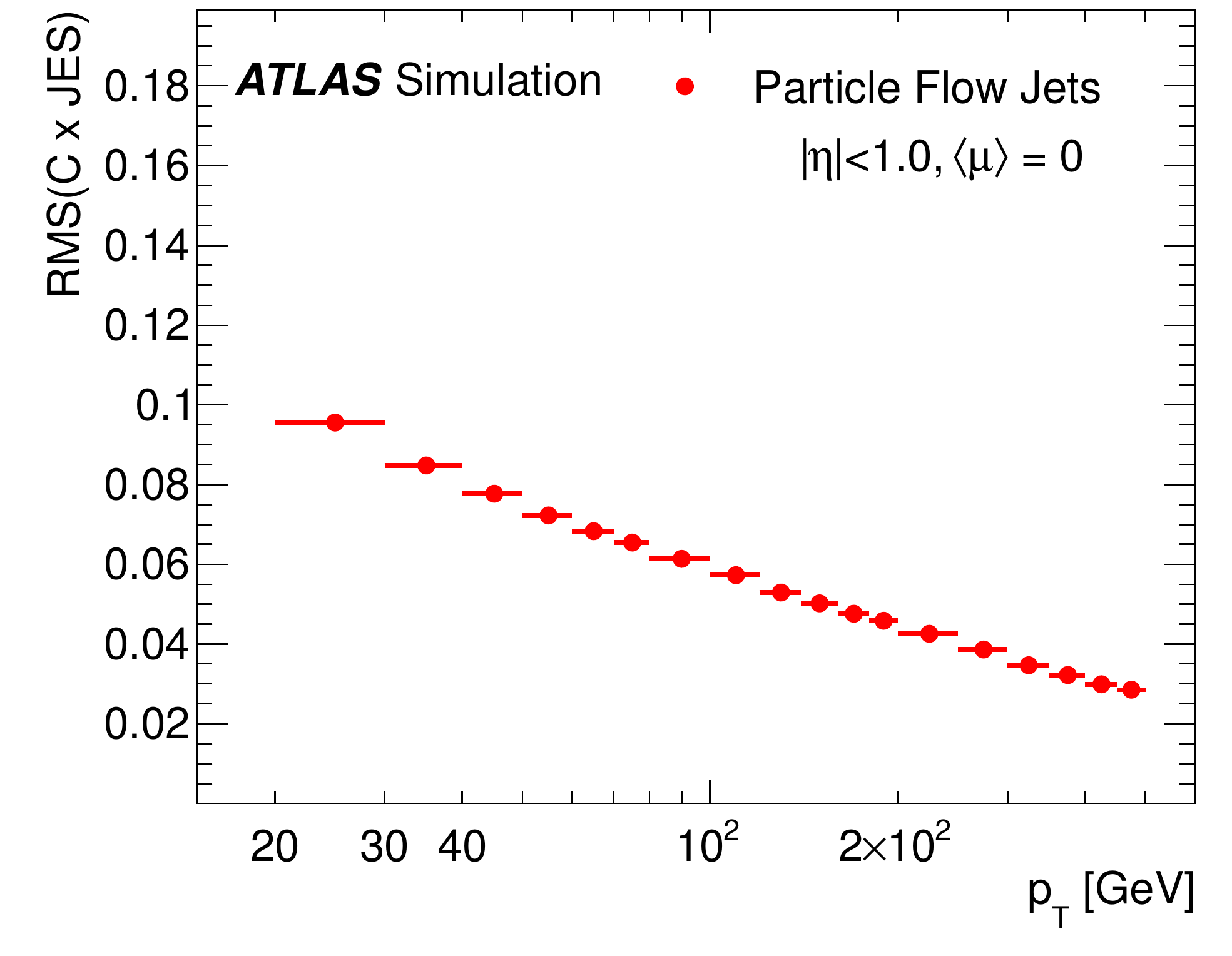}\label{fig:perf:pT:b}}
  \caption{
    As a function of the jet \pT, subfigure (a) shows the mean of the confusion term $C = R^\pm-R^0$, scaled up by the jet energy scale, derived as discussed in \Sect{\ref{sec:jet:cal}}, and (b) shows the RMS of this distribution. The error bars denote the statistical uncertainty. The MC samples used do not include pile-up.
  }
  \label{fig:perf:sub:pT}
\end{figure}

Figure~\ref{fig:perf:sub:pT} shows $C \times \text{JES}$ as a function of \pT.
The mean value of the JES weighted confusion remains close to zero and always within $\pm\SI{1.5}{\%}$,
showing that on average the algorithm removes the correct amount of energy from the calorimeter.
The RMS decreases with increasing \pT.
This is due to a combination of the particle \pT spectrum becoming harder, such that the efficiency of matching to the correct cluster increases;
the increasing difficulty of subtracting the hadronic showers in the denser environments of high-\pT jets;
and the fact that no subtraction is performed for tracks above \SI{40}{\GeV},
resulting in the fraction of the jet considered for subtraction decreasing with increasing jet \pT.

%------------------------------------------------------------------------------
\subsection{Visualising the subtraction}

\begin{figure}[htbp]
  \centering
  \subfloat[Before subtraction, no pile-up.]{\includegraphics[width=0.49\textwidth]{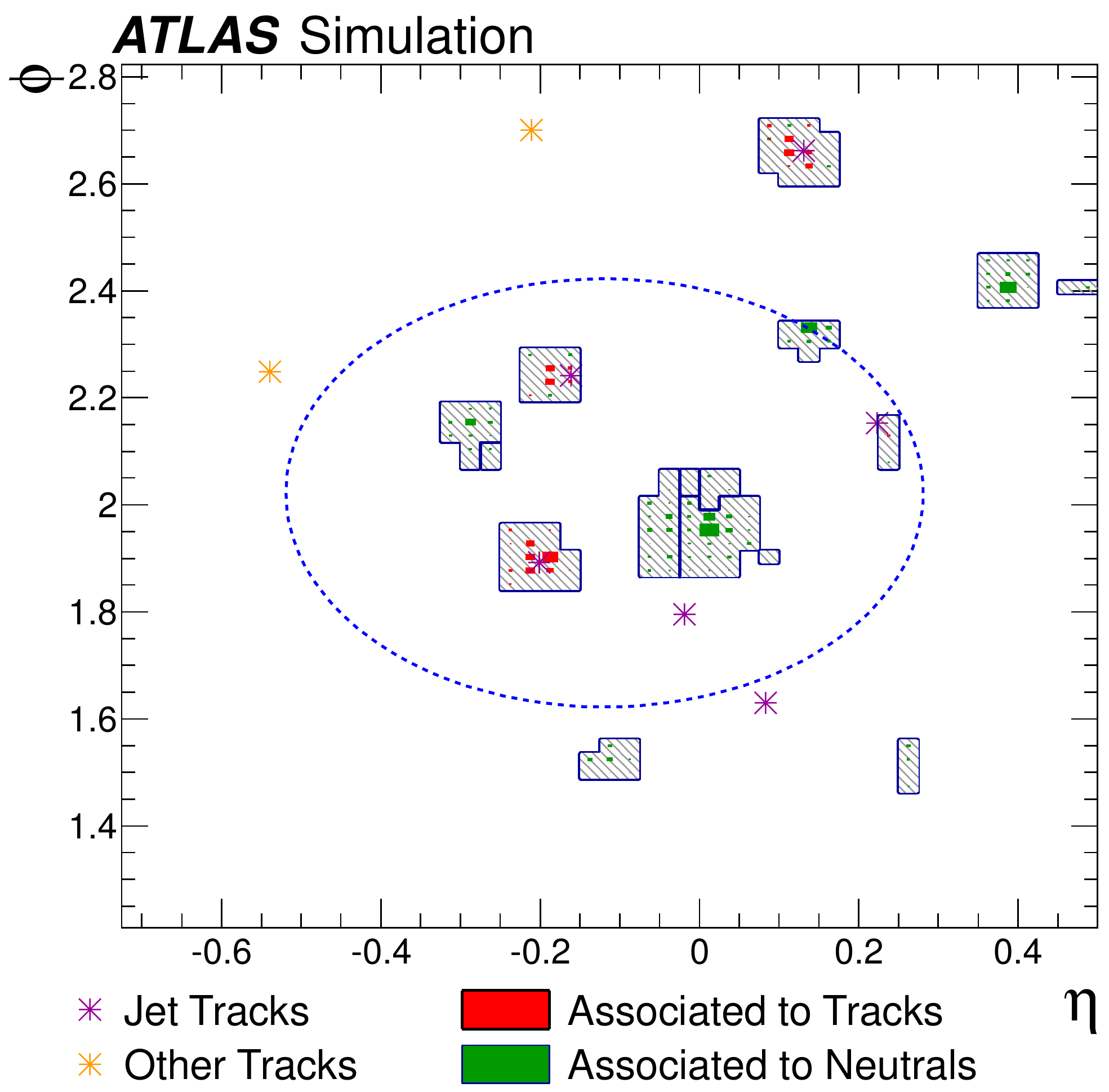}}
  \subfloat[After subtraction, no pile-up.]{\includegraphics[width=0.49\textwidth]{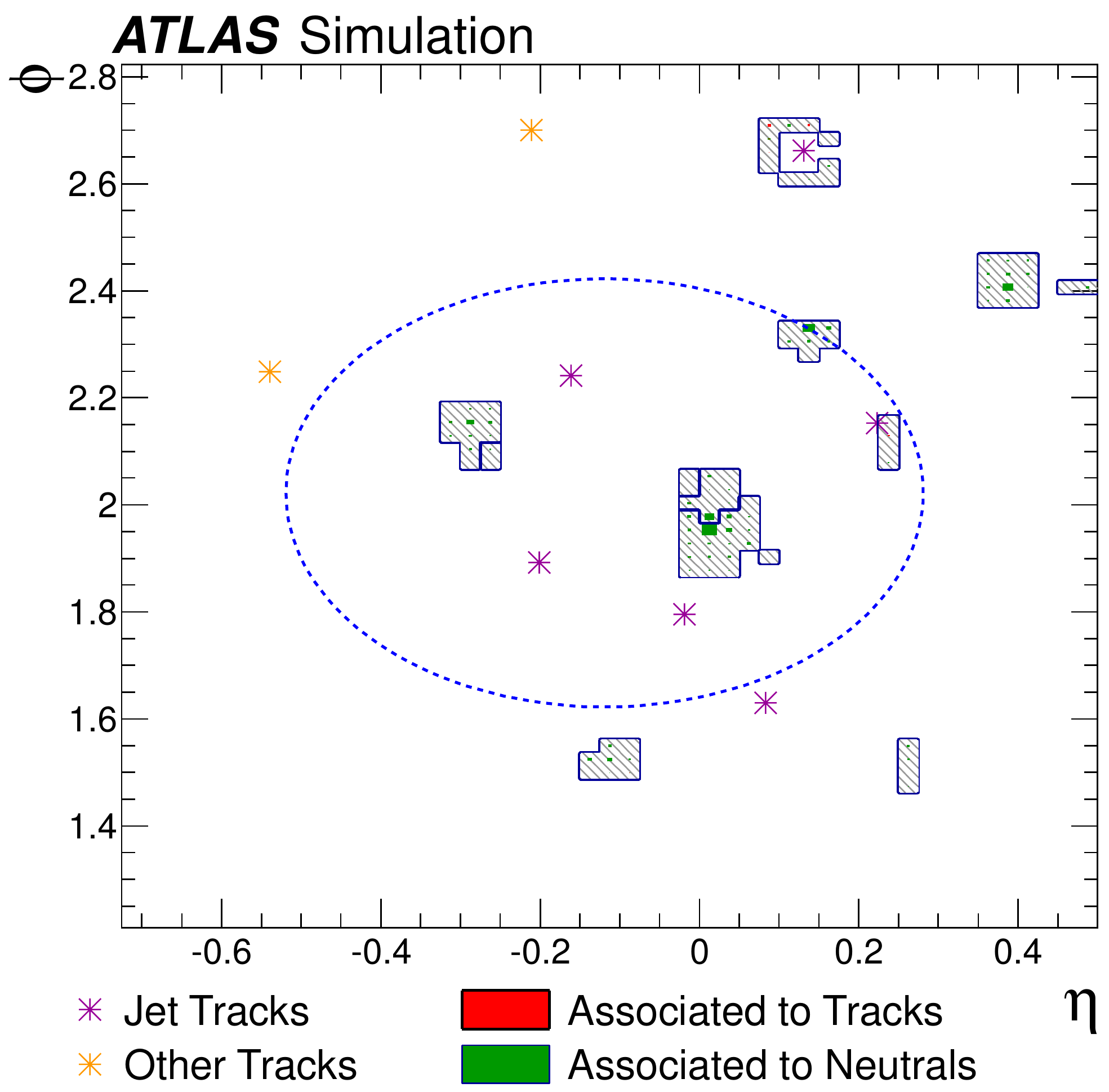}}\\
  \subfloat[Before subtraction, $\mu=40$.]{\includegraphics[width=0.49\textwidth]{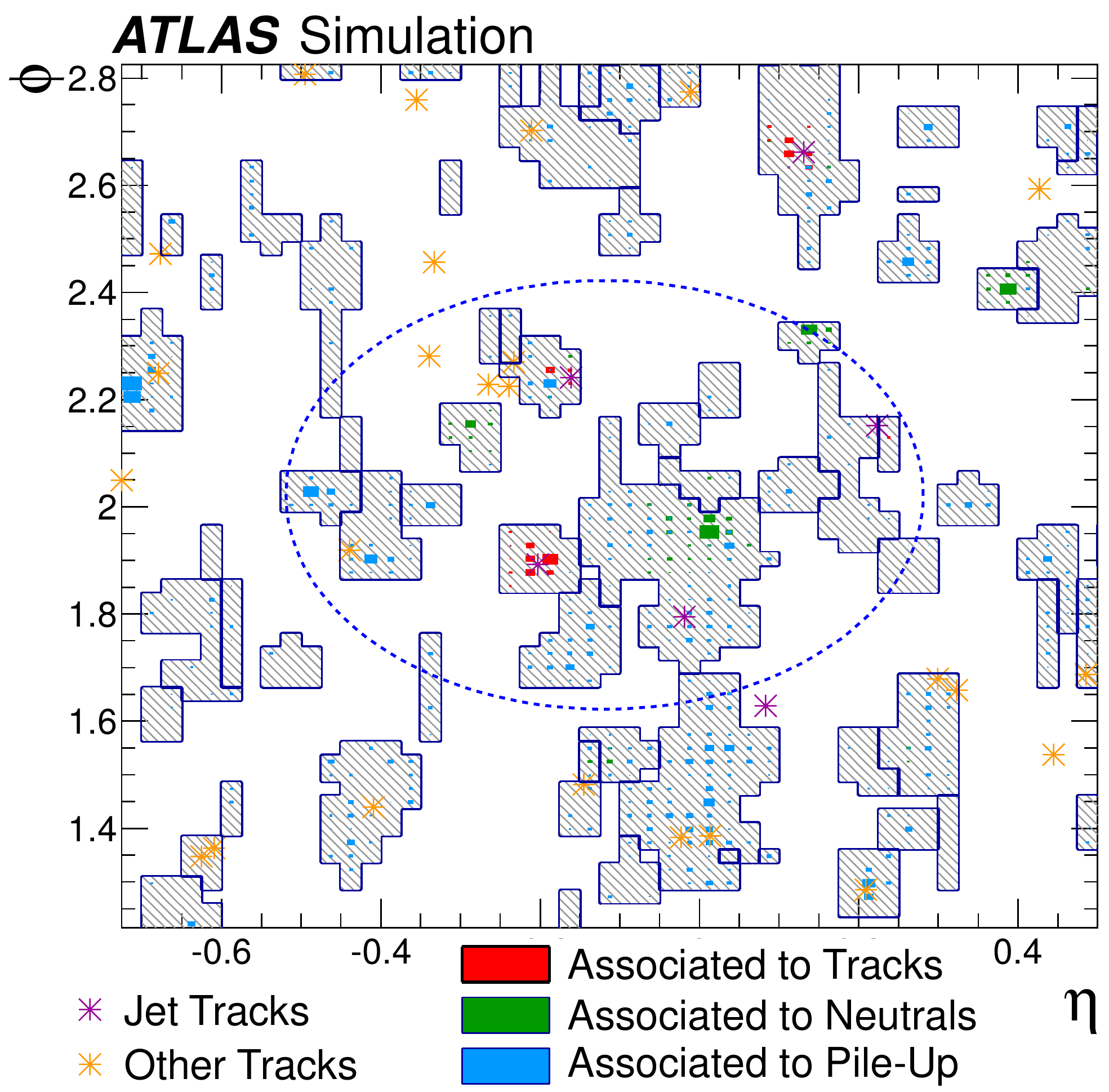}}
  \subfloat[After subtraction, $\mu=40$.]{\includegraphics[width=0.49\textwidth]{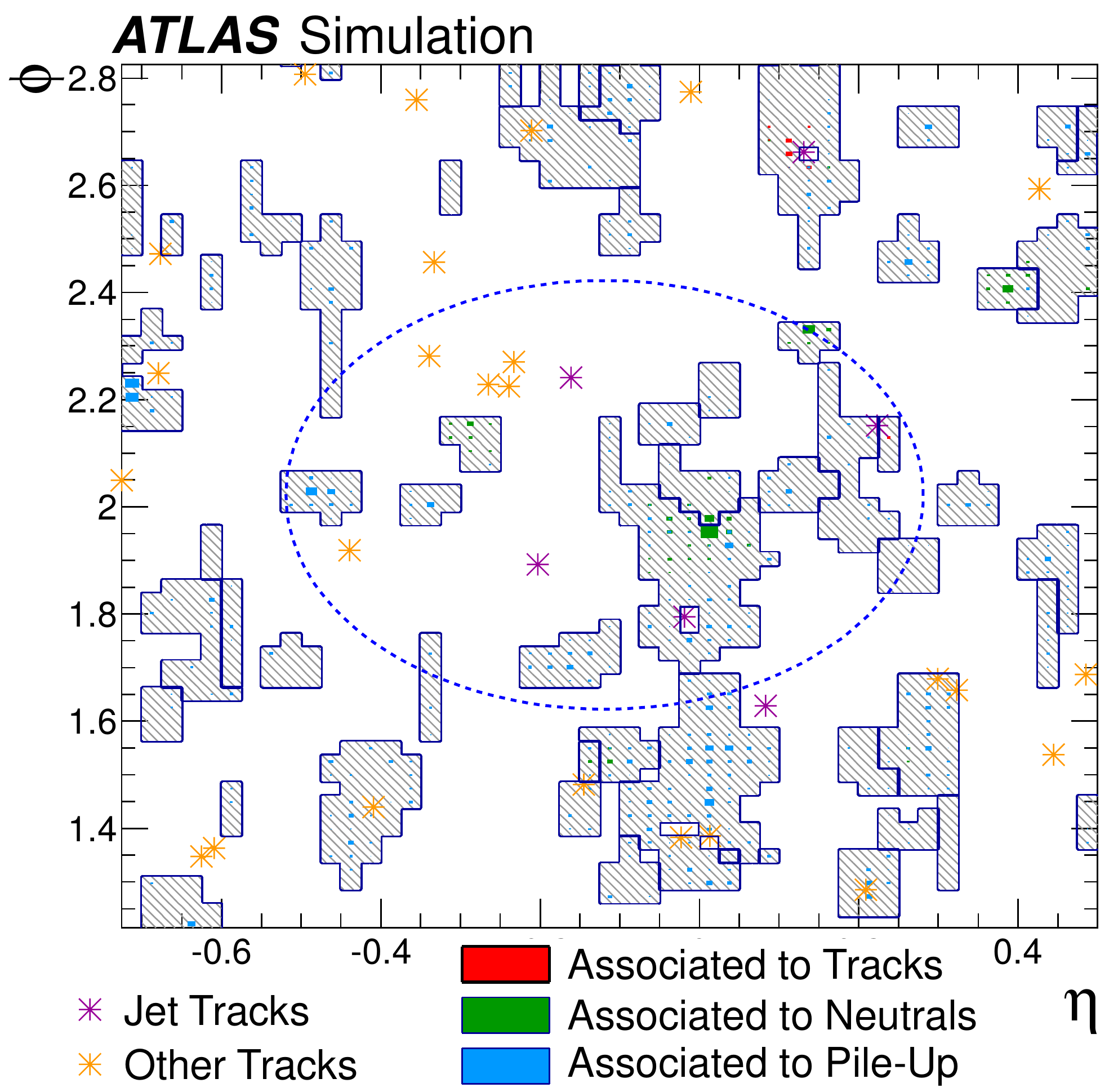}}
  \caption{A graphical display of the second layer of the EM calorimeter focusing on a \SI{30}{\GeV} truth jet, outlined by the ellipse.
  Asterisks indicate the positions of tracks extrapolated to the calorimeter, while blue framed rectangles mark the cells clustered into \topoclusters.
  The colour purple (dark) is used to indicate those tracks that are selected for particle flow jet reconstruction, 
  i.e.\ those matched to the nominal hard-scatter primary vertex (see \Sect{\ref{sec:jet:cal}}) 
  and clustered into the jet based on their momenta expressed at the perigee.
  Other tracks are shown in orange (light).
  Red and green boxes indicate reconstructed cell energies associated with truth particles where tracks have and have not been reconstructed.
  Subfigures (a) and (b) show the event without pile-up.
  Subfigures (c) and (d) show the same event with pile-up overlaid.
  Pile-up energy in the calorimeter is indicated by blue boxes.}
  \label{fig:perf:Evt}
\end{figure}

For a concrete demonstration of successes and failures of the subtraction algorithm, it is instructive to look at a specific event in the calorimeter.
Figure~\ref{fig:perf:Evt} illustrates the action of the algorithm in the second layer of the EM calorimeter,
where the majority of low-energy showers are contained.
The focus is on a region where a \SI{30}{\GeV} truth jet is present.
In general, the subtraction works well in the absence of pile-up,
as the two \topoclusters inside the jet radius with energy mainly associated with charged particles at truth level are entirely removed.
Nevertheless, examples can be seen where small mistakes are made.
For example, the algorithm additionally removes some cells containing neutral-particle energy from the \topocluster just above the track at $(\eta,\phi)=(0.0,1.8)$.

The figure also shows the same event, overlaid with pile-up corresponding to $\mu=40$.
Pile-up contributions are identified by subtracting the energy reconstructed without pile-up and are illustrated in blue.
The pile-up supplies many more energy deposits and tracks within the region under scrutiny.
However, the subtraction continues to function effectively, removing energy in the vicinity of pile-up tracks and hence the post-subtraction cell distribution more closely resembles that without pile-up, especially inside the jet radius.
Because tracks classified as originating from pile-up are ignored in jet reconstruction (see \Sect{\ref{sec:jet:cal}}), the jet energy after subtraction is mainly contaminated by neutral pile-up contributions.

%-------------------------------------------------------------------------------
% Jet reconstruction and calibration
%-------------------------------------------------------------------------------
% !TeX root = Pflow.tex
%-------------------------------------------------------------------------------
\section{Jet reconstruction and calibration}
\label{sec:jet:cal}
%-------------------------------------------------------------------------------

Improved jet performance is the primary goal of using particle flow reconstruction.
Particle flow jets are reconstructed using the \akt algorithm with radius parameter 0.4.
The inputs to jet reconstruction are the ensemble of positive energy \topoclusters surviving the energy subtraction step
and the selected tracks that are matched to the hard-scatter primary vertex.
These tracks are selected by requiring 
$|z_0\sin\theta| < \SI{2}{\mm}$, where $z_0$ is the distance of closest approach of the track to the hard-scatter primary vertex along the $z$-axis.
This criterion retains the tracks from the hard scatter,
while removing a large fraction of the tracks (and their associated calorimeter energy) from pile-up interactions~\cite{STDM-2010-01}.
Prior to jet-finding, the \topocluster $\eta$ and $\phi$ are recomputed with respect to the hard-scatter primary vertex position, rather than the detector origin.

Calorimeter jets are similarly reconstructed using the \akt algorithm with radius parameter 0.4, but take as input \topoclusters calibrated
at the LC-scale, uncorrected for the primary vertex position. For the purposes of jet calibration, truth jets are formed from stable
final-state particles excluding muons and neutrinos, using the \akt algorithm with radius parameter 0.4.\footnote{\enquote{Stable particles} are defined as those with proper lifetimes longer than \SI{30}{ps}.}

%-------------------------------------------------------------------------------
\subsection{Overview of particle flow jet calibration}

Calibration of these jets closely follows the scheme used for standard calorimeter jets described in Refs.~{\cite{ATLAS-CONF-2015-002, ATLAS-CONF-2015-017, ATLAS-CONF-2015-037, ATLAS-CONF-2015-057}} and is carried out over the range $20 < \pT < \SI{1500}{\GeV}$.
The reconstructed jets are first corrected for pile-up contamination using the jet ghost-area subtraction method~\cite{catchArea,areaSub}.
This is described in \Sect{\ref{sec:cal:pileup_corr}}.
A numerical inversion~\cite{ATLAS-CONF-2015-037} based on Monte Carlo events (see \Sect{\ref{sec:cal:jes}}) restores the jet response to match the average response at particle level.
Additional fluctuations in jet response are corrected for using a \textit{global sequential correction} process~\cite{ATLAS-CONF-2015-002}, which is detailed in \Sect{\ref{sec:jet:gsc}}.
No \textit{in situ} correction to data is applied in the context of these studies;
however, the degree of agreement between data and MC simulation is checked using the \pT balance of jets against a $Z$ boson decaying to two muons.

The features of particle flow jet calibration that differ from the calibration of calorimeter jets are discussed below,
and results from the different stages of the jet calibration are shown.

%-------------------------------------------------------------------------------
\subsection{Area-based pile-up correction}
\label{sec:cal:pileup_corr}

\begin{figure}[htbp]
  \centering
  \includegraphics[width=0.4\textwidth]{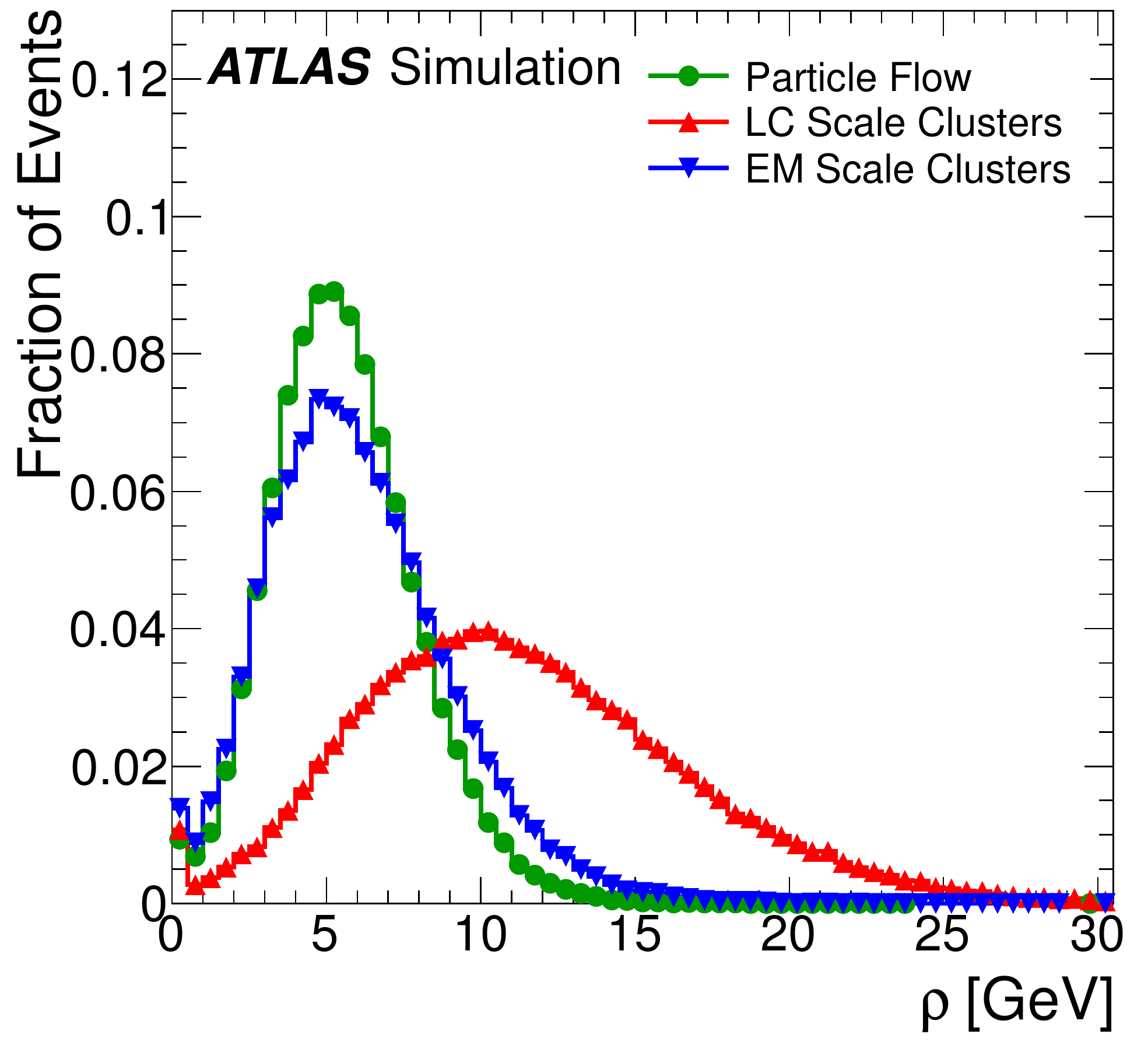}
  \caption{The distribution of the median transverse energy density $\rho$
    in dijet MC simulated events with similar pile-up to that measured in the 2012 data for different jet constituents.
  }
  \label{fig:JetRec:rho}
\end{figure}

The calorimeter jet pile-up correction uses a transverse energy density $\rho$ calculated from \topoclusters using $k_\mathrm{T}$ jets~\cite{Ellis:1993tq,Catani:1993hr},
for a correction of the form of $\rho$ multiplied by the area of the jet~\cite{areaSub}.
For particle flow jets, the transverse energy density therefore needs to be computed 
using charged and neutral particle flow objects to correctly account for the differences in the jet constituents.
As discussed above, the tracks associated to pile-up vertices are omitted,
eliminating a large fraction of the energy deposits from charged particles from pile-up interactions.
The jet-area subtraction therefore corrects for the impact of charged underlying-event hadrons,
charged particles from out-of-time interactions,
and more importantly, neutral particles from pile-up interactions.
This correction is evaluated prior to calibration of the jet energy scale.
Figure~\ref{fig:JetRec:rho} shows the distribution of the median transverse energy density $\rho$
in dijet MC events for particle flow objects and for \topoclusters.
%The particle flow $\rho$ is computed using the same particle flow objects used for jet reconstruction.
The \topocluster $\rho$ is calculated with the ensemble of clusters, calibrated either at the EM scale or LC scale, and the particle flow jets use \topoclusters calibrated at the EM scale.

The LC-scale energy density is larger than the EM-scale energy density due to the application of the cell weights to calibrate cells to the hadronic
scale. Compared to the EM- and LC-scale energy densities, $\rho$ has a lower per-event value for particle flow jets in 2012 conditions,
due to the reduced pile-up contribution.
The removal of the charged particle flow objects that are not associated with the hard-scatter primary vertex more than compensates for the higher energy scale for charged hadrons from the underlying event.

%-------------------------------------------------------------------------------
\subsection{Monte Carlo numerical inversion}
\label{sec:cal:jes}

\begin{figure}[htbp]
\centering
\includegraphics[width=0.6\textwidth]{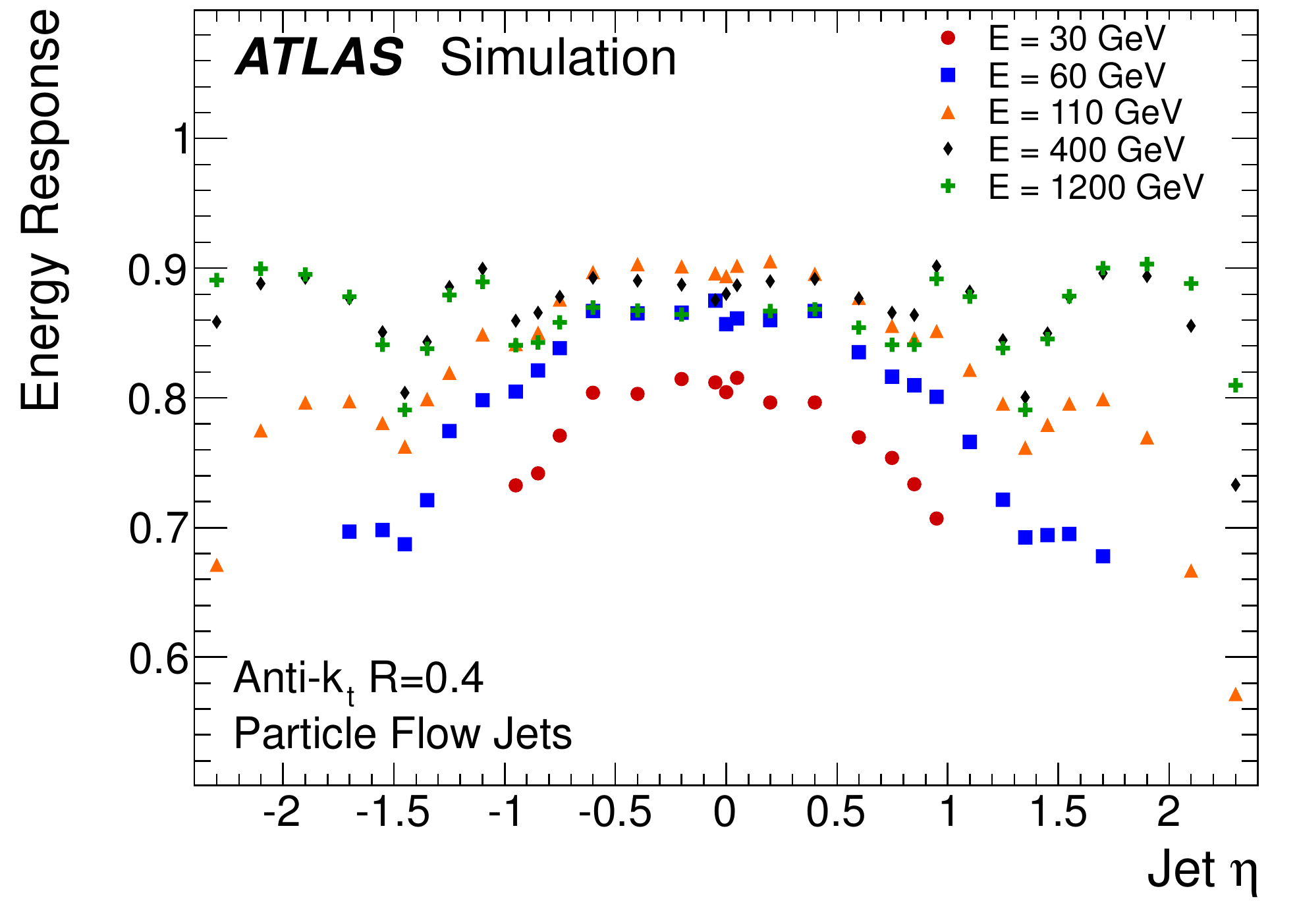}
\caption{The response $E_\text{reco}/E_\text{true}$ of \akt particle flow jets with radius parameter 0.4 from MC dijet samples with no pile-up, as a function of the jet $\eta$, measured prior to calibration, and binned in the energy of the matched truth jet.}
\label{fig:jetCal:response}
\end{figure}

Figure~\ref{fig:jetCal:response} shows the energy response $R=E_\text{reco}/E_\text{truth}$ prior to the MC-based jet energy scale correction.
The same numerical procedure as described in \Ref{\cite{ATLAS-CONF-2015-037}} is applied and
successfully corrects for the hadron response,
at a similar level to that observed in \Ref{\cite{ATLAS-CONF-2015-037}}.

%-------------------------------------------------------------------------------
\subsection{Global sequential correction}
\label{sec:jet:gsc}

The numerical inversion calibration restores the average reconstructed jet energy to the mean value of the truth jet energy,
accounting for variations in the jet response due to the jet energy and pseudorapidity.
However, other jet characteristics such as the flavour of the originating parton and the composition of the hadrons created in jet fragmentation may cause further differences in the response.
A \textit{global sequential correction}~\cite{ATLAS-CONF-2015-002} that makes use of additional observables adapts the jet energy calibration to account for such variations, thereby improving the jet resolution without changing the scale.
The variables used for particle flow jets are not the same as those used for
calorimeter jets, as tracks have already been used in the calculation
of the energy of the jet constituents.

As the name implies, the corrections corresponding to each variable are applied consecutively.
Three variables are used as inputs to the correction:
\begin{enumerate}
\item the fraction of the jet energy measured from constituent tracks (charged fraction), i.e.\ those tracks associated to the jet;
\item the fraction of jet energy measured in the third EM calorimeter layer;
\item the fraction of jet energy measured in the first Tile calorimeter layer.
\end{enumerate}
The first of these variables allows the degree of under-calibrated signal,
due to the lower energy deposit of hadrons in the hadronic calorimeter, to be determined.
The calorimeter-layer energy fractions allow corrections to be made for the energy lost in dead material between the LAr and Tile calorimeters.

%-------------------------------------------------------------------------------
\subsection{\textit{In situ} validation of JES}

\begin{figure}[htbp]
\centering
\includegraphics[width=0.6\textwidth]{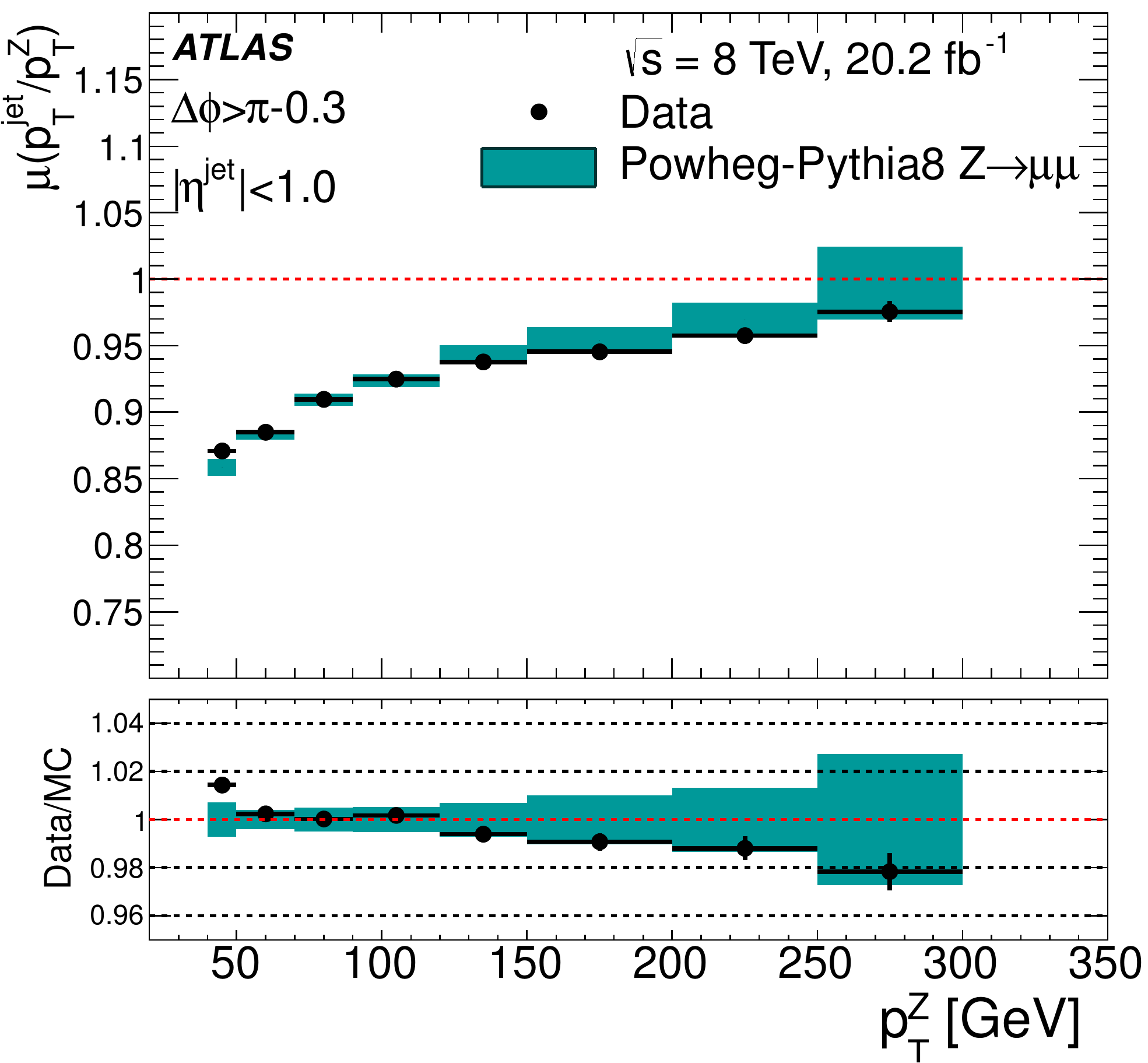}
\caption{The mean value of the ratio of the transverse momentum of a jet to that of the reconstructed $Z$ boson decaying to $\mu\mu$.
  The uncertainties shown are statistical.}
\label{fig:jetCal:Zbal}
\end{figure}

A full \textit{in situ} calibration and evaluation of the uncertainties on the JES~\cite{PERF-2012-01} is not performed for these studies.
However, to confirm that the ATLAS MC simulation describes the particle flow jet characteristics well enough
to reproduce the jet energy scale in data, similar methods are used to validate the jet calibration.
A sample of $Z \rightarrow \mu\mu$ events with a jet balancing the $Z$ boson is used for the validation.
A preselection is made using the criteria described in \Sect{\ref{sec:dataset}}.
The particle flow algorithm is run on these events and further requirements,
discussed below, are applied.
The jet with the highest \pT ($\text{j1}$) and the reconstructed $Z$ boson are required to be well separated in azimuthal angle, $\Delta\phi>\pi-0.3$.
Events with any additional jet within $|\eta|<4.5$, with $\pT^{\,\text{j2}} > \SI{20}{\GeV}$
or $\pT^{\,\text{j2}}>0.1 \pT^{\,\text{j1}}$, are vetoed,
where $\text{j2}$ denotes the jet with the second highest \pT.
In \Fig{\ref{fig:jetCal:Zbal}}, the mean value of the ratio of the transverse momentum of the jet to that of the $Z$ boson 
is shown for data and MC simulation for jets with $|\eta|<1$.
The mean value is determined using a Gaussian fit to the distribution in bins of the $Z$-boson \pT.
The double-ratio of data to MC simulation is also shown.
The simulation reproduces the data to within \SI{2}{\%}, and in general is consistent with the data points within statistical uncertainties.
At high \pT the data/MC ratio is expected to tend towards that of the calorimeter jets~\cite{ATLAS-CONF-2015-037, ATLAS-CONF-2015-057},
as a large fraction of the jet’s energy is carried by particles above the cut made on the track momentum. For \pT > 200 GeV it is observed that the 
jet energy scale of calorimeter jets in data is typically \SI{0.5}{\%} below that in simulation.

%-------------------------------------------------------------------------------
% Resolution of jets in Monte Carlo simulation
%-------------------------------------------------------------------------------
% !TeX root = Pflow.tex
%-------------------------------------------------------------------------------
\section{Resolution of jets in Monte Carlo simulation}
\label{sec:jet:res}
%-------------------------------------------------------------------------------

The largest expected benefit from using the particle flow reconstruction as input to jet-finding
is an improvement of the jet energy and angular resolution for low-\pT jets.
In this section, the jet resolution achieved with particle flow methods is compared with that attained using standard calorimeter jet reconstruction.

%-------------------------------------------------------------------------------
\subsection{Transverse momentum resolution}\label{sec:jet:pt}

\begin{figure}[htbp]
  \centering
  \subfloat[]{\includegraphics[width=0.48\textwidth]{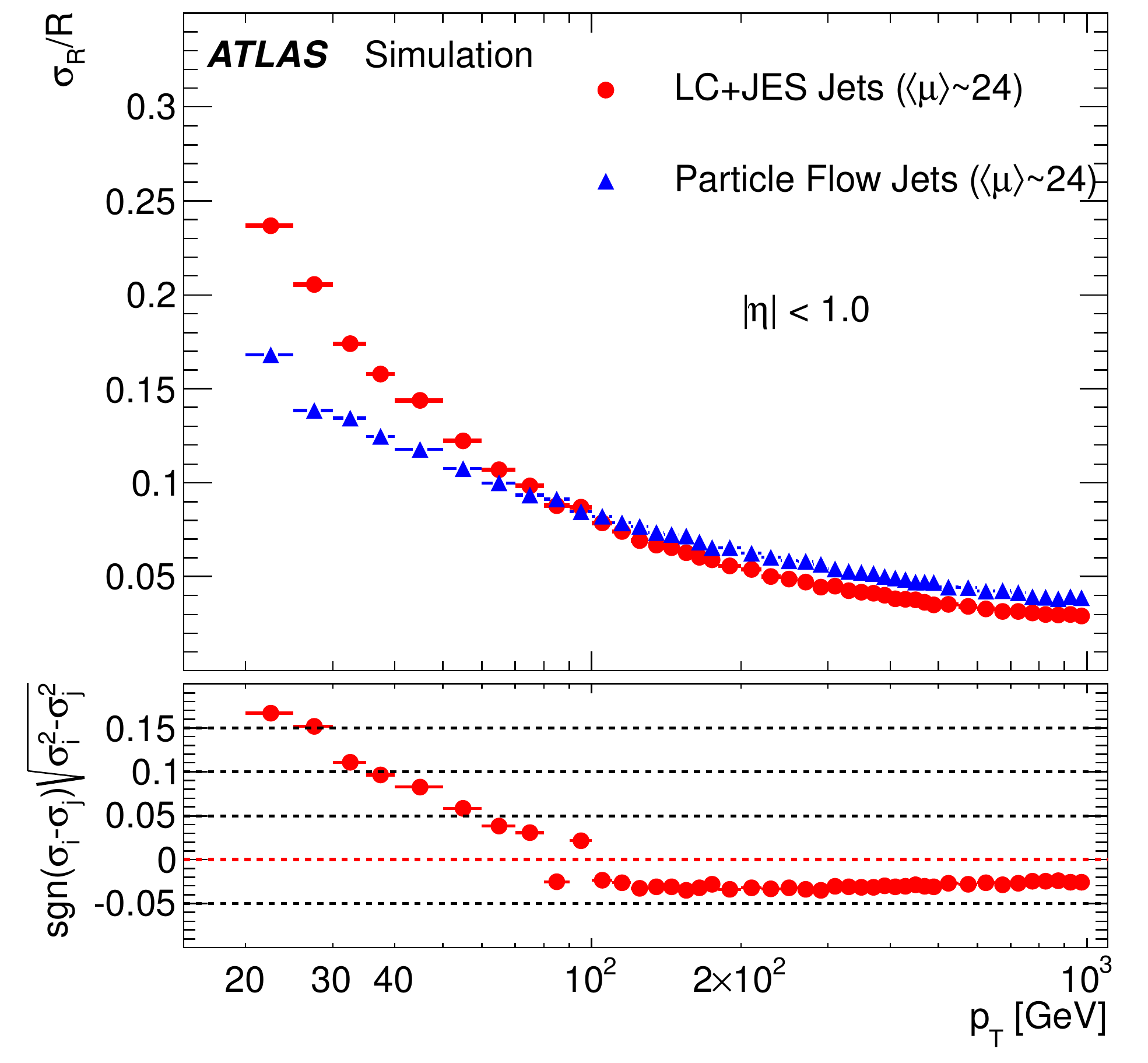}}\quad
  \subfloat[]{\includegraphics[width=0.48\textwidth]{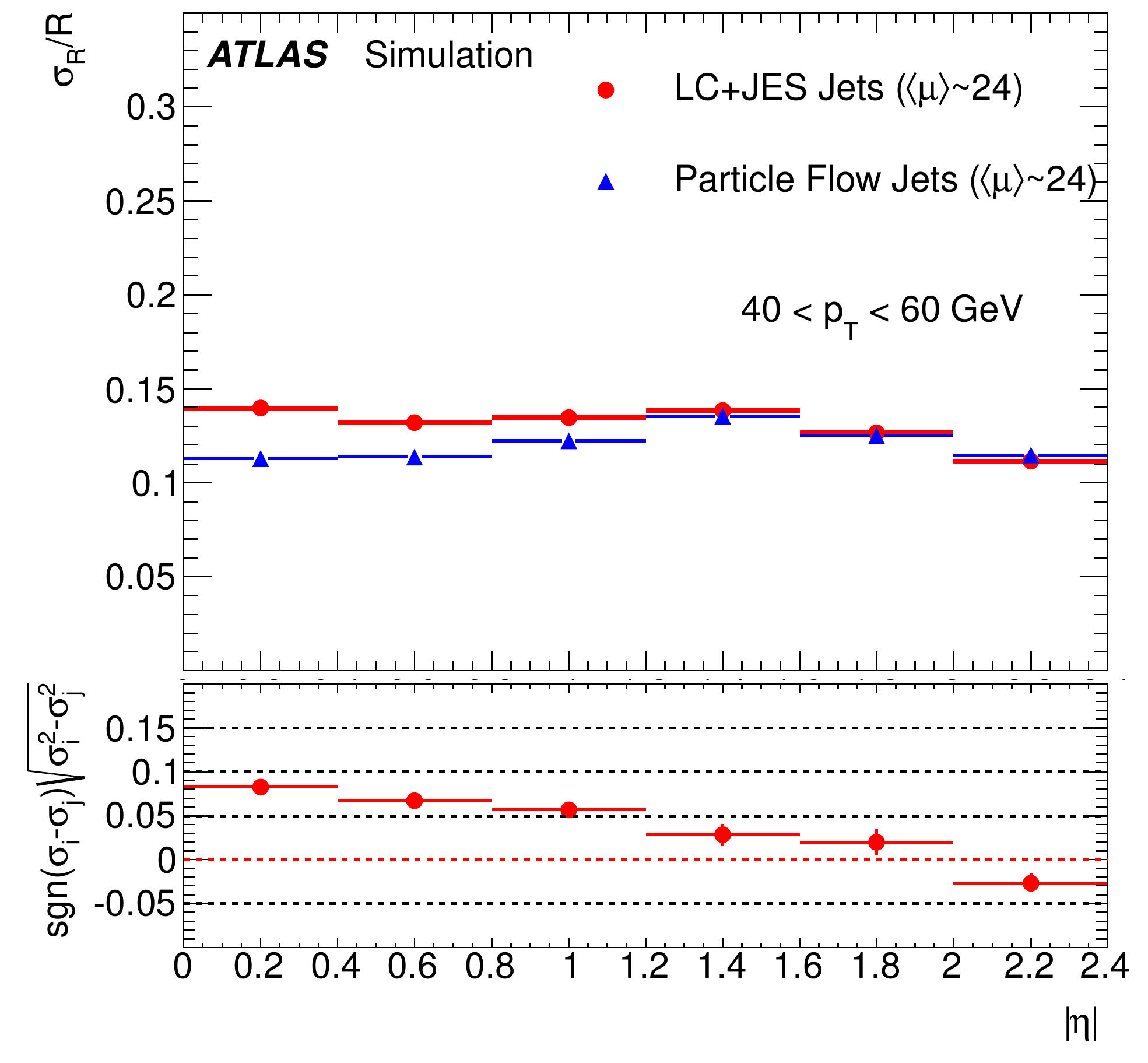}}
  \caption{The jet transverse momentum resolution as determined in dijet MC events for calorimeter jets and particle flow jets. Subfigure (a) shows the resolution as a function of \pT for jets with $|\eta|<1.0\,$ and (b) shows the resolution as a function of $|\eta|\,$ for jets with $40<\pT<\SI{60}{\GeV}$. Simulated pile-up conditions are similar to the data-taking in 2012. To quantify the difference in resolution between particle flow and calorimeter jets, the lower figure shows the square root of the difference of the squares of the resolution for the two classes of jets.
  }
  \label{fig:jetRes}
\end{figure}

In \Fig{\ref{fig:jetRes}}, the relative jet transverse momentum resolution for particle flow and calorimeter jets is shown as a function of jet transverse momentum for jets in the pseudorapidity range $|\eta|<1.0$, and as a function of $|\eta|$ for jets with $40<\pT<\SI{60}{GeV}$.
Particle flow jets are calibrated using the procedures described in \Sect{\ref{sec:jet:cal}}, while calorimeter jets use the more detailed procedure described in Refs.~{\cite{ATLAS-CONF-2015-002, ATLAS-CONF-2015-017, ATLAS-CONF-2015-037, ATLAS-CONF-2015-057}}.
The particle flow jets perform better than calorimeter jets at transverse momenta of up to $\SI{90}{GeV}$ in the central region, benefiting from the improved scale for low-\pT hadrons and intrinsic pile-up suppression (elaborated on in \Sect{\ref{sec:jet:PU}}).
However, at high transverse momenta, the particle flow reconstruction performs slightly worse than the standard reconstruction.
This is due to two effects.
The dense core of a jet poses a challenge to tracking algorithms,
causing the tracking efficiency and accuracy to degrade in high-\pT jets.
Furthermore, the close proximity of the showers within the jet increases the degree of confusion during the cell subtraction stage.
To counteract this, the track \pT used for particle flow reconstruction is required to be $< \SI{40}{\GeV}$ for the 2012 data.
Alternative solutions,
such as smoothly disabling the algorithm for individual tracks as the particle environment becomes more dense,
are expected to restore the particle flow jet performance to match that of the calorimeter jets at high energies. The benefits of particle flow also diminish toward the more forward regions as the cell granularity decreases, as shown in \Fig{\ref{fig:jetRes}(b)}

\begin{figure}[htbp]
  \centering
  \subfloat[]{\includegraphics[width=0.48\textwidth]{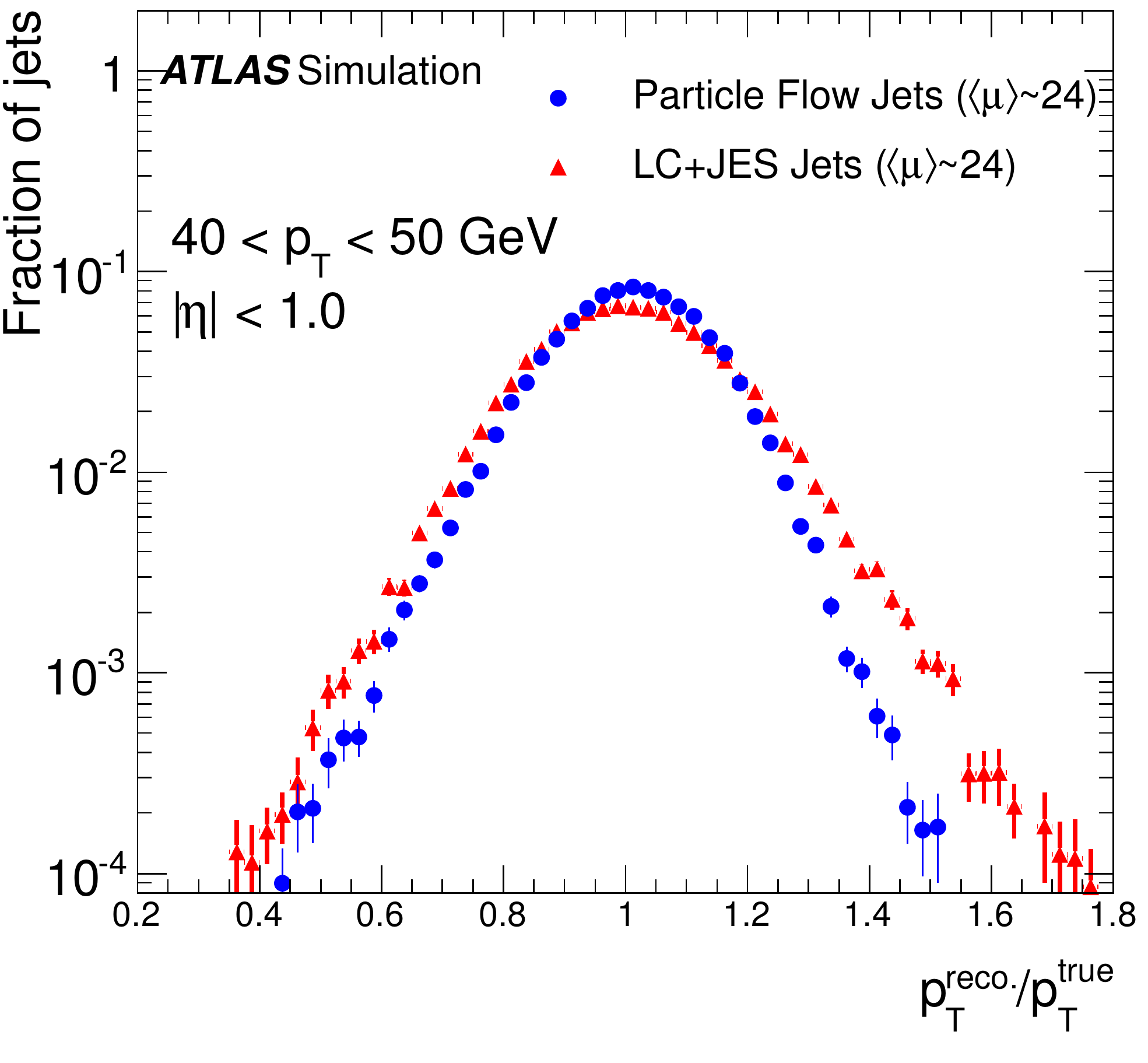}}\quad
  \subfloat[\label{fig:jetRes2:highpt}]{\includegraphics[width=0.48\textwidth]{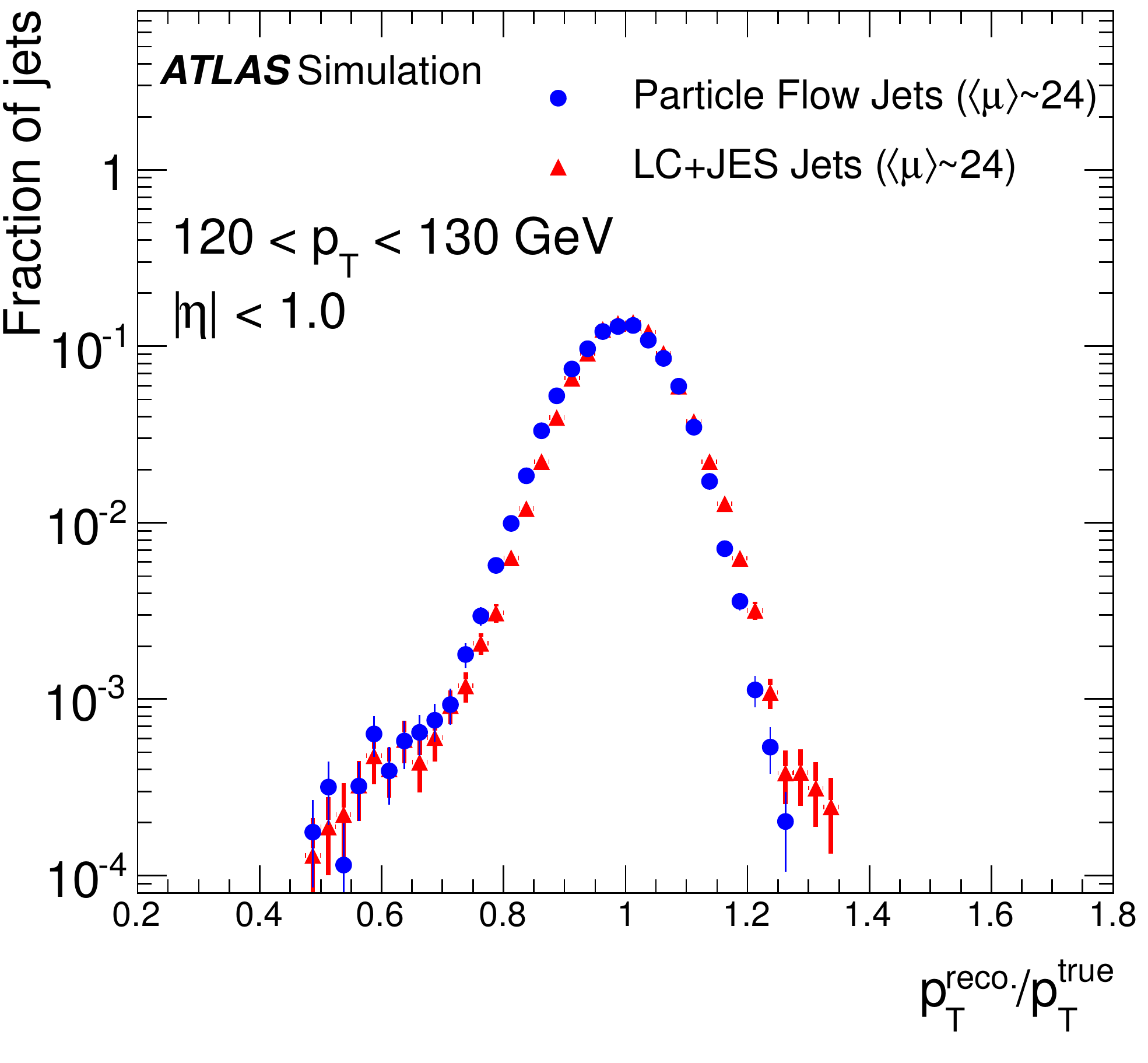}}
  \caption{The jet transverse momentum response distribution as determined in dijet MC events for calorimeter jets and particle flow jets. Two different \pT bins as shown; (a) $40<\pT<\SI{50}{\GeV}$ and (b) $120<\pT<\SI{130}{\GeV}$. Simulated pile-up conditions are similar to the data-taking in 2012. 
  %To quantify the difference in resolution between particle flow and calorimeter jets, the lower figure shows the square root of the difference of the squares of the resolution for the two classes of jets.
  }
  \label{fig:jetRes2}
\end{figure}

In \Fig{\ref{fig:jetRes2}}, the underlying distributions of the ratio of reconstructed to true \pT are shown for two different jet \pT bins.
This demonstrates that the particle flow algorithm does not introduce significant tails in the response at either low or high \pT.
The low-side tail visible in \Fig{\ref{fig:jetRes2:highpt}} is present in both calorimeter and particle flow jets and is caused by dead material and inactive detector regions.

%-------------------------------------------------------------------------------
\subsection{Angular resolution of jets}\label{sec:jet:etaphi}

Besides improving the \pT resolution of jets, the particle flow algorithm is expected to improve the angular ($\eta, \phi$) resolution of jets.
This is due to three different effects.
Firstly, usage of tracks to measure charged particles results in a much better angular resolution for individual particles than that obtained using \topoclusters,
because the tracker's angular resolution is far superior to that of the calorimeter.
Secondly, the track four-momentum can be determined at the perigee, before the charged particles have been spread out by the magnetic field, thereby improving the $\phi$ resolution for the jet.
Thirdly, the suppression of charged pile-up particles should also reduce mismeasurements of the jet direction.

\begin{figure}[htbp]
  \centering
  \subfloat[]{\includegraphics[width=0.48\textwidth]{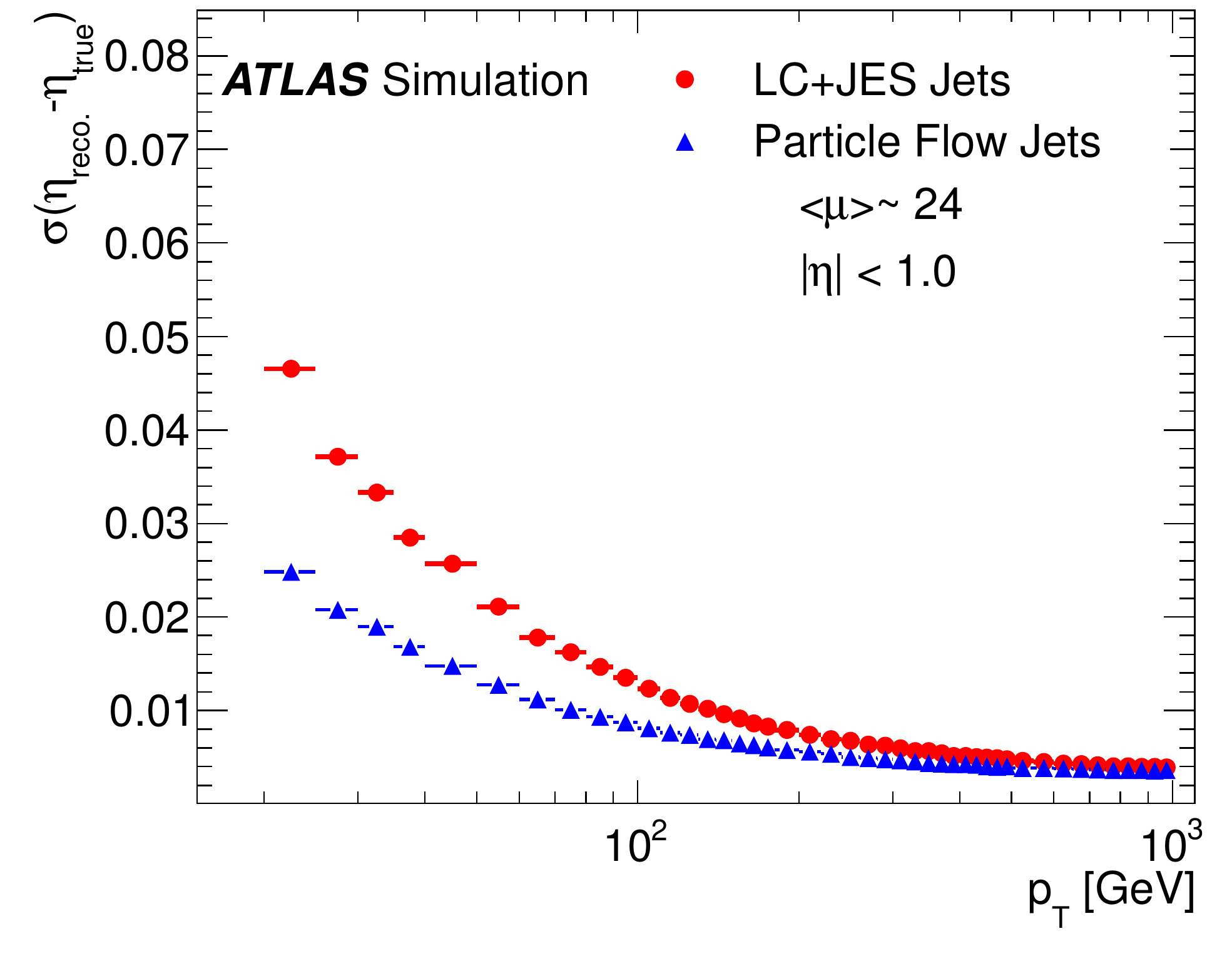}}\quad
  \subfloat[]{\includegraphics[width=0.48\textwidth]{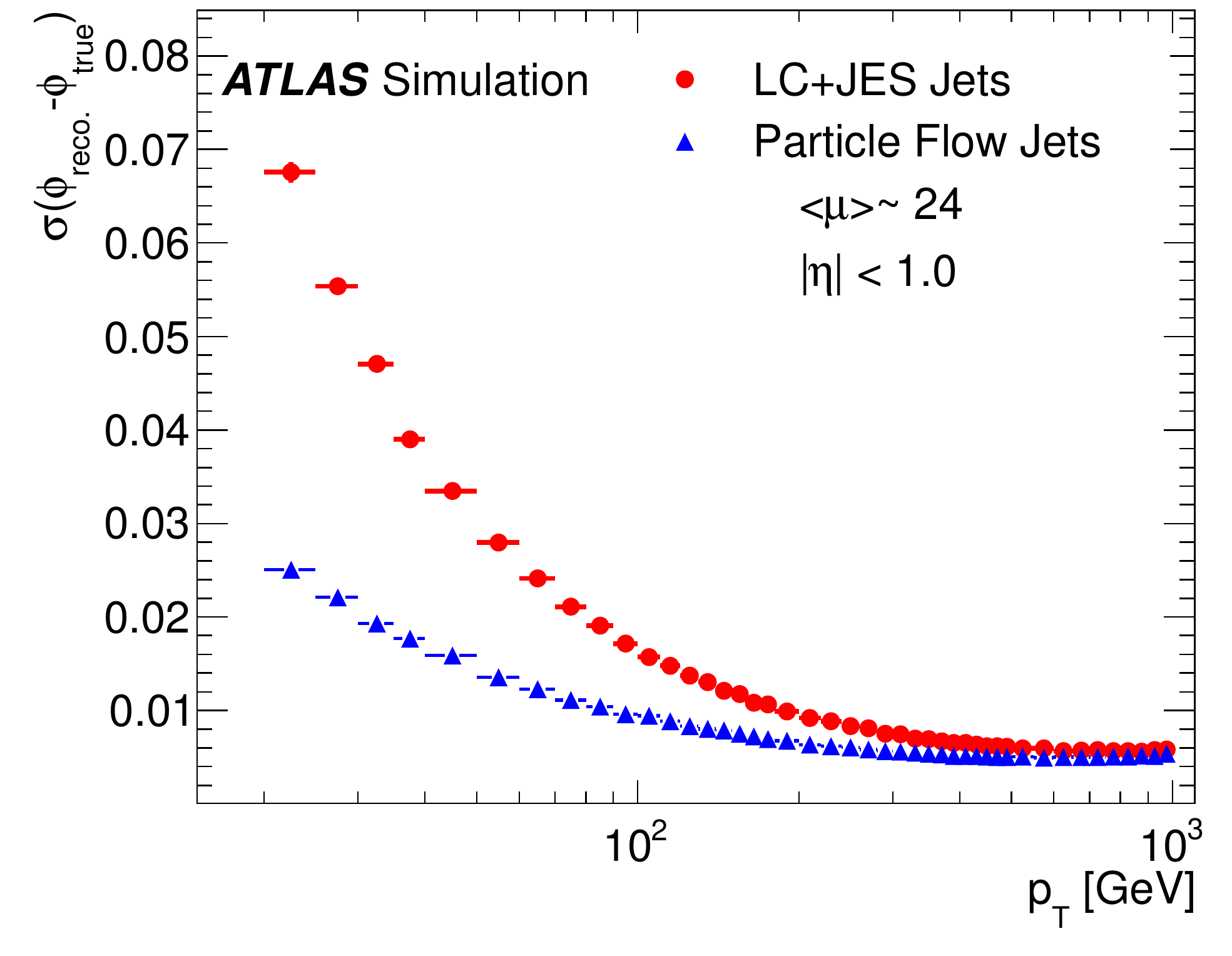}}
  \caption{The angular resolution, (a) in $\eta$ and (b) in $\phi$,
    as a function of the jet \pT, determined in dijet MC simulation
    by fitting Gaussian functions to the difference between the reconstructed and truth quantities.
    Conditions are similar to the data-taking in 2012.
  }
  \label{fig:AngularRes}
\end{figure}

Figure~\ref{fig:AngularRes} shows the angular resolution in $\eta$ and $\phi$
as a function of the reconstructed jet transverse momentum
for particle flow and calorimeter jets.
It is determined from the standard deviation of a Gaussian fit over $\pm 1.5\sigma$ to the difference between the $\eta$ and $\phi$ values for the reconstructed and matched truth ($\dR<0.3$) jets in the central region.
At low \pT, where the three effects described above are expected to be more important,
significant improvements are seen in both the $\eta$ and $\phi$ resolutions.
It is interesting to note that for particle flow jets the $\eta$ and $\phi$ resolutions are similar, while for calorimeter jets the $\phi$ resolution is worse due to the aforementioned effect of the magnetic field on charged particles.

%-------------------------------------------------------------------------------
% Effect of pile-up on the jet resolution and rejection of pile-up jets
%-------------------------------------------------------------------------------
% !TeX root = Pflow.tex
%-------------------------------------------------------------------------------
\section{Effect of pile-up on the jet resolution and rejection of pile-up jets}
\label{sec:jet:PU}
%-------------------------------------------------------------------------------

At the design luminosity of the LHC, and even in 2012 data-taking conditions,
in- and out-of-time pile-up contribute significantly to the signals measured in the ATLAS detector,
increasing the fluctuations in jet energy measurements. 
The pile-up suppression inherent in the particle flow reconstruction and 
the calibration of charged particles through the use of tracks
significantly mitigates the degradation of jet resolution from pile-up and eliminates jets reconstructed from pile-up deposits,
making the particle flow method a powerful tool,
especially as the LHC luminosity increases.

%-------------------------------------------------------------------------------
\subsection{Pile-up jet rate}

\begin{figure}[htbp]
  \centering
  \subfloat[]{%
    \includegraphics[width=0.45\textwidth]{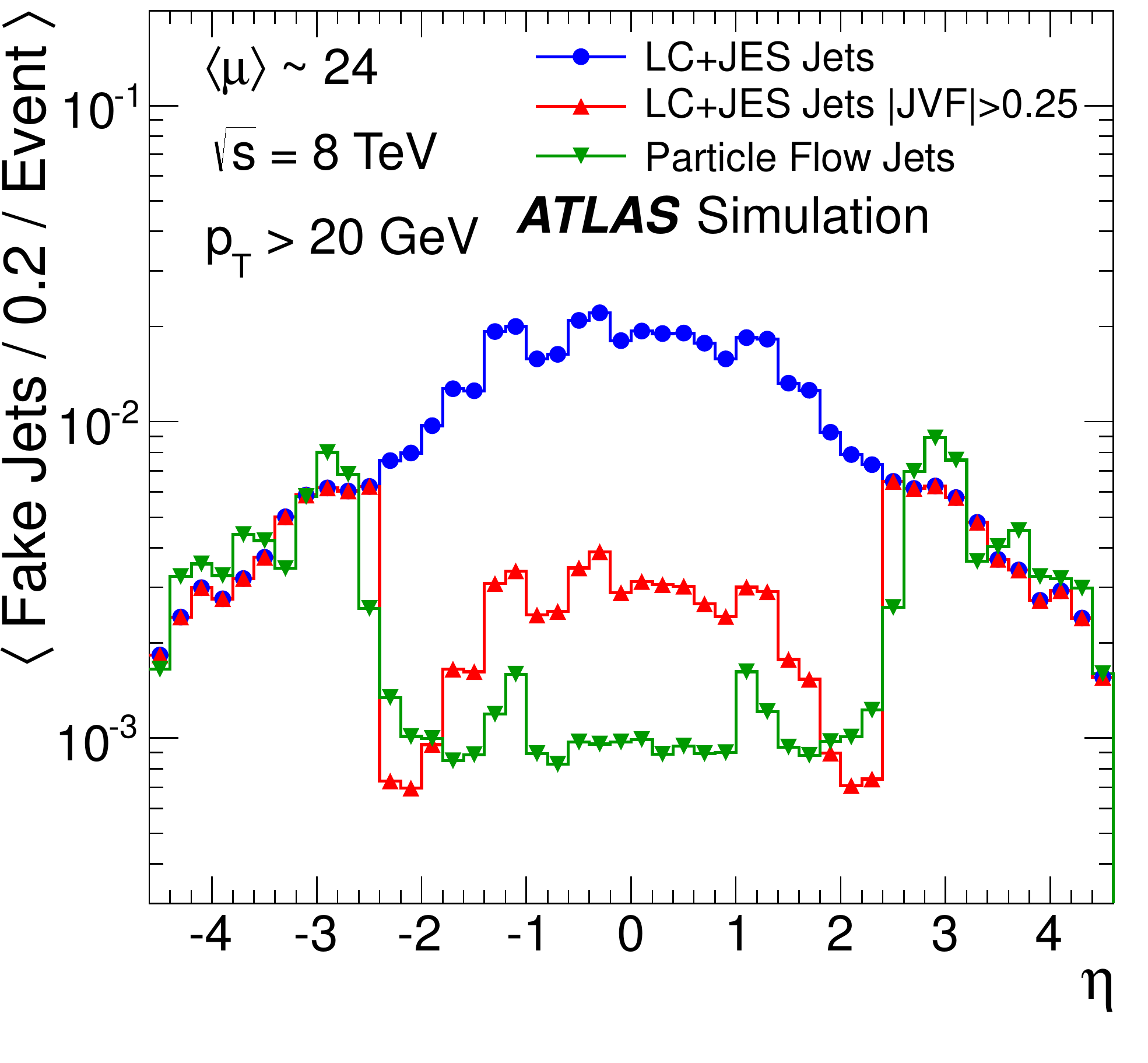}\label{fig:jetPU:fakeJetsVsEta_a}}\quad
  \subfloat[]{%
    \includegraphics[width=0.45\textwidth]{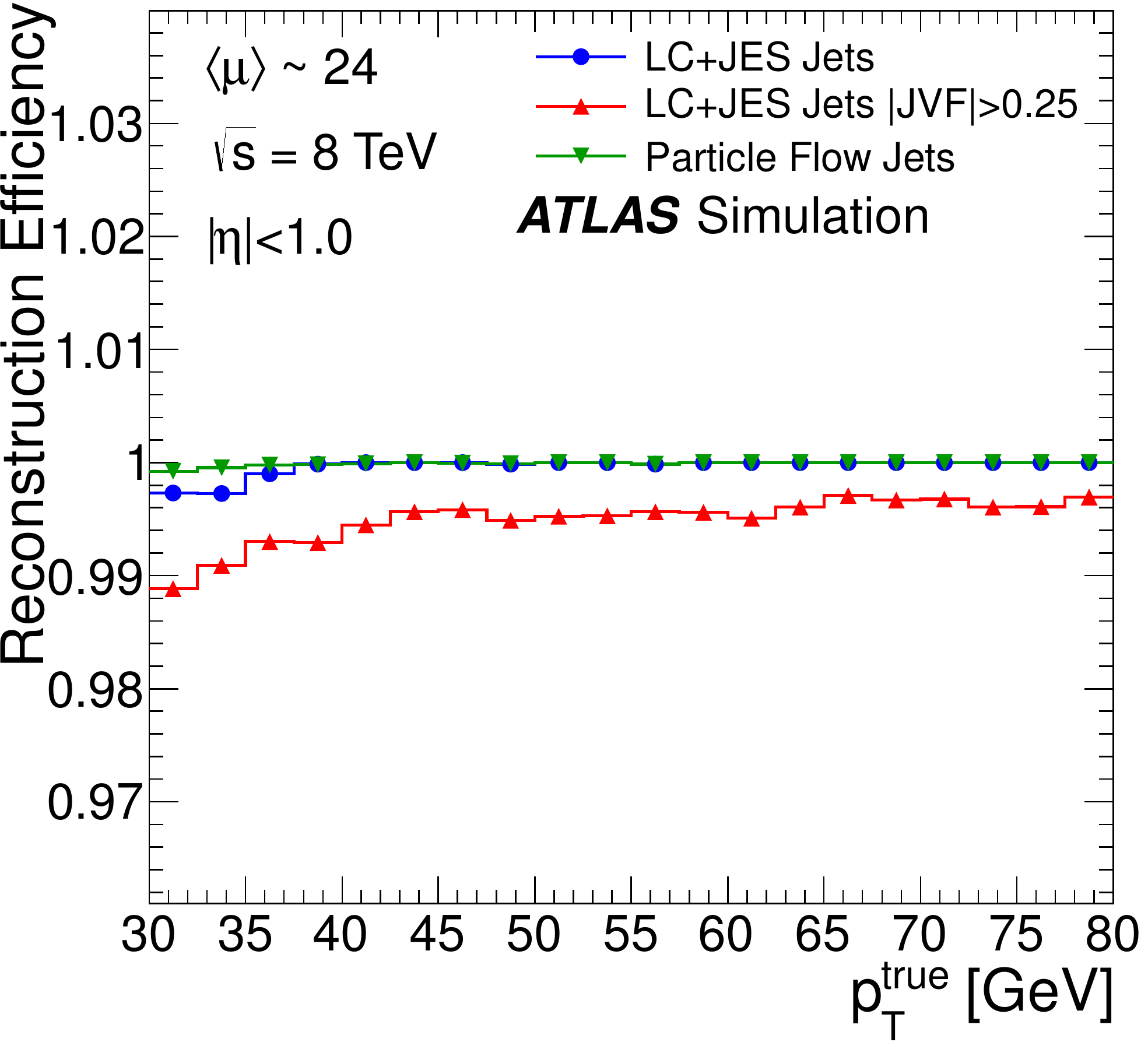}\label{fig:jetPU:fakeJetsVsEta_b}}
  \caption{ In the presence of pile-up, \enquote{fake jets} can arise from particles not produced in the hard-scatter interaction. Subfigure (a) shows the number of fake jets (jets not matched to truth jets with $\pT > \SI{4}{\GeV}$ within $\dR<0.4$)
    and (b) the efficiency of reconstructing a hard-scatter jet (reconstructed jet found within $\dR<0.4$ with $\pT > \SI{15}{\GeV}$)
    in dijet MC events.
    Simulated pile-up conditions are similar to the data-taking in 2012.}
  \label{fig:jetPU:fakeJetsVsEta}
\end{figure}

In the presence of pile-up, jets can arise from particles not produced in the hard-scatter interaction.
These jets are here referred to as \enquote{fake jets}.
Figure~\ref{fig:jetPU:fakeJetsVsEta_a} shows the fake-jet rate as a function of the jet $\eta$ for particle flow jets compared to calorimeter jets with and without track-based pile-up suppression ~\cite{PERF-2014-03}.
These rates are evaluated using a dijet MC sample
overlaid with simulated minimum-bias events approximating the data-taking conditions in 2012.
The jet vertex fraction (JVF) is defined as the ratio of two scalar sums of track momenta:
the numerator is the scalar sum of the \pT of tracks that originate from the hard-scatter primary vertex
and are associated with the jet;
the denominator is the scalar sum of the transverse momenta of all tracks associated with that jet.\footnote{%
  Jets with no tracks associated with them are assigned $\text{JVF} = -1$.}
Within the tracker coverage of $|\eta|<2.5$,
the fake rate for particle flow jets drops by an order of magnitude compared to the standard calorimeter jets.
The small increase in the rate of particle flow fake jets around $1.0 < |\eta| < 1.2$  is related
to the worse performance of the particle flow algorithm in the transition region between the barrel and extended barrel of the Tile calorimeter,
which is significantly affected by pile-up contributions~\cite{topoclustering}.

For $|\eta| > 2.5$, the jets are virtually identical, and hence the fake rate shows no differences.
This rejection rate is comparable to that achieved using the JVF discriminant,
which can likewise only be applied within the tracker coverage.
Here, the comparison is made to a $|\text{JVF}|$ threshold of 0.25 for calorimeter jets,
which is not as powerful as the particle flow fake-jet rate reduction.
The inefficiency of the particle flow jet-finding is negligible, as can be seen from \Fig{\ref{fig:jetPU:fakeJetsVsEta_b}}.
In contrast, the inefficiency generated by requiring $|\text{JVF}|>0.25$ is clearly visible
(it should be noted that in 2012 JVF cuts were only applied to calorimeter jets
up to a \pT of \SI{50}{\GeV}).
Below \SI{30}{\GeV}, the jet resolution causes some reconstructed jets to fall below the jet reconstruction energy threshold so these values are not shown.

\begin{figure}[htbp]
  \centering
  \includegraphics[width=0.3\textwidth]{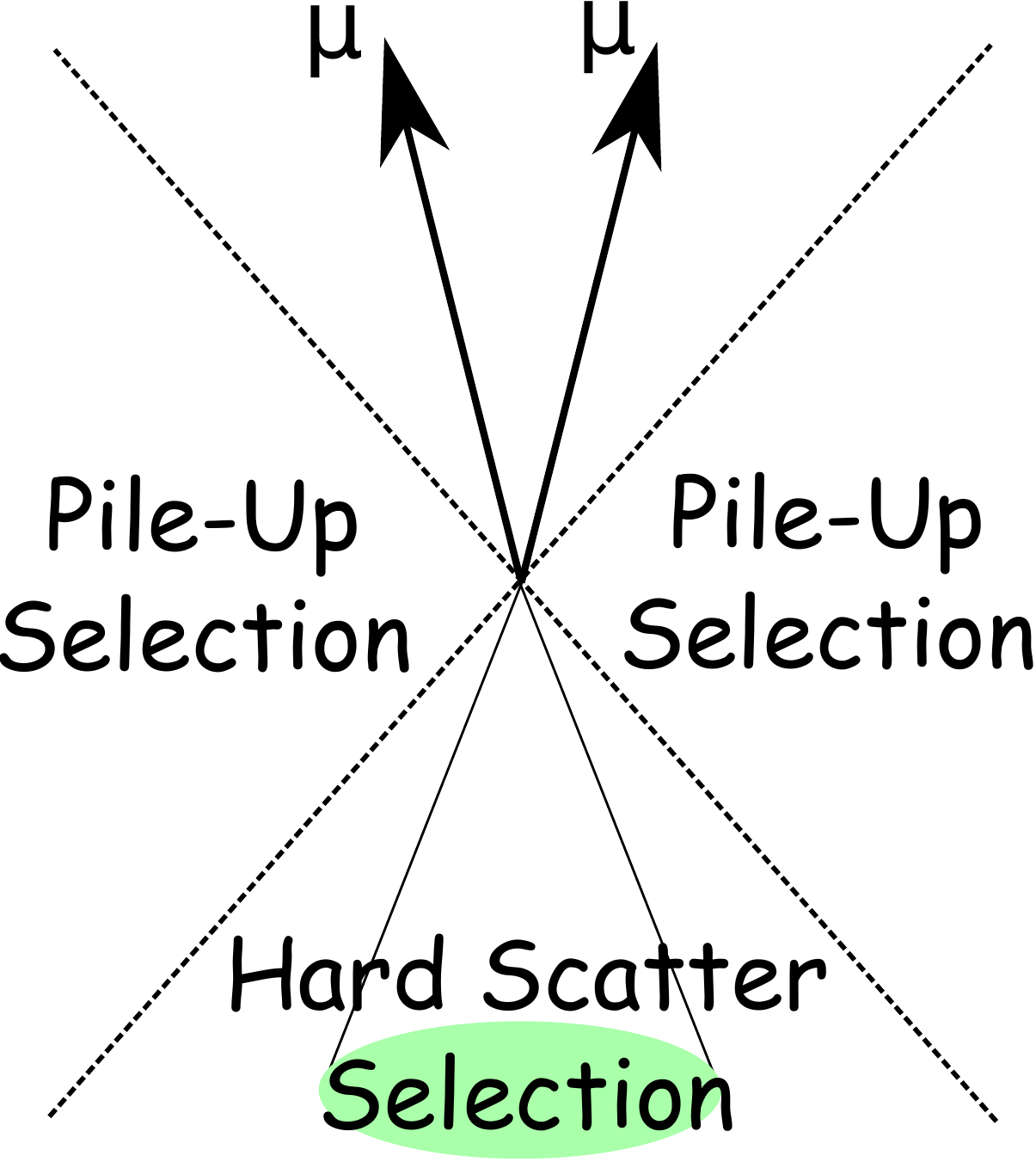}
  \caption{A diagram displaying the regions of $r$--$\phi$ phase space which are expected to be dominated by hard-scatter jets
    (opposite in the $r--\phi$ plane to the $Z \to \mu\mu$ decay)
    and where there is greater sensitivity to pile-up jet activity
    (perpendicular to the $Z \to \mu\mu$ decay).
  }
  \label{fig:jetPU:Zdiagram}
\end{figure}

A more detailed study of the pile-up jet rates is carried out in a $Z\to\mu\mu$ sample, both in data and MC simulation,
by isolating several phase-space regions that are enriched in hard-scatter or pile-up jets.
A preselection is made using the criteria described in \Sect{\ref{sec:dataset}}.
The particle flow algorithm is run on these events and further requirements are applied:
events are selected with two isolated muons, each with $\pT > \SI{25}{\GeV}$, with invariant mass $80 < m_{\mu\mu} < \SI{100}{\GeV}$ and $\pT^{\mu\mu} > \SI{32}{\GeV}$,
ensuring that the boson recoils against hadronic activity.
Figure~\ref{fig:jetPU:Zdiagram} displays two regions of phase space: one opposite the recoiling boson, where large amounts of hard-scatter jet activity are expected,
and one off-axis, which is more sensitive to pile-up jet activity.

\begin{figure}[htbp]
  \centering
  \subfloat[$|\eta|<1.0$]{\includegraphics[width=0.31\textwidth]{%
    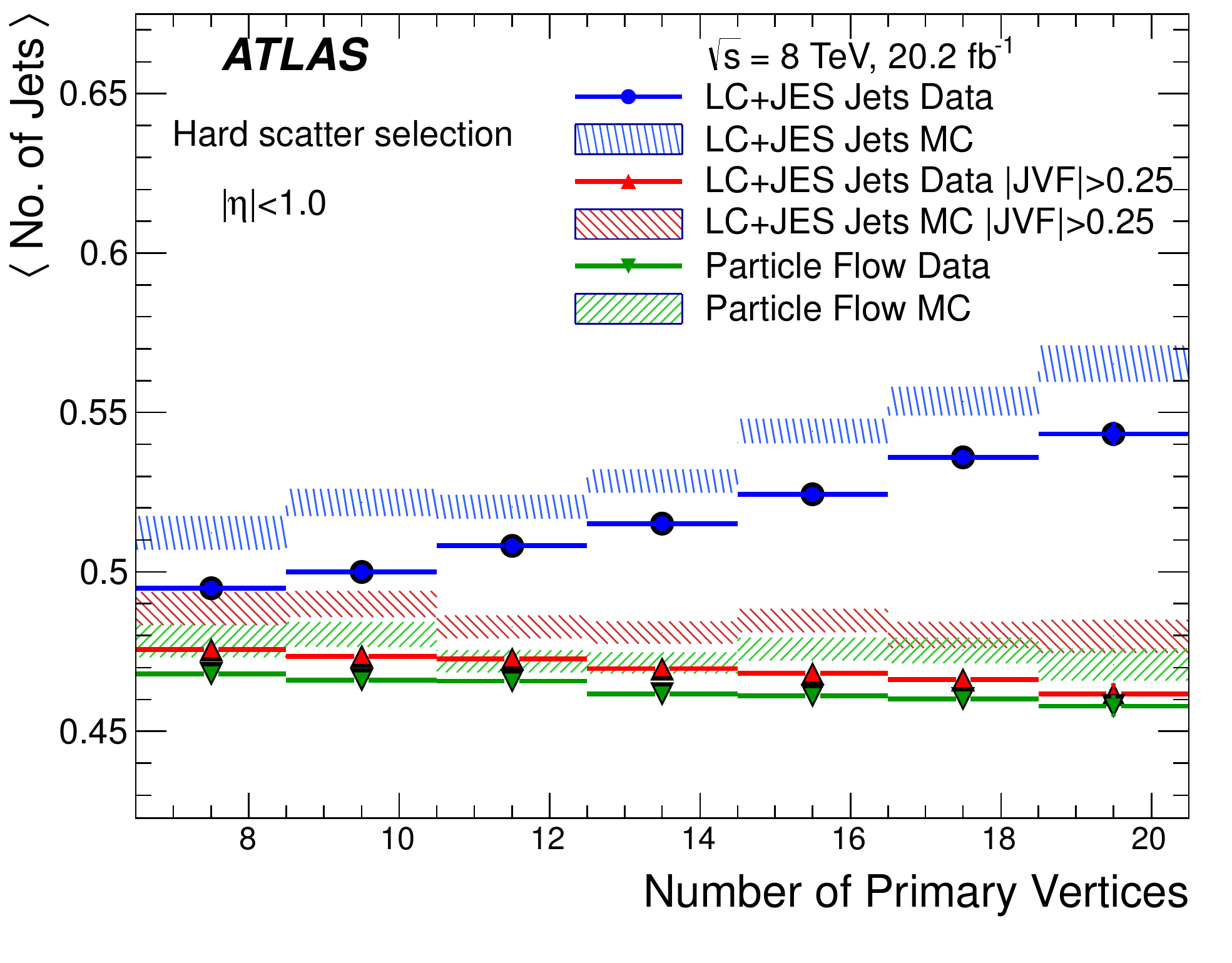}\label{fig:jetPU:HS_20_025:NPV_a}}\quad
  \subfloat[$1.0<|\eta|<2.0$]{%
    \includegraphics[width=0.31\textwidth]{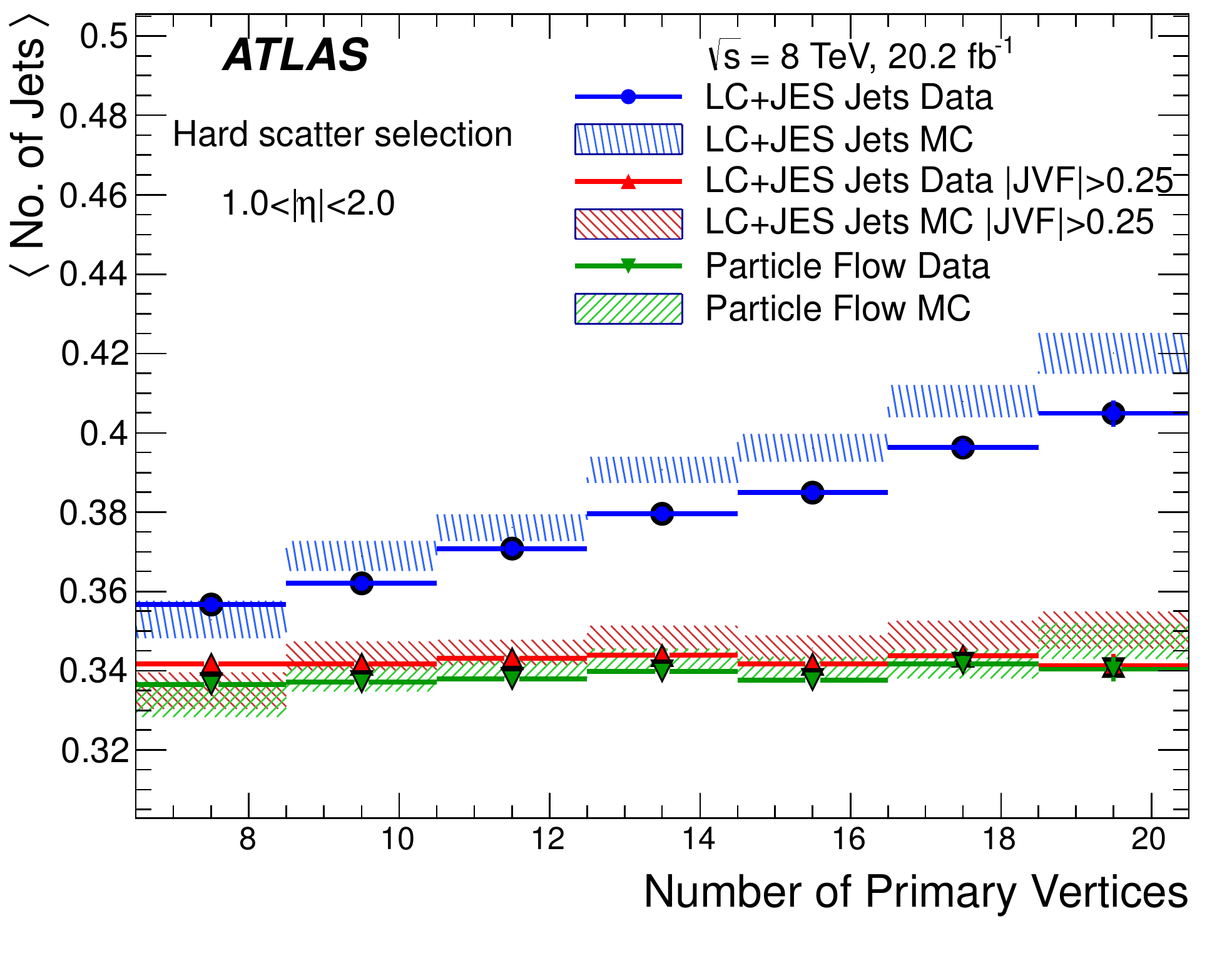}\label{fig:jetPU:HS_20_025:NPV_b}}\quad
  \subfloat[$2.0<|\eta|<2.5$]{%
    \includegraphics[width=0.31\textwidth]{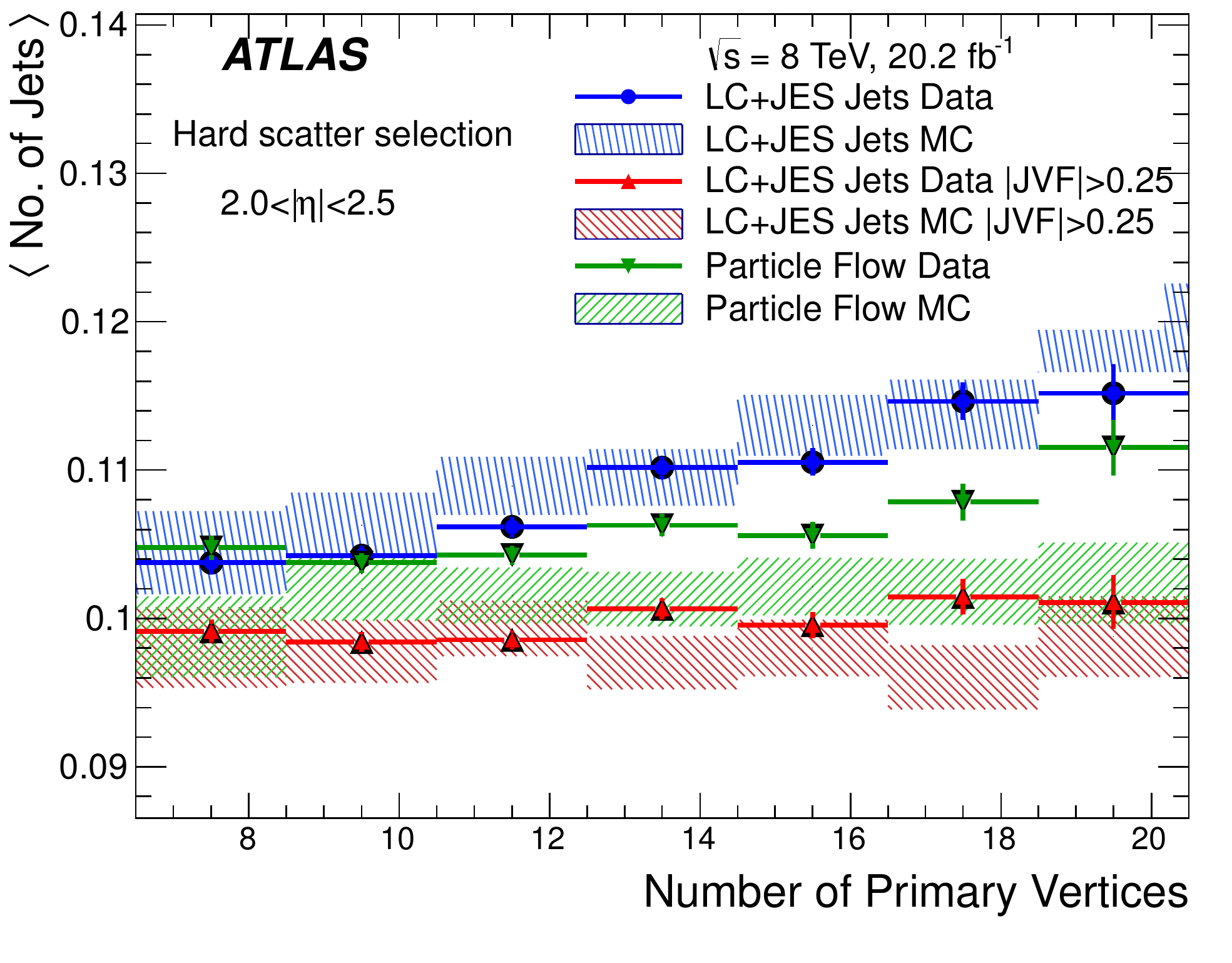}\label{fig:jetPU:HS_20_025:NPV_c}}
  \caption{The average number of jets per event, for jets with $\pT > \SI{20}{\GeV}$, as a function of the number of primary vertices
    in the $Z\to\mu\mu$ samples.
    The distributions are shown in three different $|\eta|$ regions for particle flow jets, calorimeter jets and calorimeter jets with an additional cut on JVF. 
    The jets are selected in a region of $\phi$ opposite the $Z$ boson's direction, $\Delta\phi(Z,\text{jet}) > 3\pi/4$, which is enriched in hard-scatter jets.
    The statistical uncertainties in the number of events are shown as a hatched band.
  }
  \label{fig:jetPU:HS_20_025:NPV}
\end{figure}

\begin{figure}[htbp]
  \centering
  \subfloat[$|\eta|<1.0$]{%
    \includegraphics[width=0.31\textwidth]{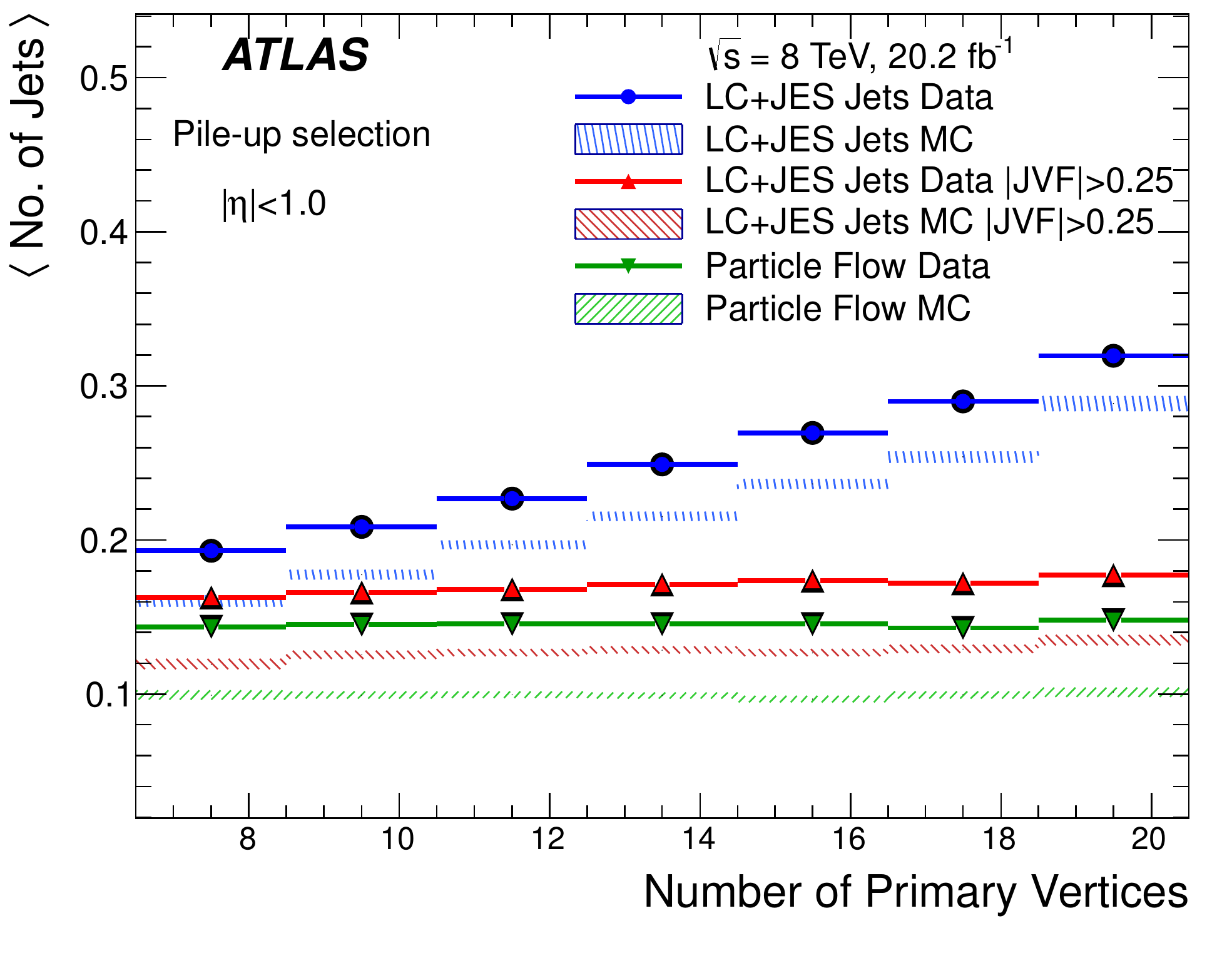}\label{fig:jetPU:PU_20_025:NPV_a}}\quad
  \subfloat[$1.0<|\eta|<2.0$]{%
    \includegraphics[width=0.31\textwidth]{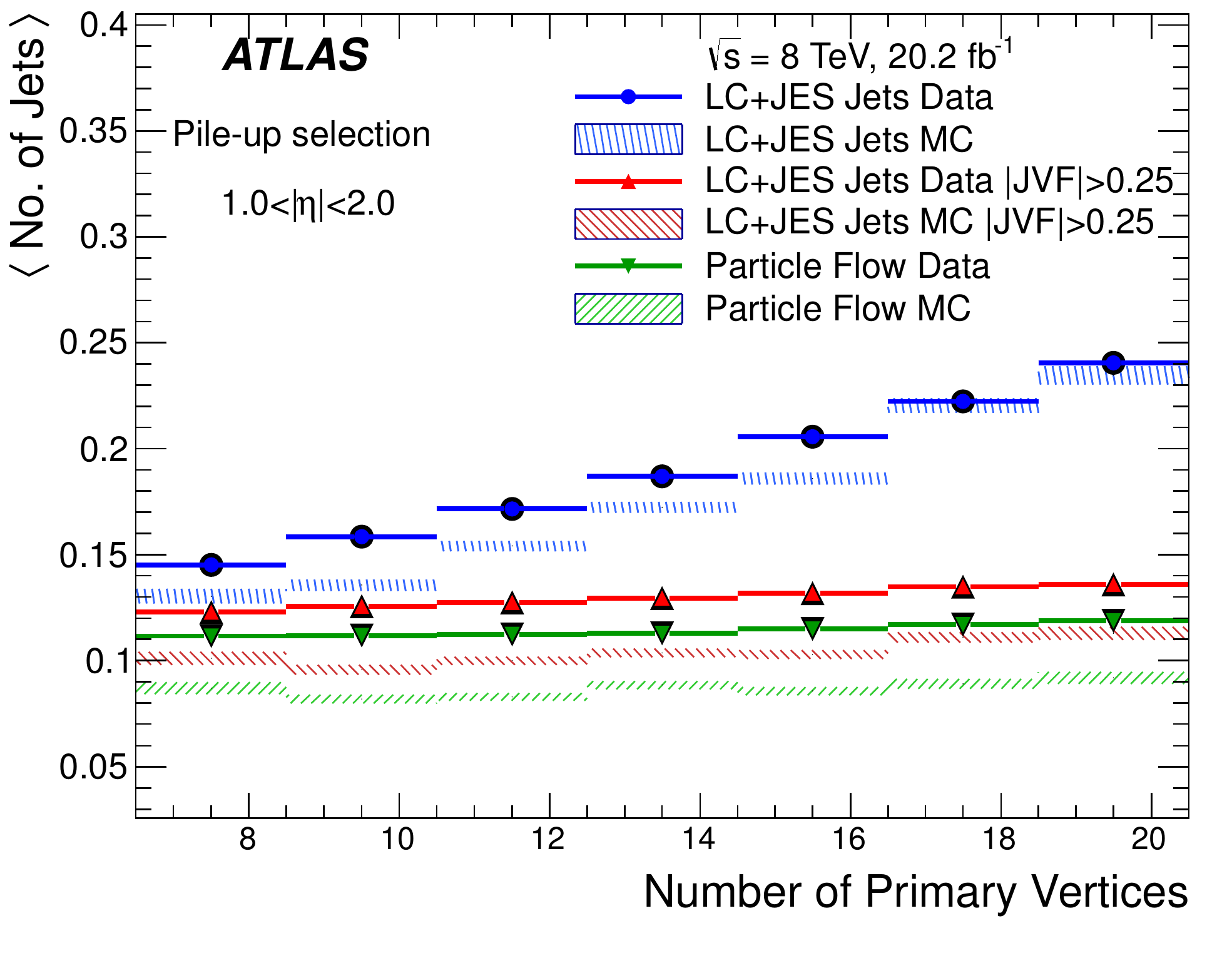}\label{fig:jetPU:PU_20_025:NPV_b}}\quad
  \subfloat[$2.0<|\eta|<2.5$]{%
    \includegraphics[width=0.31\textwidth]{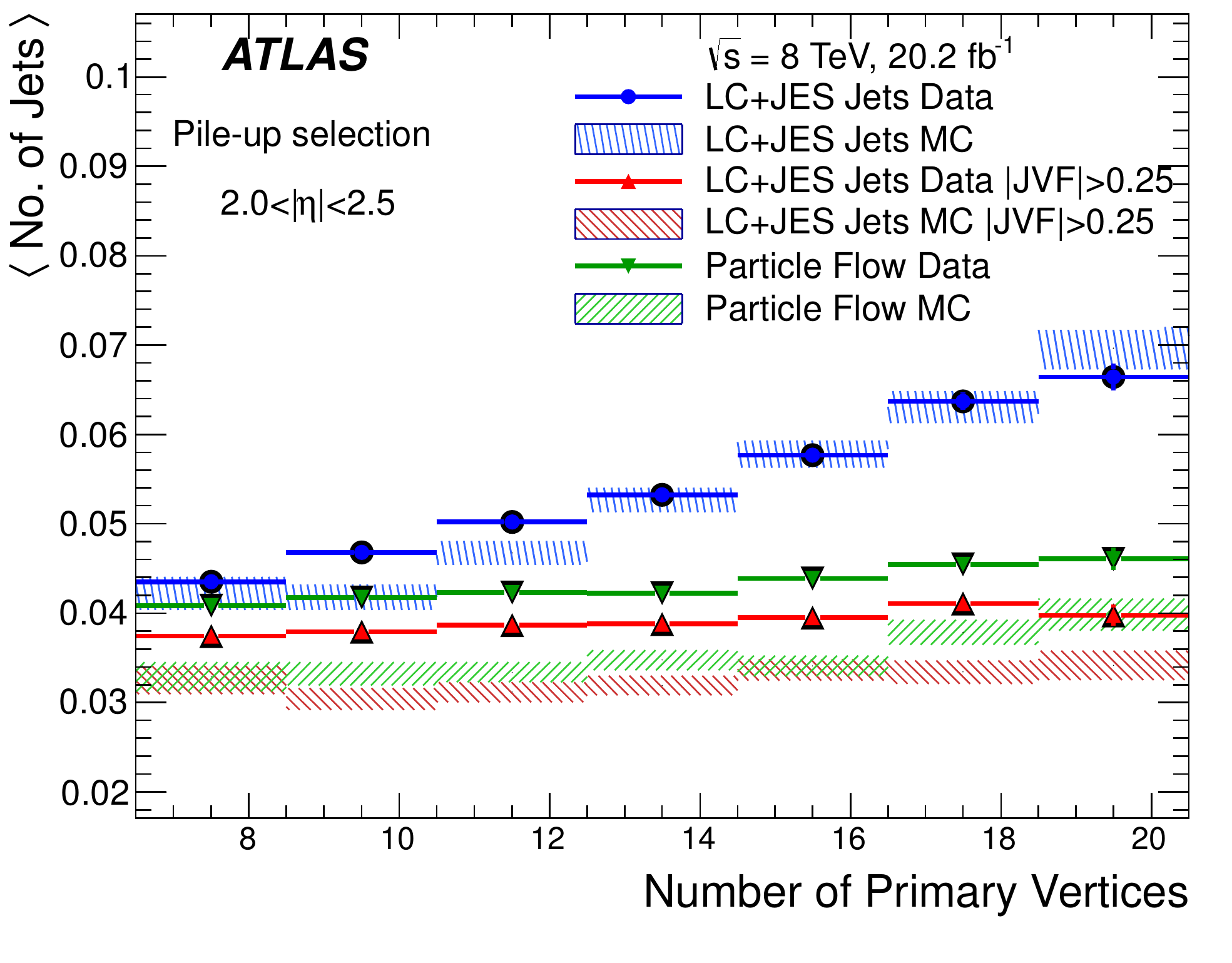}\label{fig:jetPU:PU_20_025:NPV_c}}
  \caption{The average number of jets per event, for jets with $\pT > \SI{20}{\GeV}$, as a function of the number of primary vertices
    in the $Z\to\mu\mu$ samples.
    The distributions are shown in three different $|\eta|$ regions for particle flow jets, calorimeter jets and calorimeter jets with an additional cut on JVF. 
    The jets are selected in a region of $\phi$ perpendicular to the $Z$ boson's direction, $\pi/4 < \Delta\phi(Z,\text{jet}) < 3\pi/4$, which is enriched in pile-up jets.
    The statistical uncertainties in the number of events are shown as a hatched band.
  }
  \label{fig:jetPU:PU_20_025:NPV}
\end{figure}

Figure~\ref{fig:jetPU:HS_20_025:NPV} shows the average number of jets with $\pT > \SI{20}{\GeV}$ in the hard-scatter-enriched region
for different $|\eta|$ ranges as a function of the number of primary vertices.
The distributions are stable for particle flow jets and for 
calorimeter jets with $|\text{JVF}|>0.25$ as a function the number of primary vertices
in all $|\eta|$ regions.
The only exception is in the $2.0<|\eta|<2.5$ region,
where in \Fig{\ref{fig:jetPU:fakeJetsVsEta}} a slight increase in the jet fake rate is visible for jet pseudorapidities very close to the tracker boundary.
This is due to the jet area collecting charged-particle pile-up contributions that are outside the ID acceptance.
If the JVF cut is not applied to the calorimeter jets, the jet multiplicity grows with increasing pile-up.
Figure~\ref{fig:jetPU:PU_20_025:NPV} shows that in the pile-up-enriched selection,
the particle flow and calorimeter jets with a JVF selection still show no dependence on 
the number of reconstructed vertices in all $|\eta|$ regions.
The observed difference between data and MC simulation for both jet collections is due to a poor modelling of this region of phase space.
These distributions establish the high stability of particle flow jets in varying pile-up conditions.

%-------------------------------------------------------------------------------
\subsection{Pile-up effects on jet energy resolution}

\begin{figure}[htbp]
  \centering
  \includegraphics[width=0.46\textwidth]{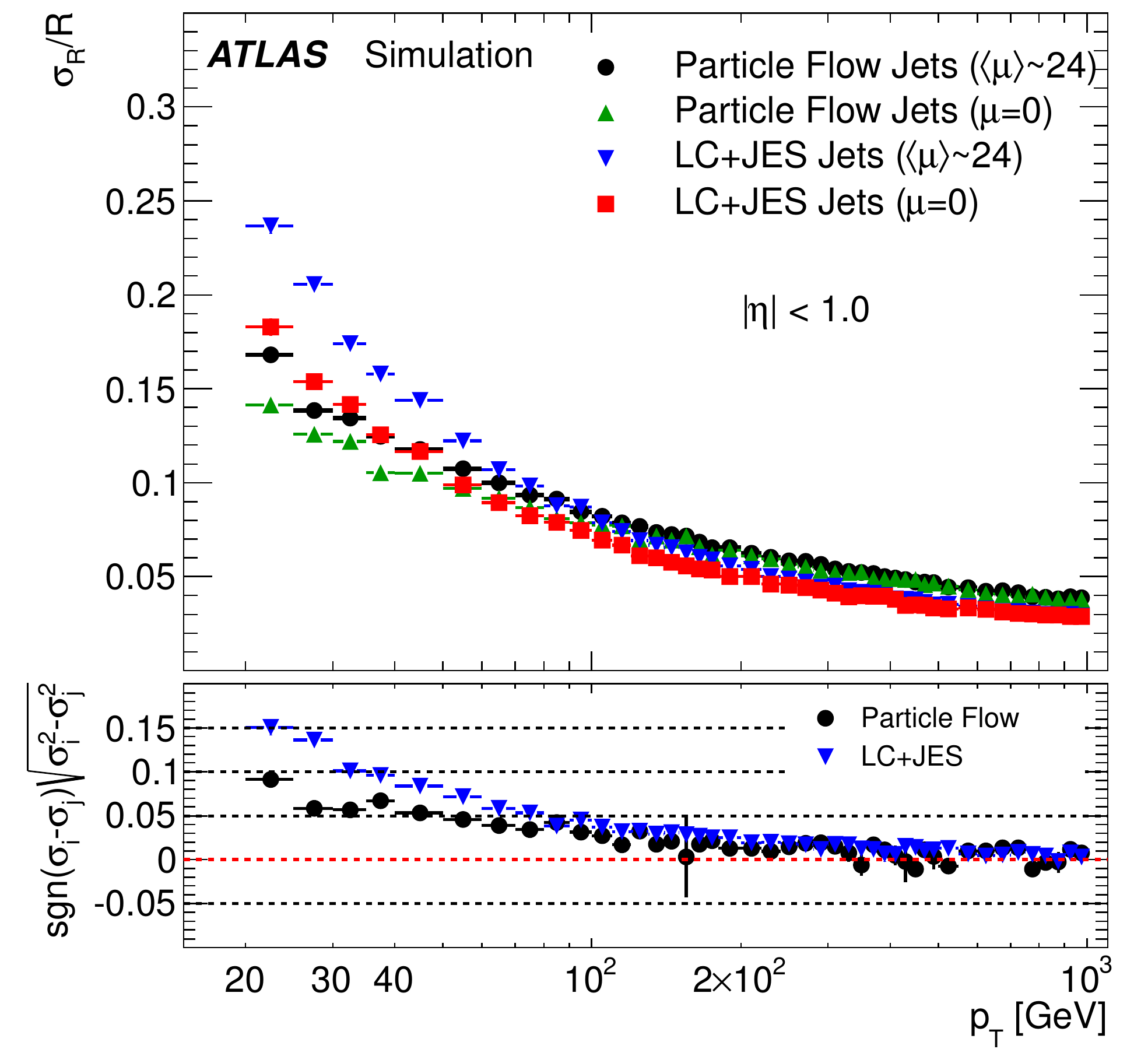}
  \caption{The resolutions of calorimeter and particle flow jets determined as a function of $\pT$ in MC dijet simulation,
    compared with no pile-up and conditions similar to those in the 2012 data.
    The quadratic difference in the resolution with and without pile-up is shown in the lower panel for LC+JES (blue) and particle flow (black) jets.
    The data are taken from a dijet sample without pile-up
      with $20 < \pTlead < \SI{500}{\GeV}$
      and the statistical uncertainties on the number of MC simulated events are shown.
  }
\label{fig:jetPU:jetReso}
\end{figure}

In addition to simply suppressing jets from pile-up,
the particle flow procedure reduces the fluctuations in the jet energy measurements due to pile-up contributions.
This is demonstrated by \Fig{\ref{fig:jetPU:jetReso}},
which compares the jet energy resolution for particle flow and calorimeter jets with and without pile-up.
Even in the absence of pile-up, the particle flow jets have a better resolution at \pT values below \SI{50}{\GeV}.
With pile-up conditions similar to those in the 2012 data, the cross-over point is at $\pT=\SI{90}{\GeV}$,
indicating that particle flow reconstruction alleviates a significant contribution from pile-up even for fairly energetic jets.
The direct effect of pile-up can be seen in the lower panel, where the difference in quadrature between the resolutions with and without pile-up is shown.
The origin of the increase in the resolution with pile-up is discussed in detail in \Ref{\cite{ATLAS-CONF-2015-037}}.
It is shown that additional energy deposits are the primary cause of the degradation of the calorimeter jet resolution.
This effect is mitigated by the particle flow algorithm in two ways.
Firstly, the subtraction of \topoclusters formed by charged particles from pile-up vertices prior to jet-finding eliminates a major source of fluctuations.
Secondly, the increase in the constituent scale of hard-scatter jets from the use of calibrated tracks, rather than energy clusters in the calorimeter, amplifies the signal,
effectively suppressing the contribution from pile-up.
This second mechanism is found to be the main contributing factor.

For \SI{40}{\GeV} jets, the total jet resolution without pile-up is \SI{10}{\%}.
Referring back to \Fig{\ref{fig:perf:sub_c}}, confusion contributes $\sim\SI{8}{\%}$ to the jet resolution in the absence of pile-up.
Since the terms are combined in quadrature, confusion contributes significantly to the overall jet resolution, although it does not totally dominate.
While additional confusion can be caused by the presence of pile-up particles,
the net effect is that pile-up affects the resolution of particle flow jets less than that of calorimeter jets.

%-------------------------------------------------------------------------------
% Comparison of data and Monte Carlo simulation
%-------------------------------------------------------------------------------
% !TeX root = Pflow.tex
%-------------------------------------------------------------------------------
\section{Comparison of data and Monte Carlo simulation}
\label{sec:DataMC}
%-------------------------------------------------------------------------------

It is crucial that the quantities used by the particle flow reconstruction are accurately described by the ATLAS detector simulation.
In this section, particle flow jet properties are compared for $Z \rightarrow \mu\mu$ and \ttbar events in data and MC simulation.
Various observables are validated, from low-level jet characteristics to derived observables relevant to physics analyses.

%-------------------------------------------------------------------------------
\subsection{Individual jet properties}

\begin{figure}[htbp]
  \centering
  \subfloat[Charged fraction]{\includegraphics[width=0.37\textwidth]{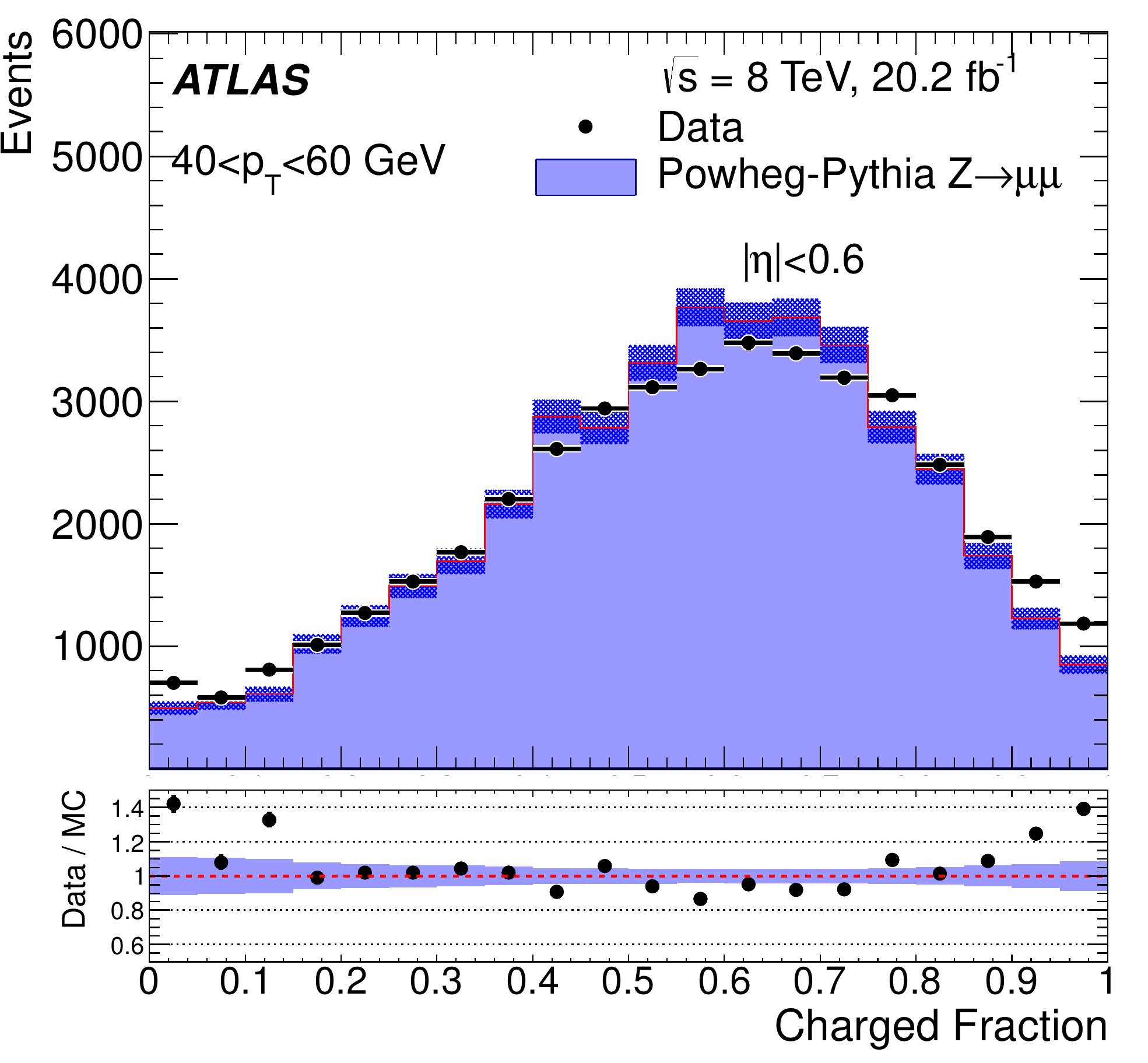}}\quad
  \subfloat[Track multiplicity]{\includegraphics[width=0.37\textwidth]{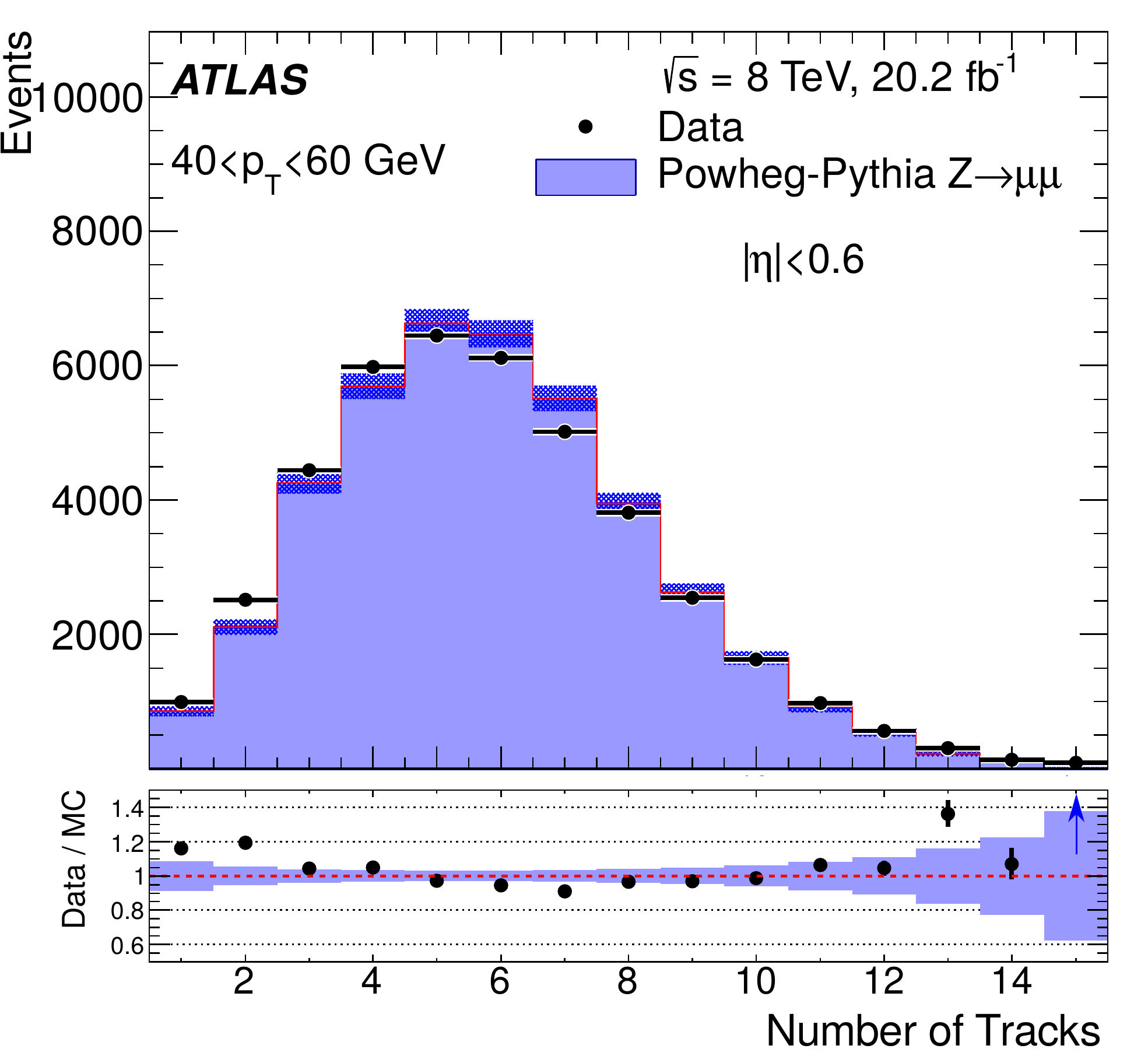}}\\
  \subfloat[Leading track $\pt$]{\includegraphics[width=0.37\textwidth]{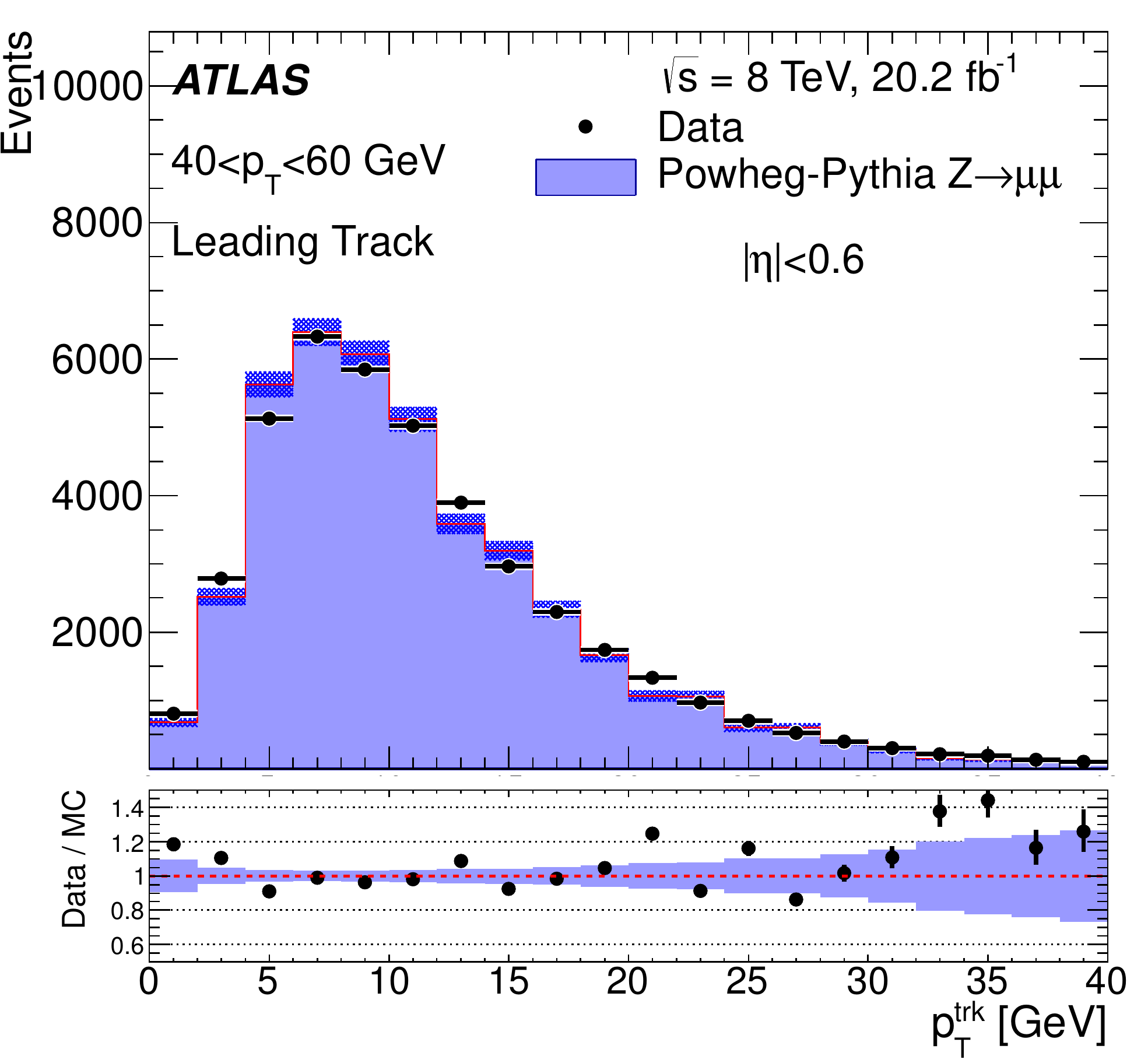}}\quad
  \subfloat[Weighted inclusive track $\pt$]{\includegraphics[width=0.37\textwidth]{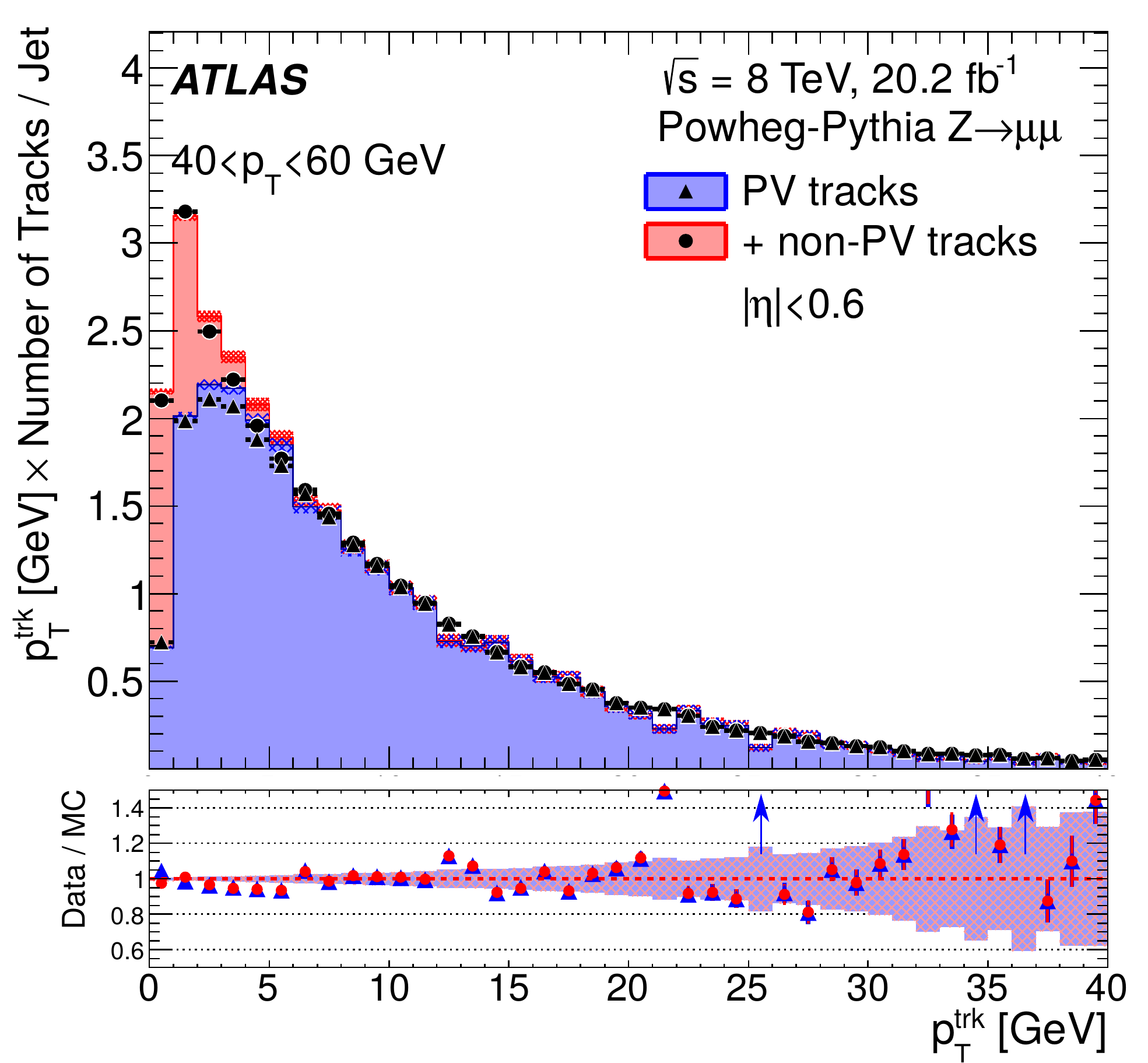}}\\
  \caption{Comparison of jet track properties, for a selection of jets with $40< \pT < \SI{60}{\GeV}$ and $|\eta|<0.6$, selected in $Z\to\mu\mu$ events from collision data and MC simulation.
  	The simulated samples are normalised to the number of events in data.
    The following distributions are shown:
    (a) the charged fraction, i.e.\ the fractional jet \pT carried by reconstructed tracks;
    (b) the number of tracks in the jet that originate from the nominal hard-scatter primary vertex;
    (c) the transverse momentum of the leading track in the jet;
    (d) the transverse momenta of all tracks in the jet weighted by the track \pT and normalised to the number of jets,
    illustrating the transverse momentum flow from particles of different \pT.
    The distribution is shown both for tracks satisfying the hard-scatter primary vertex association criteria and forming the jet
    as well as the additional tracks within $\dR=0.4$ of the jet failing to satisfy the hard-scatter primary vertex association criteria.
    The darker shaded bands represent the statistical uncertainties.
  }
  \label{fig:DataMC:TrkMomsBarrel}
\end{figure}

\begin{figure}[htbp]
  \centering
  \subfloat[EMB1, fraction removed]{\includegraphics[width=0.37\textwidth]{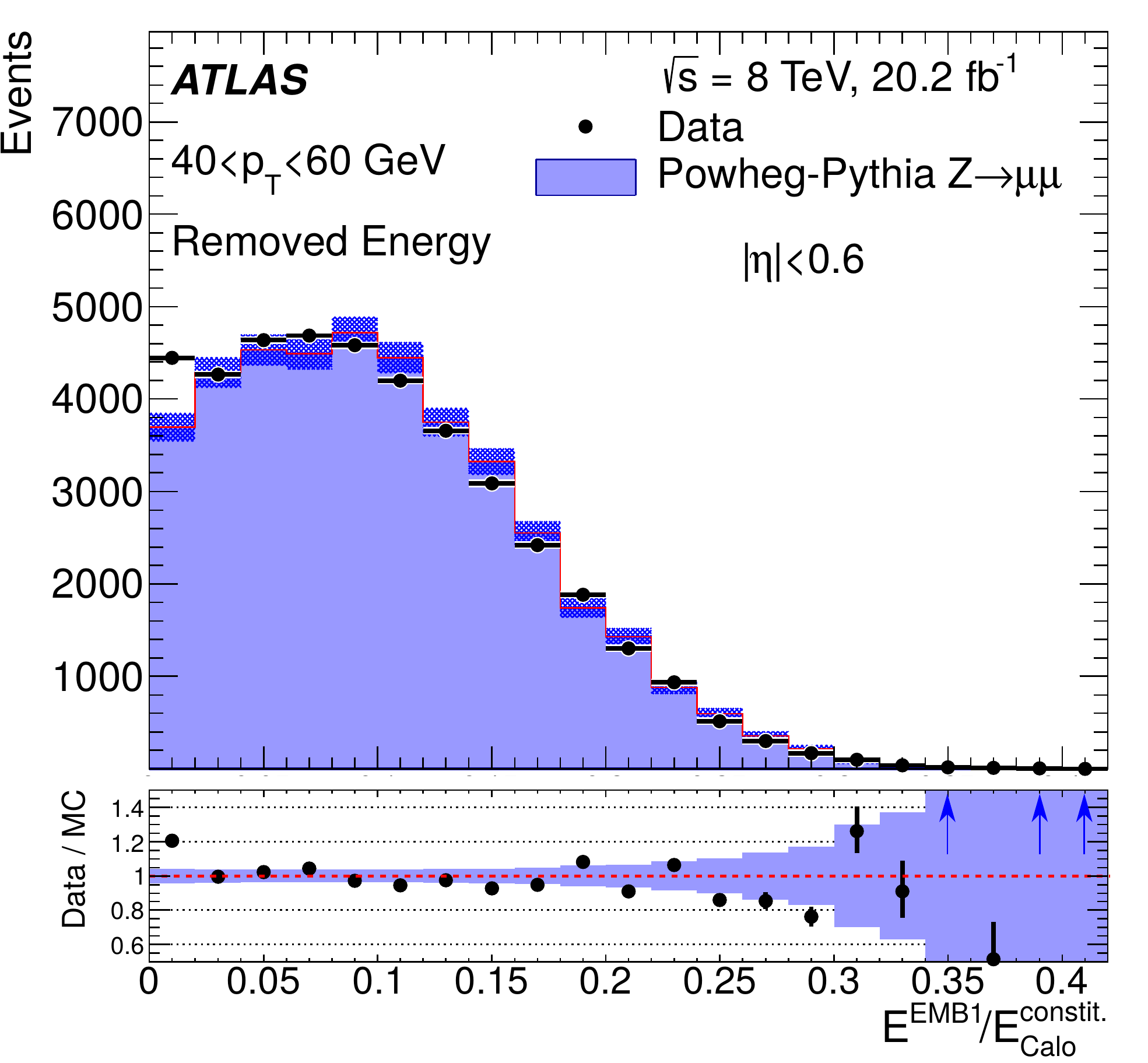}}\qquad
  \subfloat[EMB1, fraction retained]{\includegraphics[width=0.37\textwidth]{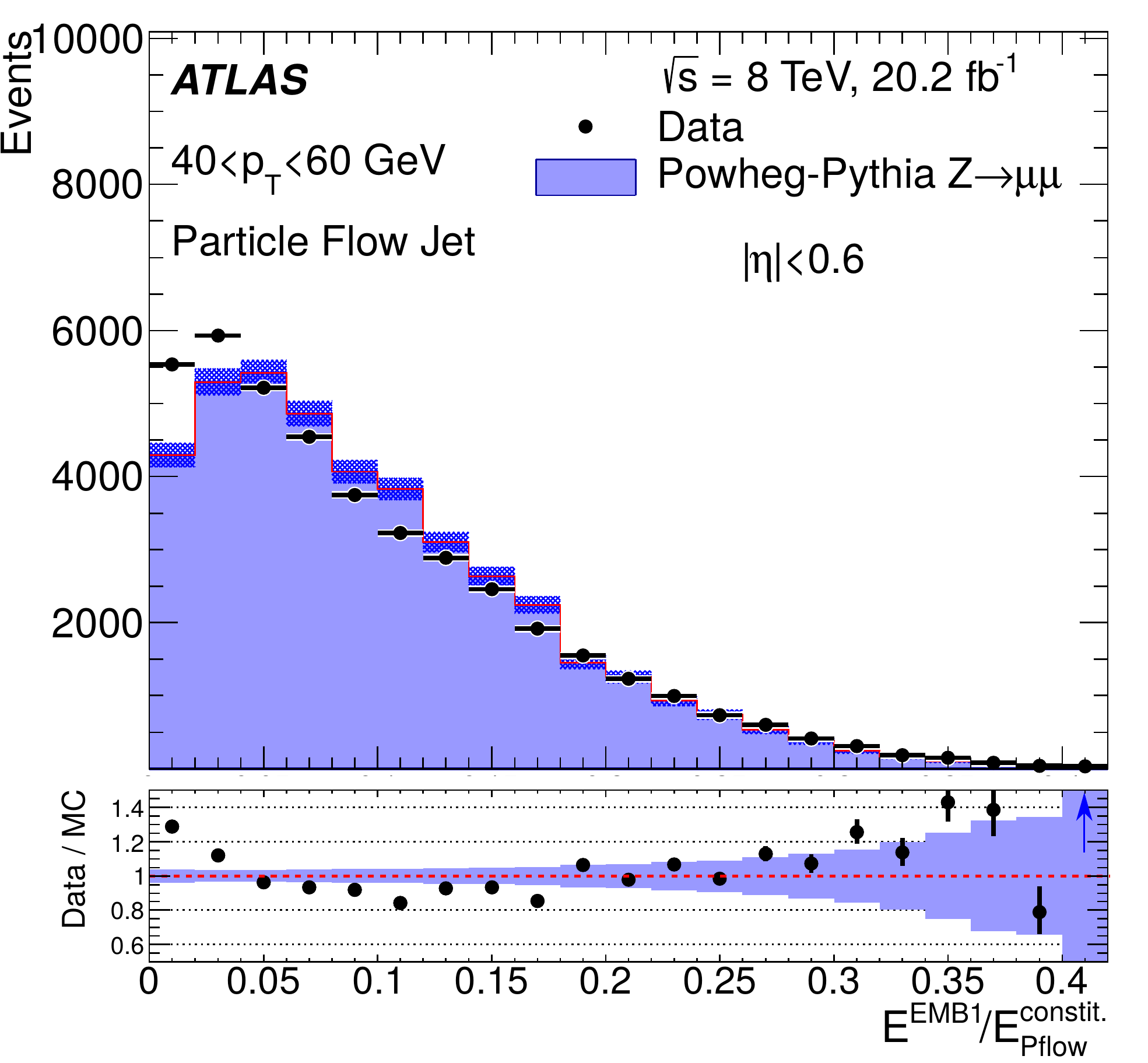}}\\
  \subfloat[EMB2, fraction removed]{\includegraphics[width=0.37\textwidth]{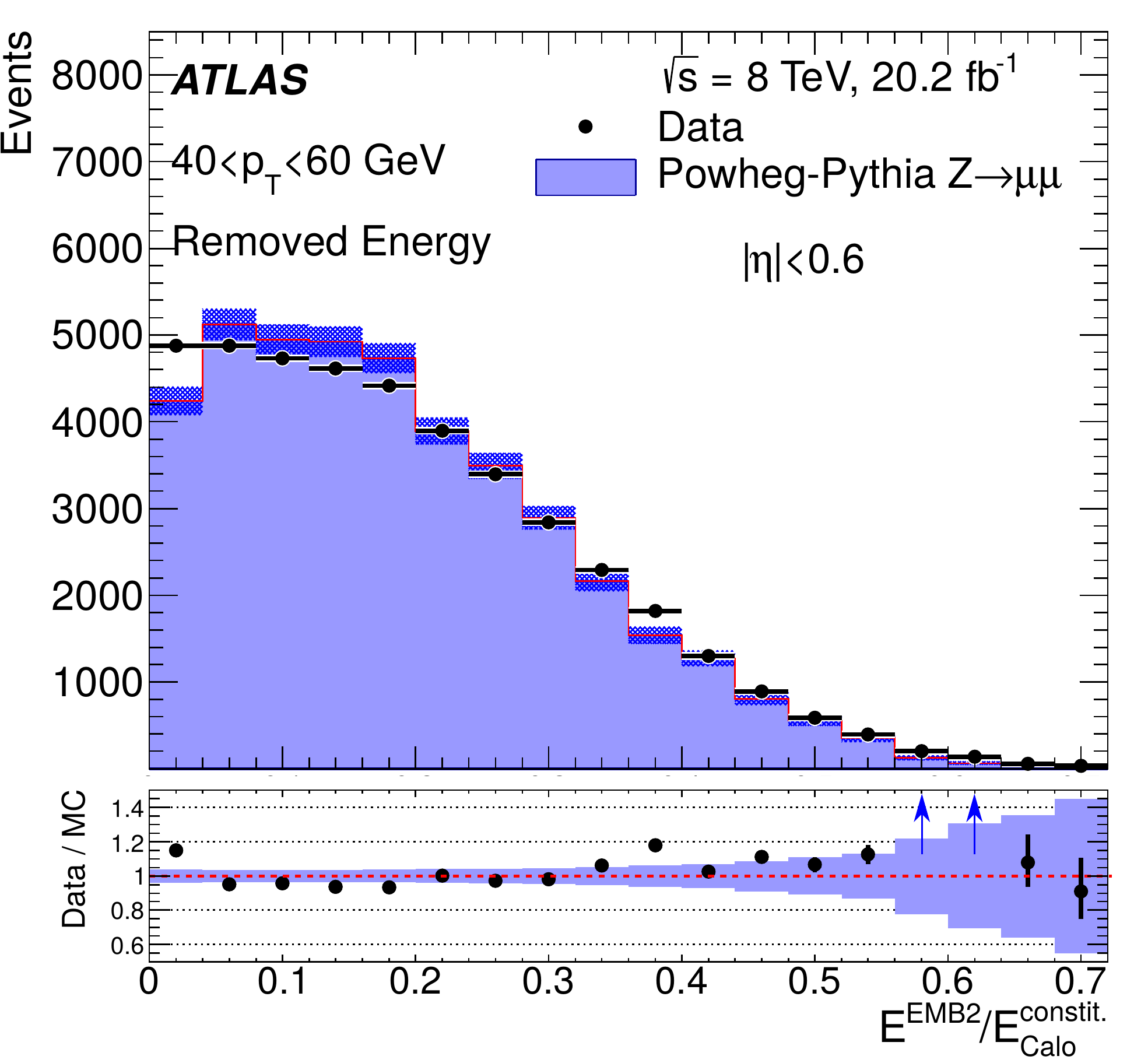}}\quad
  \subfloat[EMB2, fraction retained]{\includegraphics[width=0.37\textwidth]{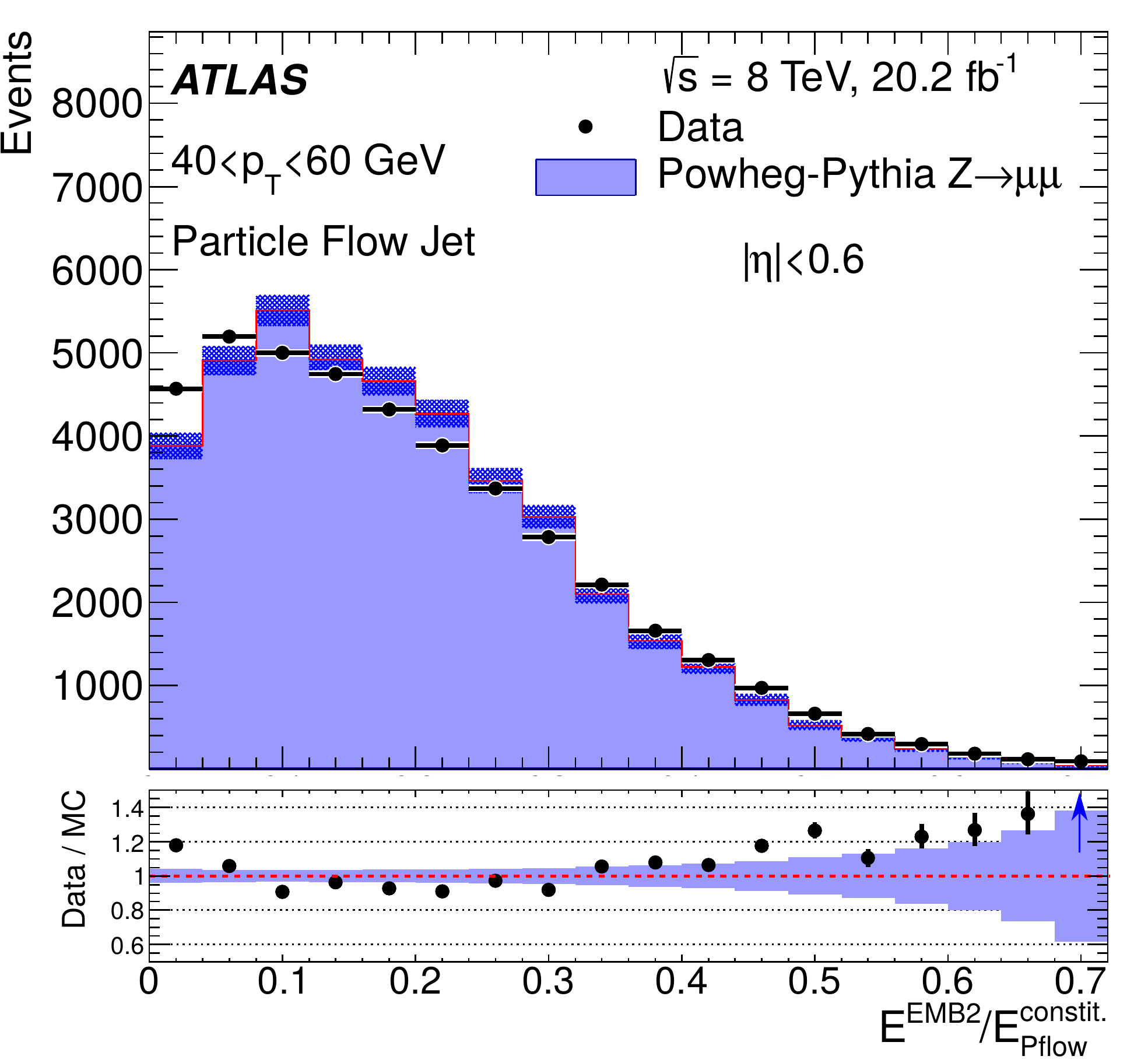}}\\
  \subfloat[EMB3, fraction removed]{\includegraphics[width=0.37\textwidth]{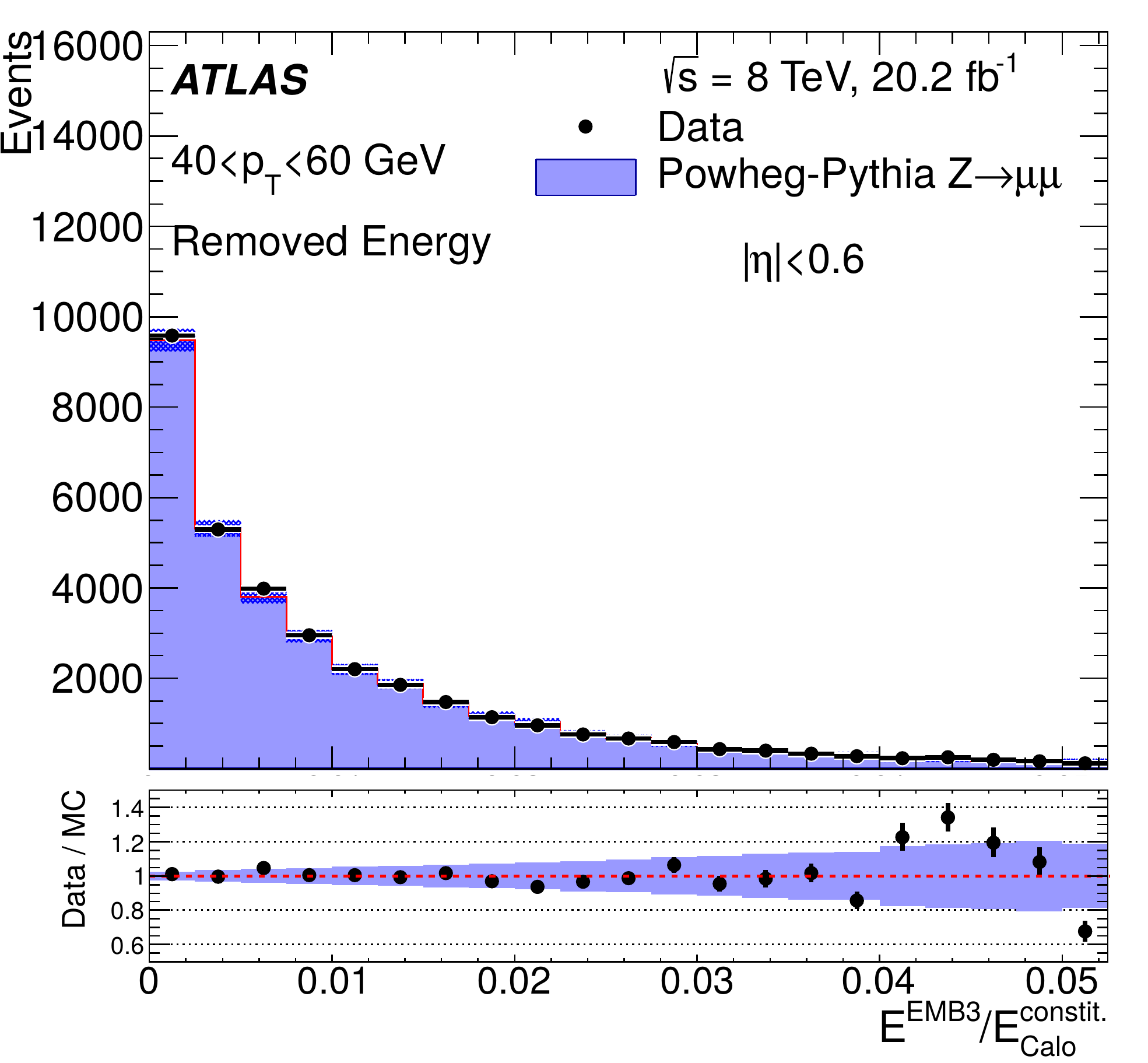}}\quad
  \subfloat[EMB3, fraction retained]{\includegraphics[width=0.37\textwidth]{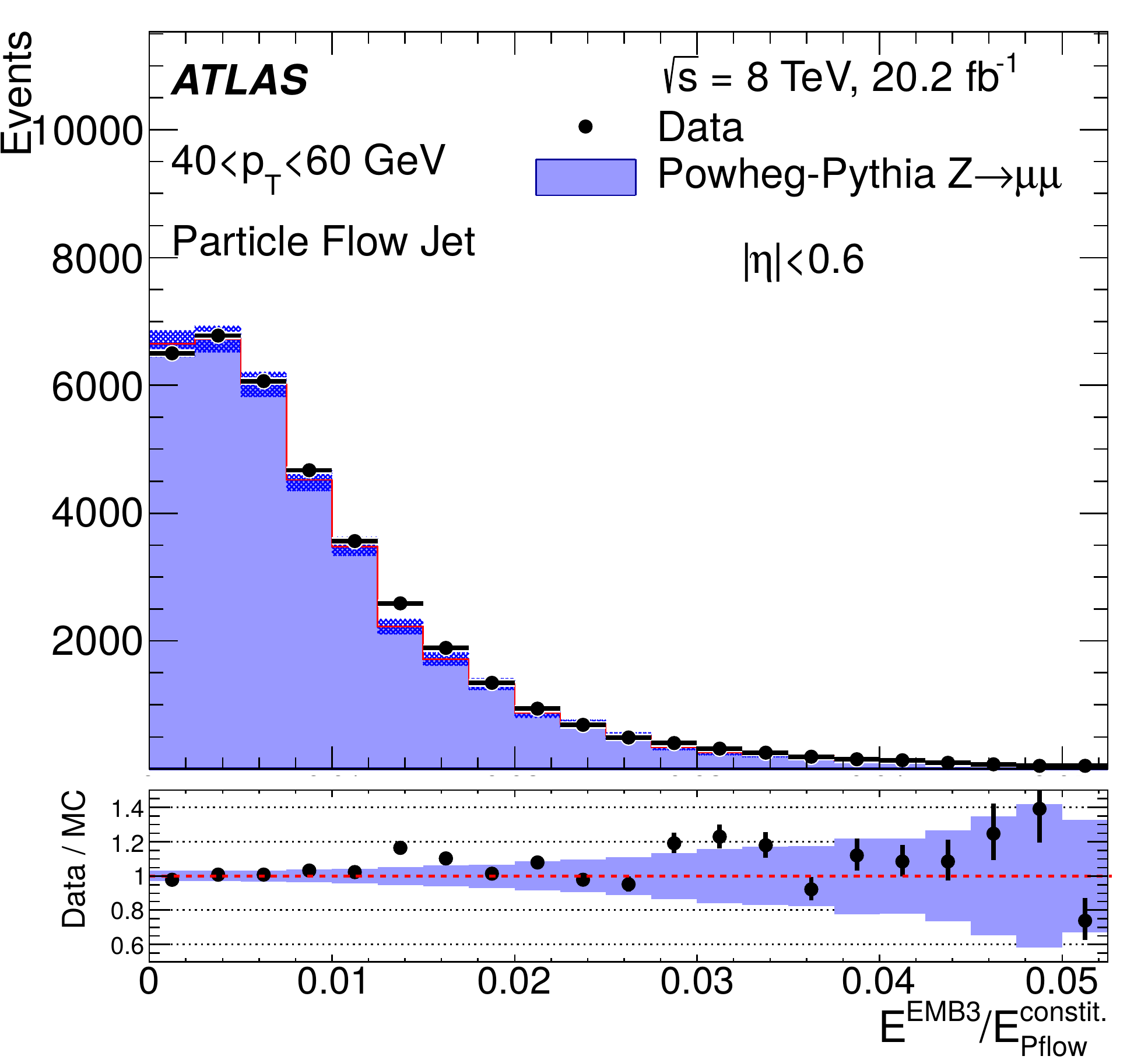}}
  \caption{Comparison of the fractions of jet energy removed from a single layer of the electromagnetic calorimeter relative to the total energy of the constituents of the matched calorimeter jet $E_{\textrm{Calo}}^{\textrm{constit.}}$ (left)
    and retained relative to the total energy of the constituents of the particle flow jet $E_{\textrm{Pflow}}^{\textrm{constit.}}$ (right)
    by the cell subtraction algorithm in different layers of the EM barrel,
    for a selection of jets with $40 < \pT < \SI{60}{\GeV}$ and $|\eta|<0.6$,
    selected in $Z\to\mu\mu$ events from collision data and MC simulation.
  	The simulated samples are normalised to the number of events in data.
    The darker shaded bands represent the statistical uncertainties.
}
\label{fig:DataMC:EfracsEMB}
\end{figure}

A sample of jets is selected in $Z \rightarrow \mu\mu$ events, as in \Sect{\ref{sec:jet:cal}}, and used for a comparison between data and MC simulation.
As the subtraction takes place at the cell level, the energy subtracted from each layer of the calorimeter demonstrates how well the subtraction procedure is modelled.
To determine the energy before subtraction the particle flow jets are matched to jets formed solely from topo-clusters at the electromagnetic scale.
A similar selection to that used to evaluate the jet energy scale is used.
The leading jet is required to be opposite a reconstructed $Z$ boson decaying to two muons with $\Delta\phi>\pi-0.4$.
The \pT of the reconstructed boson is required to be above \SI{32}{\GeV} and the reconstructed jets must have $40<\pT<\SI{60}{\GeV}$.
Figures~\ref{fig:DataMC:TrkMomsBarrel} and \ref{fig:DataMC:EfracsEMB} show the properties of central jets.
The MC simulation describes the data reasonably well for the jet track multiplicity, fraction of the jet \pT carried by tracks
as well as the amount of subtracted or surviving energy in each layer of the EM barrel.
Similar levels of agreement are observed for jets in the endcap regions of the detector.

%-------------------------------------------------------------------------------
\subsection{Event-level observables}

Finally, the particle flow performance is examined in a sample of selected \ttbar events;
a sample triggered by a single-muon trigger with a single offline reconstructed muon is used.
At least four jets with $\pT > \SI{25}{\GeV}$ and $|\eta|<2.0$ are required and two of these are required to have been $b$-tagged using the MV1 algorithm and have $\pT > 35$ and \SI{30}{\GeV}.\footnote{As the $b$-tagging algorithm has only been calibrated for calorimeter jets, the particle flow jets use the calorimeter jet information from the closest jet in $\Delta R$ in order to decide if the jet is $b$-tagged.}
This selects a \SI{95}{\%} pure sample of \ttbar events.
The event \met is reconstructed from the vector sum of the calibrated jets with $\pT > \SI{20}{\GeV}$, the muon and all remaining tracks associated with the hard-scatter primary vertex but not associated with these objects.
This is then used to form the transverse mass variable defined by $\mT = \sqrt{2 \pT^{\mu} \met (1-\cos(\Delta\phi(\mu,\met)))}$.
The invariant mass of the two leading non-$b$-tagged jets, $m_{\text{jj}}$, forms a hadronic $W$ candidate, while the invariant masses of each of the two $b$-tagged jets and these two non-$b$-tagged jets form two hadronic top quark candidates, $m_{\text{jj}b}$.

\begin{figure}[htbp]
  \centering
  \subfloat[Transverse mass, requiring $60 < m_{\text{jj}}< \SI{100}{\GeV}$.]{\includegraphics[width=0.46\textwidth]{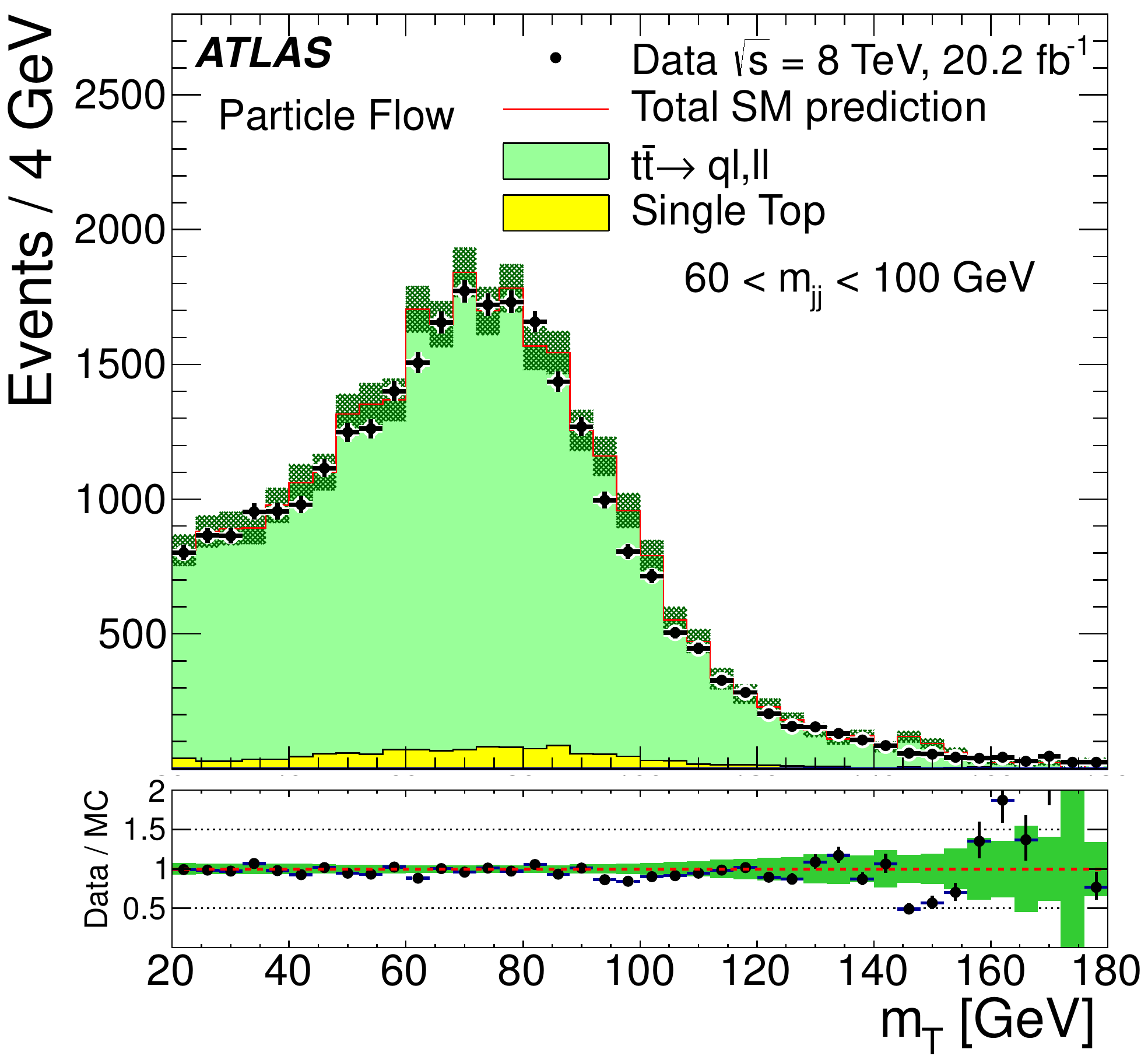}}\quad
  \subfloat[Dijet mass, requiring $\mT > \SI{45}{\GeV}$.]{\includegraphics[width=0.46\textwidth]{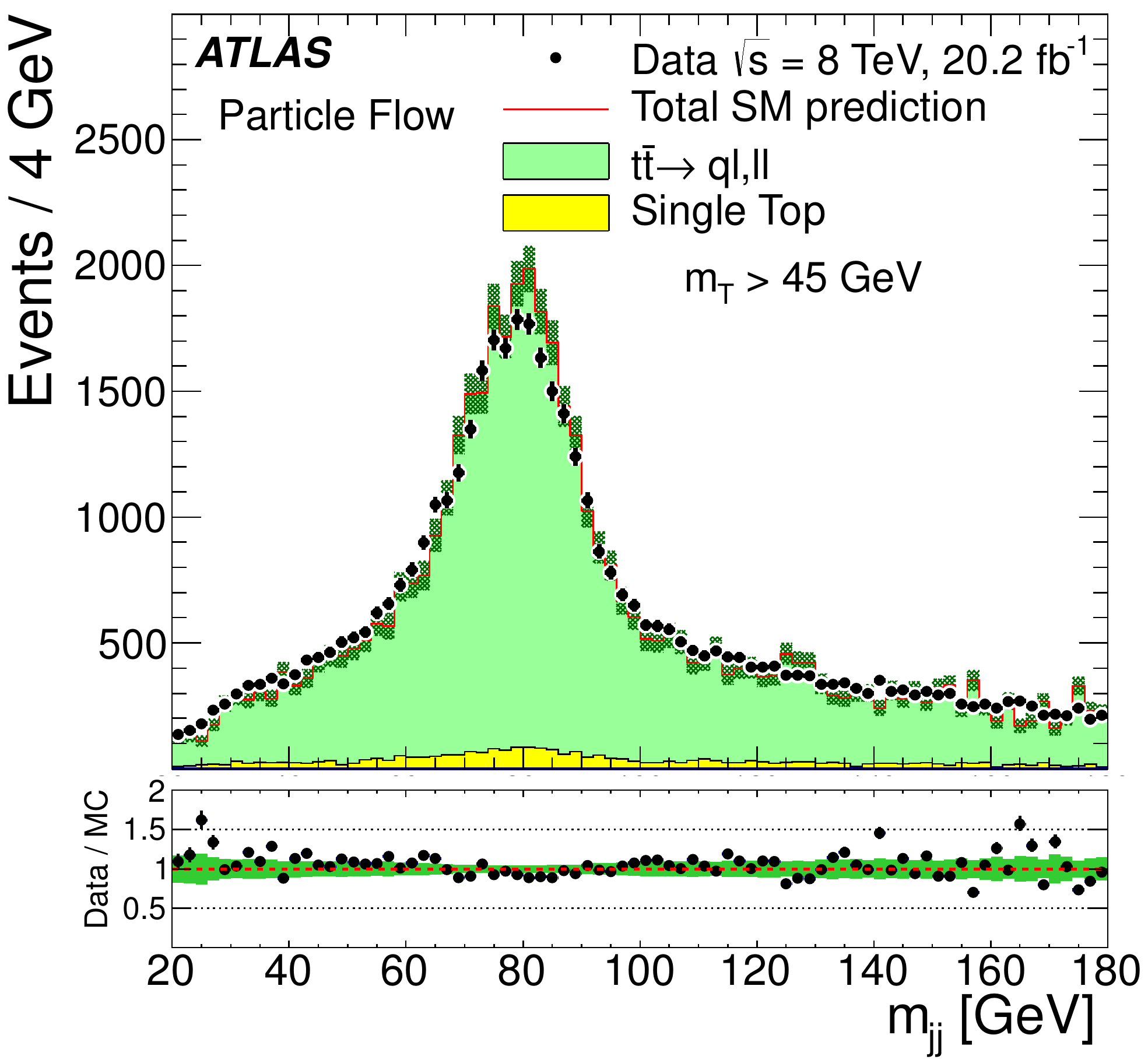}}\quad
  \subfloat[Top-quark-candidate mass,
    requiring $\mT > \SI{45}{\GeV}$ and $60 < m_{\text{jj}} < \SI{100}{\GeV}$.]{\includegraphics[width=0.46\textwidth]{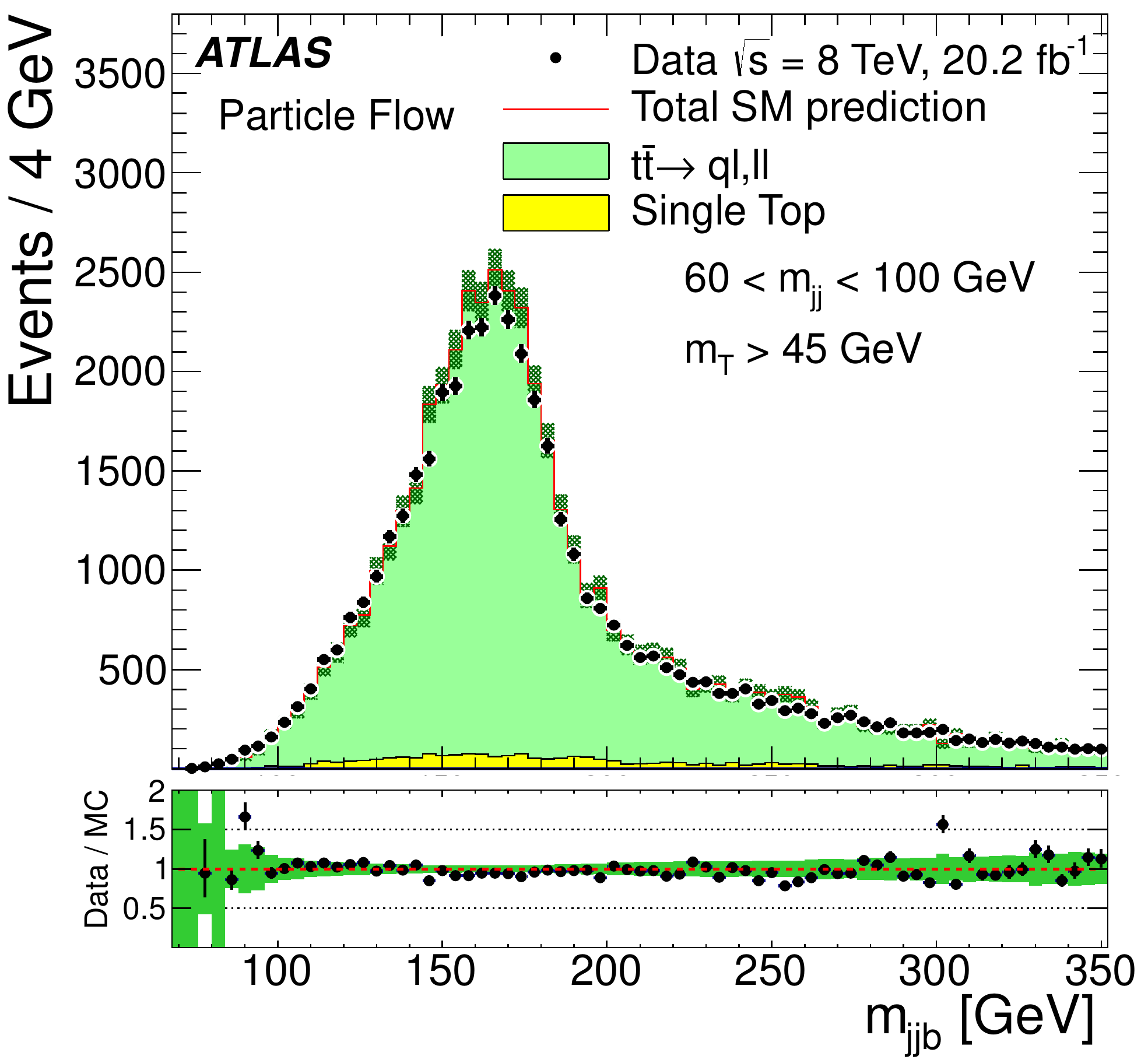}}\quad
  \caption{Comparison of the distributions of mass variables computed with particle flow jets between collision data and the MC simulation for a \ttbar event selection.
    The darker shaded bands and the errors on the collision data show the statistical uncertainties.}
\label{fig:DataMC:ttbar}
\end{figure}

\begin{figure}[htbp]
  \centering
  \subfloat[Requiring $\pT^{\text{jj}} < \SI{80}{\GeV}$.]{\includegraphics[width=0.46\textwidth]{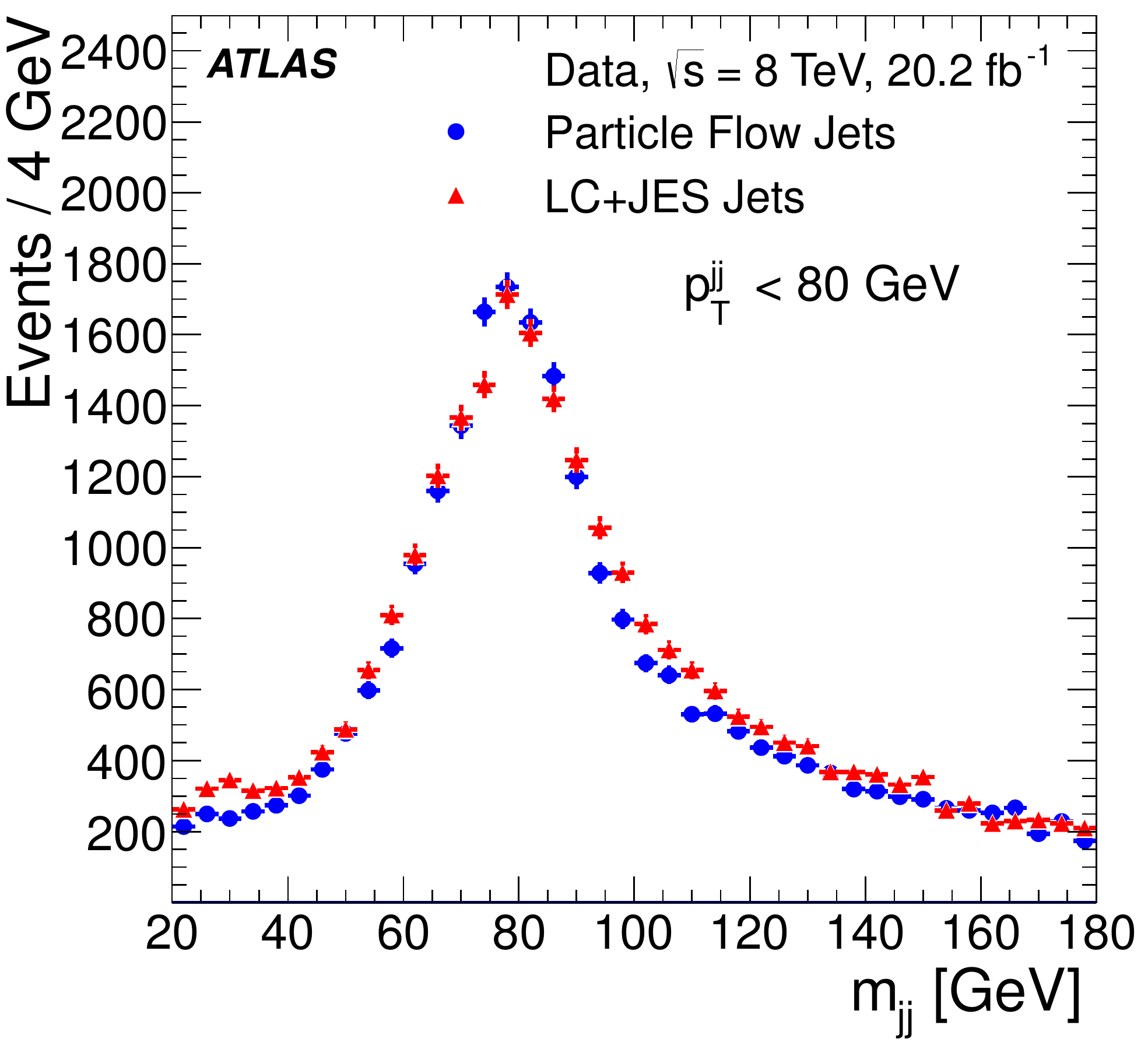}}
  \subfloat[Requiring $\pT^{\text{jj}} > \SI{80}{\GeV}$.]{\includegraphics[width=0.46\textwidth]{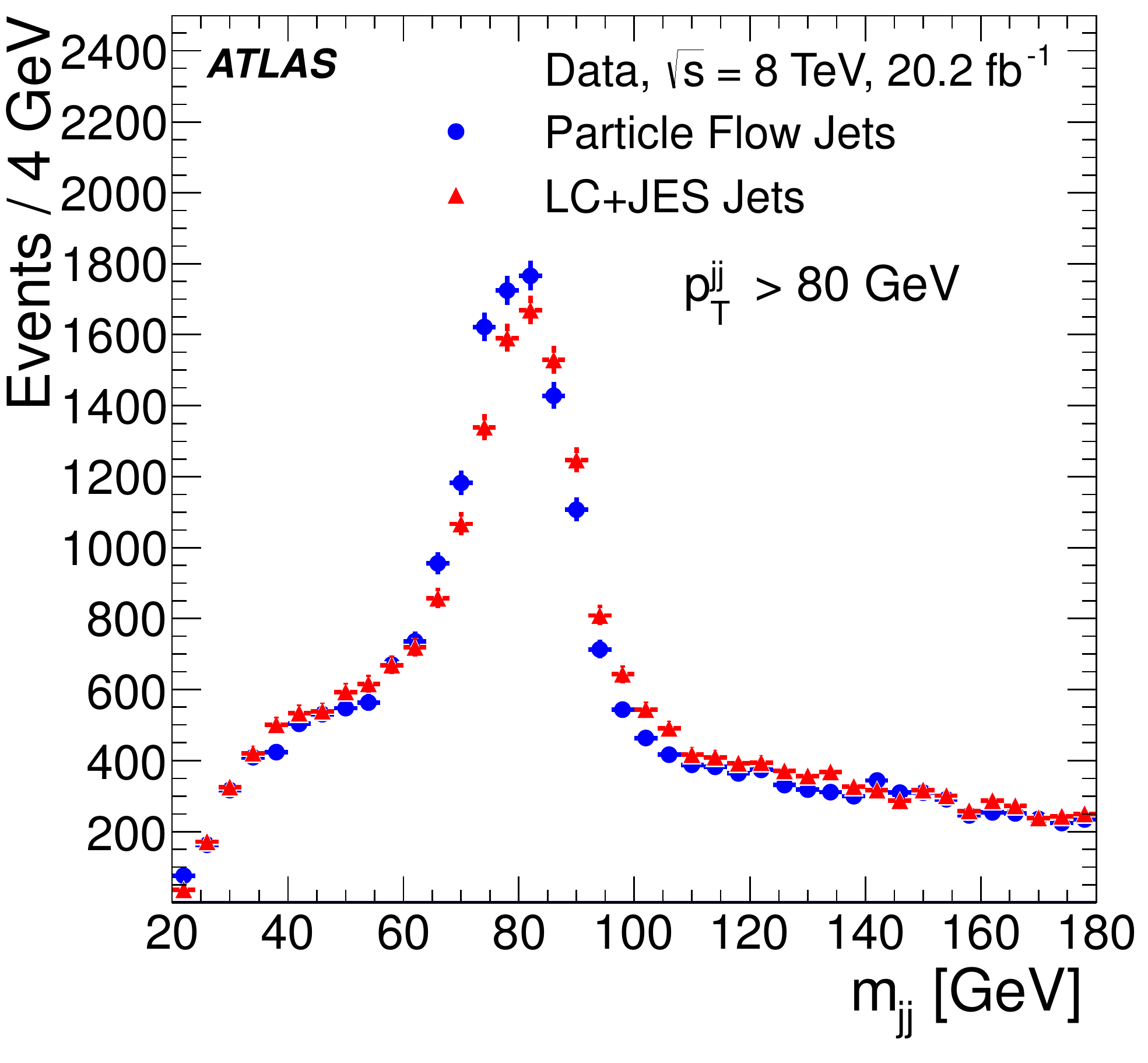}}
  \caption{Comparison between the $m_\mathrm{jj}$ distributions measured using particle flow jets and calorimeter jets with a JVF selection in data. The sample is split into those events where the reconstructed $W$ candidate has $\pT^{\text{jj}} < \SI{80}{\GeV}$ and $\pT^{\text{jj}} > \SI{80}{\GeV}$. The errors shown are purely statistical.
}
\label{fig:DataMC:mW_PFvsLC}
\end{figure}

Figure~\ref{fig:DataMC:ttbar} compares the data with MC simulation for these three variables; $\mT, m_{\text{jj}}$ and $m_{\text{jj}b}$.
The MC simulation describes the data very well in all three distributions.
Figure~\ref{fig:DataMC:mW_PFvsLC} shows the $m_{\text{jj}}$ distribution
for particle flow jets compared to the distribution obtained from the same selection applied to calorimeter jets (with $|\text{JVF}|>0.25$).
For the calorimeter jet selection, the \MET is reconstructed from the muon, jets, photons and remaining unassociated clusters~\cite{PERF-2014-04}.
The two selections are applied separately; hence the exact numbers of events in the plots differ.
The particle flow reconstruction provides a good measure and narrower width of the peak for both low and high $\pT^{\text{jj}}$.
Gaussian fits to the data in the range $65 < m_{\text{jj}} < \SI{95}{\GeV}$ give widths of $\SI{13.8 \pm 0.4}{\GeV}$ and $\SI{16.2 \pm 0.6}{\GeV}$ for particle flow reconstruction and that based on calorimeter jets, respectively, for  $\pT^{\text{jj}} < \SI{80}{\GeV}$.
For $\pT^{\text{jj}} > \SI{80}{\GeV}$, the widths were found to be $\SI{11.2 \pm 0.2}{\GeV}$ and $\SI{11.9 \pm 0.3}{\GeV}$, respectively.
At very high values of $\pT^{W}$, the gains would further diminish (see \Fig{\ref{fig:jetRes}}).
% but good performance is obtained for the ranges studied here.

% All figures and tables should appear before the summary and conclusion.
\FloatBarrier

%-------------------------------------------------------------------------------
% Conclusions
%-------------------------------------------------------------------------------
% !TeX root = Pflow.tex
%-------------------------------------------------------------------------------
\section{Conclusions}
\label{sec:conc}
%-------------------------------------------------------------------------------

The particle flow algorithm used by the ATLAS Collaboration for \SI{20.2}{\per\fb} of $pp$ collisions at \SI{8}{\TeV} at the LHC is presented.
This algorithm aims to accurately subtract energy deposited by tracks in the calorimeter,
exploiting the good calorimeter granularity and longitudinal segmentation.
Use of particle flow leads to improved energy and angular resolution of jets
compared to techniques that only use the calorimeter in the central region of the detector.

In 2012 data-taking conditions, the transverse momentum resolution of particle flow jets calibrated
with a \textit{global sequential correction} is superior up to $\pT \sim \SI{90}{\GeV}$ for $|\eta|<1.0$.
For a representative jet $\pTtrue$ of \SI{30}{\GeV}, the resolution is improved
from the \SI{17.5}{\%} resolution of calorimeter jets with local cluster weighting calibration to \SI{14}{\%}.
Jet angular resolutions are improved across the entire \pT spectrum, with $\sigma(\eta)$ and $\sigma(\phi)$ decreasing from 0.03 to 0.02 and 0.05 to 0.02, respectively,
for a jet \pT of \SI{30}{\GeV}.

Rejection of charged particles from pile-up interactions in jet reconstruction leads to substantially better jet resolution
and to the suppression of jets due to pile-up interactions by an order of magnitude within the tracker acceptance,
with negligible inefficiency for jets from the hard-scatter interaction.
This outperforms a purely track-based jet pile-up discriminant typically used in ATLAS analyses,
which achieves similar pile-up suppression at the cost of about one percent in hard-scatter jet efficiency.

The algorithm therefore achieves a better performance for hadronic observables such as reconstructed resonant particle masses.

Studies which compare data with MC simulation demonstrate that jet properties
used for energy measurement and calibration are modelled well by the ATLAS simulation,
both before and after application of the particle flow algorithm.
This translates to good agreement between data and simulation for derived physics observables, such as invariant masses of combinations of jets. 

The algorithm has been integrated into the ATLAS software framework for Run~2 of the LHC.
As demonstrated, it is robust against pile-up and should therefore perform well under the conditions encountered in Run~2.

\section*{Acknowledgements}
% Acknowledgements for papers with collision data
% Version 6-Mar-2017

% Standard acknowledgements start here
%----------------------------------------------
We thank CERN for the very successful operation of the LHC, as well as the
support staff from our institutions without whom ATLAS could not be
operated efficiently.

We acknowledge the support of ANPCyT, Argentina; YerPhI, Armenia; ARC, Australia; BMWFW and FWF, Austria; ANAS, Azerbaijan; SSTC, Belarus; CNPq and FAPESP, Brazil; NSERC, NRC and CFI, Canada; CERN; CONICYT, Chile; CAS, MOST and NSFC, China; COLCIENCIAS, Colombia; MSMT CR, MPO CR and VSC CR, Czech Republic; DNRF and DNSRC, Denmark; IN2P3-CNRS, CEA-DSM/IRFU, France; SRNSF, Georgia; BMBF, HGF, and MPG, Germany; GSRT, Greece; RGC, Hong Kong SAR, China; ISF, I-CORE and Benoziyo Center, Israel; INFN, Italy; MEXT and JSPS, Japan; CNRST, Morocco; NWO, Netherlands; RCN, Norway; MNiSW and NCN, Poland; FCT, Portugal; MNE/IFA, Romania; MES of Russia and NRC KI, Russian Federation; JINR; MESTD, Serbia; MSSR, Slovakia; ARRS and MIZ\v{S}, Slovenia; DST/NRF, South Africa; MINECO, Spain; SRC and Wallenberg Foundation, Sweden; SERI, SNSF and Cantons of Bern and Geneva, Switzerland; MOST, Taiwan; TAEK, Turkey; STFC, United Kingdom; DOE and NSF, United States of America. In addition, individual groups and members have received support from BCKDF, the Canada Council, CANARIE, CRC, Compute Canada, FQRNT, and the Ontario Innovation Trust, Canada; EPLANET, ERC, ERDF, FP7, Horizon 2020 and Marie Sk{\l}odowska-Curie Actions, European Union; Investissements d'Avenir Labex and Idex, ANR, R{\'e}gion Auvergne and Fondation Partager le Savoir, France; DFG and AvH Foundation, Germany; Herakleitos, Thales and Aristeia programmes co-financed by EU-ESF and the Greek NSRF; BSF, GIF and Minerva, Israel; BRF, Norway; CERCA Programme Generalitat de Catalunya, Generalitat Valenciana, Spain; the Royal Society and Leverhulme Trust, United Kingdom.

The crucial computing support from all WLCG partners is acknowledged gratefully, in particular from CERN, the ATLAS Tier-1 facilities at TRIUMF (Canada), NDGF (Denmark, Norway, Sweden), CC-IN2P3 (France), KIT/GridKA (Germany), INFN-CNAF (Italy), NL-T1 (Netherlands), PIC (Spain), ASGC (Taiwan), RAL (UK) and BNL (USA), the Tier-2 facilities worldwide and large non-WLCG resource providers. Major contributors of computing resources are listed in Ref.~\cite{ATL-GEN-PUB-2016-002}.
%----------------------------------------------

\clearpage
\printbibliography

\clearpage
% ATLAS Collaboration author list
% Data extracted on 13-Feb-2017 for paper reference PERF-2015-09
% \documentclass[11pt]{article}
% \usepackage{a4wide}\begin{document}
\begin{flushleft}
{\Large The ATLAS Collaboration}

\bigskip

M.~Aaboud$^\textrm{\scriptsize 137d}$,
G.~Aad$^\textrm{\scriptsize 88}$,
B.~Abbott$^\textrm{\scriptsize 115}$,
J.~Abdallah$^\textrm{\scriptsize 8}$,
O.~Abdinov$^\textrm{\scriptsize 12}$$^{,*}$,
B.~Abeloos$^\textrm{\scriptsize 119}$,
S.H.~Abidi$^\textrm{\scriptsize 161}$,
O.S.~AbouZeid$^\textrm{\scriptsize 139}$,
N.L.~Abraham$^\textrm{\scriptsize 151}$,
H.~Abramowicz$^\textrm{\scriptsize 155}$
% ...
\bigskip
\\
% ...
$^{8}$ Department of Physics, The University of Texas at Arlington, Arlington TX, United States of America\\
% ...
$^{12}$ Institute of Physics, Azerbaijan Academy of Sciences, Baku, Azerbaijan\\
% ...
$^{88}$ CPPM, Aix-Marseille Universit{\'e} and CNRS/IN2P3, Marseille, France\\
% ...
$^{115}$ Homer L. Dodge Department of Physics and Astronomy, University of Oklahoma, Norman OK, United States of America\\
% ...
$^{119}$ LAL, Univ. Paris-Sud, CNRS/IN2P3, Universit{\'e} Paris-Saclay, Orsay, France\\
% ...
$^{137}$ $^{(a)}$ Facult{\'e} des Sciences Ain Chock, R{\'e}seau Universitaire de Physique des Hautes Energies - Universit{\'e} Hassan II, Casablanca; $^{(b)}$ Centre National de l'Energie des Sciences Techniques Nucleaires, Rabat; $^{(c)}$ Facult{\'e} des Sciences Semlalia, Universit{\'e} Cadi Ayyad, LPHEA-Marrakech; $^{(d)}$ Facult{\'e} des Sciences, Universit{\'e} Mohamed Premier and LPTPM, Oujda; $^{(e)}$ Facult{\'e} des sciences, Universit{\'e} Mohammed V, Rabat, Morocco\\
% ...
$^{139}$ Santa Cruz Institute for Particle Physics, University of California Santa Cruz, Santa Cruz CA, United States of America\\
% ...
$^{151}$ Department of Physics and Astronomy, University of Sussex, Brighton, United Kingdom\\
% ...
$^{155}$ Raymond and Beverly Sackler School of Physics and Astronomy, Tel Aviv University, Tel Aviv, Israel\\
% ...
$^{161}$ Department of Physics, University of Toronto, Toronto ON, Canada\\
% ...
$^{a}$ Also at Department of Physics, King's College London, London, United Kingdom\\
$^{b}$ Also at Institute of Physics, Azerbaijan Academy of Sciences, Baku, Azerbaijan\\
$^{c}$ Also at Novosibirsk State University, Novosibirsk, Russia\\
$^{d}$ Also at TRIUMF, Vancouver BC, Canada\\
$^{e}$ Also at Department of Physics {\&} Astronomy, University of Louisville, Louisville, KY, United States of America\\
% ...
$^{*}$ Deceased
\end{flushleft}

% \end{document}
% Created with xml2latex.py

\end{document}